\documentclass[review,compress]{elsarticle}

\usepackage[margin=1in]{geometry}
\usepackage[colorlinks=true, allcolors=blue]{hyperref}
\usepackage{booktabs,caption,subcaption,booktabs,tabularray,multirow}
\usepackage{amssymb,amsmath,amsfonts,amsthm}
\usepackage{tikz}
\usepackage{lineno}
\usepackage{comment}
\usepackage{xparse}
\usepackage{epstopdf}
\usepackage{bm}
\usepackage{etoolbox}
\usepackage[utf8]{inputenc}
\usepackage[T1]{fontenc}
\usepackage[nameinlink,capitalise]{cleveref}

\usetikzlibrary{fit,positioning,backgrounds,arrows.meta}
\theoremstyle{definition}
\newtheorem{definition}{Definition}
\crefname{definition}{Definition}{Definitions}

\AppendGraphicsExtensions{.tif}
\AppendGraphicsExtensions{.jpeg}
\AppendGraphicsExtensions{.eps}

\def\R{{\mathbb R}}

\def\x{{\bm x}}

\newcommand{\bo}{\hat{\boldsymbol{\Omega}}}

\newcommand{\p}{\partial}
\newcommand{\f}{\frac}
\def\s#1{\Sigma_{#1}}

\newcommand{\TT}[1]{\mathsf{#1}}

\NewDocumentCommand{\trans}{m}{%
  \text{\textsc{trans}}_{\{#1\}}%
}
\NewDocumentCommand{\perm}{m}{%
  \text{\textsc{perm}}_{\{#1\}}%
}
\NewDocumentCommand{\diag}{m}{%
  \text{\textsc{diag}}_{\{#1\}}%
}

\journal{Nuclear Physics B}

\begin{document}

\begin{frontmatter}

\title{Tensorized Discontinuous Isogeometric Analysis Method for the 2-D Time-Independent Linearized Boltzmann Transport Equation}

\author{Patrick A. Myers$^{*}$}
\author{Joseph A. Bogdan}
\author{Majdi I. Radaideh}
\author{Brian C. Kiedrowski$^{*}$}

\cortext[mycorrespondingauthor]{Corresponding Authors: Patrick A. Myers (myerspat@umich.edu), Brian C. Kiedrowski (bckiedro@umich.edu)}

\affiliation{organization={University of Michigan, Department of Nuclear Engineering \& Radiological Sciences},
            addressline={2355 Bonisteel Boulevard}, 
            city={Ann Arbor},
            postcode={48109}, 
            state={Michigan},
            country={United States of America}}

\begin{abstract}
\par We present the novel Tensorized Discontinuous Isogeometric Analysis (TDIGA) method applied to the discontinuous Galerkin (DG) time-independent 2-D linearized Boltzmann transport equation (LBTE) with higher-order scattering, discretized with discrete ordinates in angle, multigroup in energy, and isogeometric analysis (IGA) in space. We formulate operator assembly in the tensor train (TT) format, producing seven-dimensional operators for both fixed-source and $k$-eigenvalue neutron transport problems solved using the restarted Generalized Minimum Residual Method (GMRES) and power iteration with an uncompressed solution vector. Our results on single-patch homogeneous and multi-patch heterogeneous problems, including a cruciform-shaped fuel array inspired by advanced reactor fuel designs, demonstrate the TT format's ability to compress interior operators from petabytes to megabytes, whereas the Compressed Sparse Row (CSR) matrix format requires gigabytes of storage. However, highly coupled boundary operators present a significant challenge for TT. Despite the storage savings, TT formatted operators increase time-to-solution relative to CSR as an uncompressed solution vector forces operator-vector product scaling of $\mathcal{O}(dr^2N^d\log(N))$ for TT while CSR scales at $\mathcal{O}(\text{nnz})$. We mitigate this discrepancy by using mixed formats with interior operators in TT, while high-rank boundary operators remain in CSR format. We compare all results to Monte Carlo (MC) and analytic reference solutions. While CSR remains $<10\times$ faster than this mixed format, the TDIGA method enables high-fidelity transport for expensive high-order IGA meshes. 
\end{abstract}



\end{frontmatter}


\section{Introduction}\label{sec:intro}

\par This paper presents the Tensorized Discontinuous Isogeometric Analysis (TDIGA) discrete ordinates method applied to the 2-D time-independent linearized Boltzmann transport equation (LBTE). We demonstrate assembly in the tensor train (TT) format and compare results solved with Compressed Sparse Row (CSR) matrix operators and estimated with Monte Carlo (MC) using the OpenMC code \cite{ROMANO201590} on both fixed-source and $k$-eigenvalue neutron transport problems. 

\par Traditional Finite Element Analysis (FEA) imposes a chosen basis for the \textit{unknown} solution field onto the \textit{known} geometry, often represented as a computer-aided design (CAD) model. FEA discretizes these geometries using basic geometric primitives such as triangles and quadrilaterals, which may fail to exactly represent the original CAD geometry and result in gaps between connected but separate domains \cite{MeshGen, OWENS2016501, LATIMER2020103238}. In the modeling of neutron transport in fission reactors, these issues often lead to inaccuracies in surface-area or volume calculations, fail to preserve fissile mass, and introduce numerical errors in quantities of interest, such as neutron leakage \cite{LATIMER2020103238, HALL2012160}. These problems are exacerbated by highly curvilinear geometries proposed for advanced nuclear fission or fusion reactors. Examples of contemporary systems are the Advanced Test Reactor and ITER. Resolving problems with FEA discretizations typically consumes $\sim 80\%$ of analysis time, as iterating on these meshes can be computationally expensive and require geometric approximations \cite{HUGHES20054135, IGABook}.

\par Isogeometric Analysis (IGA) \cite{HUGHES20054135, IGABook} flips the FEA paradigm by imposing the basis of the CAD model, often some spline such as a B-Spline or Non-Uniform Rational B-Spline (NURBS), onto the solution field. The approach, known as the \textit{isoparametric concept}, produces refinement-invariant meshes that are exact representations of the original CAD model with higher continuity $\left(\ge C^0\right)$. By forgoing geometric approximation, IGA enables geometric consistency between neutronics and other relevant physics (e.g., heat transport). IGA has seen application in a wide variety of engineering fields, including computational fluid dynamics (CFD) \cite{HUGHES20054135, Das_Gautam_2024, doi:10.1142/S0218202522020018}, solid mechanics \cite{HUGHES20054135, Das_Gautam_2024, NGUYEN201589}, and coupled fluid-structure interactions \cite{Das_Gautam_2024, BAZILEVS201228}. Within the neutron transport literature, IGA has been applied to the discontinuous Galerkin (DG) first-order \cite{OWENS2016501, OWENS2017215, OWENS2017352}, self-adjoint angular flux (SAAF) \cite{LATIMER2020107049, WILSON2024117414, LATIMER2021109941}, and weighted least squares (WLS) \cite{LATIMER2020103238, LATIMER2021109941} forms of the BTE. Researchers in \cite{OWENS2016501} laid the foundation for DG discrete-ordinates neutron transport with an IGA spatial discretization for both patch- and element-discontinuous formulations, known as the Patch-DG (PDG) and Fully-DG (FDG) approaches, respectively. They extended their element-discontinuous formulation to support hanging node adaptive mesh refinement (AMR) in \cite{OWENS2017215, OWENS2017352}. In all \cite{OWENS2016501, OWENS2017215, OWENS2017352}, they showcase a cycle-breaking sweeping algorithm as opposed to a Krylov solver. In \cite{LATIMER2020107049}, researchers developed an IGA discretization for the weak form SAAF equations, a linearly elliptic partial differential equation (PDE), which was later extended to a symmetric interior-penalty DG method in \cite{WILSON2024117414}. They found that further work was needed to develop a high-order, applicable preconditioner, as FEA preconditioners were less effective for high-order NURBS. Researchers in \cite{LATIMER2020103238} compared discontinuous IGA applied to the WLS to that with SAAF in \cite{LATIMER2020107049} and extended both to accommodate AMR with a continuous Budnov-Galerkin IGA spatial discretization in \cite{LATIMER2021109941}. In addition to discrete ordinates neutron transport, IGA has been applied to neutron diffusion in \cite{WELCH2017465, 10.1115/ICONE26-81316, Wilson06062024} using geometry-exact multi-level preconditioning and refinement.

\par The higher continuity of IGA meshes produces denser discretized PDE operators. As noted by \cite{OWENS2016501}, both their PDG and FDG approaches suffer from densification due to increased continuity and require the development of efficient algorithms. The size of PDE operators already scales exponentially with the number of dimensions, $\mathcal{O}(N^d)$. This phenomenon, known as the \textit{curse of dimensionality} \cite{dynamicprogramming}, necessitates the use of matrix-free inversion algorithms or sparse matrix formats, such as the Compressed Sparse Row (CSR) format. These sparse formats scale at $\mathcal{O}\left(\text{number of non-zeros (nnz)}\right)$, alleviating the curse of dimensionality; however, the higher continuity of IGA meshes and subsequent densification may challenge the efficiency of these formats. 

\par To break the curse of dimensionality, researchers in quantum mechanics \cite{annurev:/content/journals/10.1146/annurev-conmatphys-040721-022705, Orús_2019}, stochastic systems \cite{Chertkov_Oseledets_2021, doi:10.1177/0278364917753994}, CFD \cite{math12203277}, and others \cite{ION2022115593, CORONA2017145, MATVEEV2016164} are exploring tensor networks (TNs), which factorize a large tensor into a contraction network of many smaller tensors. These applications demonstrate TN factorization's ability to exploit low-rank structure, yielding a highly compressed representation of the original tensor with a prescribed error bound. Some specific TN topologies, such as the canonical \cite{Hars1972a}, Tucker \cite{Lathauwer, Tucker_1966}, and hierarchical Tucker (HT) \cite{HT} formats, form the foundation of TN research. The tensor train (TT) format \cite{Oseledets}, a special case of the HT format, enables linear scaling with respect to the number of dimensions, $\mathcal{O}(dNr^2)$. TT and its quantized-TT (QTT) variant have been applied to both multigroup neutron \cite{TRUONG2024112943} and gray thermal radiative \cite{gorodetsky2025thermalradiationtransporttensor} transport with discrete-ordinates and diamond-differencing approximations. The neutron transport application focused exclusively on homogenized Cartesian meshes with isotropic scattering; however, \cite{TRUONG2024112943} hypothesized diminishing returns as the number of unique materials increases for heterogeneous neutron transport problems. Both applications achieved exceptional speedups of $>60\times$ and high compression of $>1000\times$, especially in the angular dimensions. The success of the TT format depended on the problem, with geometrically complex problems suffering from rank explosions. 

\par Despite the efficiency of the TT format for representing high-dimensional data, solving linear systems in TT format remains difficult due to the unknown TT ranks of the solution, which must be estimated or adapted during the computation. TT optimization algorithms, such as the Alternating Linear Scheme (ALS) \cite{doi:10.1137/100818893} and the Alternating Minimal Energy (AMEn) method \cite{doi:10.1137/140953289, 62842}, solve a linear system by iteratively projecting and solving each mode of the system. ALS does not feature rank adaptivity, whereas AMEn introduces it via a residual enrichment step. This enrichment step can be costly and temporary because the algorithm must discover missing subspace directions. Alternative methods include TT-GMRES \cite{Dolgov+2013+149+172, doi:10.1137/100799010} and TT-cross \cite{Chertkov_Oseledets_2021, OSELEDETS201070, doi:10.1137/17M1138881, GHAHREMANI2024117385, doi:10.1137/22M1498401}. However, these solvers assume that the solution vector is compressible in the TT format, which is not necessarily the case even if the operators compress well in the TT format \cite{math12203277, TRUONG2024112943, gorodetsky2025thermalradiationtransporttensor}.

\par In this work, we present the Tensorized Discontinuous Isogeometric Analysis (TDIGA) method applied to the discontinuous Galerkin (DG) time-independent 2-D LBTE (DG-LBTE) for fixed source and $k$-eigenvalue discrete ordinates multigroup neutron transport. We extend the PDG neutron transport with isotropic scattering from \cite{OWENS2016501} to higher-order scattering and address the densification of LBTE operators via a TT decomposition. This work is also directly applicable to their FDG approach. We formulate operator assembly in TT format and use the restarted Generalized Minimal Residual Method (GMRES) to solve linear systems with TT or CSR operators and an uncompressed solution vector. In addition to purely academic, homogenized examples, we demonstrate TDIGA on practical problems, including a shielded cruciform fixed-source problem and an infinite array of cruciform-shaped fuel based on the Lightbridge advanced nuclear fuel design \cite{Malone01122012}. We compare solutions using MC or analytic reference solutions whose geometric representation is exactly consistent with the NURBS representation, and discuss how resolution and operator format affect compression, TT-rank, time-to-solution, and solution accuracy. We demonstrate two to three orders of magnitude greater compression for the TT format than for CSR for interior operators, whereas curvalinear boundaries introduce higher coupling and thus higher TT-rank between angle and space for boundary operators. We find that the time-to-solution for operators in TT format is limited by the operator-vector product scaling when using an uncompressed solution vector, which itself is only marginally compressible in TT format for regular meshes.

\par The remainder of this paper is organized as follows. In \cref{sec:background}, we discuss the background of multi-patch IGA, two common splines used in CAD, the discretization of the DG-LBTE, and the TT format. In \cref{sec:methods} we outline the TDIGA assembly and numerical schemes. In \cref{sec:results}, we demonstrate and discuss TDIGA on four fixed source and three $k$-eigenvalue neutron transport problems. Finally, we provide our concluding remarks in \cref{sec:conc}, discussing the implications of this work, future extensions to AMR and domain decomposition, and the potential of alternative TN formats to provide compression for both the operators and the solution vector. 
\section{Background}\label{sec:background}

\par In this section, we provide an overview of the concepts of IGA and tensor trains. We begin with \cref{sec:iga} and discuss multi-patch IGA and its applicability to the weak DG form of PDEs. We then discuss B-Splines and NURBS as the foundation of CAD models in \cref{sec:bspline,sec:nurbs}, respectively. We conclude the IGA background by discussing the various spaces and mappings required in \cref{sec:spaces}. In \cref{sec:discretized}, we derive the PDG approach for the discretized 2-D LBTE with higher-order scattering. Finally, in \cref{sec:tt}, we provide background on the TT decomposition and relevant TT operations.

\subsection{Multi-Patch Isogeometric Analysis}\label{sec:iga}

\par
\begin{figure}
    \centering
    \begin{tikzpicture}
        \node (fig1) {\includegraphics[width=0.3\textwidth]{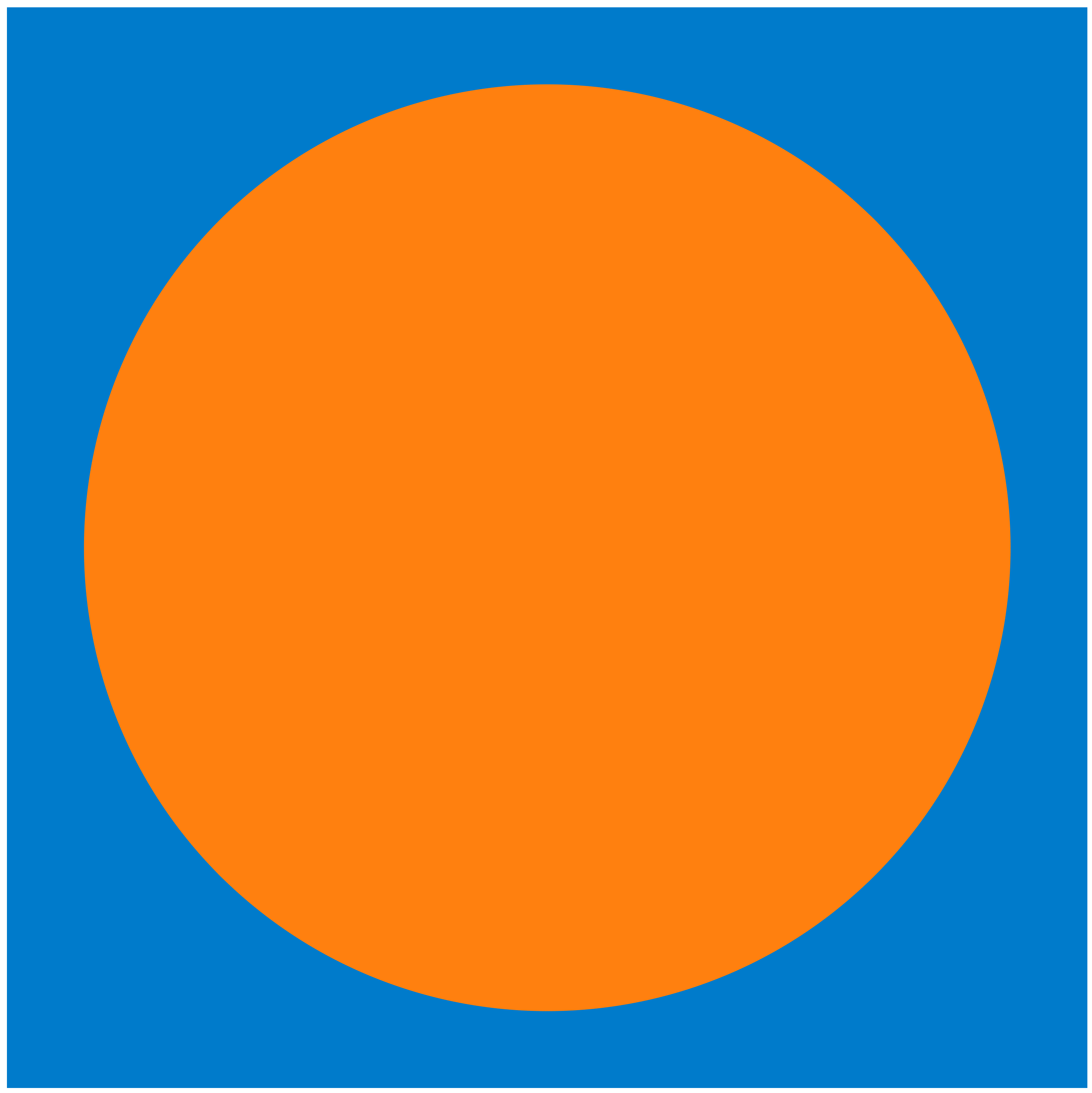}};
        
        \node[right=2cm of fig1] (fig2) {\includegraphics[width=0.35\textwidth]{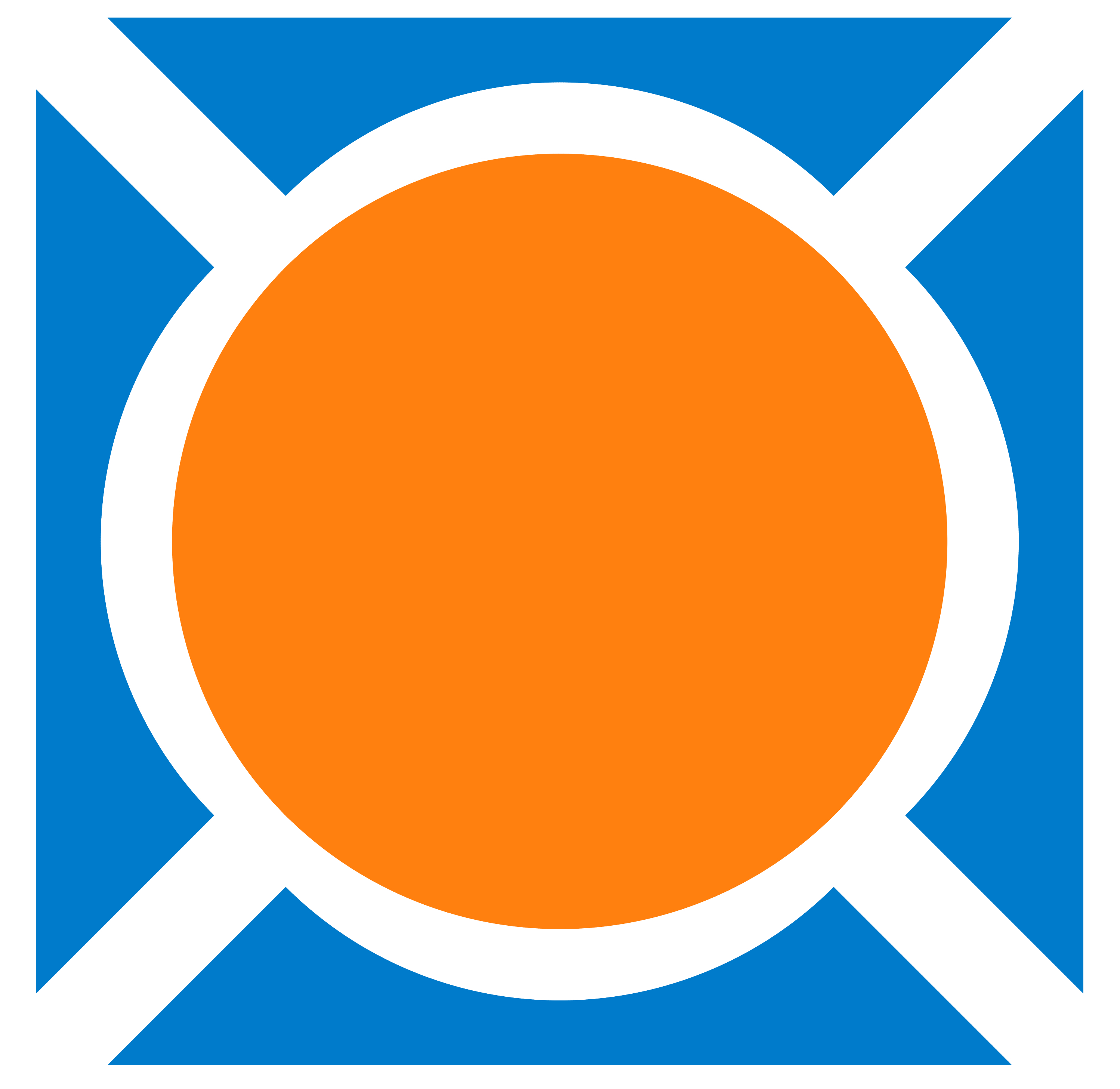}};
        
        \draw[->, line width=1mm, >=Latex, scale=1] (fig1.east) -- (fig2.west);
    \end{tikzpicture}
    \caption{NURBS patches for a pinell CAD model.}
    \label{fig:pincell_patches}
\end{figure}

 Complex CAD models incorporate multiple \textit{patches} to represent geometries due to topological differences, varying material properties, or for parallel assembly \cite{IGABook}. A patch is a parametric subdomain equipped with a set of basis functions, knot vectors, and control points that map the parametric space $\hat \Gamma$ to the physical space $\Gamma$ to represent a curve, surface, or volume. Within a given patch, physical properties are assumed to be piecewise constant. As an example, we show a pincell (a typical unit cell in a nuclear fission reactor lattice) defined by five patches in \cref{fig:pincell_patches}. This separation of the geometry into subdomains naturally accommodates the weak DG form of PDEs, in which basis functions are allowed to be discontinuous across patch interfaces. DG methods enable high parallelization through techniques such as domain decomposition and are simple to implement due to the straightforward treatment of boundary conditions \cite{Cockburn_2000}. They readily accommodate AMR strategies without the continuity restrictions imposed by continuous meshes, and their order of accuracy depends solely on the exact solution \cite{Cockburn_2000}. 

\subsubsection{B-Spline}\label{sec:bspline}

\par Many splines used in CAD are built off B-Splines, which is therefore a natural starting point for our discussion of spline curves, surfaces, and volumes \cite{10.5555/265261}. B-Spline parametric curves are given by
\begin{equation}
    \mathbf{C}(\hat x) = \sum_{i_{\hat x} = 1}^{N_{\hat{x}}}A_{i_{\hat{x}}, p_{\hat{x}}}(\hat x) \mathbf{B}_{i_{\hat{x}}},
\end{equation}
where $\mathbf{B} = \{\mathbf{B}_1, ..., \mathbf{B}_{N_{\hat{x}}}\}\subset \mathbb{R}$ are the control points and $A_{i_{\hat{x}}, p_{\hat{x}}}(\hat x)$ is the $i_{\hat{x}}$th degree-$p_{\hat{x}}$ B-Spline basis function parameterized by $0 \le \hat x < 1$. This defines a one-dimensional B-Spline patch. The Cox-de Boor recursion formula \cite{DEBOOR197250} gives the basis functions for $p_{\hat x} = 0$,
\begin{equation}
    A_{i_{\hat x}, 0}(\hat x) = \begin{cases}
        1 & \hat{\mathbf{x}}_{i_{\hat x}} \le \hat x < \hat{\mathbf{x}}_{i_{\hat x} + 1}\\
        0 & \text{otherwise}
    \end{cases},
\end{equation}
and for $p_{\hat x} \ge 1$,
\begin{equation}
    A_{i_{\hat{x}}, p_{\hat{x}}}(\hat x) = \frac{\hat x - \hat{\mathbf{x}}_{i_{\hat{x}}}}{\hat{\mathbf{x}}_{i_{\hat{x}} + p_{\hat{x}}} - \hat{\mathbf{x}}_{i_{\hat{x}}}}A_{i_{\hat{x}}, p_{\hat{x}} - 1}(\hat x) + \frac{\hat{\mathbf{x}}_{i_{\hat{x}} + p_{\hat{x}} + 1} - \hat x}{\hat{\mathbf{x}}_{i_{\hat{x}} + p_{\hat{x}} + 1} - \hat{\mathbf{x}}_{i_{\hat{x}} + 1}}A_{i_{\hat{x}} + 1, p_{\hat{x}} - 1}(\hat x),
\end{equation}
where $\hat{\mathbf{x}}$ is a vector of non-decreasing, potentially repeated real numbers known as the knot vector. The unique numbers, known as knots, in the knot vector partition the B-Spline patch into knot spans, which are similar to traditional FEA elements due to reduced continuity. The number of times a knot is repeated defines that knot's multiplicity $m$, in which the continuity between neighboring knot spans is $C^{p_{\hat x}-m}$. The continuity within knot spans is $C^\infty$. A special case, and one relevant to this work, is the open knot vector in which the first and last unique knot value is repeated $p_{\hat x} + 1$ times. The basis functions at the endpoints are therefore interpolatory. In the interior knots, we can increase their multiplicity to $m = p_{\hat x} + 1$, resulting in a discontinuity, where the original B-Spline patch splits into two patches. However, we impose a stricter formulation of the open knot vector where $m = 1$ for all interior knots, resulting in $C^{p_{\hat x}-1}$ continuity and basis function support spanning $p_{\hat x}+1$ control points. The localized non-zero basis functions at a given point $\hat x$ are \textit{active} basis functions. The open knot vector $\hat{\mathbf{x}}\in\mathbb{R}^{N_{\hat{x}} + p_{\hat{x}} + 1}$ has a unique counterpart $\hat{\mathbf{x}}^u\in\mathbf{R}^{N_{\hat x}^u + 1}$ partitioning the parametric space into knot spans with indices $i_{\hat{x}}^u\in\{1, 2, \dots, N_{\hat x}^u\}$ where $N_{\hat x}^u = N_{\hat x} - p_{\hat x}$ is the total number of knot spans along $\hat x$. The basis functions form a partition of unity for all $\hat x \in [0, 1]$,
\begin{equation}
    \sum_{i_{\hat x} = 1}^{N_{\hat x}}A_{i_{\hat{x}}, p_{\hat{x}}}(\hat x) = 1.
\end{equation}
Due to the recursive nature of B-Splines, we can exactly calculate their derivatives by
\begin{equation}
    \frac{d}{d\hat x}A_{i_{\hat{x}}, p_{\hat{x}}}(\hat x) = \frac{p_{\hat x}}{\hat{\mathbf{x}}_{i_{\hat{x}} + p_{\hat x}} - \hat{\mathbf{x}}_{i_{\hat{x}}}}A_{i_{\hat{x}}, p_{\hat x} - 1}(\hat x) - \frac{p_{\hat x}}{\hat{\mathbf{x}}_{i_{\hat{x}} + p_{\hat x} + 1} - \hat{\mathbf{x}}_{i_{\hat{x}} + 1}}A_{i_{\hat{x}} + 1, p_{\hat x} - 1}(\hat x).
\end{equation}

\par These properties carry over into B-Spline surfaces and volumes, defined respectively by
\begin{subequations}
    \begin{gather}
        \mathbf{S}(\hat x, \hat y) = \sum_{i_{\hat x} = 1}^{N_{\hat x}}\sum_{i_{\hat y} = 1}^{N_{\hat y}}A_{i_{\hat x}, p_{\hat x}}(\hat x)B_{i_{\hat y}, p_{\hat y}}(\hat y) \mathbf{B}_{i_{\hat x}, i_{\hat y}},\\
        \mathbf{V}(\hat x, \hat y, \hat z) =\sum_{i_{\hat x} = 1}^{N_{\hat x}}\sum_{i_{\hat y} = 1}^{N_{\hat y}}\sum_{i_{\hat z} = 1}^{N_{\hat z}}A_{i_{\hat x}, p_{\hat x}}(\hat x)B_{i_{\hat y}, p_{\hat y}}(\hat y)C_{i_{\hat z}, p_{\hat z}}(\hat z)\mathbf{B}_{i_{\hat x}, i_{\hat y}, i_{\hat z}},
    \end{gather}
\end{subequations}
where $B_{i_{\hat y}, p_{\hat y}}(\hat y)$ and $C_{i_{\hat z}, p_{\hat z}}(\hat z)$ are basis functions for $0 \le \hat y < 1$ and $0 \le \hat z < 1$, with polynomial degrees $p_{\hat y}$ and $p_{\hat z}$, and knot vectors $\hat{\mathbf{y}}\in\mathbb{R}^{N_{\hat y} + p_{\hat y} + 1}$ and $\hat{\mathbf{z}}\in\mathbb{R}^{N_{\hat z} + p_{\hat z} + 1}$, respectively. The control points constitute a matrix $\mathbf{B}\in\mathbb{R}^{N_{\hat x}\times N_{\hat y}}$ for surfaces and a three-dimensional tensor $\mathbf{B}\in\mathbb{R}^{N_{\hat x}\times N_{\hat y}\times N_{\hat z}}$ for volumes. Other properties include pointwise non-negativity and affine invariance. For all cases of B-Splines, we can perform knot insertion and order elevation to refine the curve \cite{IGABook, 10.5555/265261}.

\subsubsection{Non-Uniform Rational B-Spline (NURBS)}\label{sec:nurbs}

\par A NURBS is constructed by summing weighted B-Spline basis functions \cite{10.5555/265261}. A NURBS curve basis function is
\begin{equation}\label{eq:nurbs_basis_1d}
    R_{i_{\hat x}}^{p_{\hat x}}(\hat x) = \frac{A_{i_{\hat x}, p_{\hat x}}(\hat x)w_{i_{\hat x}}}{\sum_{i_{\hat x}' = 1}^{N_{\hat x}}A_{i_{\hat x}', p_{\hat x}}(\hat x)w_{i_{\hat x}'}},
\end{equation}
where $w_{i_{\hat x}}\in\{w_1, ..., w_{N_{\hat x}}\}$ are the weights. Surface and volume NURBS basis functions use weight $w_{i_{\hat x}, i_{\hat y}}$ and $w_{i_{\hat x}, i_{\hat y}, i_{\hat z}}$ in their definitions,
\begin{subequations}
    \begin{gather}
        R_{i_{\hat x}, i_{\hat y}}^{p_{\hat x}, p_{\hat y}}(\hat x, \hat y) = \frac{A_{i_{\hat x}, p_{\hat x}}(\hat x)B_{i_{\hat y}, p_{\hat y}}(\hat y)w_{i_{\hat x}, i_{\hat y}}}{\sum_{i_{\hat x}' = 1}^{N_{\hat x}}\sum_{i_{\hat y}' = 1}^{N_{\hat y}}A_{i_{\hat x}', p_{\hat x}}(\hat x)B_{i_{\hat y}', p_{\hat y}}(\hat y)w_{i_{\hat x}', i_{\hat y}'}}, \label{eq:nurbs_basis_2d}\\
        R_{i_{\hat x}, i_{\hat y}, i_{\hat z}}^{p_{\hat x}, p_{\hat y}, p_{\hat z}}(\hat x, \hat y, \hat z) = \frac{A_{i_{\hat x}, p_{\hat x}}(\hat x)B_{i_{\hat y}, p_{\hat y}}(\hat y)C_{i_{\hat z}, p_{\hat z}}(\hat z)w_{i_{\hat x}, i_{\hat y}, i_{\hat z}}}{\sum_{i_{\hat x}' = 1}^{N_{\hat x}}\sum_{i_{\hat y}' = 1}^{N_{\hat y}}\sum_{i_{\hat z}' = 1}^{N_{\hat z}}A_{i_{\hat x}', p_{\hat x}}(\hat x)B_{i_{\hat y}', p_{\hat y}}(\hat y)C_{i_{\hat z}', p_{\hat z}}(\hat z)w_{i_{\hat x}', i_{\hat y}', i_{\hat z}'}}. \label{eq:nurbs_basis_3d}
    \end{gather}
\end{subequations}
NURBS curves, surfaces, and volumes are then given by
\begin{subequations}\label{eq:NURBS}
\begin{gather}
    \mathbf{C}(\hat x) = \sum_{i_{\hat x} = 1}^{N_{\hat x}}R_{i_{\hat x}}^{p_{\hat x}}(\hat x)\mathbf{B}_{i_{\hat x}},\\
    \mathbf{S}(\hat x, \hat y) = \sum_{i_{\hat x} = 1}^{N_{\hat x}}\sum_{i_{\hat y} = 1}^{N_{\hat y}}R_{i_{\hat x}, i_{\hat y}}^{p_{\hat x}, p_{\hat y}}(\hat x, \hat y)\mathbf{B}_{i_{\hat x}, i_{\hat y}},\\
    \mathbf{V}(\hat x, \hat y, \hat z) = \sum_{i_{\hat x} = 1}^{N_{\hat x}}\sum_{i_{\hat y} = 1}^{N_{\hat y}}\sum_{i_{\hat z} = 1}^{N_{\hat z}}R_{i_{\hat x}, i_{\hat y}, i_{\hat z}}^{p_{\hat x}, p_{\hat y}, p_{\hat z}}(\hat x, \hat y, \hat z)\mathbf{B}_{i_{\hat x}, i_{\hat y}, i_{\hat z}}.
\end{gather}
\end{subequations}
Many of the properties of B-Splines also follow as NURBS form partitions of unity, have the same continuity as B-Splines, have analytic derivatives, have local support, are non-negative, and are affine invariant. We can also utilize knot insertion and order elevation to refine the NURBS.

\subsubsection{Physical, parametric, and parent element spaces}\label{sec:spaces}

\begin{figure}
\centering
\begin{subfigure}{0.32\textwidth}
    \includegraphics[width=\textwidth]{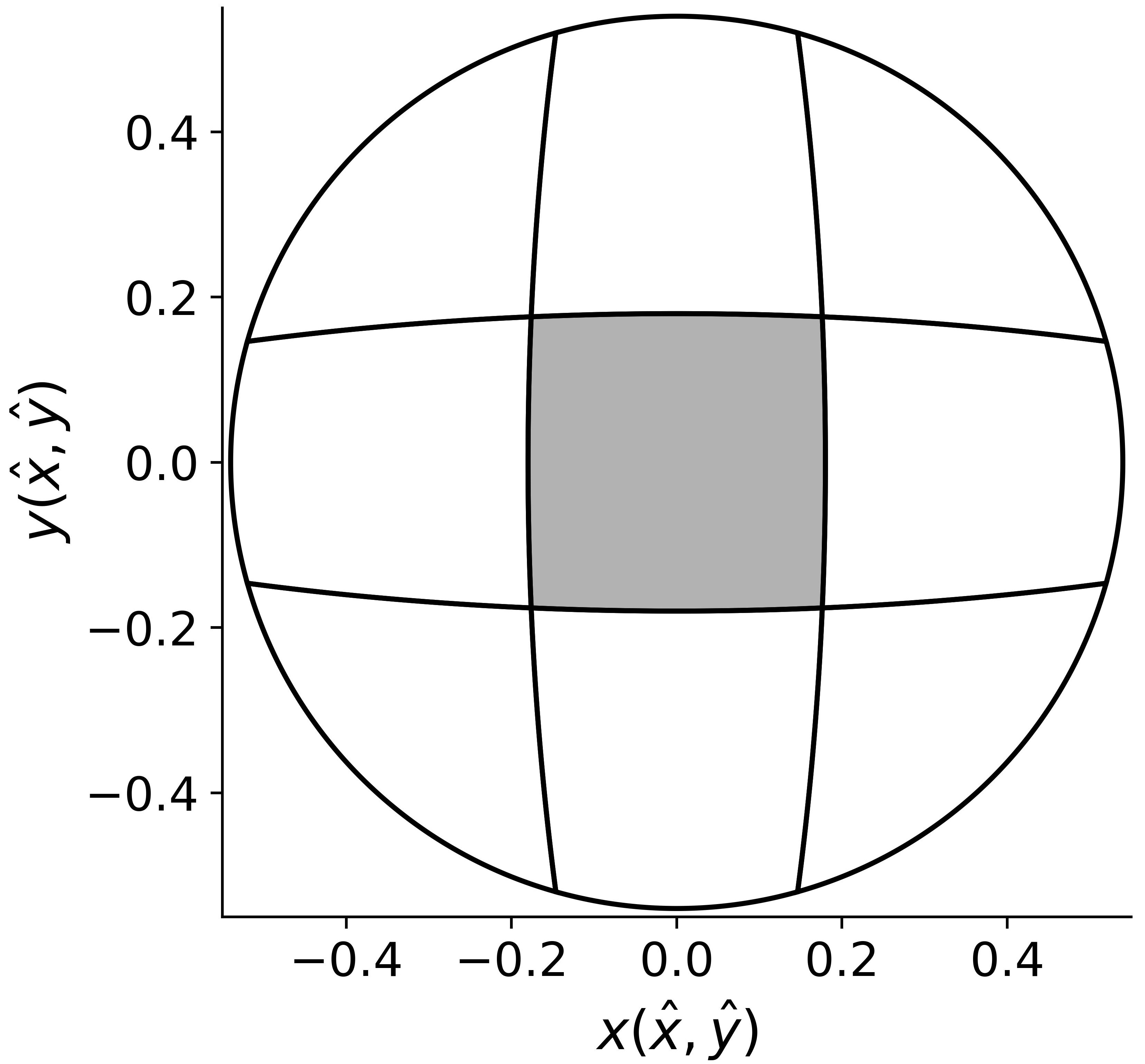}
    \caption{}
\end{subfigure}
\hfill
\begin{subfigure}{0.32\textwidth}
    \includegraphics[width=\textwidth]{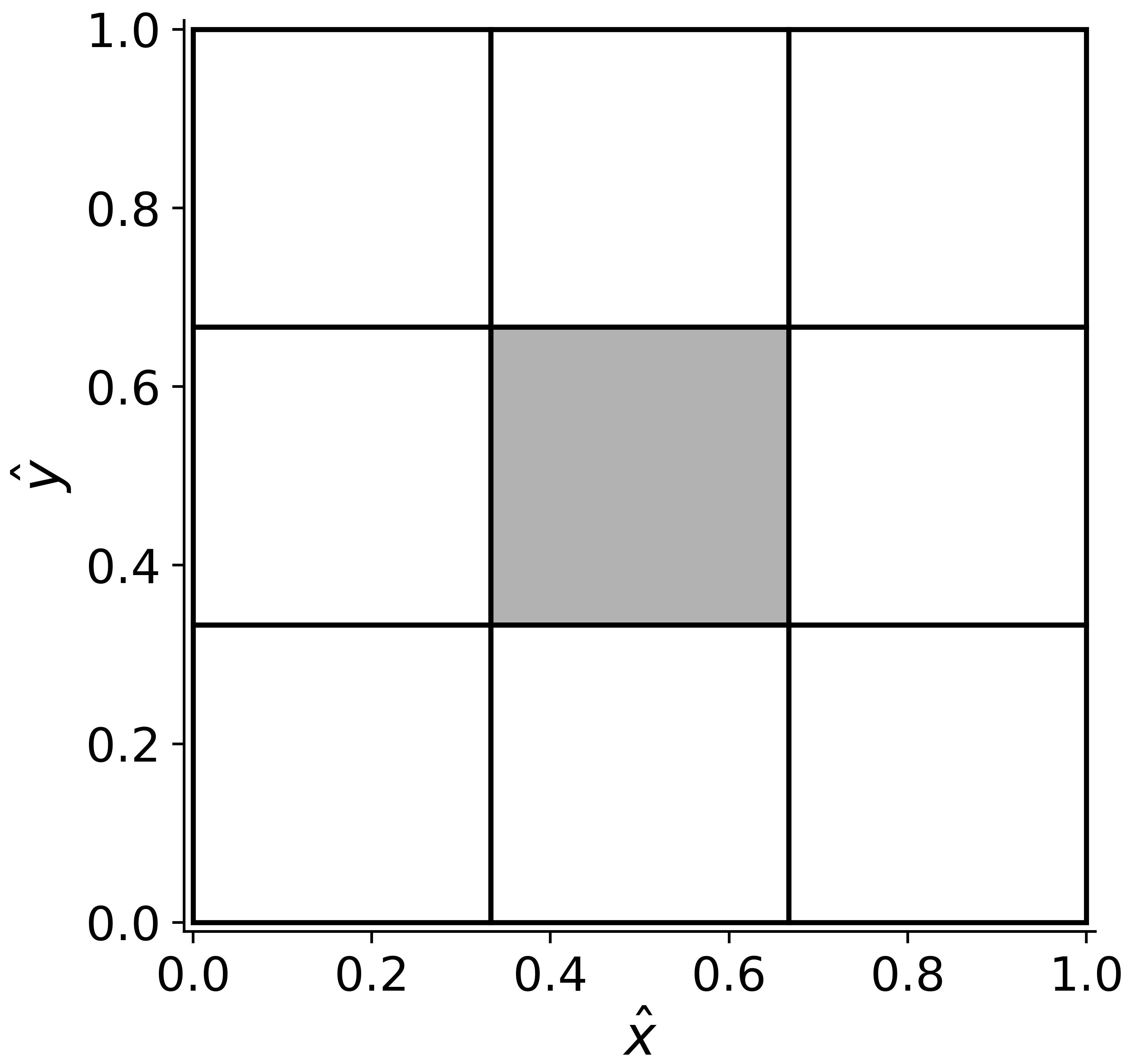}
    \caption{}
\end{subfigure}
\hfill
\begin{subfigure}{0.32\textwidth}
    \includegraphics[width=\textwidth]{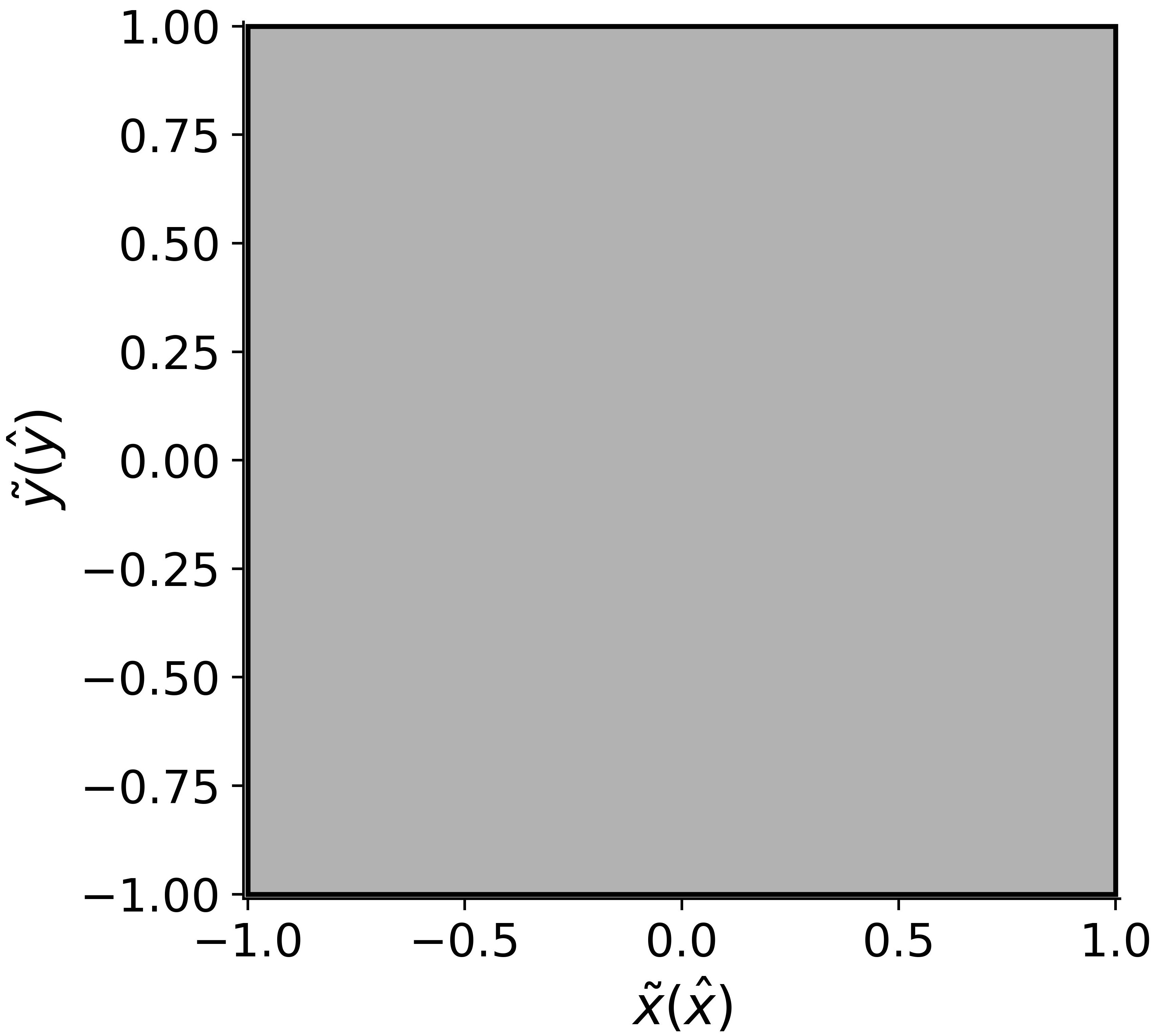}
    \caption{}
\end{subfigure}
\caption{Mapping of a knot span in a NURBS circle for the (a) physical, (b) parametric, and (c) parent element spaces.}\label{fig:mapping}
\end{figure}

\par IGA uses three spaces: the physical space, the parametric space, and the parent element space. A spline patch with index $i_e\in\{1, 2, \dots, N_e\}$ maps from the parametric space $\hat\Gamma_{i_e} = [0, 1]^3$, with coordinates $\hat{\x} = (\hat x, \hat y, \hat z)$, to the physical space $\Gamma_{i_e}$, with coordinates $\x = (x, y, z)$, where the original geometry is defined. This mapping is given by \cref{eq:NURBS} for NURBS patches. The parametric space is discretized into knot spans $\left[\hat{\mathbf{x}}_{i_{\hat x}^u}^u, \hat{\mathbf{x}}_{i_{\hat x}^u + 1}^u\right]\times \left[\hat{\mathbf{y}}_{i_{\hat y}^u}^u, \hat{\mathbf{y}}_{i_{\hat y}^u + 1}^u\right] \times \left[\hat{\mathbf{z}}_{i_{\hat z}^u}^u, \hat{\mathbf{z}}_{i_{\hat z}^u + 1}^u\right]$ which we map to the parent element space $\tilde \Gamma_{i_e}^{i_{\hat x}^u, i_{\hat y}^u} = \left[-1, 1\right]^3$ with coordinates $\tilde \x = (\tilde x, \tilde y, \tilde z)$ by
\begin{subequations}
\begin{align}\label{eq:parent_mapping}
    \hat x(\tilde x) &= \hat{\mathbf{x}}_{i_{\hat x}^u}^u + (\tilde x + 1)\frac{(\hat{\mathbf{x}}_{i_{\hat x}^u + 1}^u - \hat{\mathbf{x}}_{i_{\hat x}^u}^u)}{2},\\
    \hat y(\tilde y) &= \hat{\mathbf{y}}_{i_{\hat y}^u}^u + (\tilde y + 1)\frac{(\hat{\mathbf{y}}_{i_{\hat y}^u + 1}^u - \hat{\mathbf{y}}_{i_{\hat y}^u}^u)}{2},\\
    \hat z(\tilde z) &= \hat{\mathbf{z}}_{i_{\hat z}^u}^u + (\tilde z + 1)\frac{(\hat{\mathbf{z}}_{i_{\hat z}^u + 1}^u - \hat{\mathbf{z}}_{i_{\hat z}^u}^u)}{2}.
\end{align}
\end{subequations}
These mappings are shown in \cref{fig:mapping}. In the parent element space, we perform all arithmetic operations, such as numerical integration using a Gaussian quadrature with $(p_{\hat x} + 1) \times (p_{\hat y} + 1) \times (p_{\hat z} + 1)$ abscissae and weights $\mathbf{w}^{\tilde \x} = \left\{w_{i_{\tilde x}}^{\tilde x}w_{i_{\tilde y}}^{\tilde y}w_{i_{\tilde z}}^{\tilde z}\right\}_{i_{\tilde x} = 1, i_{\tilde y} = 1, i_{\tilde z} = 1}^{p_{\hat x} + 1, p_{\hat y} + 1, p_{\hat z} + 1}$. For NURBS weights that are not equal to one, this numerical integration is approximate; however, as the mesh is refined, it is better approximated by non-rational polynomials. This is due to the denominator of \cref{eq:nurbs_basis_1d,eq:nurbs_basis_2d,eq:nurbs_basis_3d} as well as the Jacobian mapping between the parametric and physical spaces being invariant to knot insertion and order elevation.

\subsection{Discretized Discontinuous Galerkin Time-Independent 2-D Linearized Boltzmann Transport Equation}\label{sec:discretized}

\par The time-independent 2-D LBTE is given by
\begin{equation}\label{eq:nte}
\begin{aligned}
    \bo\cdot \nabla&\psi(\x, \bo, E) + \s{t}(\x, E)\psi(\x,\bo, E) \\
    &= \int_{0}^{\infty}\int_{4\pi}\s{s}(\x,\bo'\cdot \bo, E'\rightarrow E)\psi(\x, \bo', E')d\Omega'dE' + Q(\x, \bo, E),\\
    &~~~~\x \in \Gamma,~~ \bo\in 4\pi,~~ 0 < E < \infty,
    \end{aligned}
\end{equation}
where $\x = (x, y)$ is a position in space within $\Gamma$, $\bo = (\Omega_x, \Omega_y, \Omega_z) = (\mu, \sqrt{1 - \mu^2}\cos\gamma, \sqrt{1 - \mu^2}\sin\gamma)$ is the direction the neutron is traveling defined by a unit vector on the unit sphere, $E$ is the neutron's energy, $\psi(\x, \bo, E)$ is the angular flux, $\s{t}(\x, E)$ is the macroscopic total cross section, and $\s{s}(\x,\bo'\cdot \bo, E'\rightarrow E)$ is the macroscopic double differential scattering cross section from $(\bo', E')$ to $(\bo, E)$. The source term $Q(\x, \bo, E)$ may be any internal source term but for our purposes may be an isotropic source $Q(\x, \bo, E) = \f{1}{4\pi}Q(\x, E)$ or, in the $k$-eigenvalue case, a fission source,
\begin{equation}
    Q(\x, \bo, E) = \f{\chi(E)}{4\pi k}\int_0^\infty\int_{4\pi}\nu(E') \s{f}(E')\psi(\x, \bo', E')d\Omega'dE',
\end{equation}
where $\chi(E)$ is the fission spectrum, $\nu(E)$ is the average number of neutrons per fission, $\s{f}(E)$ is the macroscopic fission cross section, and $k$ is the neutron multiplication factor. The boundary condition,
\begin{equation}
    \psi(\x, \bo, E) = \psi^b(\x, \bo, E),~~\x \in \p\Gamma,~~ \bo\cdot \mathbf{n} < 0,~~ 0 < E < \infty,
\end{equation}
with outward normal vector $\mathbf{n}$, completes the boundary value problem.

\par \Cref{eq:nte} is a five-dimensional integro-differential PDE which describes the gains and losses of neutrons in the system. We aim to find the neutron angular flux that satisfies this equation. For practical engineering problems, we must approximate this solution using numerical methods. To do so, we apply an $L$th-order Legendre expansion to the angular dimensions of $\s{s}(\x,\bo'\cdot \bo, E'\rightarrow E)$ and apply the multigroup and discrete-ordinates approximations on the energy and direction dimensions, respectively, resulting in
\begin{equation}\label{eq:xcont}
\begin{gathered}
    \begin{aligned}\bo_{i_\Omega}\cdot \nabla&\psi_{i_E, i_\Omega}(\x) + \s{t, i_E}(\x)\psi_{i_E,i_\Omega}(\x) \\=& \sum_{i_E' = 1}^{N_E}\sum_{l = 0}^{L}\s{sl, i_E'\rightarrow i_E}(\x)\sum_{i_\Omega' = 1}^{N_\Omega}w^{\Omega}_{i_\Omega'}Y_{l}(\bo_{i_\Omega}, \bo_{i_\Omega'})\psi_{i_E', i_\Omega'}(\x)+ Q_{i_E,i_\Omega}(\x),\end{aligned}\\
    \x\in\Gamma, ~~i_\Omega \in\{ 1, ..., N_\Omega\},~~ i_E\in\{1, ..., N_E\},
\end{gathered}
\end{equation}
where $w_{i_\Omega}^{\Omega}$ is the weight for ordinate $i_\Omega$. The angular quadrature satisfies $\sum_{i_\Omega = 1}^{N_\Omega}w_{i_\Omega}^{\Omega} = 1$. The spherical harmonic operator, $Y_{l}(\bo, \bo')$, is defined by
\begin{subequations}
\begin{align}
    Y_l(\bo, \bo') &= Y_{e, l}^0(\bo)Y_{e, l}^0(\bo') + 2\sum_{m = 1}^lY_{e,l}^m(\bo)Y_{e,l}^m(\bo'),\\ 
    Y_{e, l}^m(\bo) &= Y_{e, l}^m(\mu, \gamma) = \left[(2l + 1)\f{(l - |m|)!}{(l + |m|)!}\right]^{1/2} \cos(m\gamma)(1 - \mu^2)^{|m|/2}\f{d^{|m|}}{d\mu^{|m|}}P_l(\mu),
\end{align}
\end{subequations}
with even spherical harmonic function $Y_{e, l}^m(\bo)$ and Legendre polynomial $P_l(\mu)$. The formulation for $Y_l(\bo, \bo')$ was derived using the symmetry across the $xy$-plane for the 2-D LBTE. For the 3-D LBTE, there is another term for the odd spherical harmonic function in the summation. The scalar flux is
\begin{equation}
    \phi_{i_E}(\x) = \sum_{i_\Omega = 1}^{N_\Omega}w_{i_\Omega}^\Omega\psi_{i_E, i_\Omega}(\x).
\end{equation}

\par This system of equations now only depends on $\x$ for which we can apply the PDG scheme developed in \cite{OWENS2016501}. We multiply \cref{eq:xcont} by a test function $v(\x)\in H^1(\Gamma_{i_e})$ for element $i_e\in\{1, 2, \dots, N_e\}$ and integrate over the element,
\begin{equation}\label{eq:pre_parts}
\begin{aligned}
    \int_{\Gamma_{i_e}}v(\x)\bo_{i_\Omega}\cdot \nabla&\psi_{i_E, i_\Omega}(\x)dV + \s{t, i_E}\int_{\Gamma_{i_e}}v(\x)\psi_{i_E,i_\Omega}(\x)dV\\
    =& ~\sum_{i_E' = 1}^{N_E}\sum_{l = 0}^{L}\s{sl, i_E'\rightarrow i_E}\sum_{i_\Omega' = 1}^{N_\Omega}w_{i_\Omega'}^\Omega Y_l(\bo_{i_E}, \bo_{i_E'})\int_{\Gamma_{i_e}}v(\x)\psi_{i_E', i_\Omega'}(\x)dV \\
    &~+ \int_{\Gamma_{i_e}}v(\x)Q_{i_E,i_\Omega}(\x)dV
\end{aligned}
\end{equation}
We then apply integration by parts to the streaming term and split the boundary term into inflow $\p \Gamma_{i_e}^-$ for $\bo_{i_\Omega}\cdot \mathbf{n} < 0$ and outflow $\p\Gamma_{i_e}^+$ for $\bo_{i_\Omega}\cdot \mathbf{n} > 0$,
\begin{equation}
\begin{aligned}
    \int_{\Gamma_{i_e}}v(\x)&\bo_{i_\Omega}\cdot \nabla\psi_{i_E, i_\Omega}(\x)dV \\
    =& \int_{\p\Gamma_{i_e}^+}(\bo_{i_\Omega}\cdot \mathbf{n})v(s)\psi_{i_E, i_\Omega}(s)dS - \int_{\p\Gamma_{i_e}^-}|\bo_{i_\Omega}\cdot \mathbf{n}|v(s)\psi_{i_E, i_\Omega}^{\text{adj}}(s)dS \\
    &- \bo_{i_\Omega}\cdot \int_{\Gamma_{i_e}}\psi_{i_E, i_\Omega}(\x)\nabla v(\x)dV,
\end{aligned}
\end{equation}
where $\psi^{\text{adj}}_{i_E, i_\Omega}(s)$ is the incident neutron flux from neighboring elements or boundary conditions. Note that in the PDG scheme, the element $i_e$ is a parameterized NURBS patch given by \cref{eq:NURBS} for which the cross sections are assumed to be constant. We map the parametric space $\hat \Gamma_{i_e}$ to the physical space $\Gamma_{i_e}$ using the Jacobian $\mathbf{J}_{\hat{\x}}^{i_e} = \frac{\p(x, y)}{\p(\hat x, \hat y)}$, its determinant $\det\mathbf{J}_{\hat{\x}}^{i_e}$, and its inverse $\left(\mathbf{J}_{\hat{\x}}^{i_e}\right)^{-1}$. To simplify the notation, we define three inner products:
\begin{subequations}
    \begin{align}
        \{a(\hat{\x}), b(\hat{\x})\}&:= \int_{\hat{\Gamma}_{i_e}}a(\hat{\x})b(\hat{\x})|\det(\mathbf{J}_{\hat{\x}}^{i_e})|d\hat V,\\
        \langle a(\hat s), b(\hat s)\rangle^- &:= \int_{\p \hat{\Gamma}_{i_e}^-}a(\hat s)b(\hat s)|\hat{\x}'|d\hat S,\\
        \langle a(\hat s), b(\hat s)\rangle^+ &:= \int_{\p \hat{\Gamma}_{i_e}^+}a(\hat s)b(\hat s)|\hat{\x}'|d\hat S.
    \end{align}
\end{subequations}
The resulting system is then
\begin{equation}\label{eq:nte_inter}
\begin{aligned}
    \langle (\bo_{i_\Omega}\cdot\mathbf{n})v(\hat s), &\psi_{i_E, i_\Omega}(\hat s) \rangle^+ - \bo_{i_\Omega} \cdot \{\left(\mathbf{J}_{\hat{\x}}^{i_e}\right)^{-1}\nabla_{\hat \x} v(\hat \x), \psi_{i_E, i_\Omega}(\hat \x)\} + \s{t, i_E}\{v(\hat\x), \psi_{i_E,i_\Omega}(\hat\x)\}\\
    =& \sum_{i_E' = 1}^{N_E}\sum_{l = 0}^L\s{sl, i_E'\rightarrow i_E}\sum_{i_\Omega' = 1}^{N_\Omega}w_{i_\Omega'}^\Omega Y_l(\bo_{i_\Omega}, \bo_{i_\Omega'})\{v(\hat\x), \psi_{i_E',i_\Omega'}(\hat\x)\}\\& + \{v(\hat\x), Q_{i_E, i_\Omega}(\hat{\x})\} + \langle |\bo_{i_\Omega}\cdot \mathbf{n}|v(\hat s),\psi_{i_E, i_\Omega}^{\text{adj}}(\hat s)\rangle^-
\end{aligned}
\end{equation}
where $\nabla_{\hat{\x}} = \left(\frac{\p}{\p \hat{x}}, \frac{\p}{\p \hat{y}}\right)$. We then seek a solution $\psi_{g, n}(\x)\in H^1(\Gamma_{i_e})$ such that \cref{eq:nte_inter} holds $\forall v(\x)\in H^1(\Gamma_{i_e})$, and with the isoparametric concept, we assume the solution and test function have the same form as the NURBS patch defining the geometry,
\begin{align}
    v(\hat x, \hat y) &= \sum\limits_{i_{\hat x} = 1}^{N_{\hat x}}\sum\limits_{i_{\hat y} = 1}^{N_{\hat y}}R_{i_{\hat z}, i_{\hat y}}^{p_{\hat x}, p_{\hat y}}(\hat x, \hat y)\mathbf{v}_{i_{\hat x}, i_{\hat y}} = \mathbf{R}^T\mathbf{v},\\
    \psi_{i_E, i_\Omega}(\hat x, \hat y) &= \sum\limits_{i_{\hat x} = 1}^{N_{\hat x}}\sum\limits_{i_{\hat y} = 1}^{N_{\hat y}}R_{i_{\hat z}, i_{\hat y}}^{p_{\hat x}, p_{\hat y}}(\hat x, \hat y)\mathbf{\Psi}_{i_E, i_\Omega,i_{\hat x}, i_{\hat y}} = \mathbf{R}^T\mathbf{\Psi}_{i_E, i_\Omega},
\end{align}
where $\mathbf{R}$ is a vector of NURBS basis functions of shape $N_{\hat x}N_{\hat y}\times 1$ with control variables $\mathbf{v},\mathbf{\Psi}_{g, n}\in\R^{N_{\hat x}N_{\hat y}\times 1}$. The resulting system of $N_\Omega\times N_E\times N_{\hat x}\times N_{\hat y}$ linear equations for element $i_e$ is
\begin{subequations}\label{eq:operator}
\begin{equation}
    \mathcal H \mathbf{\Psi} + \mathcal{B}_{\text{out}}\mathbf{\Psi} = \mathcal S \mathbf{\Psi} + \mathcal{B}_{\text{in}}\mathbf{\Psi} + \mathbf{Q},
\end{equation}
\begin{align}
        (\mathcal H\mathbf{\Psi})_{i_E, i_\Omega} =& ~- \bo_{i_\Omega} \cdot \{\left(\mathbf{J}_{\hat{\x}}^{i_e}\right)^{-1}\nabla_{\hat \x} \mathbf{R}, \mathbf{R}^T\}\mathbf{\Psi}_{i_E, i_\Omega} + \s{t, i_E}\{\mathbf{R}, \mathbf{R}^T\}\mathbf{\Psi}_{i_E, i_\Omega}, \\ 
        (\mathcal S\mathbf{\Psi})_{i_E, i_\Omega} =& ~\sum_{i_E' = 1}^{N_E}\sum_{l = 0}^L\s{sl, i_E'\rightarrow i_E}\sum_{i_\Omega' = 1}^{N_\Omega}w_{i_\Omega'}^{\Omega}Y_l(\bo_{i_\Omega}, \bo_{i_\Omega'})\{\mathbf{R}, \mathbf{R}^T\}\mathbf{\Psi}_{i_E', i_\Omega'},\\
        (\mathcal B_{\text{out}}\mathbf{\Psi})_{i_E, i_\Omega} = &~ \langle (\bo_{i_\Omega}\cdot\mathbf{n})\mathbf{R},\mathbf{R}^T \rangle^+\mathbf{\Psi}_{i_E, i_\Omega},\\
        (\mathcal B_{\text{in}}\mathbf{\Psi})_{i_E, i_\Omega} = &~\langle |\bo_{i_\Omega}\cdot \mathbf{n}|\mathbf{R},\mathbf{R}^T\rangle^-\mathbf{\Psi}_{i_E, i_\Omega}^{\text{adj}},\\
        \mathbf Q_{i_E, i_\Omega} = &~\{\mathbf{R}, Q_{i_E, i_\Omega}(\hat{\x})\},
\end{align}
\end{subequations}
where $\mathcal{H}$ is the streaming and collision operator, $\mathcal{B}_{\text{out}}$ is the outflow boundary operator, $\mathcal{S}$ is the scattering operator, $\mathcal{B}_{\text{in}}$ is the inflow boundary operator, and $Q_{g,n}(\hat \x)$ is either a fixed source, in which $\mathbf{Q}$ will be a vector, or a fission source. In the fission source case, we expand $\mathbf{Q}$ to be
\begin{equation}
    \mathbf{Q}_{i_E, i_\Omega} = (\mathcal{F}\mathbf{\Psi})_{i_E, i_\Omega} = \f{\chi_{i_E}}{k}\sum_{i_E' = 1}^{N_E}\nu\s{f, i_E'}\sum_{i_\Omega' = 1}^{N_\Omega}w_{i_\Omega'}^\Omega\{\mathbf{R}, \mathbf{R}^T\}\mathbf{\Psi}_{i_E', i_\Omega'}
\end{equation}
with fission operator $\mathcal{F}$.

\par For the 2-D LBTE, we define the outflow boundary integral as four integrals over each side of the parametric square $\hat \Gamma_{i_e} = [0, 1]^2$,
\begin{equation}
\begin{aligned}
    \langle (\bo_{i_\Omega}\cdot\mathbf{n})\mathbf{R},\mathbf{R}^T \rangle^+ &= \int_{\p \hat\Gamma_{i_e}^+}(\bo_{i_\Omega}\cdot \mathbf{n})\mathbf{R}\mathbf{R^T}|\hat{\x}'|d\hat S\\
    &= \int_0^1\mathbf{f}_{i_\Omega}^{\Omega}(\hat x, 0)d\hat x + \int_0^1\mathbf{f}_{i_\Omega}^{\Omega}(\hat x, 1)d\hat x + \int_0^1\mathbf{f}_{i_\Omega}^{\Omega}(0, \hat y)d\hat y + \int_0^1\mathbf{f}_{i_\Omega}^{\Omega}(1, \hat y)d\hat y,
\end{aligned}
\end{equation}
where
\begin{subequations}
\begin{align}
    \mathbf{f}_{i_\Omega}^{\Omega}(\hat x, \hat y) &= \begin{cases}
        (\bo_{i_\Omega}\cdot \mathbf{n})\mathbf{R}(\hat x, \hat y)\mathbf{R}^T(\hat x, \hat y)|\hat{\x}'(\hat x, \hat y)| & \bo_{i_\Omega}\cdot\mathbf{n} > 0 \\
        0 & \text{otherwise}
    \end{cases},\\
    |\hat{\x}'(\hat x, \hat y)| &= \begin{cases}
        \sqrt{\left(\frac{\p x}{\p \hat x}|_{(\hat x, 0)}\right)^2 + \left(\frac{\p y}{\p \hat x}|_{(\hat x, 0)}\right)^2} & \hat y = 0\\
        \sqrt{\left(\frac{\p x}{\p \hat x}|_{(\hat x, 1)}\right)^2 + \left(\frac{\p y}{\p \hat x}|_{(\hat x, 1)}\right)^2} & \hat y = 1\\
        \sqrt{\left(\frac{\p x}{\p \hat y}|_{(0, \hat y)}\right)^2 + \left(\frac{\p y}{\p \hat y}|_{(0, \hat y)}\right)^2} & \hat x = 0\\
        \sqrt{\left(\frac{\p x}{\p \hat y}|_{(1, \hat y)}\right)^2 + \left(\frac{\p y}{\p \hat y}|_{(1, \hat y)}\right)^2} & \hat x = 1
    \end{cases}.
\end{align}
\end{subequations}
The outflow boundary integral is computed in the same manner. 

\par To numerically evaluate each spatial integral, we map each knot span $\left[\hat{\mathbf{x}}_{i_{\hat x}^u}^u, \hat{\mathbf{x}}_{i_{\hat x}^u + 1}^u\right]\times \left[\hat{\mathbf{y}}_{i_{\hat y}^u}^u, \hat{\mathbf{y}}_{i_{\hat y}^u + 1}^u\right]$ to the parent element space and use a $(p_{\hat x} + 1)\times(p_{\hat y} + 1)$ tensor product grid of Gauss-Legendre quadrature points, $\tilde x_{i_{\tilde x}}\in\{\tilde x_1, \tilde x_2, ..., \tilde x_{p_{\hat x} + 1}\}$ and $\tilde y_{i_{\tilde y}}\in\{\tilde y_1, \tilde y_2, ..., \tilde y_{p_{\hat y} + 1}\}$ with weights $\mathbf{w}^{\tilde \x}=\left\{w_{i_{\tilde x}}^{\tilde x}w_{i_{\tilde y}}^{\tilde y}\right\}_{i_{\tilde x} = 1, i_{\tilde y} = 1}^{p_{\hat x} + 1, p_{\hat y} + 1}$. Each integral over $\hat \Gamma_{i_e}$ is then
\begin{equation}
\begin{aligned}
    \int_{\hat \Gamma_{i_e}}\mathbf{f}(\hat x, \hat y)d\hat V &= \sum_{i_{\hat x}^u = 1}^{N_{\hat x}^u}\sum_{i_{\hat y}^u = 1}^{N_{\hat y}^u}\int_{\hat x_{i_{\hat x}^u}}^{\hat x_{i_{\hat x}^u + 1}}\int_{\hat y_{i_{\hat y}^u}}^{\hat y_{i_{\hat y}^u + 1}}\mathbf{f}(\hat x, \hat y)d\hat x d\hat y\\
    &= \sum_{i_{\hat x}^u = 1}^{N_{\hat x}^u}\sum_{i_{\hat y}^u = 1}^{N_{\hat y}^u}\int_{-1}^{1}\int_{-1}^{1}\mathbf{f}(\hat x(\tilde x), \hat y(\tilde y))\left|\det\mathbf{J}_{\tilde{\x}}^{i_{\hat x}^u, i_{\hat y}^u}(\tilde x, \tilde y)\right|d\tilde x d\tilde y\\
    &\approx \sum_{i_{\hat x}^u = 1}^{N_{\hat x}^u}\sum_{i_{\hat y}^u = 1}^{N_{\hat y}^u}\sum_{i_{\tilde x} = 1}^{p_{\hat x} + 1}\sum_{i_{\tilde y} = 1}^{p_{\hat y} + 1}w_{i_{\tilde x}}^{\tilde x}w_{i_{\tilde y}}^{\tilde y}\mathbf{f}(\hat x(\tilde x_{i_{\tilde x}}), \hat y(\tilde y_{i_{\tilde y}}))\left|\det\mathbf{J}_{\tilde{\x}}^{i_{\hat x}^u, i_{\hat y}^u}(\tilde x_{i_{\tilde x}}, \tilde y_{i_{\tilde y}})\right|
\end{aligned}
\end{equation}
with Jacobian $\mathbf{J}_{\tilde{\x}}^{i_{\hat x}^u, i_{\hat y}^u}(\tilde x, \tilde y) = \frac{\p(\hat x, \hat y)}{\p(\tilde x, \tilde y)}$ computed from \cref{eq:parent_mapping} and
\begin{equation*}
\mathbf{f}(\hat x, \hat y) = \mathbf{R}(\hat x, \hat y)\mathbf{R}^T(\hat x, \hat y)| \det\mathbf{J}_{\hat \x}^{i_e}(\hat x, \hat y)| \text{  or  } \mathbf{f}(\hat x, \hat y) = \left(\mathbf{J}_{\hat \x}^{i_e}(\hat x, \hat y)\right)^{-1}\nabla_{\hat \x}\mathbf{R}(\hat x, \hat y)\mathbf{R}^T(\hat x, \hat y)|\det\mathbf{J}_{\hat \x}^{i_e}(\hat x, \hat y)|.
\end{equation*}
The line integrals at the boundaries follow a similar pattern. For example, the line integral for $\hat y = 0$ is
\begin{equation}
\begin{aligned}
    \int_0^1\mathbf{f}_{\Omega}(\hat x, 0)d\hat x &= \sum_{i_{\hat x}^u = 1}^{N_{\hat x}^u}\int_{\hat x_{i_{\hat x}^u}}^{\hat x_{i_{\hat x}^u + 1}}\mathbf{f}_{\Omega}(\hat x, 0)d\hat x\\
    &= \sum_{i_{\hat x}^u = 1}^{N_{\hat x}^u}\int_{-1}^1\mathbf{f}_{\Omega}(\hat x(\tilde x), 0)\left|\frac{d\hat x}{d\tilde x}\right|d\tilde x\\
    &\approx \sum_{i_{\hat x}^u = 1}^{N_{\hat x}^u}\sum_{i_{\tilde x} = 1}^{p_{\hat x} + 1}w_{i_{\tilde x}}^{\tilde x}\mathbf{f}_\Omega\left(\hat x(\tilde x_{i_{\tilde x}}), 0\right)\left|\frac{d\hat x}{d\tilde x}\right|_{\tilde x_{i_{\tilde x}}}.
\end{aligned}
\end{equation}

\subsection{Tensor Train Format}\label{sec:tt}

\par Here we discuss the tensor train (TT) decomposition first proposed in \cite{Oseledets}. Since then, it has seen extensive application for solving PDEs in scientific computing; we refer the reader to several review papers for a comprehensive discussion of its properties, operations, and applications \cite{annurev:/content/journals/10.1146/annurev-conmatphys-040721-022705, https://doi.org/10.1002/gamm.201310004, KHOROMSKIJ20121, doi:10.1137/07070111X, refId0}.

\par
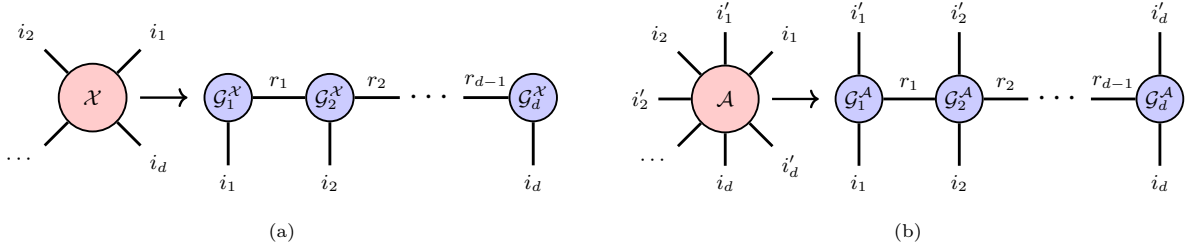
\begin{figure}
\centering

\begin{subfigure}[b]{0.47\textwidth}
\centering
\resizebox{\columnwidth}{!}{
\begin{tikzpicture}[
    thick,
    every node/.style={font=\small},
    tensor/.style={
        circle,
        draw=black,
        fill=blue!20,
        minimum size=7mm,
        inner sep=0pt
    },
    bigtensor/.style={
        circle,
        draw=black,
        fill=red!20,
        minimum size=10mm,
        inner sep=1pt
    },
    leg/.style={line width=1.2pt},
    arrowstyle/.style={->, thick, line width=1pt},
    dotstyle/.style={font=\Large}
]

\node[bigtensor] (T) at (-2,0) {$\mathcal{X}$};

\draw[leg] (T) -- ++(45:1) node[above right] {$i_1$};
\draw[leg] (T) -- ++(135:1) node[above left] {$i_2$};
\draw[leg] (T) -- ++(225:1) node[below left] {$\cdots$};
\draw[leg] (T) -- ++(315:1) node[below right] {$i_d$};

\draw[arrowstyle] (-1.3,0) -- (-0.6,0);

\node[tensor] (G1) at (0,0) {$\mathcal{G}_1^{\mathcal X}$};
\node[tensor] (G2) at (1.5,0) {$\mathcal{G}_2^{\mathcal X}$};
\node[dotstyle] (dots) at (3,0) {$\cdots$};
\node[tensor] (Gd) at (4.5,0) {$\mathcal{G}_d^{\mathcal X}$};

\draw[leg] (G1) -- ++(0,-1) node[below] {$i_1$};
\draw[leg] (G2) -- ++(0,-1) node[below] {$i_2$};
\draw[leg] (Gd) -- ++(0,-1) node[below] {$i_d$};

\draw[leg] (G1) -- (G2) node[midway, above] {$r_1$};
\draw[leg] (G2) -- (dots) node[midway, above] {$r_2$};
\draw[leg] (dots) -- (Gd) node[midway, above] {$r_{d-1}$};

\end{tikzpicture}
}
\caption{}
\label{fig:tt_vector}
\end{subfigure}
\hspace{0.02\textwidth}
\begin{subfigure}[b]{0.47\textwidth}
\centering
\resizebox{\columnwidth}{!}{
\begin{tikzpicture}[
    thick,
    every node/.style={font=\small},
    tensor/.style={
        circle,
        draw=black,
        fill=blue!20,
        minimum size=7mm,
        inner sep=0pt
    },
    bigtensor/.style={
        circle,
        draw=black,
        fill=red!20,
        minimum size=10mm,
        inner sep=1pt
    },
    leg/.style={line width=1.2pt},
    arrowstyle/.style={->, thick, line width=1pt},
    dotstyle/.style={font=\Large}
]

\node[bigtensor] (T) at (-2,0) {$\mathcal{A}$};

\draw[leg] (T) -- ++(45:1) node[above right] {$i_1$};
\draw[leg] (T) -- ++(90:1) node[above] {$i_1'$};
\draw[leg] (T) -- ++(135:1) node[above left] {$i_2$};
\draw[leg] (T) -- ++(180:1) node[left] {$i_2'$};
\draw[leg] (T) -- ++(225:1) node[below left] {$\cdots$};
\draw[leg] (T) -- ++(270:1) node[below] {$i_d$};
\draw[leg] (T) -- ++(315:1) node[below right] {$i_d'$};

\draw[arrowstyle] (-1.3,0) -- (-0.6,0);

\node[tensor] (G1) at (0,0) {$\mathcal{G}_1^{\mathcal A}$};
\node[tensor] (G2) at (1.5,0) {$\mathcal{G}_2^{\mathcal A}$};
\node[dotstyle] (dots) at (3,0) {$\cdots$};
\node[tensor] (Gd) at (4.5,0) {$\mathcal{G}_d^{\mathcal A}$};

\draw[leg] (G1) -- ++(0,1) node[above] {$i_1'$};
\draw[leg] (G1) -- ++(0,-1) node[below] {$i_1$};

\draw[leg] (G2) -- ++(0,1) node[above] {$i'_2$};
\draw[leg] (G2) -- ++(0,-1) node[below] {$i_2$};

\draw[leg] (Gd) -- ++(0,1) node[above] {$i'_d$};
\draw[leg] (Gd) -- ++(0,-1) node[below] {$i_d$};

\draw[leg] (G1) -- (G2) node[midway, above] {$r_1$};
\draw[leg] (G2) -- (dots) node[midway, above] {$r_2$};
\draw[leg] (dots) -- (Gd) node[midway, above] {$r_{d-1}$};

\end{tikzpicture}
}
\caption{}
\label{fig:tt_operator}
\end{subfigure}

\caption{Visualization of a tensor train (TT) (a) vector and (b) operator (matrix). For a tensor $\mathcal{X}\in\mathbb{R}^{N_1\times N_2\times\dots\times N_d}$ we can decompose it into $d$ three-dimensional tensors connected by tensor products over the ranks $r_j\in\{r_1, \dots,r_{d - 1}\}$. The operator tensor $\mathcal{A}\in\mathbb{R}^{N_1\times N_1'\times N_2\times N_2'\times\dots\times N_d\times N_d'}$ follows similarly with an additional dimension on each core, resulting in a linear network of $d$ four-dimensional tensors. The labeled circles represent tensors, the lines connecting the tensors indicate a tensor contraction over the contracted dimensions, and the dangling lines labeled with an index are the free indices that are open to contract with other tensor networks (TNs).}
\label{fig:tt_decompositions}
\end{figure}

 Given a vector $\mathbf{x}\in\mathbb{R}^{N_1N_2\dots N_d}$ we can fold it into a $d$-dimensional tensor $\mathcal{X}\in\mathbb{R}^{N_1\times N_2\times\cdots\times N_d}$ such that $\mathbf{x}(\overline{i_1i_2\dots i_d}) = \mathcal{X}(i_1, i_2, \dots, i_d)$ where $\overline{i_1i_2\dots i_d}$ denotes the raveled single index of $i_1, i_2, \dots, i_d$. These indices are known as \textit{free indices}. We can then decompose this vector using the tensor train (TT) decomposition given by
\begin{equation}
    \mathbf{x}(\overline{i_1i_2\dots i_d}) = \mathcal{X}(i_1, i_2, \dots, i_d) = \sum_{\alpha_1 = 1}^{r_1}\sum_{\alpha_2 = 1}^{r_2}\cdots\sum_{\alpha_{d - 1} = 1}^{r_{d - 1}}\mathcal{G}_1^{\mathcal X}(1, i_1, \alpha_1)\mathcal{G}_2^{\mathcal X}(\alpha_1, i_2, \alpha_2)\cdots\mathcal{G}_{d}^{\mathcal X}(\alpha_{d - 1}, i_d, 1),
\end{equation}
where $\alpha_j\in\{1, ..., r_j\}$ is the $j$th-rank index with TT-rank $r_j$ and TT-core $\mathcal{G}_j^{\mathcal{X}}\in\mathbb{R}^{r_{j - 1}\times N_j\times r_j}$. The result is a linear network of $d$ three-dimensional tensors coupled by tensor products across their rank dimensions, as shown in \cref{fig:tt_vector} where the labeled circular nodes represent tensors who's number of dimensions is the number of lines emanating from it, the connection between tensors represents a tensor contraction, and the dangling lines are free indices or dimensions which can be contracted with other networks. For a discussion on reading tensor network diagrams, please see \ref{app:ball_and_stick}. During the factorization procedure, we truncate each successive singular value decomposition (SVD) to a rank-$r_j$ approximation, resulting in an approximate TT representation $\tilde{\mathcal{X}}$ such that
\begin{equation}
    \|\mathcal{X} - \tilde{\mathcal{X}}\|_F \le \epsilon\|\mathcal{X}\|_F,
\end{equation}
where $\epsilon$ is the truncation tolerance and $\|\cdot \|_F$ is the Frobenius norm. This factorization procedure, known as the TT-SVD algorithm, is discussed in detail in \cite{Oseledets}. We can generalize the TT format to operators,
\begin{equation}
    \mathcal{A}(i_1, i_1', i_2, i_2', \dots, i_d, i_d') = \sum_{\alpha_1 = 1}^{r_1}\sum_{\alpha_2 = 1}^{r_2}\cdots\sum_{\alpha_{d - 1} = 1}^{r_{d - 1}}\mathcal{G}_1^{\mathcal A}(1, i_1, i_1', \alpha_1)\mathcal{G}_2^{\mathcal A}(\alpha_1, i_2, i_2', \alpha_2)\cdots\mathcal{G}_{d}^{\mathcal A}(\alpha_{d - 1}, i_d, i_d', 1),
\end{equation}
resulting in a linear contraction network of $d$ four-dimensional tensors, see \cref{fig:tt_operator}. This low-rank representation has $\mathcal{O}(dNr^2)$ and $\mathcal{O}(dN^2r^2)$ memory complexity for TT-vectors and TT-operators, respectively.

\par Common linear-algebra operations on vectors and matrices have direct analogs in the TT format. Basic operations including summation ($+$), Hadamard (element-wise) multiplication ($\odot$), and operator-operator or operator-vector products ($\cdot$) are discussed in \cite{Oseledets}. For a faster TT Hadamard multiplication, we use the algorithm proposed in \cite{michailidis2024tensortrainmultiplication}. As opposed to standard TT-operator by TT-operator or TT-vector products, we use the optimization-based AMEn algorithm to compute these products \cite{math12203277, doi:10.1137/140953289}. We also use the TT-operator-by-full-vector product in the linear solver. When converting from a full tensor to a TT, we use the TT-SVD algorithm in \cite{Oseledets}. With all these basic linear algebra operations, the resulting TT incurs a rank increase, which is recompressed to the truncation tolerance $\epsilon$ using the TT-rounding algorithm \cite{Oseledets}. For each operation, the resulting rank, complexity, and algorithm citation are shown in \cref{tbl:tt_operations}.

\begin{table}
\centering
\caption{Tensor train (TT) operations with resulting rank and complexity where $\mathcal{X},\mathcal{Y}, \mathcal{Z}\in\mathbb{R}^{N_1\times N_2\times \dots\times N_d}$ and $\mathcal{A}\in\mathbb{R}^{N_1\times N_1'\times N_2\times N_2'\times \dots\times N_d\times N_d'}$ are TTs with ranks $\mathbf{r}_{\mathcal X},\mathbf{r}_{\mathcal{Y}}, \mathbf{r}_{\mathcal{Z}}, \mathbf{r}_{\mathcal{A}}\in\mathbb{R}^{d - 1}$. We define $a$ as a constant, ($\circ$) as the vector Hadamard product, $\mathbf{x}\in\mathbb{R}^{N_1N_2\dots N_d\times 1}$ and $\mathbf{z}\in\mathbb{R}^{N_1N_2\dots N_d\times 1}$ as full vectors, and $N = \max_{j\in\{1, 2, \dots,d\}}N_j$.}\label{tbl:tt_operations}
\begin{tabular}{llll} 
\toprule
\textbf{Operation}                                  & \textbf{Resulting Rank}                             & \textbf{Complexity}           & \textbf{Citation}  \\ 
\hline\hline
$\mathcal{Z}=a\mathcal{X}$                            & $\mathbf{r}_{\mathcal Z} = \mathbf{r}_{\mathcal X}$                  & $\mathcal{O}(d\max(\mathbf{r}_{\mathcal X}))$ &   \cite{Oseledets}                 \\
$\mathcal{Z} = \mathcal{X} + \mathcal{Y}$              & $\mathbf{r}_{\mathcal Z}\le \mathbf{r}_{\mathcal X} +\mathbf{r}_{\mathcal Y}$ & $\mathcal{O}(dN\max(\mathbf{r}_{\mathcal X} + \mathbf{r}_{\mathcal Y}))^2)$ & \cite{Oseledets}                   \\
$\mathcal{Z} = \mathcal{X} \odot \mathcal{Y}$          & $\mathbf{r}_{\mathcal Z} \le \mathbf{r}_{\mathcal X}\circ \mathbf{r}_{\mathcal Y}$   & $\mathcal{O}(dN\max(\mathbf{r}_{\mathcal X}\circ\mathbf{r}_{\mathcal Y})^3)$ &     \cite{michailidis2024tensortrainmultiplication}               \\
$\mathcal{Z} = \mathcal{A} \cdot \mathcal{X}$           & $\mathbf{r}_{\mathcal Z} \le \mathbf{r}_{\mathcal A}\circ \mathbf{r}_{\mathcal X}$ & $\mathcal{O}(dN\max(\mathbf{r}_{\mathcal A}\circ\mathbf{r}_{\mathcal X})^3)$ &      \cite{math12203277, doi:10.1137/140953289}              \\
$\mathbf{z}=\mathcal{A}\cdot\mathbf{x}$                            & $-$                  & $\mathcal{O}(dN^d\max(\mathbf{r}_A)^2\log(N))$ &   \cite{Oseledets}                 \\
$\mathcal{Z} = \textsc{round}_\epsilon(\mathcal{X})$ & $\mathbf{r}_{\mathcal Z} \le \mathbf{r}_{\mathcal X}$ & $\mathcal{O}(dN\max(\mathbf{r}_{\mathcal X})^3)$                               &        \cite{Oseledets}            \\
\bottomrule
\end{tabular}
\end{table}
\section{Methods}\label{sec:methods}

\par In this section, we discuss the TDIGA method. We begin with a discussion of preliminary definitions and notation in \cref{sec:not}, followed by a discussion of operator assembly in the TT format in \cref{sec:tensorized_operator_assembly}. In \cref{sec:numerical_scheme} we discuss the numerical scheme used to solve the global system, and in \cref{sec:implementation} we briefly discuss software implementation details.

\subsection{Notation and Preliminaries}\label{sec:not}

\begin{figure}
\centering
\begin{tikzpicture}[
    thick,
    every node/.style={font=\small},
    tensor/.style={
        circle,
        draw=black,
        fill=blue!20,
        minimum size=7mm,
        inner sep=0pt
    },
    bigtensor/.style={
        circle,
        draw=black,
        fill=red!20,
        minimum size=10mm,
        inner sep=1pt
    },
    freeindex/.style={},
    leg/.style={line width=1.2pt},
    arrowstyle/.style={->, thick, line width=1pt},
    dotstyle/.style={font=\Large}
]

\node[tensor] (G1) at (0,0) {$\mathcal{G}_1^{\mathcal X}$};
\node[tensor] (G2) at (1.5,0) {$\mathcal{G}_2^{\mathcal X}$};
\node[dotstyle] (dots) at (3,0) {$\cdots$};
\node[tensor] (Gd) at (4.5,0) {$\mathcal{G}_d^{\mathcal X}$};

\node[freeindex] (N1) at (0, -1.2) {$i_1$};
\node[freeindex] (N2) at (1.5, -1.2) {$i_2$};
\node[freeindex] (Nd) at (4.5, -1.2) {$i_d$};

\draw[leg] (G1) -- (G2) node[midway, above] {$r_1$};
\draw[leg] (G2) -- (dots) node[midway, above] {$r_2$};
\draw[leg] (dots) -- (Gd) node[midway, above] {$r_{d-1}$};

\draw[leg] (G1) -- (N1);
\draw[leg] (G2) -- (N2);
\draw[leg] (Gd) -- (Nd);

\node[dotstyle] (otimes) at (5.25, 0) {$\otimes$};
\node[tensor] (c) at (6, 0) {$\mathbf{c}$};
\node[freeindex] (Nd1) at (6, -1.2) {$i_{d + 1}$};
\draw[leg] (c) -- (Nd1);

\draw[arrowstyle] (6.7,0) -- (7.4,0);

\node[tensor] (b1) at (8,0) {$\mathcal{G}_1^{\mathcal X}$};
\node[tensor] (b2) at (9.5,0) {$\mathcal{G}_2^{\mathcal X}$};
\node[dotstyle] (dots2) at (11,0) {$\cdots$};
\node[tensor] (bd) at (12.5,0) {$\mathcal{G}_d^{\mathcal X}$};
\node[tensor] (c2) at (14, 0) {$\mathbf{c}$};

\node[freeindex] (bn1) at (8, -1.2) {$i_1$};
\node[freeindex] (bn2) at (9.5, -1.2) {$i_2$};
\node[freeindex] (bnd) at (12.5, -1.2) {$i_d$};
\node[freeindex] (bnd1) at (14, -1.2) {$i_{d + 1}$};

\draw[leg] (b1) -- (bn1);
\draw[leg] (b2) -- (bn2);
\draw[leg] (bd) -- (bnd);
\draw[leg] (c2) -- (bnd1);

\draw[leg] (b1) -- (b2) node[midway, above] {$r_1$};
\draw[leg] (b2) -- (dots2) node[midway, above] {$r_2$};
\draw[leg] (dots2) -- (bd) node[midway, above] {$r_{d-1}$};
\draw[leg] (bd) -- (c2) node[midway, above] {$r_{d}=1$};

\begin{pgfonlayer}{background}
    \node[
        fit=(G1)(Nd),
        inner xsep=5pt,  
        inner ysep=10pt,   
        fill=yellow,
        fill opacity=0.3,  
        draw=red,
        thick,
        rounded corners,
        label=above:{$\mathcal{X}$}
    ] {};
\end{pgfonlayer}
\begin{pgfonlayer}{background}
    \node[
        fit=(c)(Nd1),
        inner xsep=0.5pt,  
        inner ysep=10pt,   
        fill=yellow,
        fill opacity=0.3,  
        draw=red,
        thick,
        rounded corners,
        label=above:{$\mathbf{c}$}
    ] {};
\end{pgfonlayer}
\begin{pgfonlayer}{background}
    \node[
        fit=(b1)(bnd1),
        inner xsep=4pt,  
        inner ysep=10pt,   
        fill=yellow,
        fill opacity=0.3,  
        draw=red,
        thick,
        rounded corners,
        label=above:{$\mathcal{Y} = \mathcal{X}\otimes \mathbf{c}$}
    ] {};
\end{pgfonlayer}

\end{tikzpicture}
\caption{A diagram of the tensor Kronecker product of a tensor train (TT) $\mathcal{X}\in\mathbf{R}^{N_1\times N_2\times \cdots \times N_d}$ and a vector $\mathbf{c}\in\mathbb{R}^{N_{d + 1}}$, \cref{eq:kronecker}. The result, $\mathcal{Y}\in\mathbb{R}^{N_1\times N_2\times \cdots \times N_d\times N_{d + 1}}$, is a TT with the cores of $\mathcal{X}$ coupled to $\mathbf{c}$ by a tensor product of rank $r_{d} = 1$ between $\mathcal{G}_d^{\mathcal{X}}$ and $\mathbf{c}$.}\label{fig:kronecker}
\end{figure}
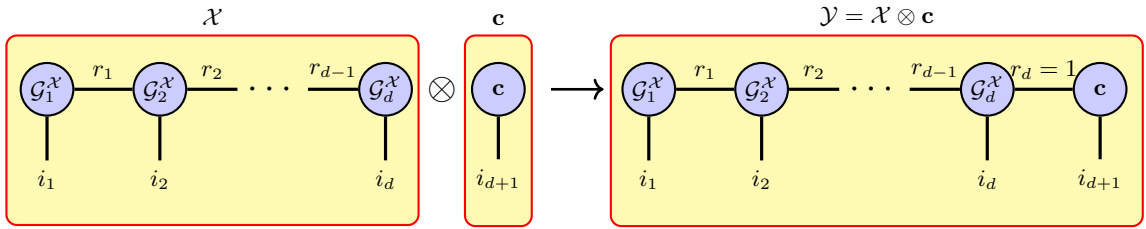

\begin{definition}[Tensor Kronecker Product]
    Let $\mathcal{X}\in\mathbb{R}^{N_1\times N_2\times\dots\times N_d}$ be a $d$-dimensional TT with free index $i_j$ for the $j$th dimension of size $N_j$. The tensor Kronecker product of $\mathcal{X}$ with a vector $\mathbf{c}\in\mathbb{R}^{N_{d + 1}}$ is
    \begin{equation}\label{eq:kronecker}
        \mathcal{Y} = \mathcal{X}\otimes \mathbf{c},
    \end{equation}
    where $\mathcal{Y}\in\mathbb{R}^{N_1\times N_2\times\dots\times N_d \times N_{d + 1}}$ is a $(d + 1)$-dimensional TT with $r_d = 1$. We show this operation in \cref{fig:kronecker}. Subsequently, the tensor Kronecker product of a tensor defines a rank-one coupling.
\end{definition}

\begin{definition}[Core Transpose]\label{def:trans}
    Given a TT $\mathcal{X}\in\mathbb{R}^{N_1\times N_2\times\dots\times N_d}$, we define the core transpose operation of the $j$th core to be,
    \begin{equation}
        \mathcal{Y} = \trans{j}(\mathcal{X}),
    \end{equation}
    where the $j$th core of $\mathcal{X}$ is transposed. For rank indices $(\alpha_{j - 1}, \alpha_j)$ the $N_j \times 1$ vector becomes a $1\times N_j$ vector,
    \begin{equation}
    \mathcal{G}_j^{\mathcal{X}}(\alpha_{j - 1}, :, \alpha_{j}) = \begin{bmatrix}
        a_1 \\ 
        a_2 \\
        a_3 \\
        \vdots \\ 
        a_{N_j}
    \end{bmatrix} \xrightarrow[]{\textsc{trans}(\mathcal{G}_j^{\mathcal X})} \mathcal{G}_j^{\mathcal{Y}}(\alpha_{j - 1}, :, :, \alpha_{j}) = \begin{bmatrix}
        a_1 & a_2 & a_3 & \cdots &a_{N_j}
    \end{bmatrix}.
    \end{equation}
    such that $\mathcal{Y}(i_1, i_2, \dots, 1, i_j, \dots, i_d) = \mathcal{X}(i_1, i_2, \dots, j, \dots, i_d)$.
\end{definition}

\begin{definition}[Core Diagonalization]\label{def:diag}
    Given a TT $\mathcal{X}\in\mathbb{R}^{N_1\times N_2\times\dots\times N_d}$, the core diagonalization operator converts a core within a TT-vector to a TT-operator with elements along the diagonal, 
    \begin{equation}
        \mathcal{Y} = \diag{j}(\mathcal{X}),
    \end{equation}
    where the $j$th core of $\mathcal{X}$ becomes a diagonal matrix for rank indices $(\alpha_{j - 1}, \alpha_j)$ resulting in a $N_j\times N_j$ matrix,
    \begin{equation}
        \mathcal{G}_j^{\mathcal{X}}(\alpha_{j - 1}, :, \alpha_{j}) = \begin{bmatrix}
            a_1 \\ 
            a_2 \\
            a_3 \\
            \vdots \\ 
            a_{N_j}
        \end{bmatrix} \xrightarrow[]{\textsc{diag}(\mathcal{G}_j^{\mathcal X})} \mathcal{G}_j^{\mathcal{X}}(\alpha_{j - 1}, :, :, \alpha_{j}) = \begin{bmatrix}
            a_1 \\ & a_2 \\ && a_3 \\ &&& \ddots \\&&&&a_{N_j}
        \end{bmatrix},
    \end{equation}
    such that $\mathcal{Y}(i_1, i_2,\dots, i_j, i_j, \dots i_d) = \mathcal{X}(i_1, i_2, \dots, i_j, \dots, i_d)$.
\end{definition}

\begin{figure}
\centering
\resizebox{\columnwidth}{!}{
\begin{tikzpicture}[
    thick,
    every node/.style={font=\small},
    tensor/.style={
        circle,
        draw=black,
        fill=blue!20,
        minimum size=7mm,
        inner sep=0pt
    },
    bigtensor/.style={
        circle,
        draw=black,
        fill=red!20,
        minimum size=10mm,
        inner sep=1pt
    },
    freeindex/.style={},
    leg/.style={line width=1.2pt},
    arrowstyle/.style={->, thick, line width=1pt},
    dotstyle/.style={font=\Large}
]

\node[dotstyle] (G0) at (-1.5,0) {$\cdots$};
\node[tensor] (G1) at (0,0) {$\mathcal{G}_{j - 1}^{\mathcal X}$};
\node[tensor] (G2) at (1.5,0) {$\mathcal{G}_{j}^{\mathcal X}$};
\node[tensor] (G3) at (3,0) {$\mathcal{G}_{j + 1}^{\mathcal X}$};
\node[tensor] (G4) at (4.5,0) {$\mathcal{G}_{j + 2}^{\mathcal X}$};
\node[dotstyle] (G5) at (6,0) {$\cdots$};

\node[freeindex] (N1) at (0, -1.2) {$i_{j - 1}$};
\node[freeindex] (N2) at (1.5, -1.2) {$i_{j}$};
\node[freeindex] (N3) at (3, -1.2) {$i_{j + 1}$};
\node[freeindex] (N4) at (4.5, -1.2) {$i_{j + 2}$};

\draw[leg] (G0) -- (G1);
\draw[leg] (G1) -- (G2);
\draw[leg] (G2) -- (G3);
\draw[leg] (G3) -- (G4);
\draw[leg] (G4) -- (G5);

\draw[leg] (G1) -- (N1);
\draw[leg] (G2) -- (N2);
\draw[leg] (G3) -- (N3);
\draw[leg] (G4) -- (N4);

\draw[arrowstyle] (6.5,0) -- (7,0);

\node[dotstyle] (G02) at (7.5,0) {$\cdots$};
\node[tensor] (G12) at (9,0) {$\tilde{\mathcal{G}}_{j - 1}^{\mathcal X}$};
\node[tensor] (G22) at (10.5,0) {$\tilde{\mathcal{G}}_{j + 1}^{\mathcal X}$};
\node[tensor] (G32) at (12,0) {$\tilde{\mathcal{G}}_{j}^{\mathcal X}$};
\node[tensor] (G42) at (13.5,0) {$\tilde{\mathcal{G}}_{j + 2}^{\mathcal X}$};
\node[dotstyle] (G52) at (15,0) {$\cdots$};

\node[freeindex] (N12) at (9, -1.2) {$i_{j - 1}$};
\node[freeindex] (N22) at (10.5, -1.2) {$i_{j + 1}$};
\node[freeindex] (N32) at (12, -1.2) {$i_{j}$};
\node[freeindex] (N42) at (13.5, -1.2) {$i_{j + 2}$};

\draw[leg] (G02) -- (G12);
\draw[leg] (G12) -- (G22);
\draw[leg] (G22) -- (G32);
\draw[leg] (G32) -- (G42);
\draw[leg] (G42) -- (G52);

\draw[leg] (G12) -- (N12);
\draw[leg] (G22) -- (N22);
\draw[leg] (G32) -- (N32);
\draw[leg] (G42) -- (N42);

\begin{pgfonlayer}{background}
    \node[
        fit=(G0)(G5)(N1),
        inner xsep=0pt,  
        inner ysep=10pt,   
        fill=yellow,
        fill opacity=0.3,  
        draw=red,
        thick,
        rounded corners,
        label=above:{$\mathcal{X}$}
    ] {};
\end{pgfonlayer}
\begin{pgfonlayer}{background}
    \node[
        fit=(G02)(G52)(N12),
        inner xsep=0pt,  
        inner ysep=10pt,   
        fill=yellow,
        fill opacity=0.3,  
        draw=red,
        thick,
        rounded corners,
        label=above:{$\mathcal{Y} = \perm{:(j - 1), j + 1, j, (j + 2):}(\mathcal{X})$}
    ] {};
\end{pgfonlayer}

\end{tikzpicture}
}
\caption{Diagram of the permutation of tensor train (TT) $\mathcal{X}\in\mathbb{R}^{N_1\times \cdots \times N_j \times N_{j + 1} \times \dots \times N_d}$ to $\mathcal{Y}\in\mathbb{R}^{N_1 \times \dots \times N_{j + 1} \times N_{j} \times\dots \times N_d}$ where the core $\mathcal{G}_j^{\mathcal{X}}$ and $\mathcal{G}_{j + 1}^{\mathcal{X}}$ are swapped. Note that physical dimensions are swapped; however, the cores themselves changed, $\mathcal{G}_j^{\mathcal{X}} \ne \tilde{\mathcal G}_j^{\mathcal{X}}$, along with the ranks.}\label{fig:perm}
\end{figure}
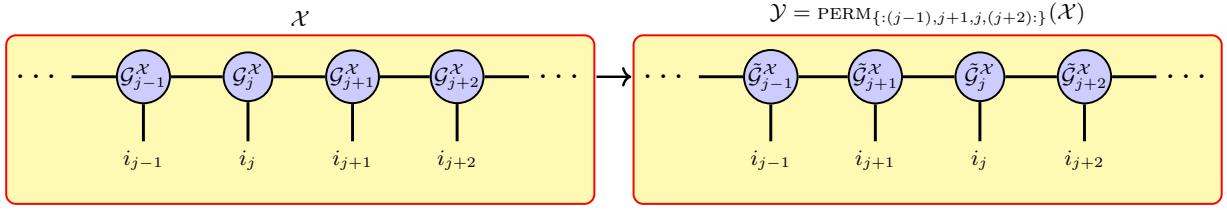

\begin{definition}[Permutation]\label{def:perm}
    Given a TT $\mathcal{X}\in\mathbb{R}^{N_1\times N_2\times\dots\times N_d}$, the permutation operator swaps cores,
    \begin{equation}
        \mathcal{Y} = \perm{:(j - 1), j + 1, j, (j + 2):}(\mathcal{X}),
    \end{equation}
    where cores $j$ and $j + 1$ are swapped, refer to \cref{fig:perm} for a visual example, resulting in $\mathcal{Y}(i_1, \dots, i_{k + 1}, i_k, \dots, i_d) = \mathcal{X}(i_1, \dots, i_{k}, i_{k + 1}, \dots, i_d)$. This operation is not exact and is subject to a given tolerance $\epsilon$, resulting in cores that do not match $\mathcal{X}$ and differ in rank.
\end{definition}

\par For the operations defined in \cref{def:trans,def:diag} we may provide a subscirpt set of indices $\{\cdots\}$ which defines the cores operatored on. If no set is provided, then all cores in the TT are operated on. The subscript set of indices defines the new order of the cores for \cref{def:perm}. We use MATLAB-like notation for indices in these operations and for indexing TTs. Some examples include $\{:j\} = \{1, 2, \dots, j\}$, $\{j:(j + 3)\} = \{j, j + 1, j + 2, j + 3\}$, and $\{j:\} = \{j, j + 1, \dots, d\}$.

\begin{definition}[TT-vector of Ones]
    A TT vector of ones of shape $N_1\times N_2\times\cdots\times N_d$ is defined as
    \begin{equation}
        1_{\{N_1, N_2, \dots, N_d\}} = \bigotimes_{i = 1}^d\mathbf{1}_{N_i},
    \end{equation}
    where $\mathbf{1}_{N_i}\in\{1\}^{N_i}$ is a vector of ones for the $i$th dimension of length $N_i$.
\end{definition}

\begin{definition}[Identity Matrix in TT Format]
    The identity TT is
    \begin{equation}
        \mathcal{I}_{\{N_1, N_2, \dots, N_d\}} = \bigotimes_{i = 1}^d \mathbf{I}_{N_i},
    \end{equation}
    where $\mathbf{I}_{N_i}\in\{0, 1\}^{N_i\times N_i}$ is an identity matrix for the $i$th dimension.
\end{definition}

\begin{definition}[Indicator Vector]
    The indicator vector $\boldsymbol{\delta}_{N}^i\in\{0, 1\}^{N}$ is a vector of length $N$ with a one at $i$ and zeros everywhere else.
\end{definition}

\subsection{Tensorized Operator Assembly}\label{sec:tensorized_operator_assembly}

\par Here we discuss operator assembly in the TT format using the concatenation procedure proposed in \cite{math12203277}. In \cref{sec:angular}, we discuss angular quadrature and integration, as well as the spherical harmonic operator in TT format. In \cref{sec:interior_operator_space}, we discuss the assembly procedure for building the spatial cores of the interior operators for each patch in TT format, followed by the construction of the interior operators in \cref{sec:interior_operators}. We discuss boundary operator assembly in \cref{sec:boundary_operators}.

\subsubsection{Tensorized angular operators}\label{sec:angular}

\par In this work, we assume a quadrant-symmetric angular quadrature set with $N_\Omega=4N_\mu N_\gamma$ ordinates. We can unravel the index such that $\bo_{i_{\Omega}} = \bo_{i_q, i_\mu, i_\gamma} = \left(\mathbf{c}^{\mu}_{i_q}\mu_{i_\mu}, \mathbf{c}_{i_q}^{\eta}\sqrt{1 - \mu_{i_\mu}^2}\cos \gamma_{i_\gamma}\right)$ where $\mathbf{c}^{\mu} = (1, 1, -1, -1)$, $\mathbf{c}^{\eta} = (1, -1, 1, -1)$, $i_q\in\{1, 2, 3, 4\}$, $i_\mu\in\{1, 2, \dots, N_\mu\}$, and $i_\gamma\in\{1, 2, \dots, N_\gamma\}$. The weights form a tensor product such that $w_{i_\Omega}^\Omega = w_{i_q, i_\mu, i_\gamma}^{\Omega} = w_{i_\mu}^\mu w_{i_\gamma}^\gamma$. In this work, we use a Chebyshev-Legendre quadrature set \cite{osti_5958402} where $N_\mu = N_\gamma$. In the TT format, we can represent this tensor product exactly,
\begin{subequations}\label{eq:directions_in_tt}
\begin{align}
    \hat{\mathsf{\Omega}} &= \begin{pmatrix}
        \mathsf{\Omega}_x & \mathsf{\Omega}_y
    \end{pmatrix},\\
    \mathsf{\Omega}_x &= \mathbf{c}^{\mu}\otimes \left\{\mu_{i_\mu}\right\}_{i_\mu = 1}^{N_\mu}\otimes \mathbf{1}_{N_\gamma},\\
    \mathsf{\Omega}_y &= \mathbf{c}^{\eta}\otimes \left\{\sqrt{1 - \mu_{i_\mu}^2}\right\}_{i_\mu = 1}^{N_\mu}\otimes \left\{\cos\gamma_{i_\gamma}\right\}_{i_{\gamma} = 1}^{N_{\gamma}}.
\end{align}
\end{subequations}
The spherical harmonic operator $\mathcal{Y}\in\mathbb{R}^{\left[4\times N_\mu \times N_\gamma  \times(L + 1)\right]^2}$ is
\begin{subequations}
\begin{align}
    \mathcal{Y} &= \trans{4}\left[\sum_{l = 0}^L\left( \trans{1}\mathcal{Y}_{e,l}\cdot\trans{2:4}\mathcal{Y}_{e,l}\right)\otimes \boldsymbol{\delta}_{L + 1}^{l + 1}\right],\\
    \mathcal{Y}_{e, l} &= \begin{pmatrix}
        Y_{e, l}^{0}\left(\hat{\mathsf{\Omega}}\right) & \sqrt{2}Y_{e, l}^1\left(\hat{\mathsf{\Omega}}\right) & \cdots & \sqrt{2}Y_{e, l}^{l}\left(\hat{\mathsf{\Omega}}\right)
    \end{pmatrix}.\label{eq:even_sphm_tt}
\end{align}
\end{subequations}
Finally, the angular integration operator is
\begin{equation}
    \mathcal{W} = \textsc{trans}\left(\mathbf{1}_4 \otimes \left\{w_{i_\mu}^{\mu}\right\}_{i_\mu = 1}^{N_\mu} \otimes \left\{w_{i_\gamma}^{\gamma}\right\}_{i_\gamma = 1}^{N_\gamma}\right).
\end{equation}

\subsubsection{Interior spatial integral operators}\label{sec:interior_operator_space}

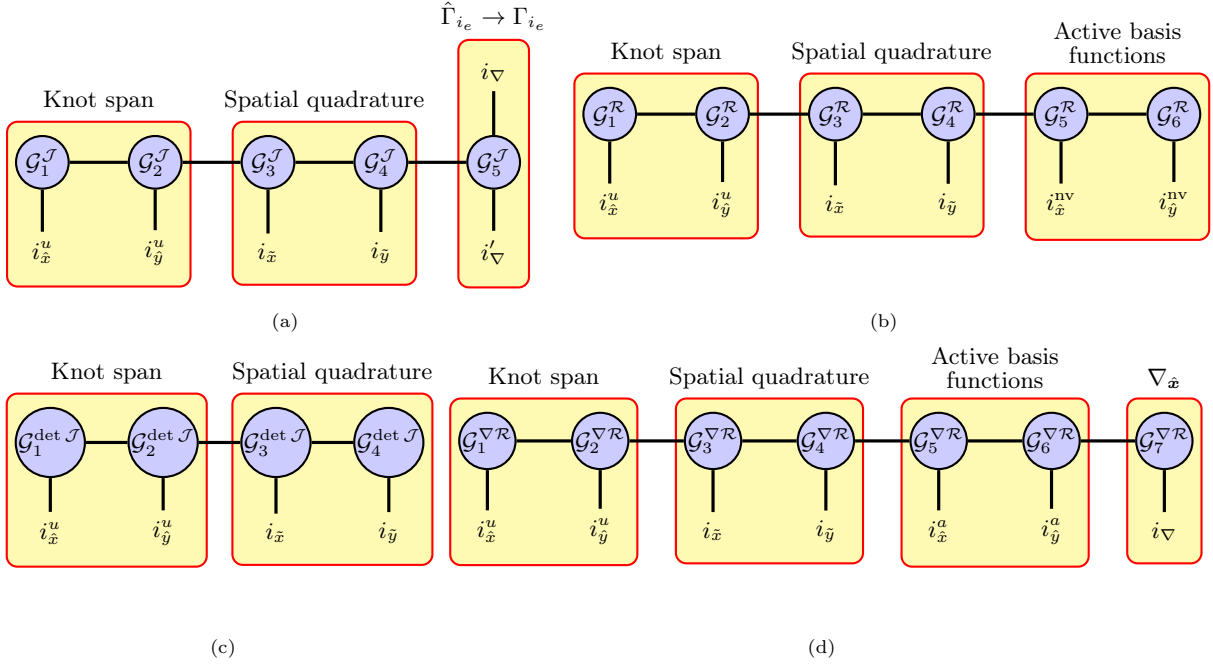
\begin{figure}
\centering
\begin{subfigure}[b]{0.45\textwidth}
\centering
\begin{tikzpicture}[
    thick,
    every node/.style={font=\small},
    tensor/.style={
        circle,
        draw=black,
        fill=blue!20,
        minimum size=7mm,
        inner sep=0pt
    },
    bigtensor/.style={
        circle,
        draw=black,
        fill=red!20,
        minimum size=10mm,
        inner sep=1pt
    },
    leg/.style={line width=1.2pt},
    freeindex/.style={},
    arrowstyle/.style={->, thick, line width=1pt},
    dotstyle/.style={font=\Large}
]

\node[tensor] (G1) at (0,0) {$\mathcal{G}_{1}^{\mathcal J}$};
\node[tensor] (G2) at (1.5,0) {$\mathcal{G}_{2}^{\mathcal J}$};
\node[tensor] (G3) at (3,0) {$\mathcal{G}_{3}^{\mathcal J}$};
\node[tensor] (G4) at (4.5,0) {$\mathcal{G}_{4}^{\mathcal J}$};
\node[tensor] (G5) at (6,0) {$\mathcal{G}_{5}^{\mathcal J}$};

\node[freeindex] (N1) at (0, -1.2) {$i_{\hat x}^u$};
\node[freeindex] (N2) at (1.5, -1.2) {$i_{\hat y}^u$};
\node[freeindex] (N3) at (3, -1.2) {$i_{\tilde x}$};
\node[freeindex] (N4) at (4.5, -1.2) {$i_{\tilde y}$};
\node[freeindex] (N5) at (6, 1.2) {$i_\nabla$};
\node[freeindex] (N6) at (6, -1.2) {$i_\nabla'$};

\draw[leg] (G1) -- (G2);
\draw[leg] (G2) -- (G3);
\draw[leg] (G3) -- (G4);
\draw[leg] (G4) -- (G5);

\draw[leg] (G1) -- (N1);
\draw[leg] (G2) -- (N2);
\draw[leg] (G3) -- (N3);
\draw[leg] (G4) -- (N4);
\draw[leg] (G5) -- (N5);
\draw[leg] (G5) -- (N6);

\begin{pgfonlayer}{background}
    \node[
        fit=(G1)(G2)(N1),
        inner xsep=3pt,  
        inner ysep=5pt,   
        fill=yellow,
        fill opacity=0.3,  
        draw=red,
        thick,
        rounded corners,
        label=above:{Knot span}
    ] {};
\end{pgfonlayer}
\begin{pgfonlayer}{background}
    \node[
        fit=(G3)(G4)(N3),
        inner xsep=3pt,  
        inner ysep=5pt,   
        fill=yellow,
        fill opacity=0.3,  
        draw=red,
        thick,
        rounded corners,
        label=above:{Spatial quadrature}
    ] {};
\end{pgfonlayer}
\begin{pgfonlayer}{background}
    \node[
        fit=(G5)(N5)(N6),
        inner xsep=3pt,  
        inner ysep=5pt,   
        fill=yellow,
        fill opacity=0.3,  
        draw=red,
        thick,
        rounded corners,
        label=above:{$\hat{\Gamma}_{i_e} \rightarrow \Gamma_{i_e}$}
    ] {};
\end{pgfonlayer}

\end{tikzpicture}
\caption{}
\label{fig:J}
\end{subfigure}
\begin{subfigure}[b]{0.5\textwidth}
\centering
\begin{tikzpicture}[
    thick,
    every node/.style={font=\small},
    tensor/.style={
        circle,
        draw=black,
        fill=blue!20,
        minimum size=7mm,
        inner sep=0pt
    },
    bigtensor/.style={
        circle,
        draw=black,
        fill=red!20,
        minimum size=10mm,
        inner sep=1pt
    },
    leg/.style={line width=1.2pt},
    freeindex/.style={},
    arrowstyle/.style={->, thick, line width=1pt},
    dotstyle/.style={font=\Large}
]

\node[tensor] (G1) at (0,0) {$\mathcal{G}_{1}^{\mathcal R}$};
\node[tensor] (G2) at (1.5,0) {$\mathcal{G}_{2}^{\mathcal R}$};
\node[tensor] (G3) at (3,0) {$\mathcal{G}_{3}^{\mathcal R}$};
\node[tensor] (G4) at (4.5,0) {$\mathcal{G}_{4}^{\mathcal R}$};
\node[tensor] (G5) at (6,0) {$\mathcal{G}_{5}^{\mathcal R}$};
\node[tensor] (G6) at (7.5,0) {$\mathcal{G}_{6}^{\mathcal R}$};

\node[freeindex] (N1) at (0, -1.2) {$i_{\hat x}^u$};
\node[freeindex] (N2) at (1.5, -1.2) {$i_{\hat y}^u$};
\node[freeindex] (N3) at (3, -1.2) {$i_{\tilde x}$};
\node[freeindex] (N4) at (4.5, -1.2) {$i_{\tilde y}$};
\node[freeindex] (N5) at (6, -1.2) {$i_{\hat x}^{\text{nv}}$};
\node[freeindex] (N6) at (7.5, -1.2) {$i_{\hat y}^{\text{nv}}$};

\draw[leg] (G1) -- (G2);
\draw[leg] (G2) -- (G3);
\draw[leg] (G3) -- (G4);
\draw[leg] (G4) -- (G5);
\draw[leg] (G5) -- (G6);

\draw[leg] (G1) -- (N1);
\draw[leg] (G2) -- (N2);
\draw[leg] (G3) -- (N3);
\draw[leg] (G4) -- (N4);
\draw[leg] (G5) -- (N5);
\draw[leg] (G6) -- (N6);

\begin{pgfonlayer}{background}
    \node[
        fit=(G1)(G2)(N1),
        inner xsep=3pt,  
        inner ysep=5pt,   
        fill=yellow,
        fill opacity=0.3,  
        draw=red,
        thick,
        rounded corners,
        label=above:{Knot span}
    ] {};
\end{pgfonlayer}
\begin{pgfonlayer}{background}
    \node[
        fit=(G3)(G4)(N3),
        inner xsep=3pt,  
        inner ysep=5pt,   
        fill=yellow,
        fill opacity=0.3,  
        draw=red,
        thick,
        rounded corners,
        label=above:{Spatial quadrature}
    ] {};
\end{pgfonlayer}
\begin{pgfonlayer}{background}
    \node[
        fit=(G5)(G6)(N6),
        inner xsep=3pt,  
        inner ysep=5pt,   
        fill=yellow,
        fill opacity=0.3,  
        draw=red,
        thick,
        rounded corners,
        label=above:{\shortstack{Active basis\\functions}},
    ] {};
\end{pgfonlayer}

\end{tikzpicture}
\caption{}
\label{fig:R}
\end{subfigure}
\begin{subfigure}[b]{0.35\textwidth}
\centering
\begin{tikzpicture}[
    thick,
    every node/.style={font=\small},
    tensor/.style={
        circle,
        draw=black,
        fill=blue!20,
        minimum size=7mm,
        inner sep=0pt
    },
    bigtensor/.style={
        circle,
        draw=black,
        fill=red!20,
        minimum size=10mm,
        inner sep=1pt
    },
    leg/.style={line width=1.2pt},
    freeindex/.style={},
    arrowstyle/.style={->, thick, line width=1pt},
    dotstyle/.style={font=\Large}
]

\node[tensor] (G1) at (0,0) {$\mathcal{G}_{1}^{\det \mathcal J}$};
\node[tensor] (G2) at (1.5,0) {$\mathcal{G}_{2}^{\det \mathcal J}$};
\node[tensor] (G3) at (3,0) {$\mathcal{G}_{3}^{\det \mathcal J}$};
\node[tensor] (G4) at (4.5,0) {$\mathcal{G}_{4}^{\det \mathcal J}$};

\node[freeindex] (N1) at (0, -1.2) {$i_{\hat x}^u$};
\node[freeindex] (N2) at (1.5, -1.2) {$i_{\hat y}^u$};
\node[freeindex] (N3) at (3, -1.2) {$i_{\tilde x}$};
\node[freeindex] (N4) at (4.5, -1.2) {$i_{\tilde y}$};

\draw[leg] (G1) -- (G2);
\draw[leg] (G2) -- (G3);
\draw[leg] (G3) -- (G4);

\draw[leg] (G1) -- (N1);
\draw[leg] (G2) -- (N2);
\draw[leg] (G3) -- (N3);
\draw[leg] (G4) -- (N4);

\begin{pgfonlayer}{background}
    \node[
        fit=(G1)(G2)(N1),
        inner xsep=3pt,  
        inner ysep=5pt,   
        fill=yellow,
        fill opacity=0.3,  
        draw=red,
        thick,
        rounded corners,
        label=above:{Knot span}
    ] {};
\end{pgfonlayer}
\begin{pgfonlayer}{background}
    \node[
        fit=(G3)(G4)(N3),
        inner xsep=3pt,  
        inner ysep=5pt,   
        fill=yellow,
        fill opacity=0.3,  
        draw=red,
        thick,
        rounded corners,
        label=above:{Spatial quadrature}
    ] {};
\end{pgfonlayer}

\end{tikzpicture}
\caption{}
\label{fig:det_J}
\end{subfigure}
\begin{subfigure}[b]{0.6\textwidth}
\centering
\begin{tikzpicture}[
    thick,
    every node/.style={font=\small},
    tensor/.style={
        circle,
        draw=black,
        fill=blue!20,
        minimum size=7mm,
        inner sep=0pt
    },
    bigtensor/.style={
        circle,
        draw=black,
        fill=red!20,
        minimum size=10mm,
        inner sep=1pt
    },
    leg/.style={line width=1.2pt},
    freeindex/.style={},
    arrowstyle/.style={->, thick, line width=1pt},
    dotstyle/.style={font=\Large}
]

\node[tensor] (G1) at (0,0) {$\mathcal{G}_{1}^{\nabla \mathcal R}$};
\node[tensor] (G2) at (1.5,0) {$\mathcal{G}_{2}^{\nabla\mathcal R}$};
\node[tensor] (G3) at (3,0) {$\mathcal{G}_{3}^{\nabla\mathcal R}$};
\node[tensor] (G4) at (4.5,0) {$\mathcal{G}_{4}^{\nabla\mathcal R}$};
\node[tensor] (G5) at (6,0) {$\mathcal{G}_{5}^{\nabla\mathcal R}$};
\node[tensor] (G6) at (7.5,0) {$\mathcal{G}_{6}^{\nabla\mathcal R}$};
\node[tensor] (G7) at (9,0) {$\mathcal{G}_{7}^{\nabla\mathcal R}$};

\node[freeindex] (N1) at (0, -1.2) {$i_{\hat x}^u$};
\node[freeindex] (N2) at (1.5, -1.2) {$i_{\hat y}^u$};
\node[freeindex] (N3) at (3, -1.2) {$i_{\tilde x}$};
\node[freeindex] (N4) at (4.5, -1.2) {$i_{\tilde y}$};
\node[freeindex] (N5) at (6, -1.2) {$i_{\hat x}^{a}$};
\node[freeindex] (N6) at (7.5, -1.2) {$i_{\hat y}^{a}$};
\node[freeindex] (N7) at (9, -1.2) {$i_\nabla$};

\draw[leg] (G1) -- (G2);
\draw[leg] (G2) -- (G3);
\draw[leg] (G3) -- (G4);
\draw[leg] (G4) -- (G5);
\draw[leg] (G5) -- (G6);
\draw[leg] (G6) -- (G7);

\draw[leg] (G1) -- (N1);
\draw[leg] (G2) -- (N2);
\draw[leg] (G3) -- (N3);
\draw[leg] (G4) -- (N4);
\draw[leg] (G5) -- (N5);
\draw[leg] (G6) -- (N6);
\draw[leg] (G7) -- (N7);

\begin{pgfonlayer}{background}
    \node[
        fit=(G1)(G2)(N1),
        inner xsep=3pt,  
        inner ysep=5pt,   
        fill=yellow,
        fill opacity=0.3,  
        draw=red,
        thick,
        rounded corners,
        label=above:{Knot span}
    ] {};
\end{pgfonlayer}
\begin{pgfonlayer}{background}
    \node[
        fit=(G3)(G4)(N3),
        inner xsep=3pt,  
        inner ysep=5pt,   
        fill=yellow,
        fill opacity=0.3,  
        draw=red,
        thick,
        rounded corners,
        label=above:{Spatial quadrature}
    ] {};
\end{pgfonlayer}
\begin{pgfonlayer}{background}
    \node[
        fit=(G5)(G6)(N6),
        inner xsep=3pt,  
        inner ysep=5pt,   
        fill=yellow,
        fill opacity=0.3,  
        draw=red,
        thick,
        rounded corners,
        label=above:{\shortstack{Active basis\\functions}},
    ] {};
\end{pgfonlayer}
\begin{pgfonlayer}{background}
    \node[
        fit=(G7)(N7),
        inner xsep=3pt,  
        inner ysep=5pt,   
        fill=yellow,
        fill opacity=0.3,  
        draw=red,
        thick,
        rounded corners,
        label=above:{$\nabla_{\hat \x}$},
    ] {};
\end{pgfonlayer}

\end{tikzpicture}
\caption{}
\label{fig:nabla_R}
\end{subfigure}

\caption{Diagram for tensor trains (a) $\mathcal{J}\in\mathbb{R}^{N_{\hat x}^u \times N_{\hat y}^u\times (p_{\hat x} + 1) \times (p_{\hat y} + 1) \times 2 \times 2}$, (b) $\mathcal{R}\in\mathbb{R}^{N_{\hat x}^u \times N_{\hat y}^u\times (p_{\hat x} + 1) \times (p_{\hat y} + 1)\times (p_{\hat x} + 1) \times (p_{\hat y} + 1)}$, (c) $\det\mathcal{J}\in\mathbb{R}^{N_{\hat x}^u \times N_{\hat y}^u\times (p_{\hat x} + 1) \times (p_{\hat y} + 1)}$, and (d) $\nabla\mathcal{R}\in\mathbb{R}^{N_{\hat x}^u \times N_{\hat y}^u\times (p_{\hat x} + 1) \times (p_{\hat y} + 1)\times (p_{\hat x} + 1) \times (p_{\hat y} + 1) \times 2}$. We label the dimensions responsible for the knot span, the spatial quadrature, the active basis function, and the Jacobian matrix or gradient vector index, $i_\nabla\in\{1, 2\}$. The active basis functions are those that evaluate to non-zero values and define the support of that basis function at the quadrature within the knot span.}
\label{fig:evaluated_tts}
\end{figure}

\par To build the spatial operators used in the interior operators of \cref{eq:operator}, we must first compute the integrals at each knot span. We require four evaluated tensors: the inverse of the Jacobian $\left[\mathbf{J}_{\hat \x}^{i_e}(\hat x, \hat y)\right]^{-1}$, the Jacobian determinant $\det\mathbf{J}_{\hat \x}^{i_e}(\hat x, \hat y)\det\mathbf{J}_{\tilde \x}^{i_{\hat x}^u, i_{\hat y}^u}(\tilde x, \tilde y)$, the basis functions $R_{i_{\hat x}, i_{\hat y}}^{p_{\hat x}, p_{\hat y}}(\hat x, \hat y)$, and the gradient of the basis functions $\nabla_{\hat\x}R_{i_{\hat x}, i_{\hat y}}^{p_{\hat x}, p_{\hat y}}(\hat x, \hat y)$ at all quadrature points within each knot span. Each tensor is then decomposed into the TT format such that,
\begin{subequations}
\begin{align}
    \mathcal{J}_{\text{inv}}^{i_{\hat x}^u,i_{\hat y}^u}(i_{\tilde x}, i_{\tilde y}, :, :) &= \left[\mathbf{J}_{\hat \x}^{i_e}(\hat x(\tilde x_{i_{\tilde x}}), \hat y(\tilde y_{i_{\tilde y}}))\right]^{-1},\\
    \det\mathcal{J}^{i_{\hat x}^u,i_{\hat y}^u}(i_{\tilde x}, i_{\tilde y}) &= w_{i_{\tilde x}}^{\tilde x}w_{i_{\tilde y}}^{\tilde y}\left|\det\mathbf{J}_{\hat \x}^{i_e}(\hat x(\tilde x_{i_{\tilde x}}), \hat y(\tilde y_{i_{\tilde y}}))\right|\left|\det\mathbf{J}_{\tilde \x}^{i_{\hat x}^u,i_{\hat y}^u}(\tilde x_{i_{\tilde x}}, \tilde y_{i_{\tilde y}})\right|,\\
    \mathcal{R}^{i_{\hat x}^u,i_{\hat y}^u}(i_{\tilde x}, i_{\tilde y}, :, :) &= \left\{R_{i_{\hat x}, i_{\hat y}}^{p_{\hat x}, p_{\hat y}}(\hat x(\tilde x_{i_{\tilde x}}), \hat y(\tilde y_{i_{\tilde y}}))\right\}_{i_{\hat x} = i_{\hat x}^u, i_{\hat y} = i_{\hat y}^u}^{i_{\hat x}^u + p_{\hat x}, i_{\hat y}^u + p_{\hat y}},\\
    \nabla\mathcal{R}^{i_{\hat x}^u,i_{\hat y}^u}(i_{\tilde x}, i_{\tilde y}, :, :, :) &=  \left\{\nabla_{\hat\x}R_{i_{\hat x}, i_{\hat y}}^{p_{\hat x}, p_{\hat y}}(\hat x(\tilde x_{i_{\tilde x}}), \hat y(\tilde y_{i_{\tilde y}}))\right\}_{i_{\hat x} = i_{\hat x}^u, i_{\hat y} = i_{\hat y}^u}^{i_{\hat x}^u + p_{\hat x}, i_{\hat y}^u + p_{\hat y}},
\end{align}
\end{subequations}
where $\left[\mathbf{J}_{\hat \x}^{i_e}(\hat x, \hat y)\right]^{-1}$ is the inverse of the Jacobian at $(\hat x, \hat y)$ which for the 2-D LBTE is the inverse of a $2\times 2$ matrix. Both $\mathcal{R}^{i_{\hat x}^u,i_{\hat y}^u}$ and $\nabla\mathcal{R}^{i_{\hat x}^u,i_{\hat y}^u}$ only include the active basis functions with indices $i_{\hat x}^{a}\in\{1, 2, \dots, p_{\hat x} + 1\}$ and $i_{\hat y}^{a}\in\{1, 2, \dots, p_{\hat y} + 1\}$. The resulting TTs, $\mathcal{J}_{\text{inv}}^{i_{\hat x}^u,i_{\hat y}^u}\in\mathbb{R}^{(p_{\hat x} + 1)\times (p_{\hat y} +1)\times 2\times 2}$, $\det\mathcal{J}^{i_{\hat x}^u,i_{\hat y}^u}\in\mathbb{R}^{(p_{\hat x} + 1)\times (p_{\hat y} +1)}$, $\mathcal{R}^{i_{\hat x}^u,i_{\hat y}^u}\in\mathbb{R}^{(p_{\hat x} + 1)\times (p_{\hat y} +1)\times (p_{\hat x} + 1)\times (p_{\hat y} +1)}$, and $\nabla\mathcal{R}^{i_{\hat x}^u,i_{\hat y}^u}\in\mathbb{R}^{(p_{\hat x} + 1)\times (p_{\hat y} +1)\times (p_{\hat x} + 1)\times (p_{\hat y} +1)\times 2}$, are then concatenated with TTs from other knot spans,
\begin{equation}
    \mathcal{X} = \sum_{i_{\hat x}^u = 1}^{N_{\hat x}^u}\sum_{i_{\hat y} = 1}^{N_{\hat y}^u}\boldsymbol{\delta}_{N_{\hat x}^u}^{i_{\hat x}^u}\otimes \boldsymbol{\delta}_{N_{\hat y}^u}^{i_{\hat y}^u}\otimes \mathcal{X}^{i_{\hat x}^u, i_{\hat y}^u},
\end{equation}
where $\mathcal{X}^{i_{\hat x}^u, i_{\hat y}^u}$ is $\mathcal{J}_{\text{inv}}^{i_{\hat x}^u, i_{\hat y}^u}$, $\det\mathcal{J}^{i_{\hat x}^u, i_{\hat y}^u}$, $\mathcal{R}^{i_{\hat x}^u, i_{\hat y}^u}$, or $\nabla\mathcal{R}^{i_{\hat x}^u, i_{\hat y}^u}$ resulting in $\mathcal{J}_{\text{inv}}$, $\mathcal{R}$, $\det\mathcal{J}$, or $\nabla\mathcal{R}$ with TT representations shown in \cref{fig:evaluated_tts}. As an alternative to the concatenation algorithm from \cite{math12203277}, we can use TT-cross \cite{OSELEDETS201070} or an incremental TT algorithm \cite{doi:10.1137/22M1537734} to build these TTs.

\begin{figure}
\centering
\begin{tikzpicture}[
    thick,
    every node/.style={font=\small},
    tensor/.style={
        circle,
        draw=black,
        fill=blue!20,
        minimum size=7mm,
        inner sep=0pt
    },
    bigtensor/.style={
        circle,
        draw=black,
        fill=red!20,
        minimum size=10mm,
        inner sep=1pt
    },
    freeindex/.style={},
    leg/.style={line width=1.2pt},
    arrowstyle/.style={->, thick, line width=1pt},
    dotstyle/.style={font=\Large}
]

\node[tensor] (G11) at (0,0) {$\mathcal{G}_{1}$};
\node[tensor] (G21) at (1.25,0) {$\mathcal{G}_{2}$};
\node[tensor] (G31) at (2.5,0) {$\mathcal{G}_{3}$};
\node[tensor] (G41) at (3.75,0) {$\mathcal{G}_{4}$};
\node[tensor] (G51) at (5,0) {$\mathcal{G}_{5}$};
\node[tensor] (G61) at (6.25,0) {$\mathcal{G}_{6}$};
\node[tensor] (G71) at (7.5,0) {$\mathcal{G}_{7}$};

\node[tensor] (G12) at (0,-1.5) {$\mathcal{G}_{1}$};
\node[tensor] (G22) at (1.25,-1.5) {$\mathcal{G}_{2}$};
\node[tensor] (G32) at (2.5,-1.5) {$\mathcal{G}_{3}$};
\node[tensor] (G42) at (3.75,-1.5) {$\mathcal{G}_{4}$};
\node[tensor] (G52) at (5,-1.5) {$\mathcal{G}_{5}$};
\node[tensor] (G62) at (6.25,-1.5) {$\mathcal{G}_{6}$};
\node[tensor] (G72) at (7.5,-1.5) {$\mathcal{G}_{7}$};

\draw[leg] (G11) -- (G21);
\draw[leg] (G21) -- (G31);
\draw[leg] (G31) -- (G41);
\draw[leg] (G41) -- (G51);
\draw[leg] (G51) -- (G61);
\draw[leg] (G61) -- (G71);

\draw[leg] (G12) -- (G22);
\draw[leg] (G22) -- (G32);
\draw[leg] (G32) -- (G42);
\draw[leg] (G42) -- (G52);
\draw[leg] (G52) -- (G62);
\draw[leg] (G62) -- (G72);

\draw[leg] (G11) -- (G12) node[midway,left] {$N_{\hat x}^u$};
\draw[leg] (G21) -- (G22) node[midway,left] {$N_{\hat y}^u$};
\draw[leg] (G31) -- (G32) node[midway,left] {$p_{\hat x} + 1$};
\draw[leg] (G41) -- (G42) node[midway,left] {$p_{\hat y} + 1$};
\draw[leg] (G51) -- (G52) node[midway,left] {$1$};
\draw[leg] (G61) -- (G62) node[midway,left] {$1$};
\draw[leg] (G71) -- (G72) node[midway,left] {$2$};

\node[freeindex] (N1) at (0, 1.2) {$i_{\hat x}^u$};
\node[freeindex] (N2) at (1.25, 1.2) {$i_{\hat y}^u$};
\node[freeindex] (N3) at (5, 1.2) {$i_{\hat x}^{a}$};
\node[freeindex] (N4) at (6.25, 1.2) {$i_{\hat y}^{a}$};
\node[freeindex] (N5) at (5, -2.7) {$\left(i_{\hat x}^{a}\right)'$};
\node[freeindex] (N6) at (6.25, -2.7) {$\left(i_{\hat y}^{a}\right)'$};
\node[freeindex] (N7) at (7.5, 1.2) {$i_{\nabla}$};

\draw[leg] (G11) -- (N1);
\draw[leg] (G21) -- (N2);
\draw[leg] (G51) -- (N3);
\draw[leg] (G61) -- (N4);
\draw[leg] (G52) -- (N5);
\draw[leg] (G62) -- (N6);
\draw[leg] (G71) -- (N7);

\draw[arrowstyle] (7.9,-0.75) -- (8.5,-0.75);

\node[tensor] (G13) at (9, -0.75) {$\mathcal{G}_1$};
\node[tensor] (G23) at (10.25, -0.75) {$\mathcal{G}_2$};
\node[tensor] (G33) at (11.5, -0.75) {$\mathcal{G}_3$};
\node[tensor] (G43) at (12.75, -0.75) {$\mathcal{G}_4$};
\node[tensor] (G53) at (14, -0.75) {$\mathcal{G}_5$};

\draw[leg] (G13) -- (G23);
\draw[leg] (G23) -- (G33);
\draw[leg] (G33) -- (G43);
\draw[leg] (G43) -- (G53);

\node[freeindex] (N13) at (9, 0.75) {$i_{\hat x}^u$};
\node[freeindex] (N23) at (10.25, 0.75) {$i_{\hat y}^u$};
\node[freeindex] (N33) at (11.5, 0.75) {$i_{\hat x}^{a}$};
\node[freeindex] (N43) at (12.75, 0.75) {$i_{\hat y}^{a}$};
\node[freeindex] (N53) at (14, 0.75) {$i_{\nabla}$};
\node[freeindex] (N63) at (11.5, -2.25) {$\left(i_{\hat x}^{a}\right)'$};
\node[freeindex] (N73) at (12.75, -2.25) {$\left(i_{\hat y}^{a}\right)'$};

\draw[leg] (G13) -- (N13);
\draw[leg] (G23) -- (N23);
\draw[leg] (G33) -- (N33);
\draw[leg] (G43) -- (N43);
\draw[leg] (G53) -- (N53);
\draw[leg] (G33) -- (N63);
\draw[leg] (G43) -- (N73);

\begin{pgfonlayer}{background}
    \node[
        fit=(N1)(G11)(N4)(G71),
        inner xsep=3pt,  
        inner ysep=3pt,   
        fill=yellow,
        fill opacity=0.3,  
        draw=red,
        thick,
        rounded corners,
        label=above:{$\diag{1, 2, 7}\trans{3, 4}\mathcal{J\nabla R}$}
    ] {};
\end{pgfonlayer}

\begin{pgfonlayer}{background}
    \node[
        fit=(G12)(N6)(G72),
        inner xsep=3pt,  
        inner ysep=3pt,   
        fill=yellow,
        fill opacity=0.3,  
        draw=red,
        thick,
        rounded corners,
        label=below:{$\left(\mathcal{JR}\otimes \mathbf{1}_2\right)$}
    ] {};
\end{pgfonlayer}

\begin{pgfonlayer}{background}
    \node[
        fit=(N13)(N73)(G13)(G53),
        inner xsep=3pt,  
        inner ysep=3pt,   
        fill=yellow,
        fill opacity=0.3,  
        draw=red,
        thick,
        rounded corners,
        label=above:{$\mathcal{V}_{\nabla R}^{\text{span}}$}
    ] {};
\end{pgfonlayer}

\end{tikzpicture}
\caption{Diagram of the product in \cref{eq:intg_spandR}. The spatial quadrature is evaluated as an inner product, and the resulting $1\times 1$ cores are absorbed into the neighboring cores. The active basis-function dimensions are combined via outer products. \Cref{eq:intg_spanR} looks the same without the last core for the gradient.}\label{fig:VR}

\end{figure}
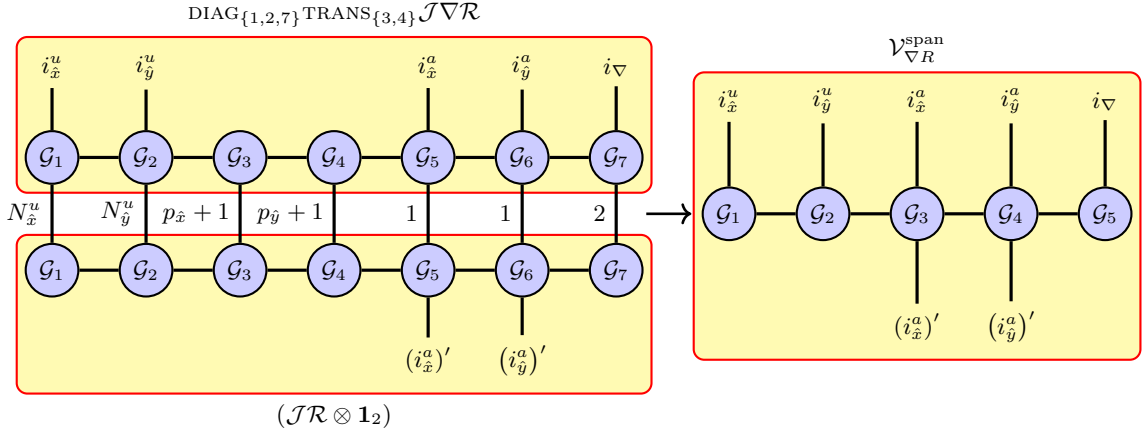
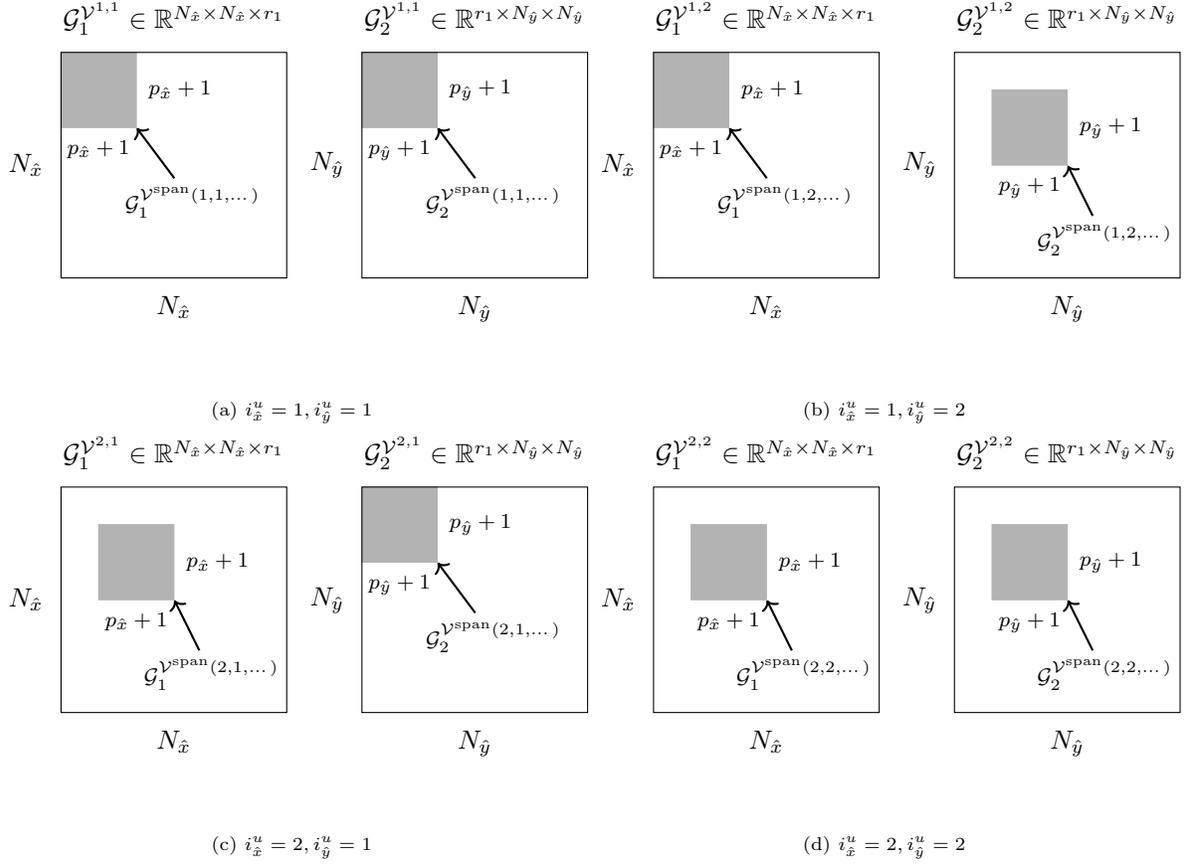
\begin{figure}
\centering
\begin{subfigure}[b]{0.47\textwidth}
\centering
\begin{tikzpicture}
    \node[draw, minimum width=3cm, minimum height=3cm] at (0, 0) (xcore) {};
    \node[above=3pt of xcore.north] {$\mathcal{G}_1^{\mathcal{V}^{1, 1}}\in\mathbb{R}^{N_{\hat x}\times N_{\hat x}\times r_1}$};
    \node[left=3pt of xcore.west] {$N_{\hat x}$};
    \node[below=3pt of xcore.south] {$N_{\hat x}$};

    \node[draw, minimum width=1cm, minimum height=1cm, fill=black, opacity=0.3] at (-1, 1) (xspan) {};
    \node[right=1pt of xspan.east] {\footnotesize$p_{\hat x} + 1$};
    \node[below=1pt of xspan.south] {\footnotesize$p_{\hat x} + 1$};

    \node[] at (0.25, -0.5) (xname) {\footnotesize$\mathcal{G}_1^{\mathcal{V}^{\text{span}}(1, 1,\dots)}$};
    \draw[->, thick] (xname) -- (xspan.south east);

    \node[draw, minimum width=3cm, minimum height=3cm] at (4, 0) (ycore) {};
    \node[above=3pt of ycore.north] {$\mathcal{G}_2^{\mathcal{V}^{1, 1}}\in\mathbb{R}^{r_1\times N_{\hat y}\times N_{\hat y}}$};
    \node[left=3pt of ycore.west] {$N_{\hat y}$};
    \node[below=3pt of ycore.south] {$N_{\hat y}$};

    \node[draw, minimum width=1cm, minimum height=1cm, fill=black, opacity=0.3] at (3, 1) (yspan) {};
    \node[right=1pt of yspan.east] {\footnotesize$p_{\hat y} + 1$};
    \node[below=1pt of yspan.south] {\footnotesize$p_{\hat y} + 1$};

    \node[] at (4.25, -0.5) (yname) {\footnotesize$\mathcal{G}_2^{\mathcal{V}^{\text{span}}(1, 1,\dots)}$};
    \draw[->, thick] (yname) -- (yspan.south east);
\end{tikzpicture}
\caption{$i_{\hat x}^u = 1,i_{\hat y}^u = 1$}
\label{fig:11}
\end{subfigure}
\begin{subfigure}[b]{0.47\textwidth}
\centering
\begin{tikzpicture}
    \node[draw, minimum width=3cm, minimum height=3cm] at (0, 0) (xcore) {};
    \node[above=3pt of xcore.north] {$\mathcal{G}_1^{\mathcal{V}^{1, 2}}\in\mathbb{R}^{N_{\hat x}\times N_{\hat x}\times r_1}$};
    \node[left=3pt of xcore.west] {$N_{\hat x}$};
    \node[below=3pt of xcore.south] {$N_{\hat x}$};

    \node[draw, minimum width=1cm, minimum height=1cm, fill=black, opacity=0.3] at (-1, 1) (xspan) {};
    \node[right=1pt of xspan.east] {\footnotesize$p_{\hat x} + 1$};
    \node[below=1pt of xspan.south] {\footnotesize$p_{\hat x} + 1$};

    \node[] at (0.25, -0.5) (xname) {\footnotesize$\mathcal{G}_1^{\mathcal{V}^{\text{span}}(1, 2,\dots)}$};
    \draw[->, thick] (xname) -- (xspan.south east);

    \node[draw, minimum width=3cm, minimum height=3cm] at (4, 0) (ycore) {};
    \node[above=3pt of ycore.north] {$\mathcal{G}_2^{\mathcal{V}^{1, 2}}\in\mathbb{R}^{r_1\times N_{\hat y}\times N_{\hat y}}$};
    \node[left=3pt of ycore.west] {$N_{\hat y}$};
    \node[below=3pt of ycore.south] {$N_{\hat y}$};

    \node[draw, minimum width=1cm, minimum height=1cm, fill=black, opacity=0.3] at (3.5, 0.5) (yspan) {};
    \node[right=1pt of yspan.east] {\footnotesize$p_{\hat y} + 1$};
    \node[below=1pt of yspan.south] {\footnotesize$p_{\hat y} + 1$};

    \node[] at (4.5, -1) (yname) {\footnotesize$\mathcal{G}_2^{\mathcal{V}^{\text{span}}(1, 2,\dots)}$};
    \draw[->, thick] (yname) -- (yspan.south east);
\end{tikzpicture}
\caption{$i_{\hat x}^u = 1,i_{\hat y}^u = 2$}
\label{fig:12}
\end{subfigure}
\begin{subfigure}[b]{0.47\textwidth}
\centering
\begin{tikzpicture}
    \node[draw, minimum width=3cm, minimum height=3cm] at (0, 0) (xcore) {};
    \node[above=3pt of xcore.north] {$\mathcal{G}_1^{\mathcal{V}^{2, 1}}\in\mathbb{R}^{N_{\hat x}\times N_{\hat x}\times r_1}$};
    \node[left=3pt of xcore.west] {$N_{\hat x}$};
    \node[below=3pt of xcore.south] {$N_{\hat x}$};

    \node[draw, minimum width=1cm, minimum height=1cm, fill=black, opacity=0.3] at (-0.5, 0.5) (xspan) {};
    \node[right=1pt of xspan.east] {\footnotesize$p_{\hat x} + 1$};
    \node[below=1pt of xspan.south] {\footnotesize$p_{\hat x} + 1$};

    \node[] at (0.5, -1) (xname) {\footnotesize$\mathcal{G}_1^{\mathcal{V}^{\text{span}}(2, 1,\dots)}$};
    \draw[->, thick] (xname) -- (xspan.south east);

    \node[draw, minimum width=3cm, minimum height=3cm] at (4, 0) (ycore) {};
    \node[above=3pt of ycore.north] {$\mathcal{G}_2^{\mathcal{V}^{2, 1}}\in\mathbb{R}^{r_1\times N_{\hat y}\times N_{\hat y}}$};
    \node[left=3pt of ycore.west] {$N_{\hat y}$};
    \node[below=3pt of ycore.south] {$N_{\hat y}$};

    \node[draw, minimum width=1cm, minimum height=1cm, fill=black, opacity=0.3] at (3, 1) (yspan) {};
    \node[right=1pt of yspan.east] {\footnotesize$p_{\hat y} + 1$};
    \node[below=1pt of yspan.south] {\footnotesize$p_{\hat y} + 1$};

    \node[] at (4.25, -0.5) (yname) {\footnotesize$\mathcal{G}_2^{\mathcal{V}^{\text{span}}(2, 1,\dots)}$};
    \draw[->, thick] (yname) -- (yspan.south east);
\end{tikzpicture}
\caption{$i_{\hat x}^u = 2,i_{\hat y}^u = 1$}
\label{fig:21}
\end{subfigure}
\begin{subfigure}[b]{0.47\textwidth}
\centering
\begin{tikzpicture}
    \node[draw, minimum width=3cm, minimum height=3cm] at (0, 0) (xcore) {};
    \node[above=3pt of xcore.north] {$\mathcal{G}_1^{\mathcal{V}^{2, 2}}\in\mathbb{R}^{N_{\hat x}\times N_{\hat x}\times r_1}$};
    \node[left=3pt of xcore.west] {$N_{\hat x}$};
    \node[below=3pt of xcore.south] {$N_{\hat x}$};

    \node[draw, minimum width=1cm, minimum height=1cm, fill=black, opacity=0.3] at (-0.5, 0.5) (xspan) {};
    \node[right=1pt of xspan.east] {\footnotesize$p_{\hat x} + 1$};
    \node[below=1pt of xspan.south] {\footnotesize$p_{\hat x} + 1$};

    \node[] at (0.5, -1) (xname) {\footnotesize$\mathcal{G}_1^{\mathcal{V}^{\text{span}}(2, 2,\dots)}$};
    \draw[->, thick] (xname) -- (xspan.south east);

    \node[draw, minimum width=3cm, minimum height=3cm] at (4, 0) (ycore) {};
    \node[above=3pt of ycore.north] {$\mathcal{G}_2^{\mathcal{V}^{2, 2}}\in\mathbb{R}^{r_1\times N_{\hat y}\times N_{\hat y}}$};
    \node[left=3pt of ycore.west] {$N_{\hat y}$};
    \node[below=3pt of ycore.south] {$N_{\hat y}$};

    \node[draw, minimum width=1cm, minimum height=1cm, fill=black, opacity=0.3] at (3.5, 0.5) (yspan) {};
    \node[right=1pt of yspan.east] {\footnotesize$p_{\hat y} + 1$};
    \node[below=1pt of yspan.south] {\footnotesize$p_{\hat y} + 1$};

    \node[] at (4.5, -1) (yname) {\footnotesize$\mathcal{G}_2^{\mathcal{V}^{\text{span}}(2, 2,\dots)}$};
    \draw[->, thick] (yname) -- (yspan.south east);
\end{tikzpicture}
\caption{$i_{\hat x}^u = 2,i_{\hat y}^u = 2$}
\label{fig:22}
\end{subfigure}

\caption{Example of the core assignment procedure in \cref{eq:interior_assignment} for $(i_{\hat x}^u, i_{\hat y}^u)\in\{1, 2\}\times \{1, 2\}$. Each active basis function core of $\mathcal{V}^{\text{span}}$ is placed on the diagonal, with the overlap between knot spans dictated by the polynomial degree $p_{\hat x}$ and $p_{\hat y}$. Note that $\mathcal{V}^{\text{span}}(i_{\hat x}^u, i_{\hat y}^u, \dots)$ becomes a two core tensor train (TT) and three core TT for $\mathcal{V}^{\text{span}}_R$ and $\mathcal{V}^{\text{span}}_{\nabla R}$, respectively, where the first core is the active basis functions along $\hat x$ and the second is the same for $\hat y$.}
\label{fig:interior_assignment}
\end{figure}

\par We define some intermediates,
\begin{subequations}
\begin{align}
    \mathcal{RJ} &= \trans{5, 6}\left[\mathcal{R} \odot \left(\det\mathcal{J}\otimes 1_{\{p_{\hat x} + 1, p_{\hat y} + 1\}}\right)\right],\\
    \mathcal{J_{\text{inv}}\nabla R} &= \diag{1, 2, 5, 6}\left[\perm{:4, 6, 7, 5}\left(\mathcal{J}_{\text{inv}}\otimes1_{\{p_{\hat x} + 1, p_{\hat y} + 1\}}\right)\right]\cdot \mathcal{\nabla R}.
\end{align}
\end{subequations}
The interior integrals for all knot spans are
\begin{subequations}\label{eq:intg_span}
\begin{align}
    \mathcal{V}_{R}^{\text{span}} &= \left(\diag{1, 2}\trans{3, 4}\mathcal{R}\right)\cdot\mathcal{RJ},\label{eq:intg_spanR}\\
    \mathcal{V}_{\nabla R}^{\text{span}} &= \left(\diag{1, 2, 7}\trans{3, 4}\mathcal{J\nabla R}\right)\cdot \left(\mathcal{RJ}\otimes \mathbf{1}_{2}\right),\label{eq:intg_spandR}
\end{align}
\end{subequations}
such that
\begin{subequations}
\begin{align}
    \mathcal{V}_{R}^{\text{span}}\left(i_{\hat x}^u, i_{\hat y}^u, :, :, :, :\right) &\approx \int_{\hat x_{i_{\hat x}^u}}^{\hat x_{i_{\hat x}^u + 1}}\int_{\hat y_{i_{\hat y}^u}}^{\hat y_{i_{\hat y}^u + 1}}\mathbf{R}\mathbf{R}^T\left|\det\left(\mathbf{J}_{\hat \x}^{i_e}\right)\right| d\hat V,\\
    \mathcal{V}_{\nabla R}^{\text{span}}\left(i_{\hat x}^u, i_{\hat y}^u, :, :, :, :, :\right) &\approx \int_{\hat x_{i_{\hat x}^u}}^{\hat x_{i_{\hat x}^u + 1}}\int_{\hat y_{i_{\hat y}^u}}^{\hat y_{i_{\hat y}^u + 1}}\left(\mathbf{J}_{\hat \x}^{i_e}\right)^{-1}\nabla_{\hat \x}\mathbf{R}\mathbf{R}^T\left|\det\left(\mathbf{J}_{\hat \x}^{i_e}\right)\right| d\hat V.
\end{align}
\end{subequations}
We depict the product for \cref{eq:intg_spandR} in \cref{fig:VR}. \Cref{eq:intg_spanR} is the same without the core for the gradient. The TTs in \cref{eq:intg_span}, are the integrals over individual knot spans. To get these into a form suitable for the interior operators in \cref{eq:operator}, we must add the contribution of each integral to the correct supported control variable location. We define $\mathcal{V}_R^{i_{\hat x}^u, i_{\hat y}^u}\in\{0\}^{N_{\hat x}\times N_{\hat x}\times N_{\hat y}\times N_{\hat y}}$ and $\mathcal{V}_{\nabla R}^{i_{\hat x}^u, i_{\hat y}^u}\in\{0\}^{N_{\hat x}\times N_{\hat x}\times N_{\hat y}\times N_{\hat y}\times 2}$ and set
\begin{subequations}\label{eq:interior_assignment}
\begin{align}
    \mathcal{V}_{R}^{i_{\hat x}^u, i_{\hat y}^u}\left(i_{\hat x}^u:\left(i_{\hat x}^u + p_{\hat x}\right), i_{\hat x}^u:\left(i_{\hat x}^u + p_{\hat x}\right), i_{\hat y}^u:\left(i_{\hat y}^u + p_{\hat y}\right), i_{\hat y}^u:\left(i_{\hat y}^u + p_{\hat y}\right)\right) &= \mathcal{V}_{R}^{\text{span}}\left(i_{\hat x}^u, i_{\hat y}^u, :, :, :, :\right),\\
    \mathcal{V}_{\nabla R}^{i_{\hat x}^u, i_{\hat y}^u}\left(i_{\hat x}^u:\left(i_{\hat x}^u + p_{\hat x}\right), i_{\hat x}^u:\left(i_{\hat x}^u + p_{\hat x}\right), i_{\hat y}^u:\left(i_{\hat y}^u + p_{\hat y}\right), i_{\hat y}^u:\left(i_{\hat y}^u + p_{\hat y}\right), :\right) &= \mathcal{V}_{\nabla R}^{\text{span}}\left(i_{\hat x}^u, i_{\hat y}^u, :, :, :, :, :\right).
\end{align}
\end{subequations}
\Cref{fig:interior_assignment} shows an example of this assignment procedure for $(i_{\hat x}^u, i_{\hat y}^u)\in\{1, 2\}\times \{1, 2\}$. We can then sum and round all elements into a global system for patch $i_e$,
\begin{equation}
    \mathcal{V}^{i_e} = \sum_{i_{\hat x}^u = 1}^{N_{\hat x}^u}\sum_{i_{\hat y}^u = 1}^{N_{\hat x}^u} \mathcal{V}^{i_{\hat x}^u, i_{\hat y}^u},
\end{equation}
for both $\mathcal{V}_{R}^{i_{\hat x}^u, i_{\hat y}^u}$ and $\mathcal{V}_{\nabla R}^{i_{\hat x}^u, i_{\hat y}^u}$ resulting in $\mathcal{V}_{R}^{i_e}\in\mathbb{R}^{N_{\hat x}\times N_{\hat x}\times N_{\hat y}\times N_{\hat y}}$ and $\mathcal{V}_{\nabla R}^{i_e}\in\mathbb{R}^{N_{\hat x}\times N_{\hat x}\times N_{\hat y}\times N_{\hat y}\times 2}$.

\subsubsection{Interior operators}\label{sec:interior_operators}

\par The streaming and collision, scattering, and fission operators for patch $i_e$ are defined as
\begin{subequations}
\begin{align}
    \mathcal{H}^{i_e} &= 1_{\{4, N_\mu, N_\gamma\}} \otimes \{\Sigma_{t, i_E}\}_{i_E = 1}^{N_E} \otimes \mathcal{V}_R^{i_e} - \hat{\TT{\Omega}}_x\otimes \mathbf{1}_{N_E}\otimes \mathcal{V}_{\nabla R}^{i_e}(:, :, :, :, 1) -\hat{\TT{\Omega}}_y\otimes \mathbf{1}_{N_E}\otimes \mathcal{V}_{\nabla R}^{i_e}(:, :, :, :, 2),\label{eq:Hu}\\
    \mathcal{S}^{i_e} &= \left[\left[1_{\{4, N_\mu, N_\gamma, N_E\}}\cdot \left(\mathcal{W}\otimes \mathbf{1}_{N_E}^T\right)\right]\odot\left[\left(\mathcal{Y}\otimes \mathbf{I}_{N_E}\right) \cdot \left(\mathcal{I}_{\{4, N_\mu, N_\gamma\}}\otimes\mathsf{\Sigma}_s\right)\right]\right]\otimes \mathcal{V}_R^{i_e},\label{eq:Su}\\
    \mathcal{F}^{i_e} &= \left(1_{\{4, N_\mu, N_\gamma\}}\cdot \mathcal{W}\right)\otimes \mathbf{\Sigma}_f\otimes \mathcal{V}_R^{i_e},\label{eq:Fu}
\end{align}
\end{subequations}
where $\mathsf{\Sigma}_s\in\mathbb{R}^{(L + 1)\times N_E\times N_E}$ is the scattering group-to-group cross section tensor decomposed into the TT format and $\mathbf{\Sigma}_f\in\mathbb{R}^{N_E\times N_E}$ is the fission cross section matrix where $\left(\mathbf{\Sigma}_f\right)_{i_E, i_E'} = \chi_{i_E}\nu\Sigma_{f, i_E'}$. The individual terms in \cref{eq:Hu} as well as \cref{eq:Fu} are rank one with respect to angle, energy, and space. The same can be said for the scattering operator if $L = 0$. If $L > 1$, then the scattering operator is only guaranteed to be rank one between energy and space as the cross section tensor and spherical harmonic operator must be approximately decomposed in the TT format according to $\epsilon$. We can then concatenate the operators over all patches $i_e\in\{1, ..., N_e\}$ to build the global operators,
\begin{subequations}
\begin{align}
    \mathcal{H}^{\text{TT}} &= \textsc{perm}_{\{2:5, 1, 6:\}}\textsc{diag}_{\{:5\}}\left(\sum_{i_e = 1}^{N_e}\boldsymbol{\delta}_{N_e}^{i_e}\otimes\mathcal{H}^{i_e}\right),\\
    \mathcal{S}^{\text{TT}} &= \textsc{perm}_{\{2:5, 1, 6:\}}\textsc{diag}_{\{1\}}\left(\sum_{i_e = 1}^{N_e}\boldsymbol{\delta}_{N_e}^{i_e}\otimes\mathcal{S}^{i_e}\right),\\
    \mathcal{F}^{\text{TT}} &= \textsc{perm}_{\{2:5, 1, 6:\}}\textsc{diag}_{\{1\}}\left(\sum_{i_e = 1}^{N_e}\boldsymbol{\delta}_{N_e}^{i_e}\otimes\mathcal{F}^{i_e}\right).
\end{align}
\end{subequations}
The final operators are TT formatted operators with shape $(4\times N_\mu\times N_\gamma \times N_E\times N_e\times N_{\hat x}\times N_{\hat y})^2$.

\subsubsection{Boundary operators}\label{sec:boundary_operators}

\begin{figure}
\centering
\begin{subfigure}[b]{0.47\textwidth}
\centering
\begin{tikzpicture}[
    thick,
    every node/.style={font=\small},
    tensor/.style={
        circle,
        draw=black,
        fill=blue!20,
        minimum size=7mm,
        inner sep=0pt
    },
    bigtensor/.style={
        circle,
        draw=black,
        fill=red!20,
        minimum size=10mm,
        inner sep=1pt
    },
    leg/.style={line width=1.2pt},
    freeindex/.style={},
    arrowstyle/.style={->, thick, line width=1pt},
    dotstyle/.style={font=\Large}
]

\node[tensor] (G1) at (0,0) {$\mathcal{G}_{1}^{\det \mathcal J_{\partial \hat \Gamma_{\hat x}}}$};
\node[tensor] (G2) at (2,0) {$\mathcal{G}_{2}^{\det \mathcal J_{\partial \hat \Gamma_{\hat x}}}$};
\node[tensor] (G3) at (4,0) {$\mathcal{G}_{3}^{\det \mathcal J_{\partial \hat \Gamma_{\hat x}}}$};

\node[freeindex] (N1) at (0, -1.5) {$i_{\hat y}^b$};
\node[freeindex] (N2) at (2, -1.5) {$i_{\hat x}^u$};
\node[freeindex] (N3) at (4, -1.5) {$i_{\tilde x}$};

\draw[leg] (G1) -- (G2);
\draw[leg] (G2) -- (G3);

\draw[leg] (G1) -- (N1);
\draw[leg] (G2) -- (N2);
\draw[leg] (G3) -- (N3);

\begin{pgfonlayer}{background}
    \node[
        fit=(G1)(N1),
        inner xsep=3pt,  
        inner ysep=5pt,   
        fill=yellow,
        fill opacity=0.3,  
        draw=red,
        thick,
        rounded corners,
        label=above:{Boundary}
    ] {};
\end{pgfonlayer}
\begin{pgfonlayer}{background}
    \node[
        fit=(G2)(N2),
        inner xsep=3pt,  
        inner ysep=5pt,   
        fill=yellow,
        fill opacity=0.3,  
        draw=red,
        thick,
        rounded corners,
        label=above:{Knot span}
    ] {};
\end{pgfonlayer}
\begin{pgfonlayer}{background}
    \node[
        fit=(G3)(N3),
        inner xsep=3pt,  
        inner ysep=5pt,   
        fill=yellow,
        fill opacity=0.3,  
        draw=red,
        thick,
        rounded corners,
        label=above:{\shortstack{Spatial\\quadrature}}
    ] {};
\end{pgfonlayer}

\end{tikzpicture}
\caption{}
\label{fig:boundary_detj}
\end{subfigure}
\begin{subfigure}[b]{0.47\textwidth}
\centering
\begin{tikzpicture}[
    thick,
    every node/.style={font=\small},
    tensor/.style={
        circle,
        draw=black,
        fill=blue!20,
        minimum size=7mm,
        inner sep=0pt
    },
    bigtensor/.style={
        circle,
        draw=black,
        fill=red!20,
        minimum size=10mm,
        inner sep=1pt
    },
    leg/.style={line width=1.2pt},
    freeindex/.style={},
    arrowstyle/.style={->, thick, line width=1pt},
    dotstyle/.style={font=\Large}
]

\node[tensor] (G1) at (0,0) {$\mathcal{G}_{1}^{\mathcal R_{\partial \hat \Gamma_{\hat x}}}$};
\node[tensor] (G2) at (2,0) {$\mathcal{G}_{2}^{\mathcal R_{\partial \hat \Gamma_{\hat x}}}$};
\node[tensor] (G3) at (4,0) {$\mathcal{G}_{3}^{\mathcal R_{\partial \hat \Gamma_{\hat x}}}$};
\node[tensor] (G4) at (6,0) {$\mathcal{G}_{4}^{\mathcal R_{\partial \hat \Gamma_{\hat x}}}$};

\node[freeindex] (N1) at (0, -1.5) {$i_{\hat y}^b$};
\node[freeindex] (N2) at (2, -1.5) {$i_{\hat x}^u$};
\node[freeindex] (N3) at (4, -1.5) {$i_{\tilde x}$};
\node[freeindex] (N4) at (6, -1.5) {$i_{\hat x}^{a}$};

\draw[leg] (G1) -- (G2);
\draw[leg] (G2) -- (G3);
\draw[leg] (G3) -- (G4);

\draw[leg] (G1) -- (N1);
\draw[leg] (G2) -- (N2);
\draw[leg] (G3) -- (N3);
\draw[leg] (G4) -- (N4);

\begin{pgfonlayer}{background}
    \node[
        fit=(G1)(N1),
        inner xsep=3pt,  
        inner ysep=5pt,   
        fill=yellow,
        fill opacity=0.3,  
        draw=red,
        thick,
        rounded corners,
        label=above:{Boundary}
    ] {};
\end{pgfonlayer}
\begin{pgfonlayer}{background}
    \node[
        fit=(G2)(N2),
        inner xsep=3pt,  
        inner ysep=5pt,   
        fill=yellow,
        fill opacity=0.3,  
        draw=red,
        thick,
        rounded corners,
        label=above:{Knot span}
    ] {};
\end{pgfonlayer}
\begin{pgfonlayer}{background}
    \node[
        fit=(G3)(N3),
        inner xsep=3pt,  
        inner ysep=5pt,   
        fill=yellow,
        fill opacity=0.3,  
        draw=red,
        thick,
        rounded corners,
        label=above:{\shortstack{Spatial\\quadrature}}
    ] {};
\end{pgfonlayer}
\begin{pgfonlayer}{background}
    \node[
        fit=(G4)(N4),
        inner xsep=3pt,  
        inner ysep=5pt,   
        fill=yellow,
        fill opacity=0.3,  
        draw=red,
        thick,
        rounded corners,
        label=above:{\shortstack{Active basis\\functions}}
    ] {};
\end{pgfonlayer}

\end{tikzpicture}
\caption{}
\label{fig:boundary_R}
\end{subfigure}
\begin{subfigure}[b]{\textwidth}
\centering
\begin{tikzpicture}[
    thick,
    every node/.style={font=\small},
    tensor/.style={
        circle,
        draw=black,
        fill=blue!20,
        minimum size=7mm,
        inner sep=0pt
    },
    bigtensor/.style={
        circle,
        draw=black,
        fill=red!20,
        minimum size=10mm,
        inner sep=1pt
    },
    leg/.style={line width=1.2pt},
    freeindex/.style={},
    arrowstyle/.style={->, thick, line width=1pt},
    dotstyle/.style={font=\Large}
]

\node[tensor] (G1) at (0,0) {$\mathcal{G}_{4}^{\mathcal D_{\partial \hat \Gamma_{\hat x}}}$};
\node[tensor] (G2) at (2,0) {$\mathcal{G}_{5}^{\mathcal D_{\partial \hat \Gamma_{\hat x}}}$};
\node[tensor] (G3) at (4,0) {$\mathcal{G}_{6}^{\mathcal D_{\partial \hat \Gamma_{\hat x}}}$};
\node[tensor] (G4) at (-6,0) {$\mathcal{G}_{1}^{\mathcal D_{\partial \hat \Gamma_{\hat x}}}$};
\node[tensor] (G5) at (-4,0) {$\mathcal{G}_{2}^{\mathcal D_{\partial \hat \Gamma_{\hat x}}}$};
\node[tensor] (G6) at (-2,0) {$\mathcal{G}_{3}^{\mathcal D_{\partial \hat \Gamma_{\hat x}}}$};

\node[freeindex] (N1) at (0, -1.5) {$i_{\hat y}^b$};
\node[freeindex] (N2) at (2, -1.5) {$i_{\hat x}^u$};
\node[freeindex] (N3) at (4, -1.5) {$i_{\tilde x}$};
\node[freeindex] (N4) at (-6, -1.5) {$i_{q}$};
\node[freeindex] (N5) at (-4, -1.5) {$i_{\mu}$};
\node[freeindex] (N6) at (-2, -1.5) {$i_{\gamma}$};

\draw[leg] (G4) -- (G5);
\draw[leg] (G5) -- (G6);
\draw[leg] (G6) -- (G1);
\draw[leg] (G1) -- (G2);
\draw[leg] (G2) -- (G3);

\draw[leg] (G1) -- (N1);
\draw[leg] (G2) -- (N2);
\draw[leg] (G3) -- (N3);
\draw[leg] (G4) -- (N4);
\draw[leg] (G5) -- (N5);
\draw[leg] (G6) -- (N6);

\begin{pgfonlayer}{background}
    \node[
        fit=(G4)(N4)(G6)(N6),
        inner xsep=3pt,  
        inner ysep=5pt,   
        fill=yellow,
        fill opacity=0.3,  
        draw=red,
        thick,
        rounded corners,
        label=above:{Angular quadrature}
    ] {};
\end{pgfonlayer}
\begin{pgfonlayer}{background}
    \node[
        fit=(G1)(N1),
        inner xsep=3pt,  
        inner ysep=5pt,   
        fill=yellow,
        fill opacity=0.3,  
        draw=red,
        thick,
        rounded corners,
        label=above:{Boundary}
    ] {};
\end{pgfonlayer}
\begin{pgfonlayer}{background}
    \node[
        fit=(G2)(N2),
        inner xsep=3pt,  
        inner ysep=5pt,   
        fill=yellow,
        fill opacity=0.3,  
        draw=red,
        thick,
        rounded corners,
        label=above:{Knot span}
    ] {};
\end{pgfonlayer}
\begin{pgfonlayer}{background}
    \node[
        fit=(G3)(N3),
        inner xsep=3pt,  
        inner ysep=5pt,   
        fill=yellow,
        fill opacity=0.3,  
        draw=red,
        thick,
        rounded corners,
        label=above:{\shortstack{Spatial\\quadrature}}
    ] {};
\end{pgfonlayer}

\end{tikzpicture}
\caption{}
\label{fig:boundary_D}
\end{subfigure}

\caption{Diagram for the tensor trains (TTs) (a) $\det\mathcal{J}_{\p\hat{\Gamma}_{\hat x}}\in\mathbb{R}^{2\times N_{\hat x}^u\times (p_{\hat x} + 1)}$, (b) $\mathcal{R}_{\partial \hat{\Gamma}_{\hat x}}\in \mathbb{R}^{2\times N_{\hat x}^u\times (p_{\hat x} + 1)\times (p_{\hat x} + 1)}$, and (c) $\mathcal{D}_{\partial \hat{\Gamma}_{\hat x}^+},\mathcal{D}_{\partial \hat{\Gamma}_{\hat x}^-}\in\mathbb{R}^{4\times N_\mu\times N_\gamma\times 2\times N_{\hat x}^u\times (p_{\hat x} + 1)}$ with labeled dimensions. The active basis functions are those that are non-zero at a given quadrature point within a knot span.}
\label{fig:boundary_tts}
\end{figure}
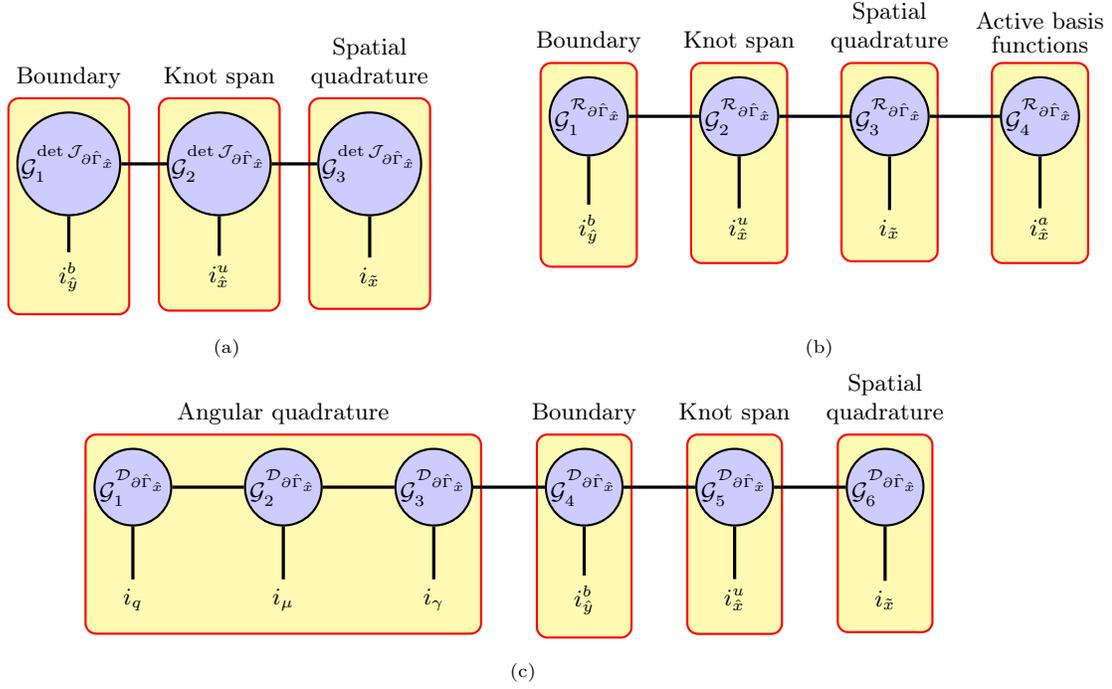

\par To build the inflow and outflow boundary operators, we need the line integral for each knot span along each of the four boundaries. We require TTs of the evaluated Jacobian, the basis functions $R_{i_{\hat x}, i_{\hat y}}^{p_{\hat x}, p_{\hat y}}(\hat x, \hat y)$, and $\bo\cdot \mathbf n$ at each quadrature point. For brevity, we only show the assembly of the boundary operator along parametric $\hat x$ for $\hat y^b_{i^b_{\hat y}} \in\{0, 1\}$. The TTs are
\begin{subequations}
\begin{align}
    \det \mathcal{J}_{\p \hat{\Gamma}_{\hat x}}^{i_{\hat x}^u, i^b_{\hat y}}(i_{\tilde x}) &= w_{i_{\tilde x}}^{\tilde x}\left|\x'\left(\hat x(\tilde x_{i_{\tilde x}}), \hat y_{i^b_{\hat y}}^b\right)\right|\left|\frac{d\hat x}{d\tilde x}\right|_{\tilde x_{i_{\tilde x}}},\\
    \mathcal{R}^{i_{\hat x}^u, i_{\hat y}(i^b_{\hat y})}_{\p \hat{\Gamma}_{\hat x}}(i_{\tilde x}, :) &= \left\{R_{i_{\hat x}, i^b_{\hat y}}^{p_{\hat x}, p_{\hat y}}(\hat x(\tilde x_{i_{\tilde x}}), \hat y_{i^b_{\hat y}}^b)\right\}_{i_{\hat x} = i_{\hat x}^u}^{i_{\hat x}^u + p_{\hat x}},\\
    \mathcal{D}_{\p \hat{\Gamma}_{\hat x}^+}^{i_{\hat x}^u, i^b_{\hat y}}\left(i_q, i_\mu, i_\gamma, i_{\tilde x}\right) &= \sqrt{\max\left[\hat{\mathsf{\Omega}}\left(i_q, i_\mu, i_\gamma\right) \cdot \mathbf{n}\left(\hat{x}\left(\tilde{x}_{i_{\tilde x}}\right), \hat y_{i^b_{\hat y}}^b\right), 0\right]},\\
    \mathcal{D}_{\p \hat{\Gamma}_{\hat x}^-}^{i_{\hat x}^u, i^b_{\hat y}}\left(i_q, i_\mu, i_\gamma, i_{\tilde x}\right) &= \sqrt{\mathbb{I}_{\text{adj}}^{i^b_{\hat y}}\max\left[-\hat{\mathsf{\Omega}}\left(i_q, i_\mu, i_\gamma\right) \cdot \mathbf{n}\left(\hat{x}\left(\tilde{x}_{i_{\tilde x}}\right), \hat y_{i^b_{\hat y}}^b\right), 0\right]},
\end{align}
\end{subequations}
given
\begin{subequations}
\begin{gather}
    \mathbb{I}_{\text{adj}}^{i^b_{\hat y}} = \begin{cases}
        1 & \text{if boundary }\mathbf{S}(\hat x, \hat{y}_{i^b_{\hat y}}^b)\text{ is an interface or a non-vacuum boundary condition}\\
        0 & \text{otherwise}
    \end{cases},\\
    i_{\hat y}(i^b_{\hat y}) = \begin{cases}
        1 & i^b_{\hat y} = 1\\
        N_{\hat x} & i^b_{\hat y} = 2
    \end{cases}.
\end{gather}
\end{subequations}
The basis functions are interpletory along $\hat y$ and only the $i_{\hat y} = 1$ and $i_{\hat y} = N_{\hat y}$ basis functions are supported for $\hat y = 0$ and $\hat y = 1$, respectively. The TTs are then $\det\mathcal{J}_{\partial \hat{\Gamma}_{\hat x}}^{i_{\hat x}^u, i^b_{\hat y}}\in\mathbb{R}^{(p_{\hat x} + 1)}$, $\mathcal{R}_{\partial \hat{\Gamma}_{\hat x}}^{i_{\hat x}^u, i^b_{\hat y}}\in\mathbb{R}^{(p_{\hat x} + 1)\times (p_{\hat x} + 1)}$, and $\mathcal{D}_{\partial \hat{\Gamma}_{\hat x}^+}^{i_{\hat x}^u, i^b_{\hat y}}, \mathcal{D}_{\partial \hat{\Gamma}_{\hat x}^-}^{i_{\hat x}^u, i^b_{\hat y}}\in\mathbb{R}^{4\times N_\mu\times N_\gamma\times (p_{\hat x} + 1)}$. Like the interior spatial integrals, we can concatenate all knot spans together and both the $\hat y = 0$ and $\hat y = 1$ boundaries,
\begin{subequations}
\begin{align}
    \det\mathcal{J}_{\partial \hat{\Gamma}_{\hat x}} &= \sum_{i_{\hat x}^u = 1}^{N_{\hat x}^u}\sum_{i^b_{\hat y} = 1}^2\boldsymbol{\delta}_{2}^{i^b_{\hat y}}\otimes \boldsymbol{\delta}_{N_{\hat x}^u}^{i_{\hat x}^u}\otimes\det\TT{J}_{\partial \hat{\Gamma}_{\hat x}}^{i_{\hat x}^u, i^b_{\hat y}},\\
    \mathcal{R}_{\partial \hat{\Gamma}_{\hat x}} &= \sum_{i_{\hat x}^u = 1}^{N_{\hat x}^u}\sum_{i^b_{\hat y} = 1}^{2}\boldsymbol{\delta}_{2}^{i^b_{\hat y}}\otimes \boldsymbol{\delta}_{N_{\hat x}^u}^{i_{\hat x}^u}\otimes\mathcal{R}_{\partial \hat{\Gamma}_{\hat x}}^{i_{\hat x}^u, i^b_{\hat y}},\\
    \mathcal{D}_{\partial \hat{\Gamma}_{\hat x}^+} &= \perm{:3, 5, 6, 4}\left(\sum_{i_{\hat x}^u = 1}^{N_{\hat x}^u}\sum_{i^b_{\hat y} = 1}^{2}\mathcal{D}_{\partial \hat{\Gamma}_{\hat x}^+}^{i_{\hat x}^u, i^b_{\hat y}}\otimes\boldsymbol{\delta}_{2}^{i^b_{\hat y}}\otimes \boldsymbol{\delta}_{N_{\hat x}^u}^{i_{\hat x}^u}\right),\\
    \mathcal{D}_{\partial \hat{\Gamma}_{\hat x}^-} &= \perm{:3, 5, 6, 4}\left(\sum_{i_{\hat x}^u = 1}^{N_{\hat x}^u}\sum_{i^b_{\hat y} = 1}^{2}\mathcal{D}_{\partial \hat{\Gamma}_{\hat x}^-}^{i_{\hat x}^u, i^b_{\hat y}}\otimes\boldsymbol{\delta}_{2}^{i^b_{\hat y}}\otimes \boldsymbol{\delta}_{N_{\hat x}^u}^{i_{\hat x}^u}\right),
\end{align}
\end{subequations}
which results in $\det\mathcal{J}_{\p\hat{\Gamma}_{\hat x}}\in\mathbb{R}^{2\times N_{\hat x}^u\times (p_{\hat x} + 1)}$, $\mathcal{R}_{\partial \hat{\Gamma}_{\hat x}}\in \mathbb{R}^{2\times N_{\hat x}^u\times (p_{\hat x} + 1)\times (p_{\hat x} + 1)}$, and $\mathcal{D}_{\partial \hat{\Gamma}_{\hat x}^+},\mathcal{D}_{\partial \hat{\Gamma}_{\hat x}^-}\in\\\mathbb{R}^{4\times N_\mu\times N_\gamma\times 2\times N_{\hat x}^u\times (p_{\hat x} + 1)}$, respectively. We show these TTs and label their dimensions in \cref{fig:boundary_tts}.

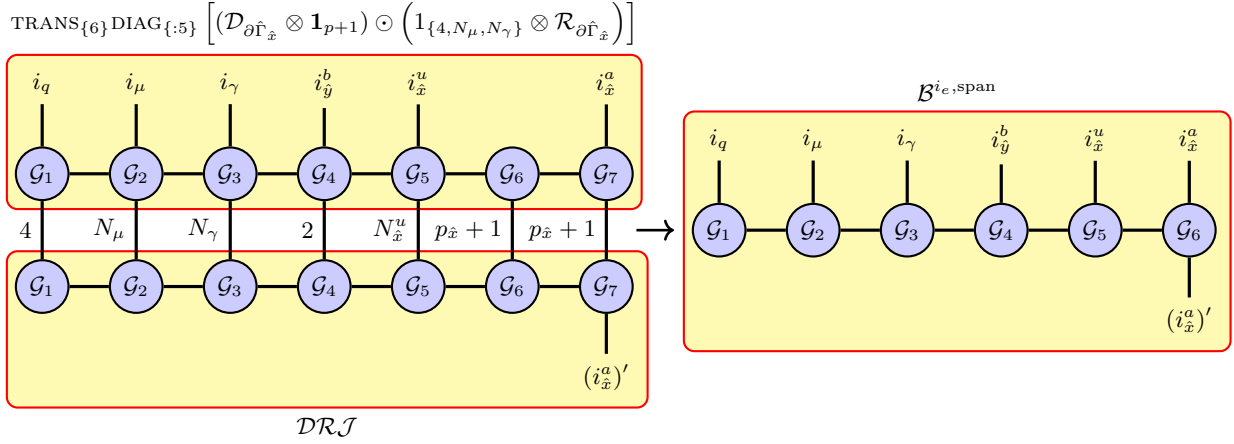
\begin{figure}
    \centering
    \begin{tikzpicture}[
    thick,
    every node/.style={font=\small},
    tensor/.style={
        circle,
        draw=black,
        fill=blue!20,
        minimum size=7mm,
        inner sep=0pt
    },
    bigtensor/.style={
        circle,
        draw=black,
        fill=red!20,
        minimum size=10mm,
        inner sep=1pt
    },
    freeindex/.style={},
    leg/.style={line width=1.2pt},
    arrowstyle/.style={->, thick, line width=1pt},
    dotstyle/.style={font=\Large}
]

\node[tensor] (G11) at (0,0) {$\mathcal{G}_{1}$};
\node[tensor] (G21) at (1.25,0) {$\mathcal{G}_{2}$};
\node[tensor] (G31) at (2.5,0) {$\mathcal{G}_{3}$};
\node[tensor] (G41) at (3.75,0) {$\mathcal{G}_{4}$};
\node[tensor] (G51) at (5,0) {$\mathcal{G}_{5}$};
\node[tensor] (G61) at (6.25,0) {$\mathcal{G}_{6}$};
\node[tensor] (G71) at (7.5,0) {$\mathcal{G}_{7}$};

\node[tensor] (G12) at (0,-1.5) {$\mathcal{G}_{1}$};
\node[tensor] (G22) at (1.25,-1.5) {$\mathcal{G}_{2}$};
\node[tensor] (G32) at (2.5,-1.5) {$\mathcal{G}_{3}$};
\node[tensor] (G42) at (3.75,-1.5) {$\mathcal{G}_{4}$};
\node[tensor] (G52) at (5,-1.5) {$\mathcal{G}_{5}$};
\node[tensor] (G62) at (6.25,-1.5) {$\mathcal{G}_{6}$};
\node[tensor] (G72) at (7.5,-1.5) {$\mathcal{G}_{7}$};

\draw[leg] (G11) -- (G21);
\draw[leg] (G21) -- (G31);
\draw[leg] (G31) -- (G41);
\draw[leg] (G41) -- (G51);
\draw[leg] (G51) -- (G61);
\draw[leg] (G61) -- (G71);

\draw[leg] (G12) -- (G22);
\draw[leg] (G22) -- (G32);
\draw[leg] (G32) -- (G42);
\draw[leg] (G42) -- (G52);
\draw[leg] (G52) -- (G62);
\draw[leg] (G62) -- (G72);

\draw[leg] (G11) -- (G12) node[midway,left] {$4$};
\draw[leg] (G21) -- (G22) node[midway,left] {$N_\mu$};
\draw[leg] (G31) -- (G32) node[midway,left] {$N_{\gamma}$};
\draw[leg] (G41) -- (G42) node[midway,left] {$2$};
\draw[leg] (G51) -- (G52) node[midway,left] {$N_{\hat x}^u$};
\draw[leg] (G61) -- (G62) node[midway,left] {$p_{\hat x} + 1$};
\draw[leg] (G71) -- (G72) node[midway,left] {$p_{\hat x} + 1$};

\node[freeindex] (N1) at (0, 1.2) {$i_{q}$};
\node[freeindex] (N2) at (1.25, 1.2) {$i_\mu$};
\node[freeindex] (N3) at (2.5, 1.2) {$i_\gamma$};
\node[freeindex] (N4) at (3.75, 1.2) {$i_{\hat y}^b$};
\node[freeindex] (N5) at (5, 1.2) {$i_{\hat x}^u$};
\node[freeindex] (N6) at (7.5, 1.2) {$i_{\hat x}^a$};
\node[freeindex] (N7) at (7.5, -2.7) {$\left(i_{\hat x}^a\right)'$};

\draw[leg] (G11) -- (N1);
\draw[leg] (G21) -- (N2);
\draw[leg] (G31) -- (N3);
\draw[leg] (G41) -- (N4);
\draw[leg] (G51) -- (N5);
\draw[leg] (G71) -- (N6);
\draw[leg] (G72) -- (N7);

\draw[arrowstyle] (7.9,-0.75) -- (8.4,-0.75);

\node[tensor] (G13) at (9,-0.75) {$\mathcal{G}_{1}$};
\node[tensor] (G23) at (10.25,-0.75) {$\mathcal{G}_{2}$};
\node[tensor] (G33) at (11.5,-0.75) {$\mathcal{G}_{3}$};
\node[tensor] (G43) at (12.75,-0.75) {$\mathcal{G}_{4}$};
\node[tensor] (G53) at (14,-0.75) {$\mathcal{G}_{5}$};
\node[tensor] (G63) at (15.25,-0.75) {$\mathcal{G}_{6}$};

\draw[leg] (G13) -- (G23);
\draw[leg] (G23) -- (G33);
\draw[leg] (G33) -- (G43);
\draw[leg] (G43) -- (G53);
\draw[leg] (G53) -- (G63);

\node[freeindex] (N13) at (9, 0.45) {$i_{q}$};
\node[freeindex] (N23) at (10.25, 0.45) {$i_\mu$};
\node[freeindex] (N33) at (11.5, 0.45) {$i_\gamma$};
\node[freeindex] (N43) at (12.75, 0.45) {$i_{\hat y}^b$};
\node[freeindex] (N53) at (14, 0.45) {$i_{\hat x}^u$};
\node[freeindex] (N63) at (15.25, 0.45) {$i_{\hat x}^a$};
\node[freeindex] (N73) at (15.25, -1.95) {$\left(i_{\hat x}^a\right)'$};

\draw[leg] (G13) -- (N13);
\draw[leg] (G23) -- (N23);
\draw[leg] (G33) -- (N33);
\draw[leg] (G43) -- (N43);
\draw[leg] (G53) -- (N53);
\draw[leg] (G63) -- (N63);
\draw[leg] (G63) -- (N73);

\begin{pgfonlayer}{background}
    \node[
        fit=(G11)(N1)(G71)(N6),
        inner xsep=3pt,  
        inner ysep=3pt,   
        fill=yellow,
        fill opacity=0.3,  
        draw=red,
        thick,
        rounded corners,
        label=above:{$ \trans{6}\diag{:5}\left[(\mathcal{D}_{\partial \hat{\Gamma}_{\hat x}}\otimes \mathbf{1}_{p + 1})\odot \left(1_{\{4, N_\mu, N_\gamma\}}\otimes \mathcal{R}_{\partial \hat{\Gamma}_{\hat x}}\right)\right]$}
    ] {};
\end{pgfonlayer}

\begin{pgfonlayer}{background}
    \node[
        fit=(G12)(G72)(N7),
        inner xsep=3pt,  
        inner ysep=3pt,   
        fill=yellow,
        fill opacity=0.3,  
        draw=red,
        thick,
        rounded corners,
        label=below:{$\mathcal{DRJ}$}
    ] {};
\end{pgfonlayer}

\begin{pgfonlayer}{background}
    \node[
        fit=(G13)(G63)(N13)(N73),
        inner xsep=3pt,  
        inner ysep=3pt,   
        fill=yellow,
        fill opacity=0.3,  
        draw=red,
        thick,
        rounded corners,
        label=above:{$\mathcal{B}^{i_e, \text{span}}$}
    ] {};
\end{pgfonlayer}

\end{tikzpicture}
    \caption{Diagram for the products in \cref{eq:b_spans}. The spatial quadrature vanishes in the inner product. An outer product combines the active basis functions.}
    \label{fig:b_spans}
\end{figure}

\par We define
\begin{subequations}
\begin{align}
    \mathcal{DRJ}^+ &= \trans{7}\left[\left(\mathcal{D}_{\partial \hat{\Gamma}_{\hat x}^+}\otimes \mathbf{1}_{p_{\hat x} + 1}\right)\odot\left[1_{\{4, N_\mu, N_\gamma\}}\otimes\left(\mathcal{R}_{\p\hat{\Gamma}_{\hat x}}\odot \left(\det\mathcal{J}_{\partial \hat{\Gamma}_{\hat x}}\otimes \mathbf{1}_{p_{\hat x} + 1}\right)\right)\right]\right],\\
    \mathcal{DRJ}^- &= \trans{7}\left[\left(\mathcal{D}_{\partial \hat{\Gamma}_{\hat x}^-}\otimes \mathbf{1}_{p_{\hat x} + 1}\right)\odot\left[1_{\{4, N_\mu, N_\gamma\}}\otimes\left(\mathcal{R}_{\p\hat{\Gamma}_{\hat x}}\odot \left(\det\mathcal{J}_{\partial \hat{\Gamma}_{\hat x}}\otimes \mathbf{1}_{p_{\hat x} + 1}\right)\right)\right]\right],
\end{align}
\end{subequations}
and compute the boundary integral for each knot span along both boundaries,
\begin{subequations}\label{eq:b_spans}
\begin{align}
    \mathcal{B}_{\text{out}}^{i_e, \text{span}} = \trans{6}\diag{:5}\left[(\mathcal{D}_{\partial \hat{\Gamma}_{\hat x}^+}\otimes \mathbf{1}_{p_{\hat x} + 1})\odot \left(1_{\{4, N_\mu, N_\gamma\}}\otimes \mathcal{R}_{\partial \hat{\Gamma}_{\hat x}}\right)\right]\cdot\mathcal{DRJ}^+,\\
    \mathcal{B}_{\text{in}}^{i_e, \text{span}} = \trans{6}\diag{:5}\left[(\mathcal{D}_{\partial \hat{\Gamma}_{\hat x}^-}\otimes \mathbf{1}_{p + 1})\odot \left(1_{\{4, N_\mu, N_\gamma\}}\otimes \mathcal{R}_{\partial \hat{\Gamma}_{\hat x}}\right)\right]\cdot\mathcal{DRJ}^-,
\end{align}
\end{subequations}
where $\mathcal{B}_{\text{out}}^{i_e, \text{span}},\mathcal{B}_{\text{in}}^{i_e, \text{span}}\in\mathbb{R}^{4\times N_\mu\times N_\gamma\times 2\times N_{\hat x}^u\times(p_{\hat x} + 1)\times(p_{\hat x} + 1)}$. We show \cref{eq:b_spans} in \cref{fig:b_spans}. These are local integrals over each span; we must sum their contributions over the supported control variables. We define $\mathcal{B}_{\text{out}}^{i_e, i_{\hat x}^u, i^b_{\hat y}},\mathcal{B}_{\text{in}}^{i_e, i_{\hat x}^u, i^b_{\hat y}}\in\{0\}^{4\times N_\mu\times N_\gamma\times N_{\hat x}\times N_{\hat x}}$ and set
\begin{equation}
    \mathcal{B}^{i_e, i_{\hat x}^u, i^b_{\hat y}}\left(:, :, :, i_{\hat x}^u:(i_{\hat x}^u + p_{\hat x}), i_{\hat x}^u:(i_{\hat x}^u + p_{\hat x})\right) = \mathcal{B}^{i_e, \text{span}}(:, :, :, i^b_{\hat y}, i_{\hat x}^u, :, :),
\end{equation}
for both inflow and outflow. The procedure for this assignment follows in the same fashion for the $\hat y$ core. We can then sum over the knot spans in the boundary,
\begin{equation}
    \mathcal{B}^{i_e, i^b_{\hat y}} = \sum_{i_{\hat x}^u = 1}^{N_{\hat x}^u} \mathcal{B}^{i_e, i_{\hat x}^u, i_{\hat y}^b}\otimes \textsc{diag}\left(\boldsymbol{\delta}_{N_{\hat y}}^{i_{\hat y}(i^b_{\hat y})}\right).
\end{equation}
We now have the inflow and outflow TTs within patch $i_e$ for each boundary $i^b_{\hat y}$, $\mathcal{B}_{\text{in}}^{i_e, i^b_{\hat y}}, \mathcal{B}_{\text{out}}^{i_e, i^b_{\hat y}}\in\mathbb{R}^{4\times N_\mu\times N_\gamma\times N_{\hat x}\times N_{\hat x}\times N_{\hat y}\times N_{\hat y}}$. The global outflow boundary integral is then
\begin{equation}
    \mathcal{B}_{\text{out}}^{\text{TT}} = \perm{3:5, 1, 2, 6:}\left[\mathbf{I}_{N_E}\otimes\diag{:4}\left(\sum_{i_e = 1}^{N_e}\boldsymbol{\delta}_{N_e}^{i_e}\otimes \left(\sum_{i_{\hat x}^b = 1}^2\mathcal{B}_{\text{out}}^{i_e, i_{\hat x}^b} + \sum_{i_{\hat y}^b = 1}^2\mathcal{B}_{\text{out}}^{i_e, i_{\hat y}^b}\right)\right)\right],
\end{equation}
where $\mathcal{B}_{\text{out}}^{i_e, i_{\hat x}^b}$ is the outgoing boundary TT for patch $i_e$ and boundary $i_{\hat x}^b$ along $\hat y$. 

\par The inflow boundary operator within a patch $i_e$ may be coupled to a patch $i_e^c$ across $i_{\hat y}^b$ or be a boundary condition. In the case that the boundary is a vacuum boundary condition, then $\mathcal{B}^{i_e, i^b_{\hat y}}_{\text{in}} = 0$. In the case that the boundary at $i_{\hat x}^b$ is reflective, then we map from the reflected quadrant,
\begin{equation}
    i_q^{\text{ref}}\left(i_q, i_{\hat y}^b\right) = \begin{cases}
        3 & i_q = 1; i_{\hat y}^b = 1\\
        4 & i_q = 2; i_{\hat y}^b = 1\\
        1 & i_q = 3; i_{\hat y}^b = 2\\
        2 & i_q = 4; i_{\hat y}^b = 2
    \end{cases}. 
\end{equation}
Note that this mapping is only valid for axis-aligned reflective boundaries. The reflected inflow boundary operator is 
\begin{equation}
    \mathcal{B}_{\text{in}}^{i_e, i_{\hat y}^b,  i_q^{\text{ref}}\rightarrow i_q}\left(i_q, i_q^{\text{ref}}\left(i_q, i_{\hat y}^b\right), :, :, :, :, :, :\right) = \mathcal{B}_{\text{in}}^{i_e, i_{\hat y}^b}\left(i_q, :, :, :, :, :, :\right),
\end{equation}
where $\mathcal{B}_{\text{in}}^{i_e, i_{\hat y}^b, i_q^{\text{ref}}\rightarrow i_q}\in\mathbb{R}^{4\times 4\times N_\mu\times N_\gamma\times N_{\hat x}\times N_{\hat x}\times N_{\hat y}\times N_{\hat y}}$. For patch interfaces, we ensure that, upon operator application, upwind angular fluxes are transferred from one patch to the other. The inflow boundary for patch $i_e$ coupled to $i_e^c$ across $i_{\hat y}^b$ is
\begin{equation}
    \mathcal{B}_{\text{in}}^{i_e^c \rightarrow i_e, i_{\hat y}^b}(:, :, :, :, :, :, :, :, i_{\hat y}) = \perm{3:5, :2, 6:}\left(\boldsymbol{\delta}^{i_e, i_e^c}_{N_e}\otimes \mathcal{B}_{\text{in}}^{i_e, i_{\hat y}^b}(:, :, :, :, :, :, N_{\hat y} - i_{\hat y})\right),
\end{equation}
where $\boldsymbol{\delta}^{i_e, i_e^c}_{N_e}\in\{0, 1\}^{N_e\times N_e}$ is a matrix of zeros with a one located at $(i_e, i_e^c)$. For both the reflective and interface conditions, we add an identity matrix for energy and permute the TT. For the reflective case, we also add an indicator vector for patch $i_e$, diagonalize that core, and permute. We then sum all inflow boundaries to obtain $\mathcal{B}_{\text{in}}^{\text{TT}}\in\mathbb{R}^{\left[4\times N_\mu\times N_\gamma \times N_E\times N_e\times N_{\hat x}\times N_{\hat y}\right]^2}$. There is no guarantee of a rank-one coupling, other than in energy, for these operators. Given the close coupling between angle and space in $\bo\cdot \mathbf{n}$, we expect higher ranks for highly-curvilinear boundaries. 

\subsection{Numerical Scheme}\label{sec:numerical_scheme}

\par We implement the operators in \cref{eq:operator} in both the TT and Compressed Sparse Row (CSR) formats. The streaming and collision and boundary operators are fully CSR; however, only the spatial and energy dimensions of the scattering and fission operators are CSR, whereas the angular dimensions are applied separately. The scattering operator in CSR is
\begin{equation}\label{eq:S_CSR}
    \mathcal{S}^{\text{CSR}}\mathbf{\Psi} = \sum_{l = 0}^L\sum_{m = 0}^l\mathbf{Y}_{l, m, i_\Omega}\mathbf{V}^{\text{CSR}}_{l}\sum_{i_\Omega' = 1}^{N_{\Omega}}\mathbf{Y}_{l, m, i_\Omega'}\mathbf{\Psi}_{i_\Omega'} w_{i_\Omega'}^\Omega,
\end{equation}
where $\mathbf{Y}_{l}\in\mathbb{R}^{(l + 1) \times N_\Omega}$ is a dense matrix of evaluated even spherical harmonic functions matching that in \cref{eq:even_sphm_tt} and $\mathbf{V}_{l}^{\text{CSR}}\in\mathbb{R}^{N_EN_eN_{\hat x}N_{\hat y}\times N_EN_eN_{\hat x}N_{\hat y}}$ is a CSR matrix including the interior spatial integration operator $\mathcal{V}_R$ in CSR format with the $l$th scattering moment cross sections. The fission operator in CSR format follows similarly,
\begin{equation}\label{eq:F_CSR}
    \mathcal{F}^{\text{CSR}}\mathbf{\Psi} = \mathbf{Y}_{0, 0, i_\Omega}\mathbf{V}_0^{\text{CSR}}\sum_{i_\Omega' = 1}^{N_{\Omega}}\mathbf{Y}_{0, 0, i_\Omega'}\mathbf{\Psi}_{i_\Omega'} w_{i_\Omega'}^\Omega.
\end{equation}

\par All operators represent a $4\times N_\mu\times N_\gamma\times N_E\times N_e\times N_{\hat x}\times N_{\hat y}$ linear or eigenvalue system. We can sum and recompress all operators in the TT format using the TT-rounding algorithm \cite{Oseledets}. The streaming and collision and boundary operators may be combined into a single CSR operator; however, the scattering and fission operators in \cref{eq:S_CSR,eq:F_CSR} must be applied separately. For eigenvalue problems, we apply power iteration over the fission source,
\begin{equation}
    \left(\mathcal{H} - \mathcal{S} + \mathcal{B}_{\text{out}} - \mathcal{B}_{\text{in}}\right)\mathbf{\Psi}^{(\tau + 1)} = \frac{1}{k^{(\tau)}}\mathcal{F}\mathbf{\Psi}^{(\tau)},
\end{equation}
with iteration index $\tau$ and assume a unit angular flux $\mathbf{\Psi}^{(0)}$ with initial eigenvalue
\begin{equation}
k^{(0)} = \frac{\left(\mathbf{\Psi}^{(0)}\right)^T\mathcal{F}\mathbf{\Psi}^{(0)}}{\left(\mathbf{\Psi}^{(0)}\right)^T\left(\mathcal{H} - \mathcal{S} + \mathcal{B}_{\text{out}} - \mathcal{B}_{\text{in}}\right)\mathbf{\Psi}^{(0)}}.
\end{equation}
We solve each resulting linear system using unpreconditioned restarted GMRES with a convergence criterion $\|\mathbf{r}\|_2 < \epsilon_{\text{GMRES}}\left\|\frac{1}{k^{(\tau)}}\mathcal{F}\mathbf{\Psi}^{(\tau)}\right\|_2$ with some acceptable error $\epsilon_{\text{GMRES}}$ and residual $\mathbf{r}$. GMRES without preconditioning is inefficient, but we leave the extension of GMRES with a tensorized multilevel preconditioner for future work. Once GMRES solves for $\mathbf{\Psi}^{(\tau + 1)}$, we calculated an updated eigenvalue 
\begin{equation}
    k^{(\tau + 1)} = k^{(\tau)}\frac{\sum\mathcal{F}\mathbf{\Psi}^{(\tau + 1)}}{\sum\mathcal{F}\mathbf{\Psi}^{(\tau)}}.
\end{equation}
We repeat until a maximum number of power iterations or we converge according to $\left\|\mathbf{\Psi}^{(\tau + 1)} - \mathbf{\Psi}^{(\tau)}\right\|_2 < \epsilon_{\text{PI}}\left\|\mathbf{\Psi}^{(\tau)}\right\|_2$ with some acceptable error $\epsilon_{\text{PI}}$.

\par We explore five cases:
\begin{subequations}\label{eq:cases}
\begin{align}
    \text{CSR: }&\left(\mathcal{H}^{\text{CSR}} + \mathcal{B}^{\text{CSR}}_{\text{out}} - \mathcal{B}^{\text{CSR}}_{\text{in}}\right)\mathbf{\Psi}^{(\tau + 1)} - \mathcal{S}^{\text{CSR}}\mathbf{\Psi}^{(\tau + 1)} = \frac{1}{k^{(\tau)}}\mathcal{F}^{\text{CSR}}\mathbf{\Psi}^{(\tau)},\label{eq:cases_csr}\\
    \text{TT: }&\mathcal{H}^{\text{TT}}\mathbf{\Psi}^{(\tau + 1)} - \mathcal{S}^{\text{TT}}\mathbf{\Psi}^{(\tau + 1)} + \mathcal{B}_{\text{out}}^{\text{TT}}\mathbf{\Psi}^{(\tau + 1)} - \mathcal{B}_{\text{in}}^{\text{TT}}\mathbf{\Psi}^{(\tau + 1)} = \frac{1}{k^{(\tau)}}\mathcal{F}^{\text{TT}}\mathbf{\Psi}^{(\tau)},\label{eq:cases_tt}\\
    \text{Mixed: }&\mathcal{H}^{\text{TT}}\mathbf{\Psi}^{(\tau + 1)} - \mathcal{S}^{\text{TT}}\mathbf{\Psi}^{(\tau + 1)} + (\mathcal{B}_{\text{out}}^{\text{CSR}} - \mathcal{B}_{\text{in}}^{\text{CSR}})\mathbf{\Psi}^{(\tau + 1)} = \frac{1}{k^{(\tau)}}\mathcal{F}^{\text{TT}}\mathbf{\Psi}^{(\tau)},\label{eq:cases_mixed}\\
    \text{TT (rounded): }&\left(\mathcal{H}^{\text{TT}} - \mathcal{S}^{\text{TT}} + \mathcal{B}_{\text{out}}^{\text{TT}} - \mathcal{B}_{\text{in}}^{\text{TT}}\right)\mathbf{\Psi}^{(\tau + 1)} = \frac{1}{k^{(\tau)}}\mathcal{F}^{\text{TT}}\mathbf{\Psi}^{(\tau)},\label{eq:cases_tt_rounded}\\
    \text{Mixed (rounded): }&\left(\mathcal{H}^{\text{TT}} - \mathcal{S}^{\text{TT}}\right)\mathbf{\Psi}^{(\tau + 1)} + \left(\mathcal{B}_{\text{out}}^{\text{CSR}} - \mathcal{B}_{\text{in}}^{\text{CSR}}\right)\mathbf{\Psi}^{(\tau + 1)} = \frac{1}{k^{(\tau)}}\mathcal{F}^{\text{TT}}\mathbf{\Psi}^{(\tau)},\label{eq:cases_mixed_rounded}
\end{align}
\end{subequations}
where $\mathbf{\Psi}$ is a full vector. Tensor train summation is rounded to a truncation error $\epsilon$. Each TT operator is applied at $\mathcal{O}(dr^2N^d\log(N))$ while the CSR operators are $\mathcal{O}(\text{nnz})$. We explore mixed cases in which interior operators are in TT format and boundary operators are in CSR format, as we expect a larger rank increase for the boundary operators in highly curvilinear domains. 

\subsection{Implementation}\label{sec:implementation}

\par Our implementation of the TDIGA method is located in our repository \href{https://github.com/myerspat/ttnte}{\texttt{ttnte}}\footnote{\href{https://github.com/myerspat/ttnte}{https://github.com/myerspat/ttnte}} and is written in C++ and Python. We define NURBS patches using \verb|igakit| \cite{igakit} and \verb|geomdl| \cite{bingol2019geomdl} and pass these to our \verb|IGAMesh| class. We refine the mesh using knot insertion and order elevation. We pass this and the material cross section information to the \verb|MatrixAssembler| or \verb|TTAssembler| to build the operators and solve the resulting linear system using the \verb|gmres| method, or eigenvalue problems using the \verb|eig| method. The cores within each TT and the CSR matrices are PyTorch \cite{DBLP:journals/corr/abs-1912-01703} tensors, allowing us to solve these systems on both CPU and GPU. We use \verb|cotengra| \cite{Gray_2021} for a FLOP optimal contraction order when applying TTs to the solution vector.
\section{Numerical Results}\label{sec:results}

\par In this section, we apply the TDIGA method to several fixed source and $k$-eigenvalue examples. In all examples, we compare TDIGA to a reference analytic or Monte Carlo (MC) scalar flux solution, except for the homogenized circle in \cref{sec:fixed_homo}, which only has a reference leakage fraction. Monte Carlo reference solutions were generated with OpenMC \cite{ROMANO201590} with either a leakage fraction or eigenvalue tally and a scalar flux regular mesh tally yielding $\mathbf{\Phi}^{\text{MC}}, \boldsymbol{\sigma}^{\text{MC}}\in\mathbb{R}^{N_E\times 128\times 128}$, where $\boldsymbol{\sigma}^{\text{MC}}$ is the standard deviation for each scalar flux mesh element. We compare this mesh tally by averaging the TDIGA solution onto the same regular mesh using trapezoidal integration and computing an $L_2$ error metric,
\begin{subequations}\label{eq:eigenvector_error}
\begin{align}
    \epsilon_2^{i_E}\left(\mathbf{\Phi}^{\text{X}}, \mathbf{\Phi}^{\text{Y}}\right) &= \frac{\sqrt{\sum_{i_{x} = 1}^{128}\sum_{i_{y} = 1}^{128}(\mathbf{\Phi}_{i_E, i_{x}, i_{y}}^{\text{X}} - \mathbf{\Phi}^{\text{Y}}_{i_E, i_{x}, i_{y}})^2}}{\sqrt{\sum_{i_{x}= 1}^{128}\sum_{i_{y} = 1}^{128}\left(\mathbf{\Phi}^{\text{Y}}_{i_E, i_{x}, i_{y}}\right)^2}},\\
    \epsilon_2\left(\mathbf{\Phi}^{\text{X}}, \mathbf{\Phi}^{\text{Y}}\right) &= \sqrt{\sum_{i_E}^{N_E}\left(\epsilon^{i_E}_2\left(\mathbf{\Phi}^{\text{X}}, \mathbf{\Phi}^{\text{Y}}\right)\right)^2},
\end{align}
\end{subequations}
where X is one of the five cases in \cref{eq:cases} and Y is either MC or CSR. We evaluate statistical significance using a $z$-score,
\begin{equation}
    \mathbf{z}_{i_E, i_x, i_y} = \frac{|\mathbf{\Phi}^{\text{X}}_{i_E, i_x, i_y} - \mathbf{\Phi}^{\text{MC}}_{i_E, i_x, i_y}|}{\boldsymbol{\sigma}^{\text{MC}}_{i_E, i_x, i_y}}.
\end{equation}
For quantifying the effectiveness with respect to memory of each case in \cref{eq:cases}, we define the compression ratio (CR),
\begin{equation}
    \text{CR}(\mathcal{X}) = \frac{\text{Number of elements in the full tensor of } \mathcal{X}}{\text{Number of elements in the compressed representation of } \mathcal{X}}.
\end{equation}
All results were generated on a cluster with a $64$-core AMD Ryzen Threadripper PRO processor, an NVIDIA RTX PRO 6000 Blackwell Max-Q GPU, and 512 GB of RAM. For all calculations, we use double-precision floating-point format, which degrades GPU tensor core performance. Further research into mixed-precision may yield improved performance for TT operator-vector products. For brevity, we show only a subset of the results, but all data, figures, and scripts can be found on our repository \href{https://github.com/myerspat/ttnte_tdiga_jcp2025}{\texttt{ttnte\_tdiga\_jcp2025}}\footnote{\href{https://github.com/myerspat/ttnte_tdiga_jcp2025}{https://github.com/myerspat/ttnte\_tdiga\_jcp2025}}.

\subsection{Fixed Source}

\par We explore four fixed source problems: homogenized square and circular sources in \cref{sec:fixed_homo}, a quarter circular source surrounded by void in \cref{sec:fixed_quarter_circle}, and a shielded cruciform source surrounded by void in \cref{sec:fixed_cruciform}. We compare the solution with TDIGA to references generated with OpenMC for both the leakage fraction,
\begin{equation}
    f_{\text{leak}} = \frac{\sum_{i_E = 1}^{N_E}\sum_{i_\Omega = 1}^{N_\Omega}w_{i_\Omega}^\Omega\int_{\p \Gamma^+}(\bo_{i_\Omega}\cdot \mathbf{n})\psi_{i_E, i_\Omega}(\x)dS}{\sum_{i_E = 1}^{N_E}\sum_{i_\Omega = 1}^{N_\Omega}w_{i_\Omega}^\Omega\int_{\Gamma}Q_{i_E, i_\Omega}(\x)dV},
\end{equation}
and scalar flux. The leakage fraction relative error is
\begin{equation}\label{eq:leakage_fraction_relative_error}
    \delta f\left(f_{\text{leak}}^\text{X}, f_{\text{leak}}^\text{Y}\right) = \frac{f_{\text{leak}}^{\text{X}} - f_{\text{leak}}^{\text{Y}}}{f_{\text{leak}}^{\text{Y}}}.
\end{equation}

\par For both \cref{sec:fixed_homo,sec:fixed_quarter_circle}, we conduct direction and mesh resolution scaling studies to examine the effects of boundary curvilinearity on operator compression, solution time, and solution error as fidelity increases. For the angular-resolution study, we use $N_\Omega\in\{16, 64, 256, 1024, 4096, 16384, 65536, 262144\}$ ordinates while fixing $N_{\hat x}^u = N_{\hat y}^u = 10$ for the square and circle in \cref{sec:fixed_homo} and $N_{\hat x}^u = N_{\hat y}^u = 8$ for the quarter circle in \cref{sec:fixed_quarter_circle}. For the mesh-resolution study, we use $N_\Omega = 256$ ordinates while the number of knot spans along each parametric direction vary as $N_{\hat x}^u = N_{\hat y}^u\in\{5, 6, 8, 11, 14, 19, 25, 33, 44, 58, 76, 100\}$. The total number of spatial degrees of freedom is $\left(N_{\hat x}^u + p_{\hat x}\right)\times \left(N_{\hat y}^u + p_{\hat y}\right) = N_{\hat x}\times N_{\hat y}$. In both studies, we use polynomial degrees $p_{\hat x} = p_{\hat y} \in\{2, 3, 4, 6\}$ for the NURBS basis functions. We round all TTs to a truncation tolerance of $\epsilon \in\{10^{-8}, 10^{-5}, 10^{-3}\}$ and run GMRES until $\epsilon_{\text{GMRES}} = 10^{-6}$ or we hit a maximum of $1000$ restarts with $100$ iterations.

\subsubsection{Homogeneous square and circle}\label{sec:fixed_homo}

\begin{figure}
\centering
\begin{subfigure}{0.47\textwidth}
    \centering
    \includegraphics[width=\textwidth]{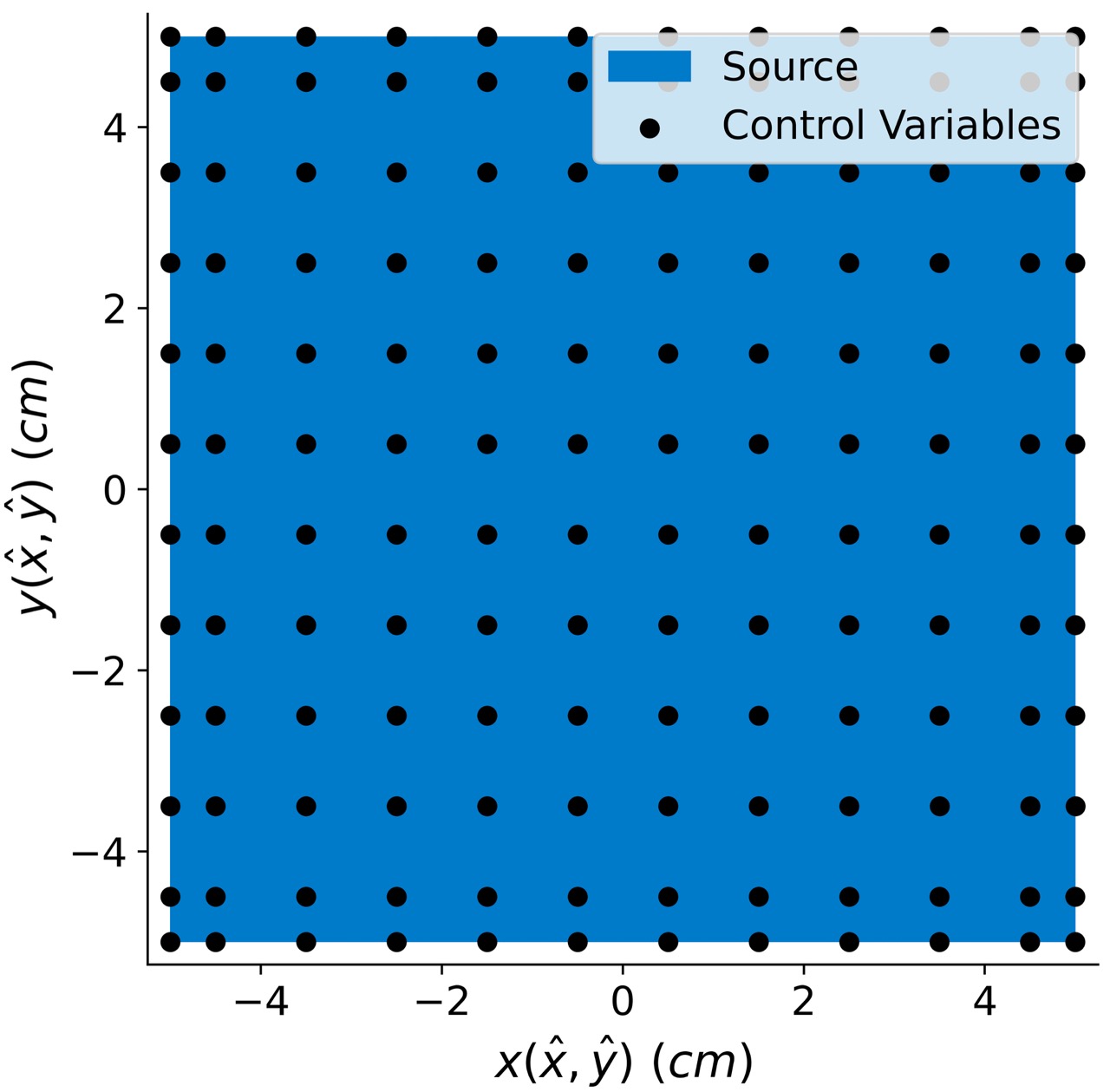}
    \caption{}
    \label{fig:square}
\end{subfigure}
\begin{subfigure}{0.47\textwidth}
    \centering
    \includegraphics[width=\textwidth]{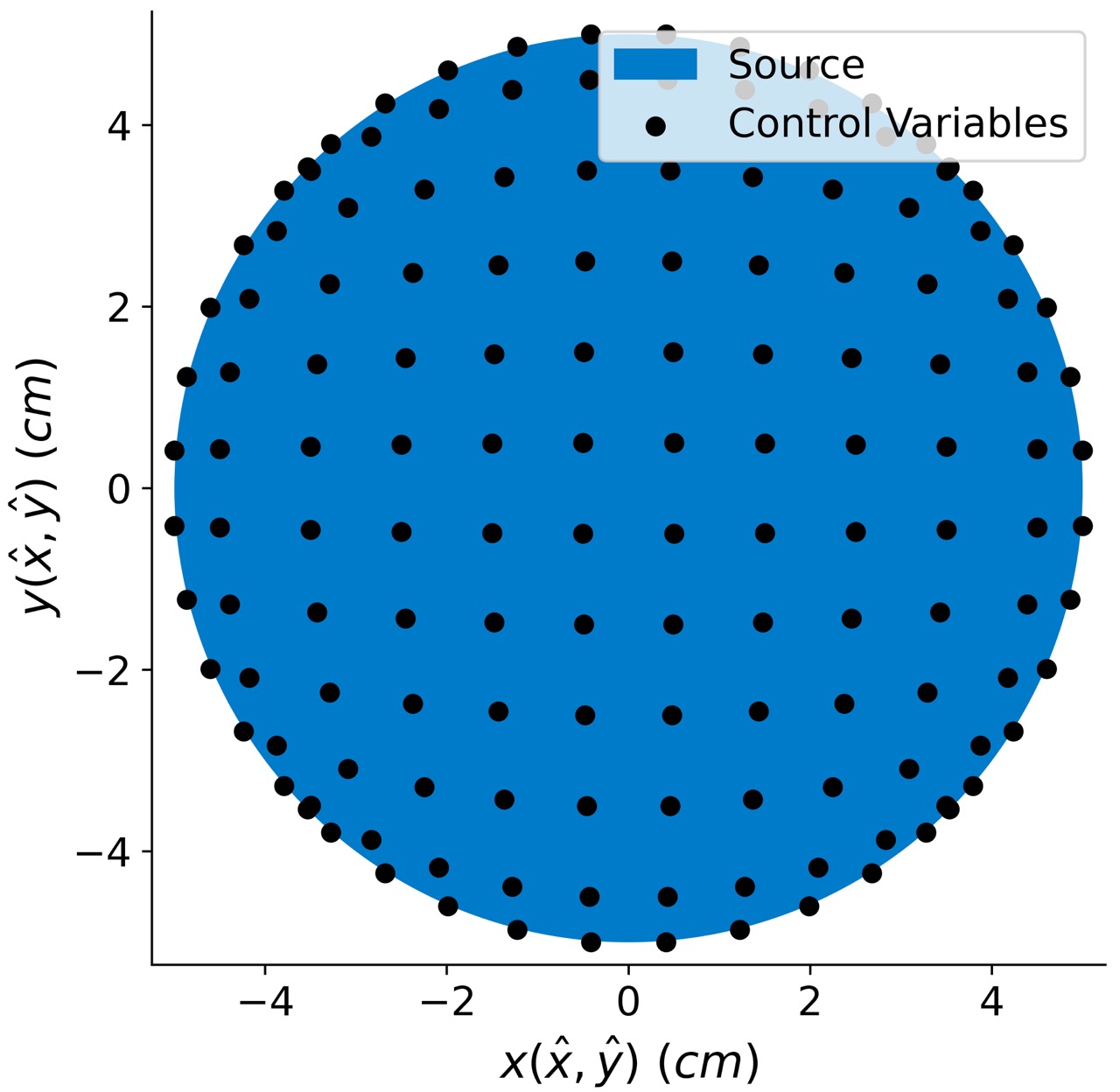}
    \caption{}
    \label{fig:circle}
\end{subfigure}
\begin{subfigure}{0.47\textwidth}
    \centering
    \includegraphics[width=\textwidth]{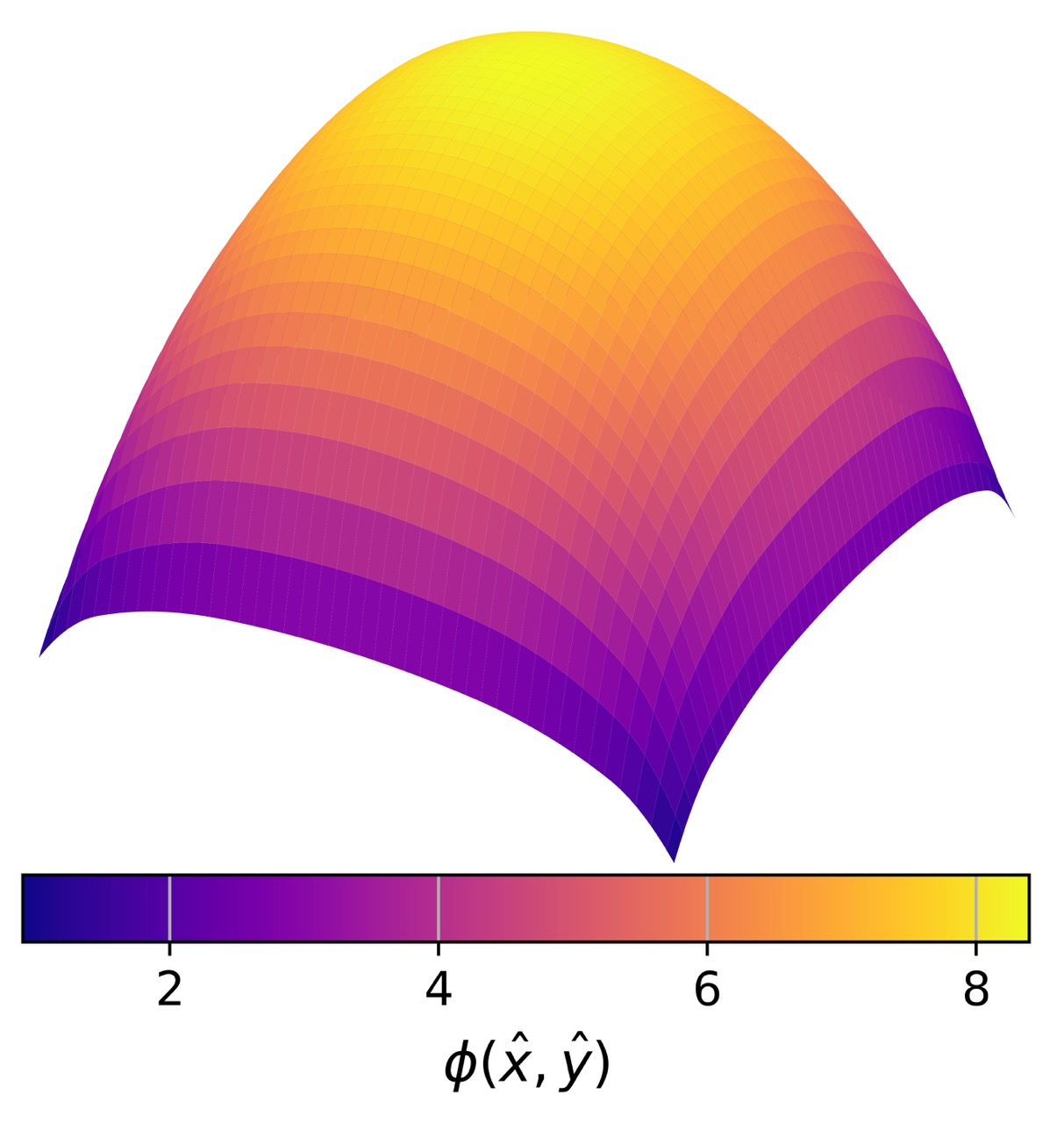}
    \caption{}
    \label{fig:square_phi}
\end{subfigure}
\begin{subfigure}{0.47\textwidth}
    \centering
    \includegraphics[width=0.90\textwidth]{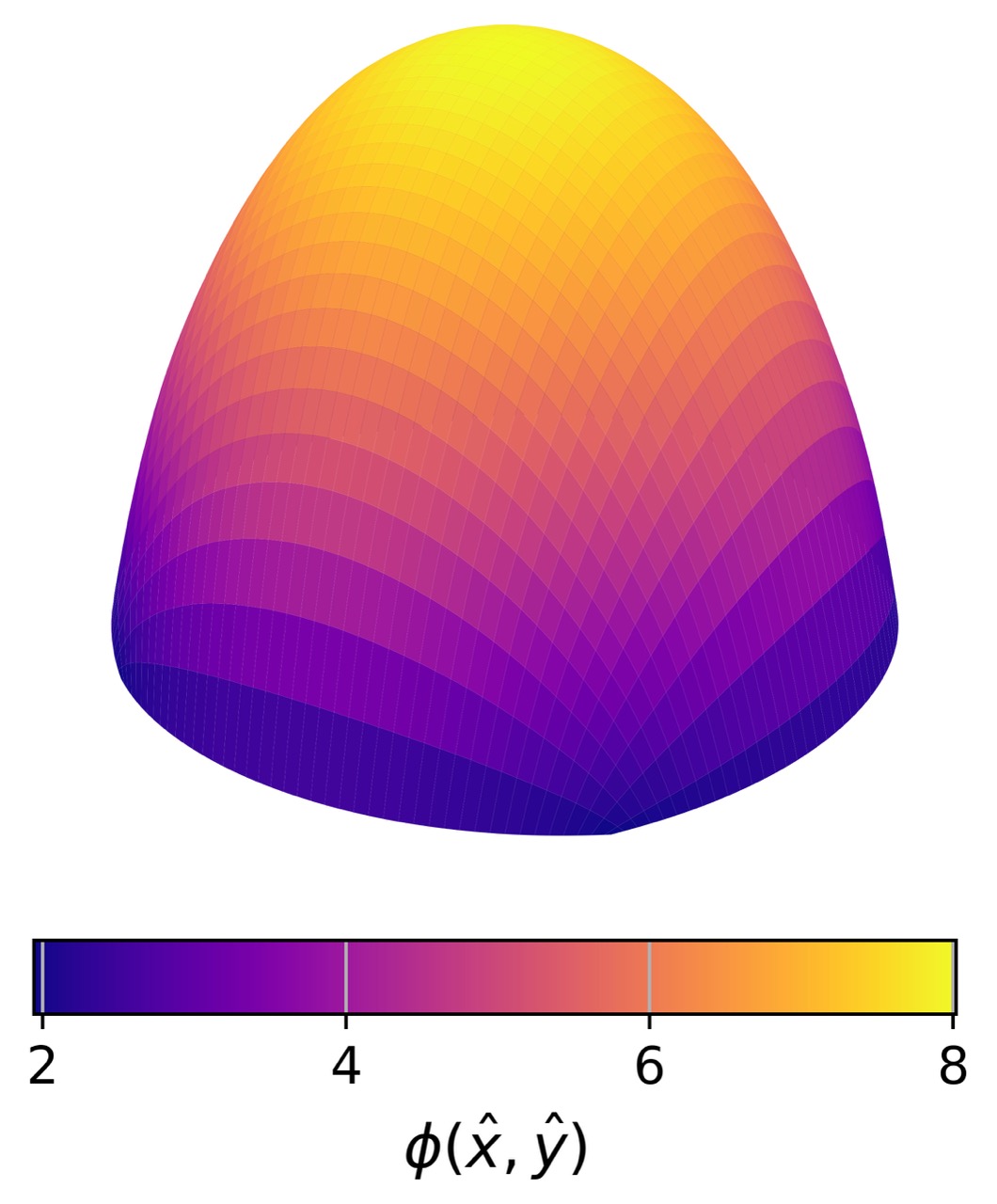}
    \caption{}
    \label{fig:circle_phi}
\end{subfigure}
        
\caption{Fixed source homogenized (a) square and (b) circle geometries with a spatial mesh of $N_{\hat x}^u = N_{\hat y}^u = 10$ knot spans along each parametric direction and polynomial degrees $p_{\hat x} = p_{\hat y} = 2$ for the NURBS basis functions. We show the scalar flux solution for the (c) square and (d) circle computed using the same spatial mesh with CSR (\cref{eq:cases_csr}) for $N_\Omega = 16384$ ordinates.}
\label{fig:square_and_circle}
\end{figure}

\par Our first test examines the effects of curvilinearity on operator compression as the angular and mesh resolutions increase. We compare the compression performance and resulting rank for two homogenized geometries: a square of side length $10~cm$ and a circle of radius $5~cm$, both with vacuum boundary conditions. Each geometry uses single-group ($N_E = 1$) cross sections $\Sigma_t = 1~cm^{-1}$ and $\Sigma_s = 0.9~cm^{-1}$ and internal source $Q = 1~cm^{-3}s^{-1}$. \Cref{fig:square,fig:circle} show the two geometries, each represented by a single patch ($N_e = 1$) with an example mesh consisting of $N_{\hat x}^u = N_{\hat y}^u = 10$ knot spans in each parametric direction and polynomial degrees $p_{\hat x} = p_{\hat y} = 2$ for the NURBS basis functions. We show their respective solutions in \cref{fig:square_phi,fig:circle_phi} computed using the CSR case with $N_\Omega=16384$ ordinates. Note that the inflow boundary operator, $\mathcal{B}_{\text{in}}$, is zero as both problems use a single patch with only vacuum boundary conditions, hence it is not included. We generated a reference solution using OpenMC. Both have reference leakage fractions, $0.42095\pm 0.00002$ for the square and $0.43995\pm 0.00002$ for the circle, while only the square has a reference scalar flux $\mathbf{\Phi}^{\text{MC}}, \boldsymbol{\sigma}^{\text{MC}}\in\mathbb{R}^{1\times 128\times 128}$.

\begin{table}
\centering
\caption{Maximum tensor train (TT) rank for operators in the fixed source homogenized square and circle problems. All max ranks were observed to be constant for the angular resolution study, except for the outflow boundary operator of the circle, and are ordered by polynomial degree $p_{\hat x} = p_{\hat y} \in\{2, 3, 4, 6\}$. The operator TT-ranks for the homogenized square are the same for the mesh resolution study as well.}\label{tbl:fixed_source_constant_ranks}
\begin{tabular}{ccccccc} 
\toprule
\multirow{2}{*}{\textbf{Operator }}    & \multicolumn{3}{c}{Homogenized Square}                                                              & \multicolumn{3}{c}{Homogenized Circle}                                                               \\ 
\cline{2-7}
                                       & $\boldsymbol{\epsilon=10^{-8}}$ & $\boldsymbol{\epsilon=10^{-5}}$ & $\boldsymbol{\epsilon=10^{-3}}$ & $\boldsymbol{\epsilon=10^{-8}}$ & $\boldsymbol{\epsilon=10^{-5}}$ & $\boldsymbol{\epsilon=10^{-3}}$  \\ 
\hline\hline
$r_{\max}\left(\mathcal{H}^{\text{TT}}\right)$              & $\{3, 3, 3, 3\}$                & $\{3, 3, 3, 3\}$                & $\{3, 3, 3, 3\}$                & $\{40, 44, 45, 44\}$            & $\{21, 21, 21, 22\}$            & $\{9, 9, 9, 10\}$                \\
$r_{\max}\left(\mathcal{S}^{\text{TT}}\right)$              & $\{1, 1, 1, 1\}$                & $\{1, 1, 1, 1\}$                & $\{1, 1, 1, 1\}$                & $\{14, 14, 14, 14\}$            & $\{7, 7, 7, 7\}$                & $\{3, 3, 3, 4\}$                 \\
$r_{\max}\left(\mathcal{B}_{\text{out}}^{\text{TT}}\right)$ & $\{4, 4, 4, 4\}$                & $\{4, 4, 4, 4\}$                & $\{4, 4, 4, 4\}$                & \multicolumn{3}{c}{Refer to \cref{fig:fixed_square_ranks_B_out}.}                         \\
\bottomrule
\end{tabular}
\end{table}

\begin{figure}
    \centering
    \includegraphics[width=0.7\linewidth, trim=0.5cm 0.5cm 0.5cm 0.5cm, clip]{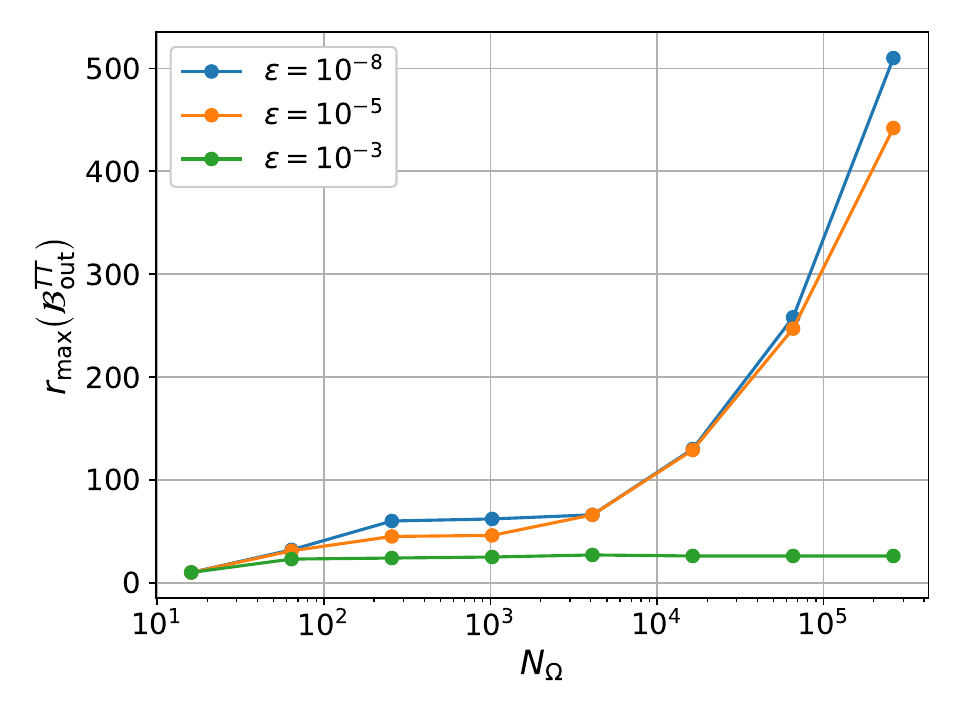}
    \caption{Max tensor train (TT) ranks of the outflow boundary operator for increasing angular resolution for $p_{\hat x} = p_{\hat y} = 2$ NURBS basis functions for the fixed source homogenized circle. NURBS basis functions of polynomial degree $p_{\hat x} = p_{\hat y}\in\{3, 4, 6\}$ follow similarly.}
    \label{fig:fixed_square_ranks_B_out}
\end{figure}

\paragraph{Angular Resolution Study}
Below, we discuss the compression of the operators in CSR and TT formats; the effect of increased angular resolution, basis function polynomial degree, and TT truncation on leakage fraction and scalar flux error; the computational scaling of CSR and TT operator-vector products; and the compressibility of the angular flux in TT format. The main observations include:
\begin{itemize}
    \item TT interior operator ($\mathcal{H}^{\text{TT}}$ and $\mathcal{S}^{\text{TT}}$) max rank is independent of angular resolution; however, it does increase from flat axis-aligned boundaries to curvalinear boundaries. The boundary operator max rank is also independent of angular resolution for regular meshes; however, as shown in \cref{fig:fixed_square_ranks_B_out}, the boundary ranks increase for increasing angular resolution for patches with curvalinear boundaries.
    \item The TT-ranks vary only marginally with basis function polynomial degree.
    \item Between truncation tolerances $\epsilon\in\{10^{-8}, 10^{-5}, 10^{-3}\}$, the max rank of TT interior operators decreases by roughly half for the circular problem, while they remain constant for the square. Only the $\epsilon = 10^{-3}$ truncation tolerance yields a plateau in the maximum rank as a function of angular resolution for the outflow boundary operator in the circular problem.
    \item TT offers superior compression in the interior operators compared to CSR. In contrast, its advantage for boundary operators depends on the degree of boundary curvilinearity and the truncation tolerance. This observation holds across polynomial basis-function degrees.
    \item A larger number of ordinates is needed to mitigate ray effects in the leakage fraction relative error for the flat boundaries and corners of the square compared to the smooth boundary of the circle. However, with enough ordinates basis function degrees $p_{\hat x} = p_{\hat y} \in\{3, 4, 6\}$ come within $\sigma$ of $f_{\text{leak}}^{\text{MC}}$ for both geometries.
    \item Leakage fraction relative error of TT, Mixed, TT (rounded), and Mixed (rounded) cases to CSR are negligible with $10^{-16} <\delta f < 10^{-6}$ for the square and $10^{-11} < \delta f < 10^{-3}$ for the circle. The square demonstrates an increase in $\delta f$ as the polynomial degree increases, while the circle does not. For increasing truncation tolerance, the leakage fraction relative error increases by two to three orders of magnitude.
    \item Only the $p_{\hat x} = p_{\hat y} = 6$ comes within $2\boldsymbol{\sigma}^{\text{MC}}$ with CSR for the square with local cell-by-cell scalar flux error being potentially statistically significant. Comparing TT, Mixed, TT (rounded), and Mixed (rounded) to CSR shows a decreasing $L_2$ error over angular resolution ranging $10^{-15} < \epsilon_2 < 10^{-4}$ depending on basis function degree. Truncation tolerances follow closely except for TT and TT (rounded) for $\epsilon = 10^{-3}$, which is two to three orders of magnitude off $\epsilon\in\{10^{-8}, 10^{-5}\}$. Again, increasing the degree of the basis functions increases the $L_2$ error for the other cases relative to CSR.
    \item The operator-vector products show that the GPU outperforms the CPU, with CSR consistently the fastest, followed by Mixed and Mixed (rounded). For increasing basis function degree, the gap between CSR and Mixed/Mixed (rounded) decreases at higher fidelity angular discretizations. The angular-resolution dependence of boundary operators for curvilinear boundaries leads to substantially unfavorable scaling.
    \item The angular flux is compressible for regular meshes for all truncation tolerances, while only compressible at $\epsilon = 10^{-3}$ for patches with curvilinear boundaries.
    \item CR for the angular flux in TT format decreases with increasing NURBS basis function degree.
\end{itemize}
With increasing angular resolution, TT offers better compression for interior operators than CSR; however, the unfavorable scaling of TT operator-vector products significantly increases time-to-solution. This is partially resolved in the Mixed and Mixed (rounded) formats, particularly for higher-order basis functions. Curvilinear boundaries drastically degrade the performance of boundary operator compression.

\subparagraph{Operator TT-Ranks and Compression}
In \cref{tbl:fixed_source_constant_ranks} we show the maximum rank for each operator. In both the square and circle problems, the ranks remain constant across all angular discretizations for the interior operators ($\mathcal{H}^{\text{TT}}$ and $\mathcal{S}^{\text{TT}}$). This is also true for the outflow boundary operator for the homogenized square, implying that rank and angular resolution are independent for regular meshes. The operators in the square case have the same ranks across truncation tolerances and polynomial order. For varying polynomial degree, $\mathcal{H}^{\text{TT}}$ and $\mathcal{S}^{\text{TT}}$ have maximum ranks that only marginally vary ($r_{\text{max}}\in[40, 45]$ for $\mathcal{H}^{\text{TT}}$ and $r_{\text{max}}\in[21, 22]$ for $\mathcal{S}^{\text{TT}}$) with angular discretization. Their max rank reduces by roughly half from $\epsilon = 10^{-8}$ to $\epsilon = 10^{-5}$ and again for $\epsilon = 10^{-5}$ to $\epsilon = 10^{-3}$.

\Cref{fig:fixed_square_ranks_B_out} shows a high degree of coupling between angular resolution and curvilinear boundaries as the max rank increases with increasing angular fidelity. Across polynomial degrees, we see a leveling off of max rank for $\epsilon=10^{-3}$ while the max rank of $\epsilon\in\{10^{-8}, 10^{-5}\}$ continues to increase. Initially, the maximum rank occurs between the spatial cores, which is overtaken by the ranks in the angular cores for $N_\Omega > 4096$, resulting in a brief plateau for $256 \le N_\Omega \le 4096$. The ranks between polynomial orders for a given truncation tolerance are similar and do not indicate an increase in rank with NURBS polynomial order.

\begin{figure}
\centering
\begin{subfigure}{0.47\textwidth}
    \includegraphics[width=\textwidth, trim=0.5cm 0.5cm 0.5cm 0.5cm, clip]{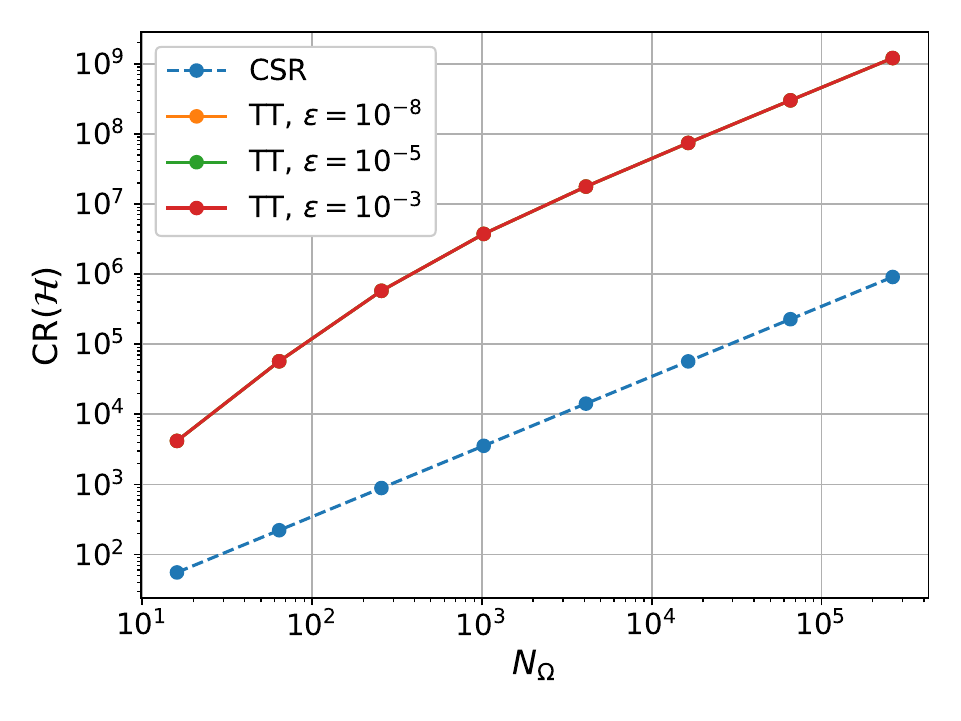}
    \caption{}
    \label{fig:fixed_square_comp_p2_H}
\end{subfigure}
\begin{subfigure}{0.47\textwidth}
    \includegraphics[width=\textwidth, trim=0.5cm 0.5cm 0.5cm 0.5cm, clip]{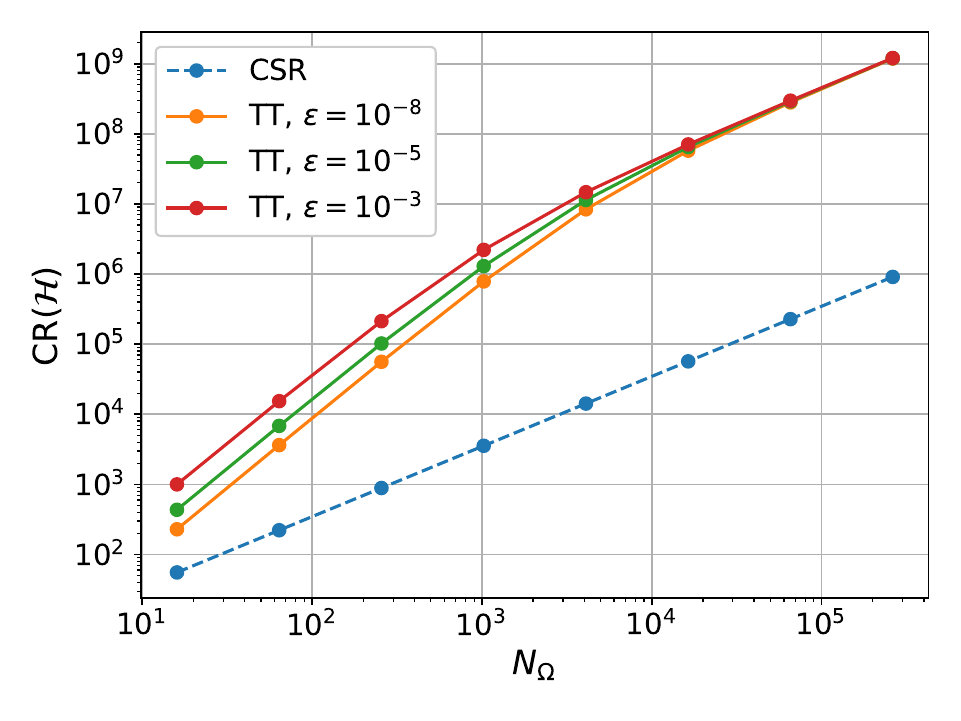}
    \caption{}
    \label{fig:fixed_circle_comp_p2_H}
\end{subfigure}
\begin{subfigure}{0.47\textwidth}
    \includegraphics[width=\textwidth, trim=0.5cm 0.5cm 0.5cm 0.5cm, clip]{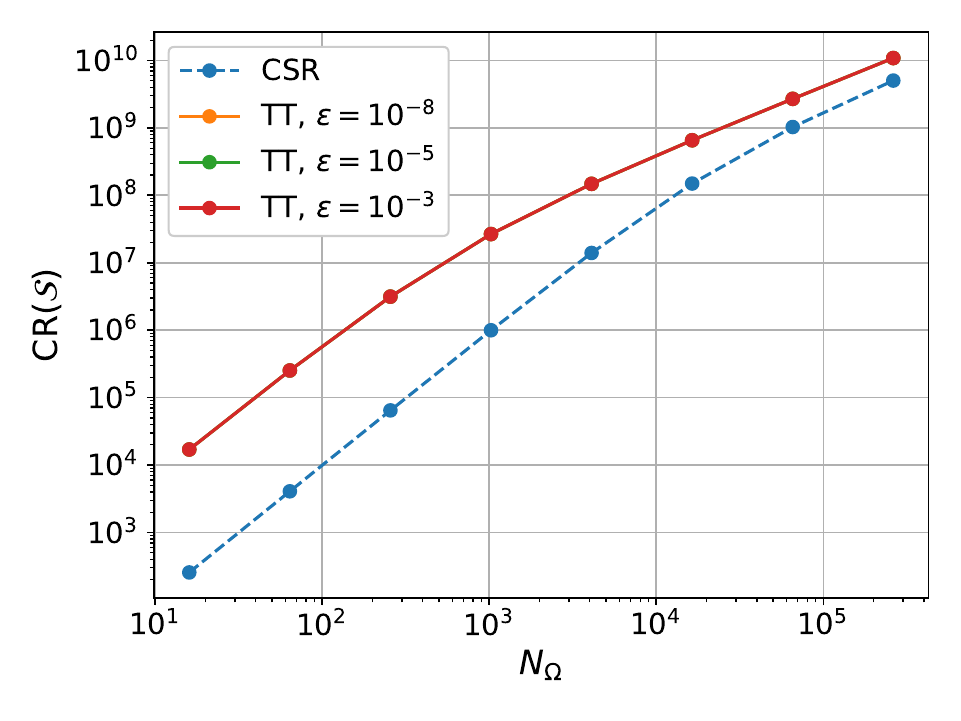}
    \caption{}
    \label{fig:fixed_square_comp_p2_S}
\end{subfigure}
\begin{subfigure}{0.47\textwidth}
    \includegraphics[width=\textwidth, trim=0.5cm 0.5cm 0.5cm 0.5cm, clip]{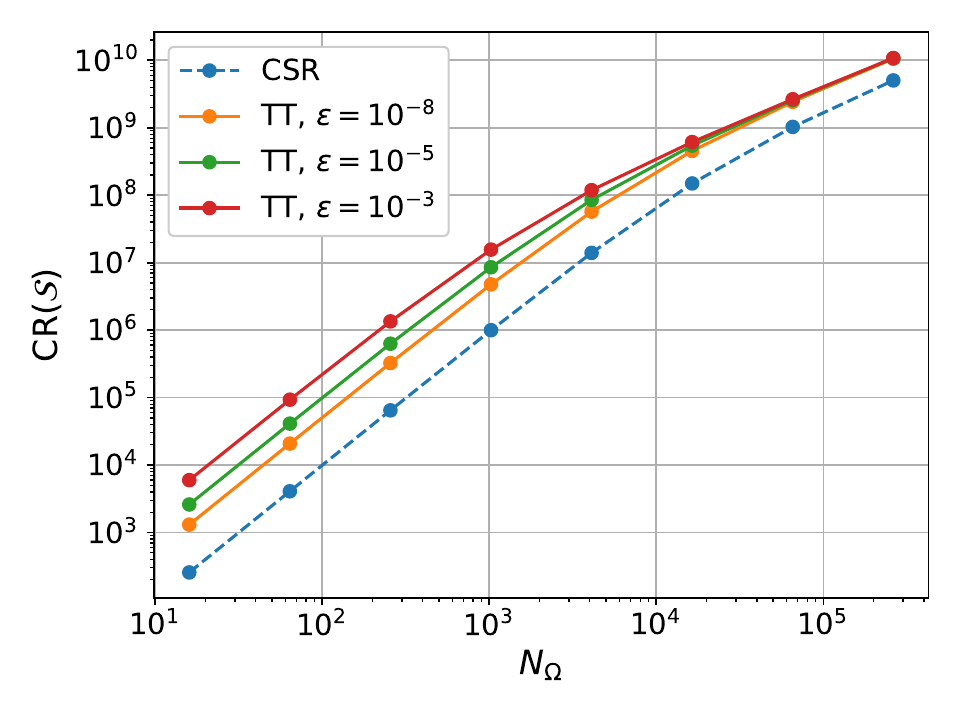}
    \caption{}
    \label{fig:fixed_circle_comp_p2_S}
\end{subfigure}
\begin{subfigure}{0.47\textwidth}
    \includegraphics[width=\textwidth, trim=0.5cm 0.5cm 0.5cm 0.5cm, clip]{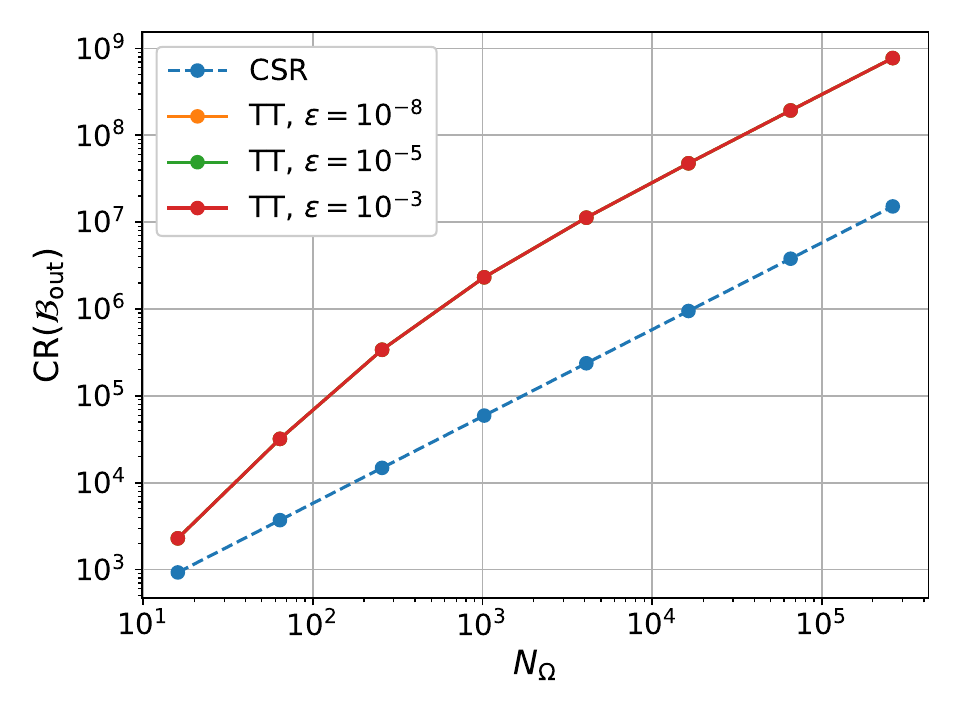}
    \caption{}
    \label{fig:fixed_square_comp_p2_B_out}
\end{subfigure}
\begin{subfigure}{0.47\textwidth}
    \includegraphics[width=\textwidth, trim=0.5cm 0.5cm 0.5cm 0.5cm, clip]{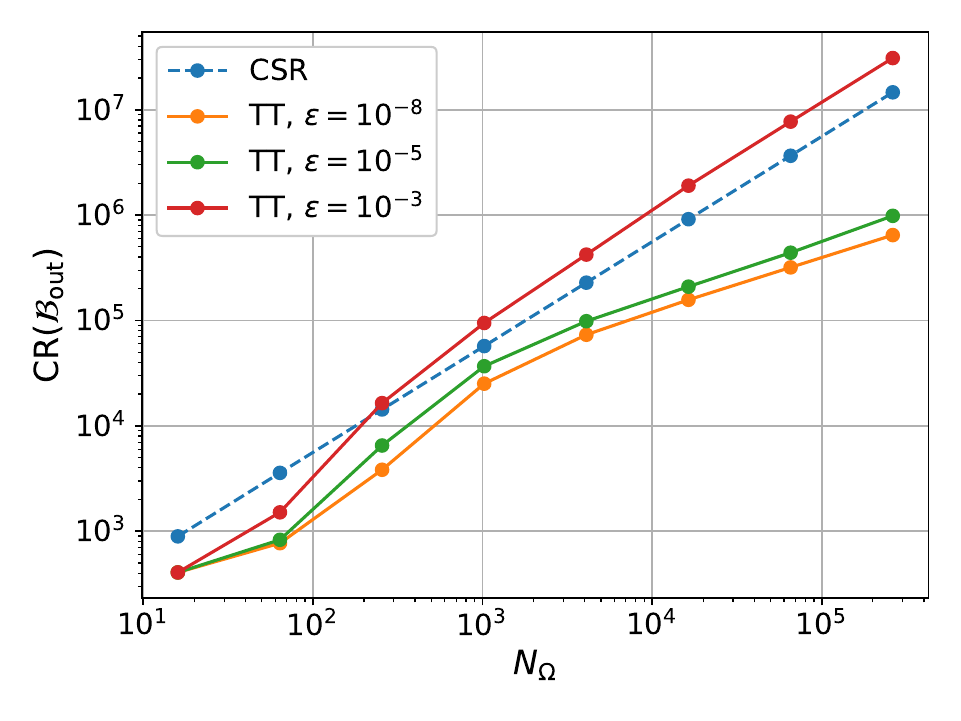}
    \caption{}
    \label{fig:fixed_circle_comp_p2_B_out}
\end{subfigure}
        
\caption{Compression ratios (CRs) for the (a, b) streaming and collision, (c, d) scattering, and (e, f) outflow boundary operators for the fixed source homogenized square (left) and circle (right) problems. The above figures only show $p_{\hat x} = p_{\hat y} = 2$; however, the CR plots of $p_{\hat x} = p_{\hat y}\in\{3, 4, 6\}$ follow the same trends with the CR of the tensor train (TT) operators relative to Compressed Sparse Row (CSR) increasing for increasing basis function degree. We show TT truncation tolerances $\epsilon \in\{10^{-8}, 10^{-5}, 10^{-3}\}$ and CSR. Note the CR for the square shows no difference between truncation tolerances.}
\label{fig:fixed_square_and_circle_comp_p2}
\end{figure}

\begin{table}
\centering
\caption{Exponent $\alpha$ for $\text{compression ratio (CR)}\propto N_{\Omega}^\alpha$ of each operator of the fixed source square and circle fit to the first and last three data points of \cref{fig:fixed_square_and_circle_comp_p2} where $p_{\hat x} = p_{\hat y} = 2$. We show the exponent for the Compressed Sparse Row (CSR) and tensor train (TT) formats.}\label{tbl:fixed_square_and_circle_comp_p2_scale}
\begin{tabular}{ccccccccc} 
\toprule
\multirow{2}{*}{\textbf{Operator }} & \multicolumn{4}{c}{First Three Data Points}                                                                              & \multicolumn{4}{c}{Last Three Data Points}                                                                                \\ 
\cmidrule{2-9}
                                    & \textbf{CSR} & $\boldsymbol{\epsilon = 10^{-8}}$ & $\boldsymbol{\epsilon = 10^{-5}}$ & $\boldsymbol{\epsilon = 10^{-3}}$ & \textbf{CSR} & $\boldsymbol{\epsilon = 10^{-8}}$ & $\boldsymbol{\epsilon = 10^{-5}}$ & $\boldsymbol{\epsilon = 10^{-3}}$  \\ 
\hline\hline
\multicolumn{9}{c}{Homogenized Square}                                                                                                                                                                                                                                                     \\ 
\midrule
$\mathcal{H}$                       & 1.000        & 1.778                             & 1.778                             & 1.778                             & 1.000        & 1.005                             & 1.005                             & 1.005                              \\
$\mathcal{S}$                       & 2.000        & 1.882                             & 1.882                             & 1.882                             & 1.268        & 1.012                             & 1.012                             & 1.012                              \\
$\mathcal{B}_{\text{out}}$          & 1.000        & 1.825                             & 1.825                             & 1.825                             & 1.000        & 1.029                             & 1.029                             & 1.029                              \\ 
\midrule
\multicolumn{9}{c}{Homogenized Circle}                                                                                                                                                                                                                                                     \\ 
\midrule
$\mathcal{H}$                       & 1.000        & 1.984                             & 1.969                             & 1.933                             & 1.000        & 1.091                             & 1.051                             & 1.023                              \\
$\mathcal{S}$                       & 2.000        & 1.989                             & 1.979                             & 1.954                             & 1.268        & 1.134                             & 1.074                             & 1.034                              \\
$\mathcal{B}_{\text{out}}$          & 1.000        & 0.809                             & 1.000                             & 1.400                             & 1.000        & 0.510                             & 0.558                             & 1.025                              \\
\bottomrule
\end{tabular}
\end{table}

\par \Cref{fig:fixed_square_and_circle_comp_p2} shows the CR as a function of angular resolution for the square and circle for $p_{\hat x} = p_{\hat y} = 2$. For the square, all three operators compress more in the TT format than CSR, and truncation tolerance does not affect compression. Assuming the first and last three points follow a power law $\text{CR}\propto N_{\Omega}^\alpha$, we fit the following exponents listed in \cref{tbl:fixed_square_and_circle_comp_p2_scale}. All three operators in TT format begin at $\alpha\approx 2$ and approach linear ($\alpha = 1$) for $N_\Omega \gg 1$. At low angular discretizations, the spatial cores consume a larger fraction of the TT's total memory footprint. In contrast, at higher-fidelity angular discretizations, the spatial cores occupy a negligible fraction of the memory footprint, resulting in linear scaling. Excluding the boundary operator, this carries over into the homogenized circle. The CSR scattering operator also exhibits a transition from superlinear to linear behavior. This is due to the treatment of the angular components of $\mathcal{S}^{\text{CSR}}$, which is functionally the same as TT, as mentioned in \cref{sec:numerical_scheme}. The interior operators of the circle in \cref{fig:fixed_circle_comp_p2_H,fig:fixed_circle_comp_p2_S} follow similarly to their counterparts in the square case; however, they show increased compression with increased truncation tolerance. Due to the higher coupling of angle and space in the outflow boundary operator, \cref{fig:fixed_circle_comp_p2_B_out} shows CSR outperforming TT for $\epsilon\in\{10^{-8}, 10^{-5}\}$. Moving to higher-order basis-function polynomial degrees follows the same trends shown here, with TT compression relative to CSR increasing as the number of nonzeros per degree of freedom increases. 

\begin{figure}
\centering
\begin{subfigure}{0.47\textwidth}
    \includegraphics[width=\textwidth, trim=0.5cm 0.5cm 0.5cm 0.5cm, clip]{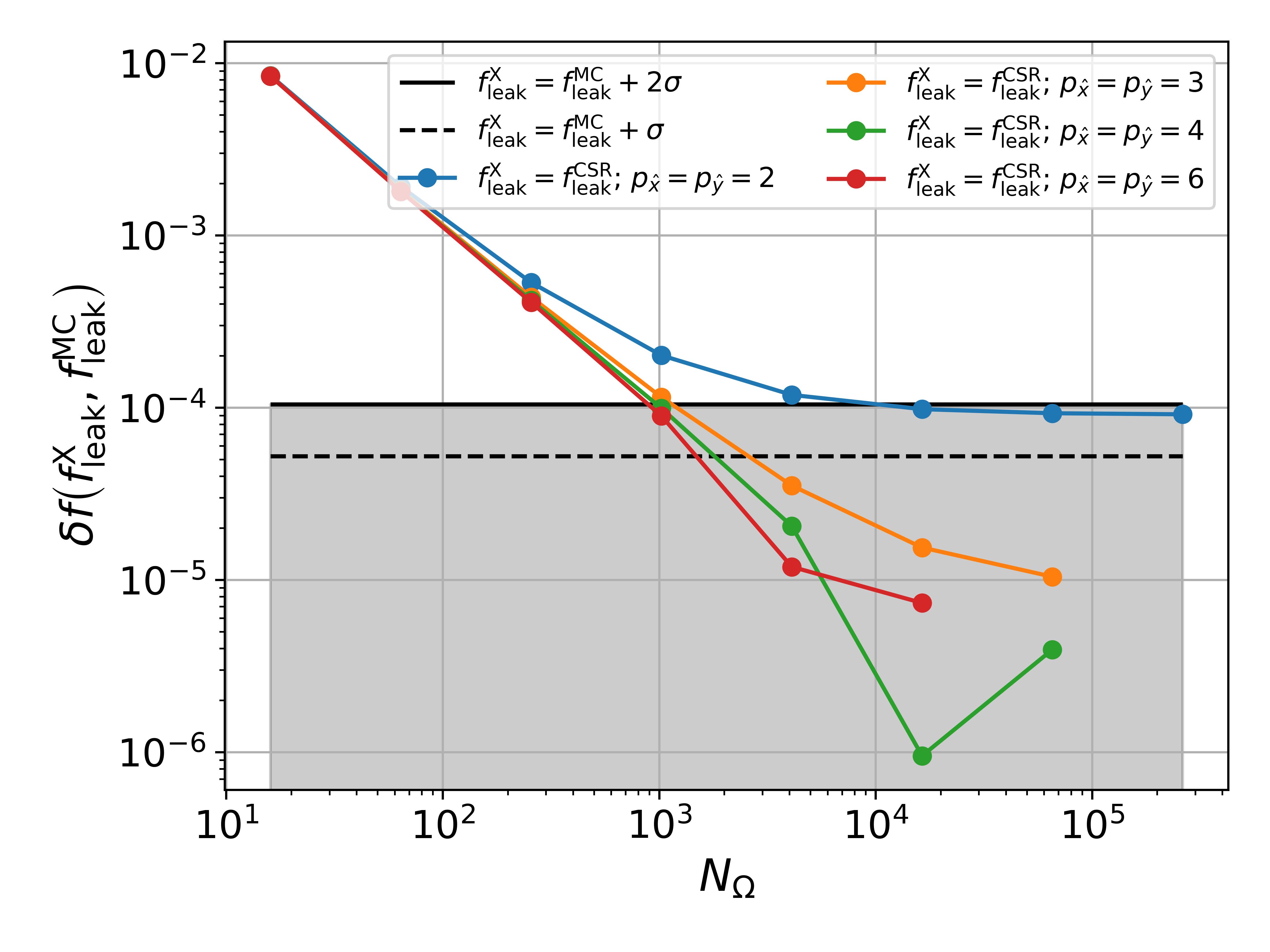}
    \caption{}
    \label{fig:fixed_square_leakage}
\end{subfigure}
\begin{subfigure}{0.47\textwidth}
    \includegraphics[width=\textwidth, trim=0.5cm 0.5cm 0.5cm 0.5cm, clip]{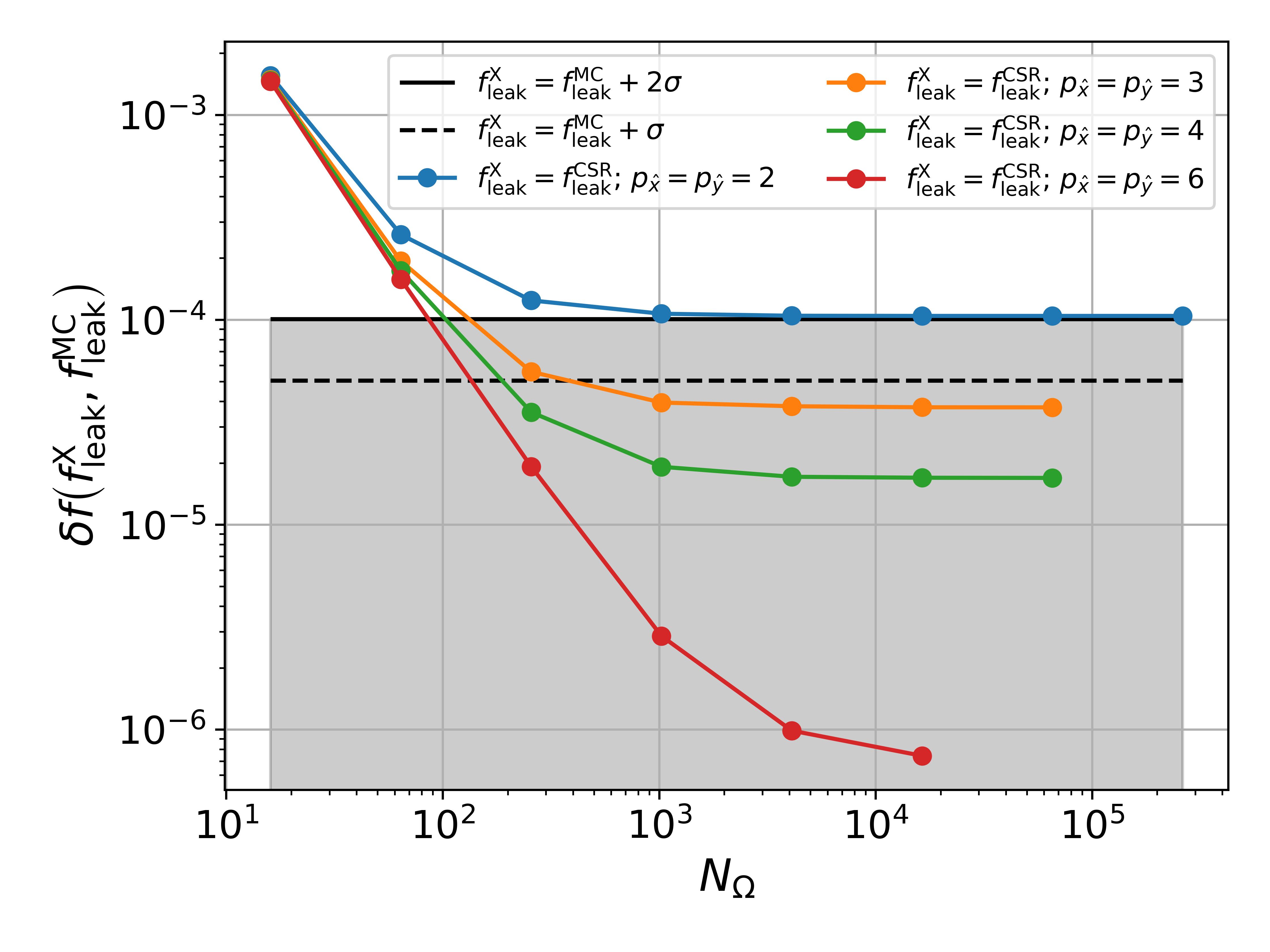}
    \caption{}
    \label{fig:fixed_circle_leakage}
\end{subfigure}
        
\caption{Leakage fraction relative error (\cref{eq:leakage_fraction_relative_error}) for the (a) square and (b) circle fixed source problems with increasing angular resolution. In both figures, we show $\sigma$ and $2\sigma$ from the Monte Carlo (MC) reference solution and CSR (\cref{eq:cases_csr}) for $p_{\hat x} = p_{\hat y}\in\{2, 3, 4, 6\}$.}
\label{fig:fixed_square_and_circle_leakage}
\end{figure}

\begin{figure}
\centering
\begin{subfigure}{0.47\textwidth}
    \includegraphics[width=\textwidth, trim=0.5cm 0.5cm 0.5cm 0.5cm, clip]{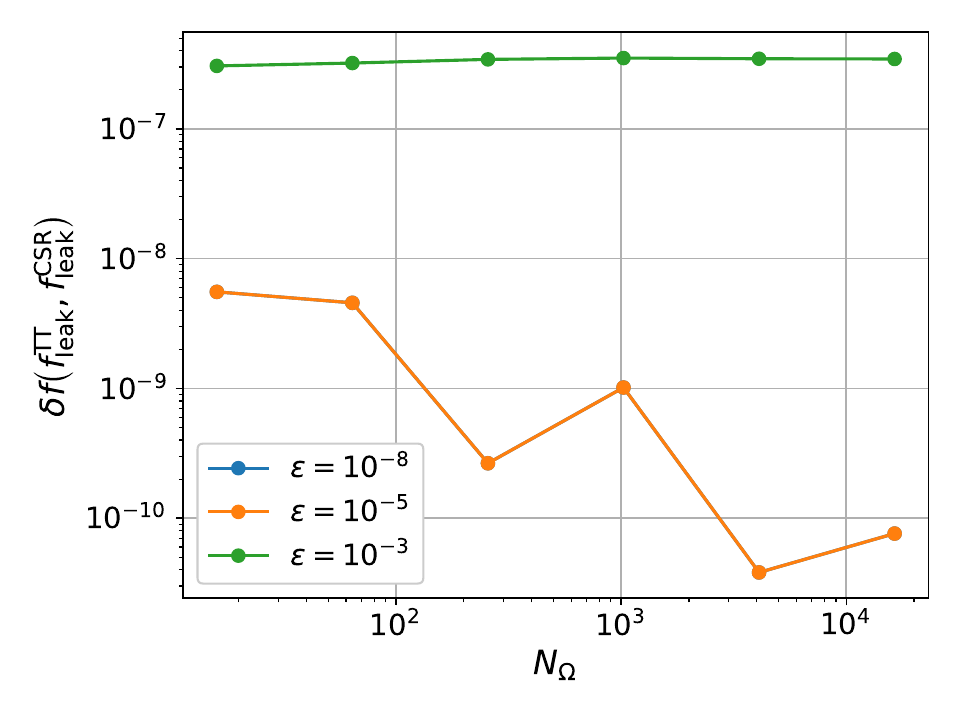}
    \caption{}
    \label{fig:fixed_square_leakage_TT}
\end{subfigure}
\begin{subfigure}{0.47\textwidth}
    \includegraphics[width=\textwidth, trim=0.5cm 0.5cm 0.5cm 0.5cm, clip]{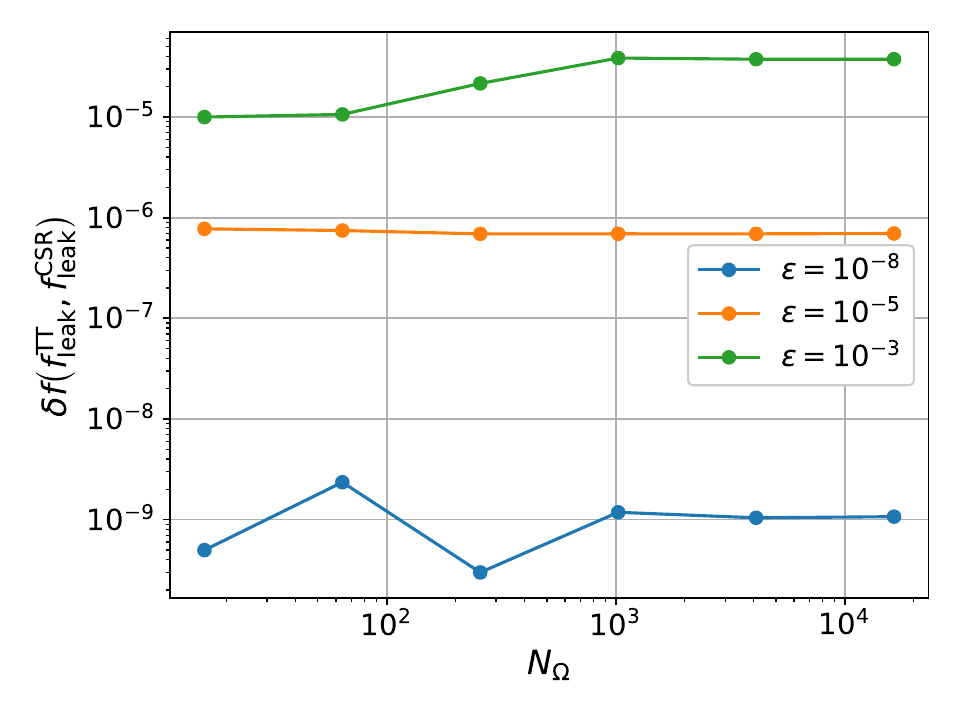}
    \caption{}
    \label{fig:fixed_circle_leakage_TT}
\end{subfigure}
\begin{subfigure}{0.47\textwidth}
    \includegraphics[width=\textwidth, trim=0.5cm 0.5cm 0.5cm 0.5cm, clip]{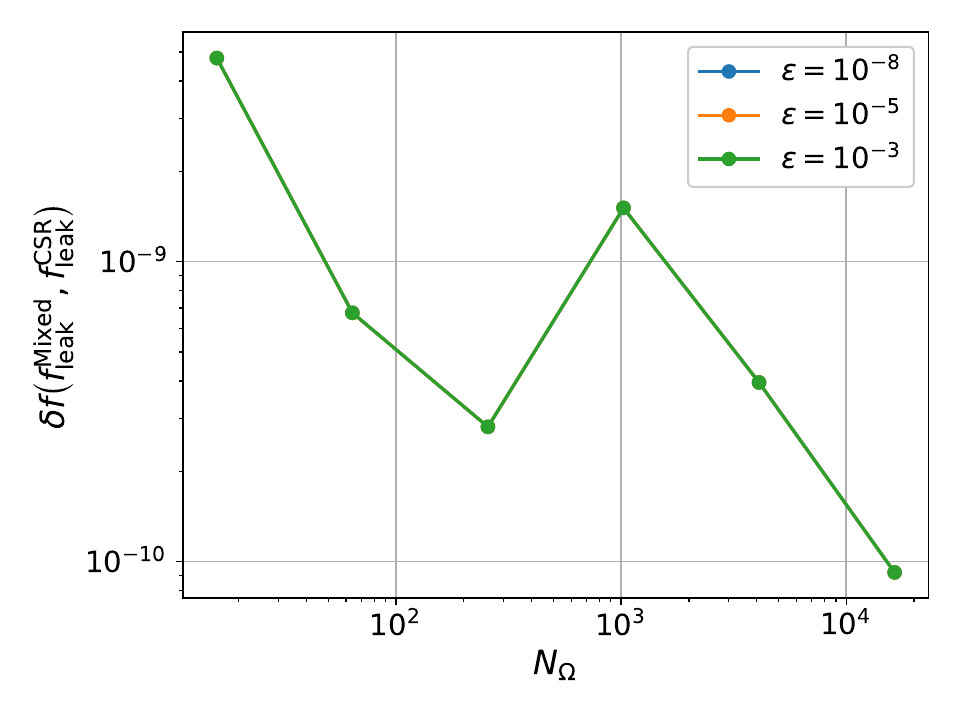}
    \caption{}
    \label{fig:fixed_square_leakage_Mixed}
\end{subfigure}
\begin{subfigure}{0.47\textwidth}
    \includegraphics[width=\textwidth, trim=0.5cm 0.5cm 0.5cm 0.5cm, clip]{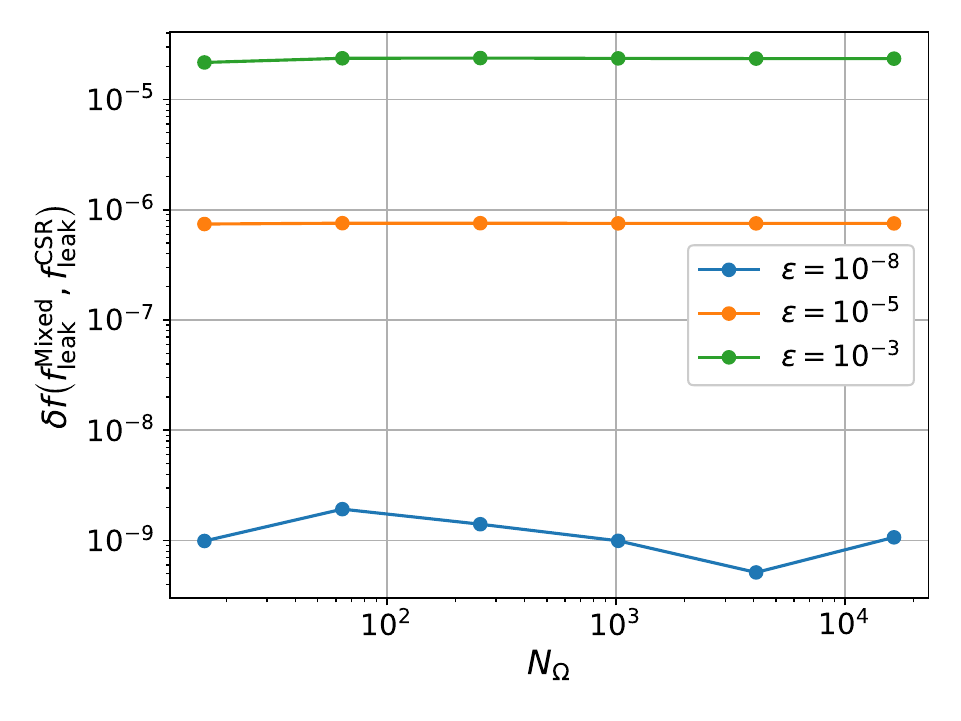}
    \caption{}
    \label{fig:fixed_circle_leakage_Mixed}
\end{subfigure}
        
\caption{Leakage fraction relative error (\cref{eq:leakage_fraction_relative_error}) with respect to CSR (\cref{eq:cases_csr}) for increasing angular resolution for (a, b) TT (\cref{eq:cases_tt}) and (c, d) Mixed (\cref{eq:cases_mixed}) representations at $\epsilon\in\{10^{-8}, 10^{-5}, 10^{-3}\}$ and $p_{\hat x} = p_{\hat y} = 6$ for the homogenized (left) square and (right) circle fixed source problems. Note the $\epsilon\in\{10^{-8}, 10^{-5}\}$ share the same line in (a) while $\epsilon\in\{10^{-8}, 10^{-5}, 10^{-3}\}$ share the same lines in (c). The TT (rounded) (\cref{eq:cases_tt_rounded}) and Mixed (rounded) (\cref{eq:cases_mixed_rounded}) follow a similar pattern.}
\label{fig:fixed_square_and_circle_leakage_to_csr}
\end{figure}

\subparagraph{Leakage Fraction Error}
In \cref{fig:fixed_square_and_circle_leakage}, we show the leakage fraction relative error as a function of angular resolution. In both problems, the $p_{\hat x} = p_{\hat y} = 2$ solution plateaus around the MC reference's $2\sigma$. The $p_{\hat x} = p_{\hat y} \in\{3, 4, 6\}$ get well within $\sigma$ for increased $N_\Omega$; however, the leakage fraction is within $\sigma$ of the MC reference for $N_\Omega \ge 1024$ for the circle while $N_\Omega \ge 4096$ for the square likely due to pronounced ray-effects at flat boundaries and corners. 

\par \Cref{fig:fixed_square_and_circle_leakage_to_csr} shows the leakage fraction relative error of TT and Mixed to CSR for $\epsilon\in\{10^{-8}, 10^{-5}, 10^{-3}\}$ and $p_{\hat x} = p_{\hat y} = 6$. The Mixed shows no difference in relative error with respect to the truncation tolerance for the square, whereas TT does. The $\epsilon = 10^{-3}$ tolerance for the square maintain $10^{-7} < \delta f\left(f_{\text{leak}}^{\text{TT}}, f_{\text{leak}}^{\text{CSR}}\right) < 10^{-6}$ in \cref{fig:fixed_square_leakage_TT}. The other curves for the square demonstrate a decrease in relative error except for $N_\Omega = 1024$, where the relative error increases from $N_\Omega = 256$. For the circle, we observe a pronounced difference in error across truncation tolerances: relative error increases with tolerance. The relative error remains largely consistent across angular discretizations with $\epsilon=10^{-8}$ around $\delta f\left(f_{\text{leak}}^{\text{X}}, f_{\text{leak}}^{\text{CSR}}\right)\approx 10^{-9}$, $\epsilon = 10^{-5}$ around $\delta f\left(f_{\text{leak}}^{\text{X}}, f_{\text{leak}}^{\text{CSR}}\right)\approx 10^{-6}$, and $\epsilon = 10^{-3}$ around $\delta f\left(f_{\text{leak}}^{\text{X}}, f_{\text{leak}}^{\text{CSR}}\right)\approx 10^{-5}$ for all cases. For $p_{\hat x} = p_{\hat y} \in\{2, 3, 4\}$, we note that all cases have leakage fraction relative errors less than $10^{-6}$ for the square and $10^{-3}$ for the circle. The results follow similarly for their rounded counterparts.

\begin{figure}
    \centering
    \includegraphics[width=0.7\linewidth, trim=0.5cm 0.5cm 0.5cm 0.5cm, clip]{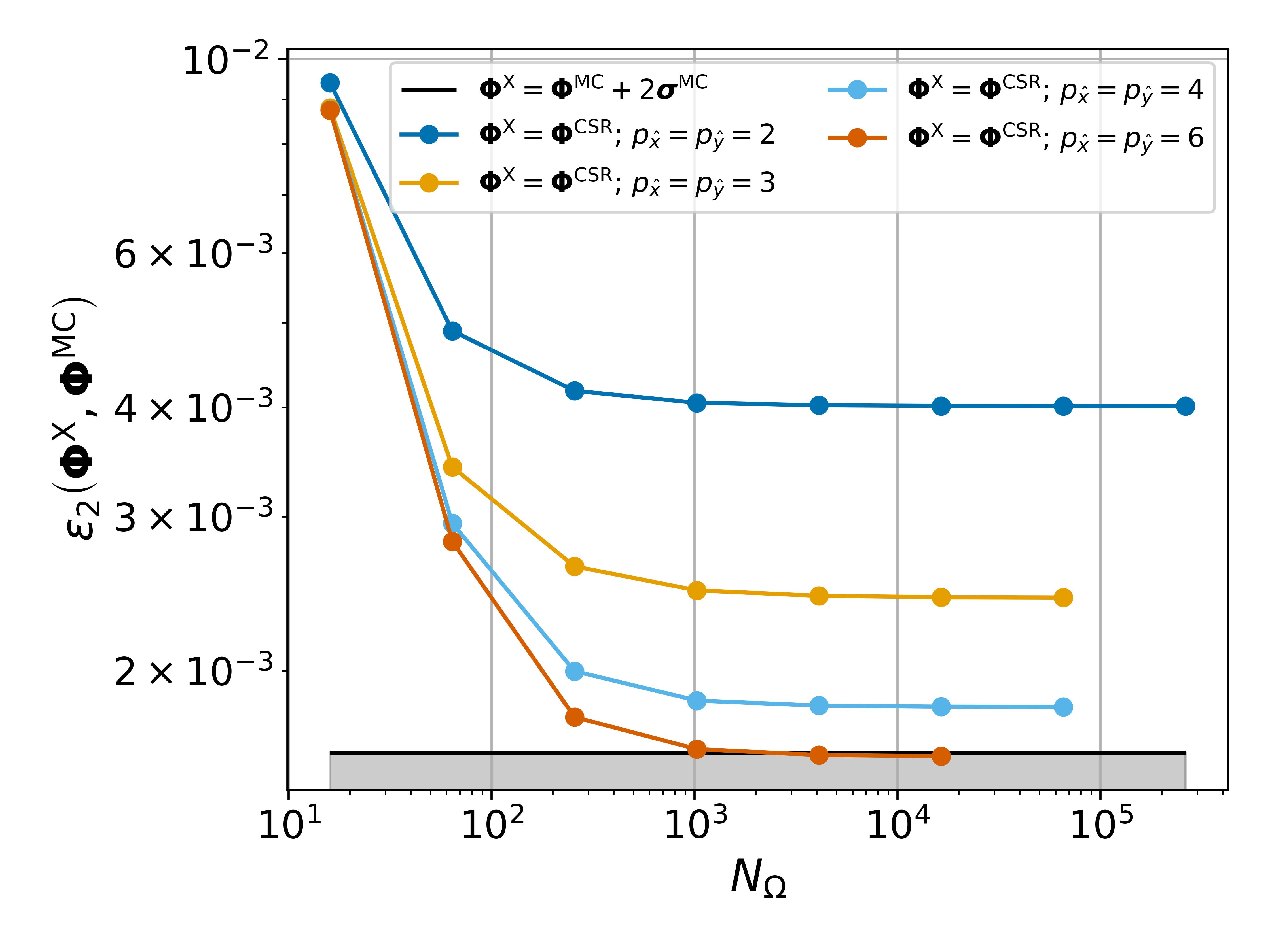}
    \caption{Scalar flux $L_2$ error (\cref{eq:eigenvector_error}) of CSR (\cref{eq:cases_csr}) to the OpenMC reference solution for the fixed source homogeneous square. We show CSR for $p_{\hat x} = p_{\hat y} \in \{2, 3, 4, 6\}$ and $L_2$ error for $2\boldsymbol{\sigma}^{\text{MC}}$ as a reference.}
    \label{fig:fixed_square_fluxl2error}
\end{figure}

\begin{figure}
\centering
\begin{subfigure}{0.47\textwidth}
    \includegraphics[width=\textwidth, trim=0.5cm 0.5cm 0.5cm 0.5cm, clip]{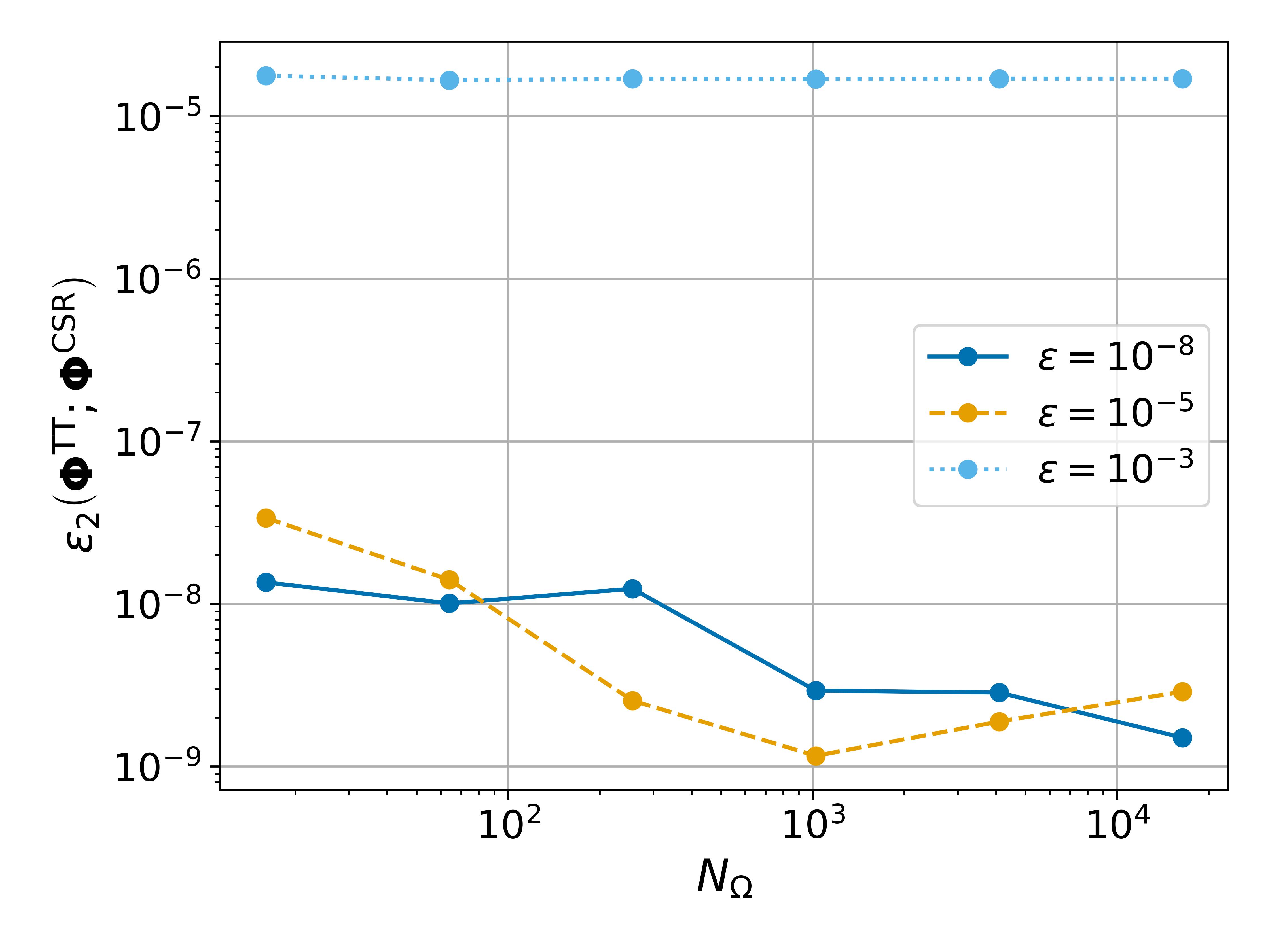}
    \caption{}
    \label{fig:fixed_square_fluxerror_TT}
\end{subfigure}
\begin{subfigure}{0.47\textwidth}
    \includegraphics[width=\textwidth, trim=0.5cm 0.5cm 0.5cm 0.5cm, clip]{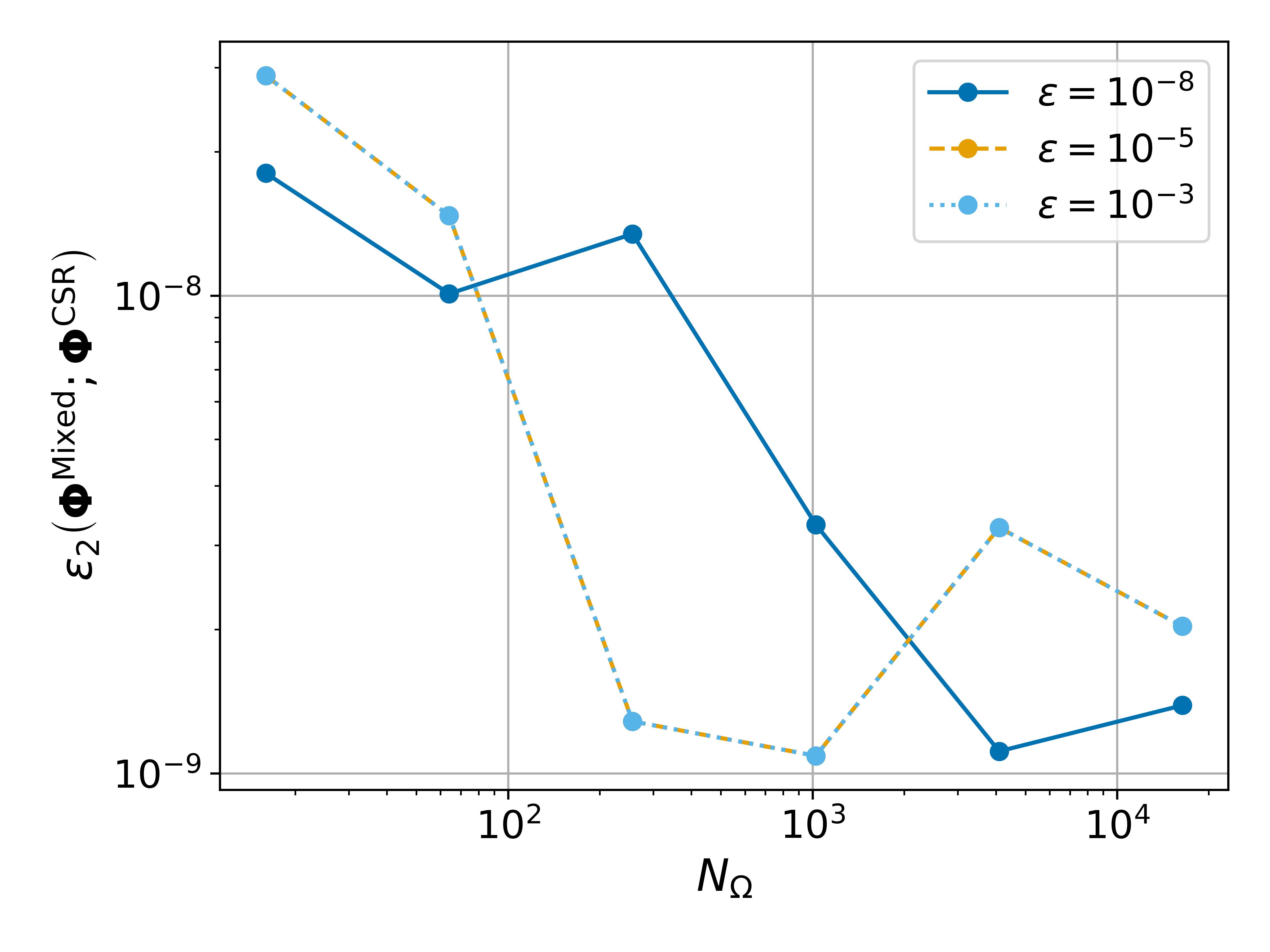}
    \caption{}
    \label{fig:fixed_square_fluxerror_Mixed}
\end{subfigure}
        
\caption{Scalar flux $L_2$ error (\cref{eq:eigenvector_error}) of (a) TT (\cref{eq:cases_tt}) and (b) Mixed (\cref{eq:cases_mixed}) to CSR (\cref{eq:cases_csr}) for the fixed source homogeneous square with $p_{\hat x} = p_{\hat y} = 6$. The TT (rounded) (\cref{eq:cases_tt_rounded}) and Mixed (rounded) (\cref{eq:cases_mixed_rounded}) cases, as well as $p_{\hat x} = p_{\hat y} \in\{2, 3, 4\}$ follow similarly.}
\label{fig:fixed_square_fluxerror_to_csr}
\end{figure}

\subparagraph{Scalar Flux Error}
In \cref{fig:fixed_square_fluxl2error} we show the $L_2$ error of the scalar flux for the CSR solution to the reference MC solution for the fixed source homogenized square problem. For reference, we also provide the $L_2$ error for $\epsilon_2\left(\mathbf{\Phi}^{\text{MC}} + 2\boldsymbol{\sigma}^{\text{MC}}, \mathbf{\Phi}^{\text{MC}}\right)$. In all these solutions, the errors on a cell-by-cell basis may be statistically significant; for example, CSR with $p_{\hat x} = p_{\hat y} = 6$ and $N_\Omega = 16384$ ordinates has a minimum $z$-score of $3.073\times 10^{-5}~\sigma$, maximum of $13.565~\sigma$, mean of $1.691~\sigma$, median of $1.129~\sigma$, $\text{Q}_1$ of $0.524~\sigma$, and $\text{Q}_2$ of $2.113~\sigma$. \Cref{fig:fixed_square_fluxerror_to_csr} shows the $L_2$ error of TT and Mixed to CSR for $p_{\hat x} = p_{\hat y} = 6$. We observe an overall decrease in $L_2$ error with increasing angular discretization, though with some oscillatory behavior. \Cref{fig:fixed_square_fluxerror_TT,fig:fixed_square_fluxerror_TT} follows \cref{fig:fixed_square_leakage_TT} where $\epsilon = 10^{-3}$ is three orders of magnitude off $\epsilon\in\{10^{-8}, 10^{-5}\}$. The $L_2$ error for the other curves decreases from $\sim10^{-7}$ to $\sim10^{-9}$ from low to high fidelity angular discretization. These trends match for $p_{\hat x}=p_{\hat y} \in\{2, 3, 4\}$ and the TT (rounded) and Mixed (rounded) cases; however, $\epsilon = 10^{-3}$ is not several orders of magnitude off $\epsilon\in\{10^{-8}, 10^{-5}\}$ for TT and TT (rounded). 

\begin{figure}
\centering
\begin{subfigure}{0.47\textwidth}
    \includegraphics[width=\textwidth, trim=0.5cm 0.5cm 0.5cm 0.5cm, clip]{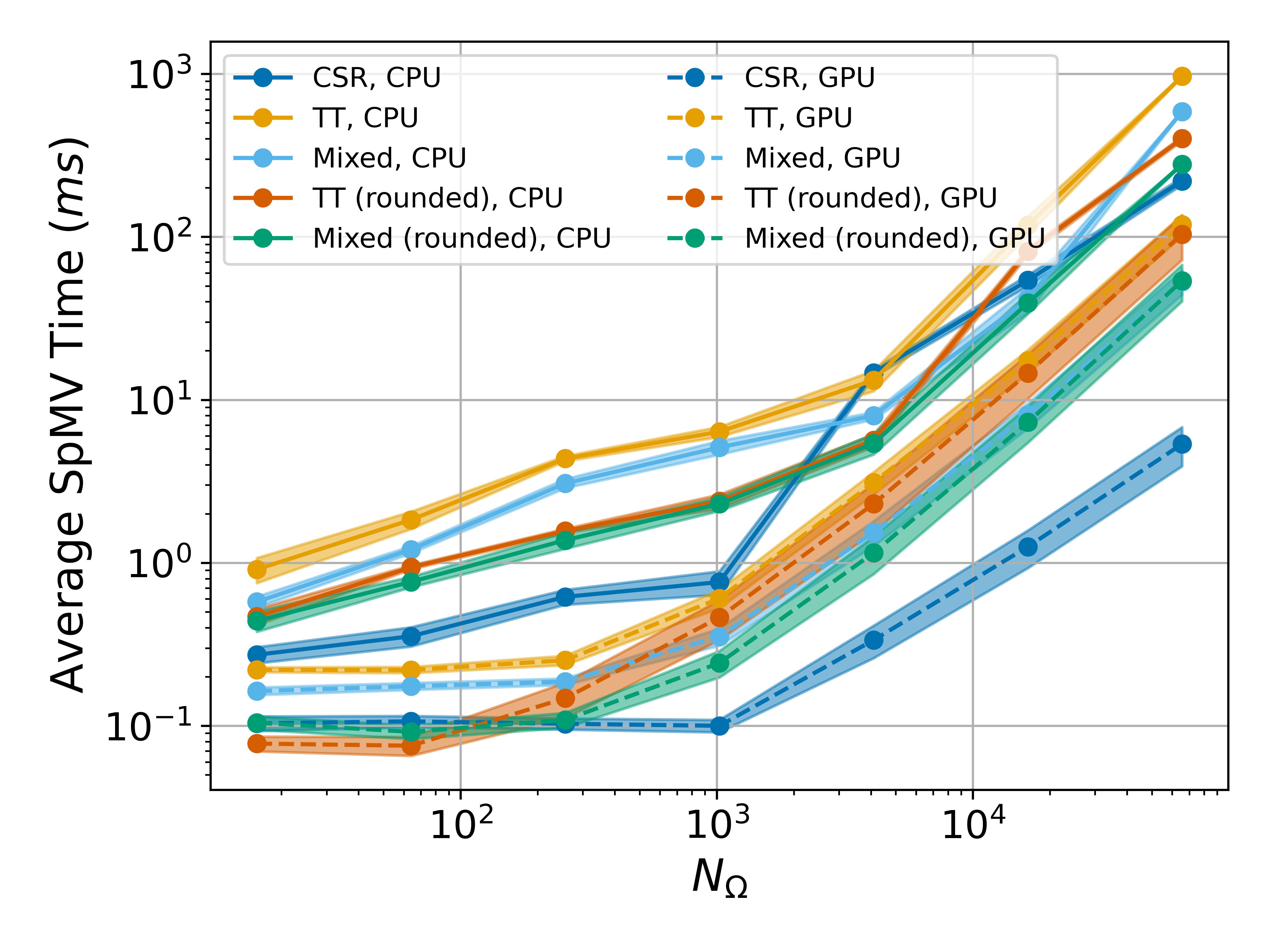}
    \caption{$p_{\hat x} = p_{\hat y} = 3$; $\epsilon = 10^{-8}$}
    \label{fig:fixed_square_matvec_p3}
\end{subfigure}
\begin{subfigure}{0.47\textwidth}
    \includegraphics[width=\textwidth, trim=0.5cm 0.5cm 0.5cm 0.5cm, clip]{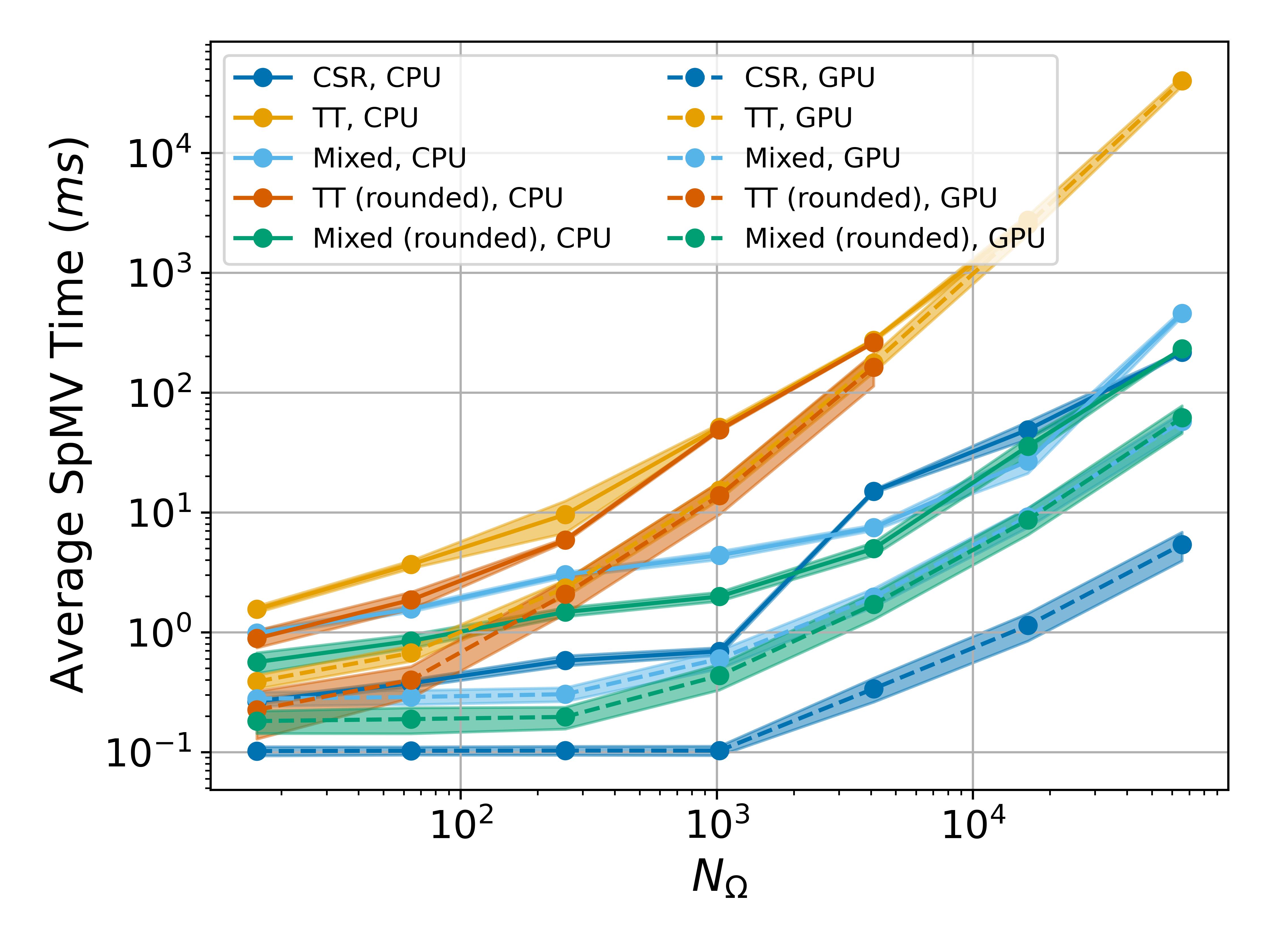}
    \caption{$p_{\hat x} = p_{\hat y} = 3$; $\epsilon = 10^{-8}$}
    \label{fig:fixed_circle_matvec_p3}
\end{subfigure}
\begin{subfigure}{0.47\textwidth}
    \includegraphics[width=\textwidth, trim=0.5cm 0.5cm 0.5cm 0.5cm, clip]{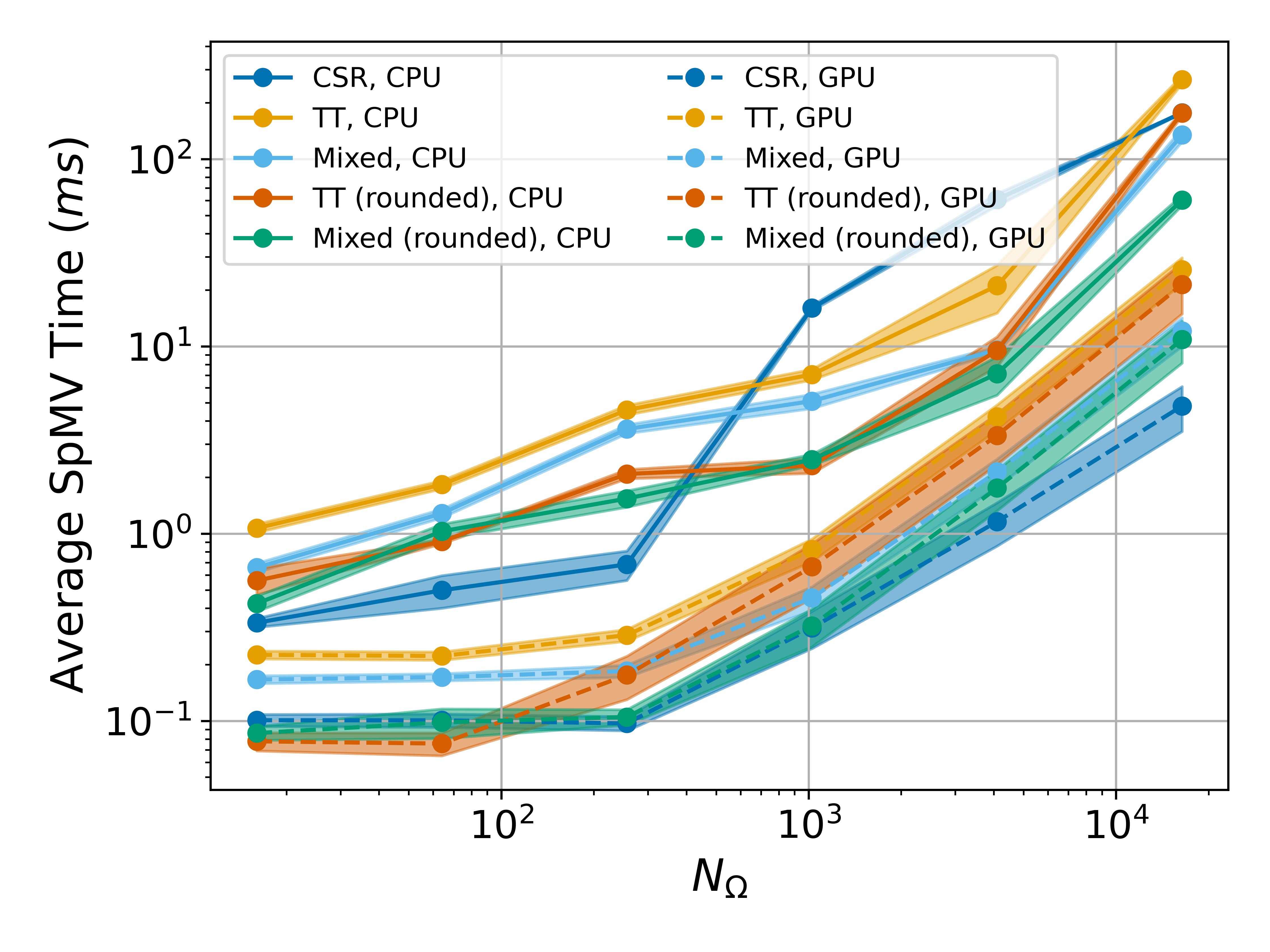}
    \caption{$p_{\hat x} = p_{\hat y} = 6$; $\epsilon = 10^{-8}$}
    \label{fig:fixed_square_matvec_p6}
\end{subfigure}
\begin{subfigure}{0.47\textwidth}
    \includegraphics[width=\textwidth, trim=0.5cm 0.5cm 0.5cm 0.5cm, clip]{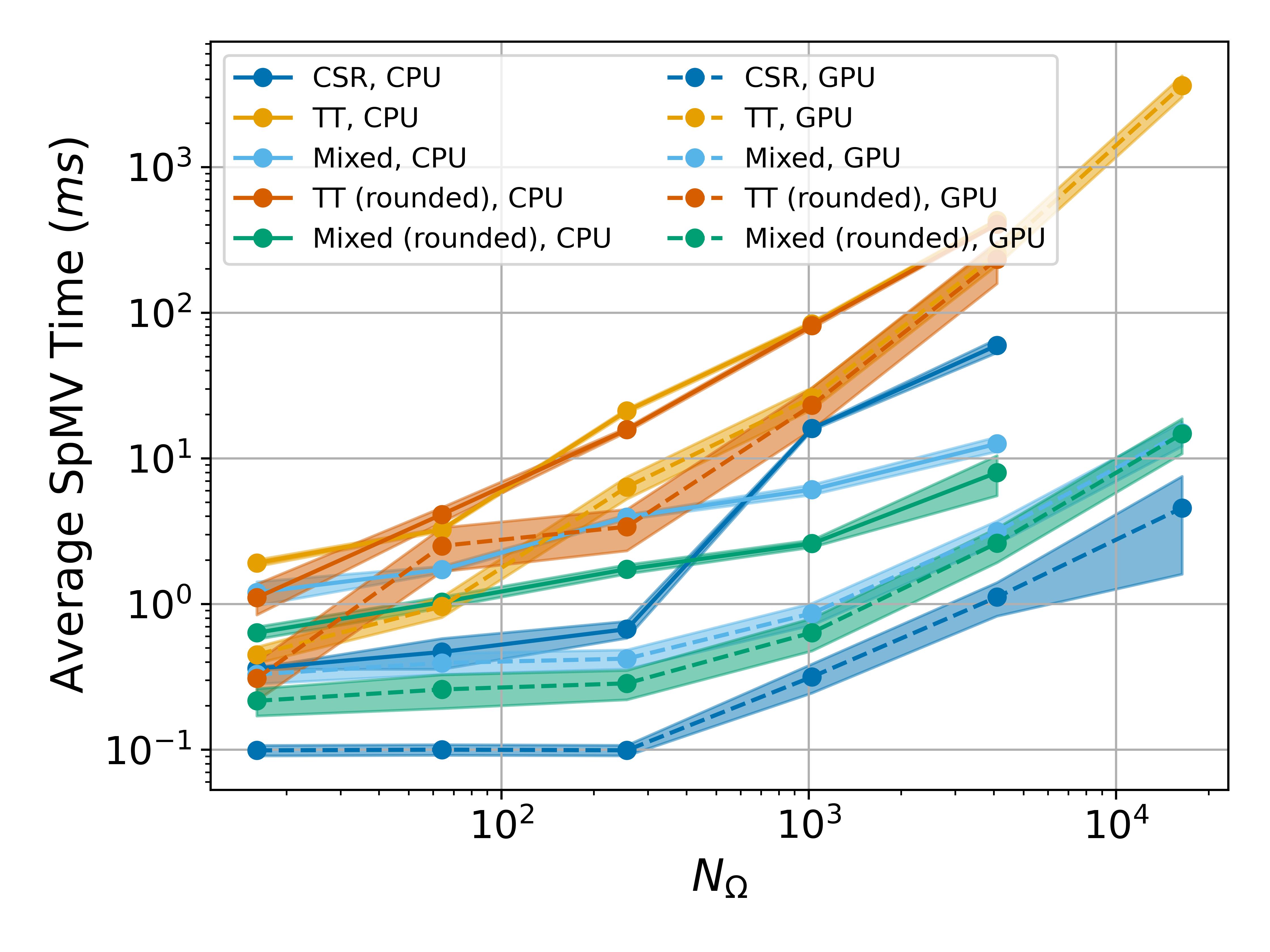}
    \caption{$p_{\hat x} = p_{\hat y} = 6$; $\epsilon = 10^{-8}$}
    \label{fig:fixed_circle_matvec_p6}
\end{subfigure}
        
\caption{Average sparse operator-vector product (SpMV) versus angular discretization for $p_{\hat x} = p_{\hat y}\in\{3, 6\}$ for the fixed source (a, c) square and (b, d) circle problems. We applied the left-hand side of \cref{eq:cases} to a random tensor 1000 times and show the average and one standard deviation of the time. Solid lines represent the CPU run times, while the dashed lines represent GPU run times.}
\label{fig:fixed_square_and_circle_matvec}
\end{figure}

\subparagraph{Operator-Vector Product Scaling}
\Cref{fig:fixed_square_and_circle_matvec} shows the average time to apply the left-hand side of \cref{eq:cases} to a random tensor computed over $1000$ iterations for varying angular discretization. We show the fixed-source square (left) and circular (right) results for $p_{\hat x} = p_{\hat y}\in\{3, 6\}$ and $\epsilon = 10^{-8}$, the least rounded TTs considered in this work. \Cref{fig:fixed_square_matvec_p3,fig:fixed_square_matvec_p6} show the GPU outperforming CPU for all cases, with the best for $N_\Omega \ge 256$ being the GPU with the CSR format, followed by the Mixed and Mixed (rounded) cases. We observed the GPU's superior performance on other GPUs, including an NVIDIA GeForce RTX 4090 and an RTX 6000 Ada Generation. The TT operator-vector products are computed via a sequence of reshapes and matrix-matrix products, which are highly optimized and massively parallel on vectorized hardware. The TT (rounded) case shows better performance for $N_\Omega \le 64$. Between polynomial degrees $3$ and $6$, the Mixed and Mixed (rounded) cases approach the CSR run time. The $2$- and $4$-degree times also show that the Mixed and Mixed (rounded) cases converge to the CSR run time as the polynomial degree increases for increasing angular discretization. This is also supported in the circular problem figures, \cref{fig:fixed_circle_matvec_p3,fig:fixed_circle_matvec_p6}. However, the GPU TT and TT (rounded) cases have longer run times than the CPU CSR, Mixed, and Mixed (rounded) cases. This indicates that the increasing ranks of the boundary operators, driven by high curvilinearity, dominate the run times for the TT and TT (rounded) cases. Further rounding to $\epsilon\in\{10^{-5}, 10^{-3}\}$ reduces TT-rank in the operators of the circle and thus reduces operator-vector run times, but still, as seen by the low rank operators in the square, does not surpass CSR for higher fidelity discretizations. These observations carry into GMRES.

\begin{figure}
\centering
\begin{subfigure}{0.47\textwidth}
    \includegraphics[width=\textwidth, trim=0.5cm 0.5cm 0.5cm 0.5cm, clip]{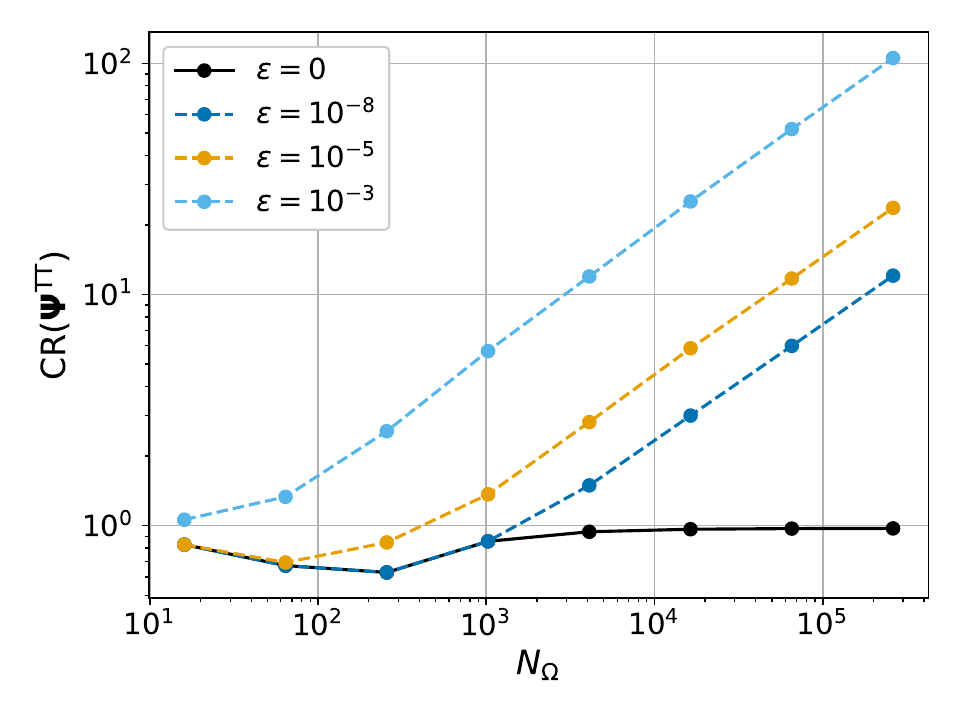}
    \caption{$p_{\hat x} = p_{\hat y} = 2$}
    \label{fig:fixed_square_compression_psi}
\end{subfigure}
\begin{subfigure}{0.47\textwidth}
    \includegraphics[width=\textwidth, trim=0.5cm 0.5cm 0.5cm 0.5cm, clip]{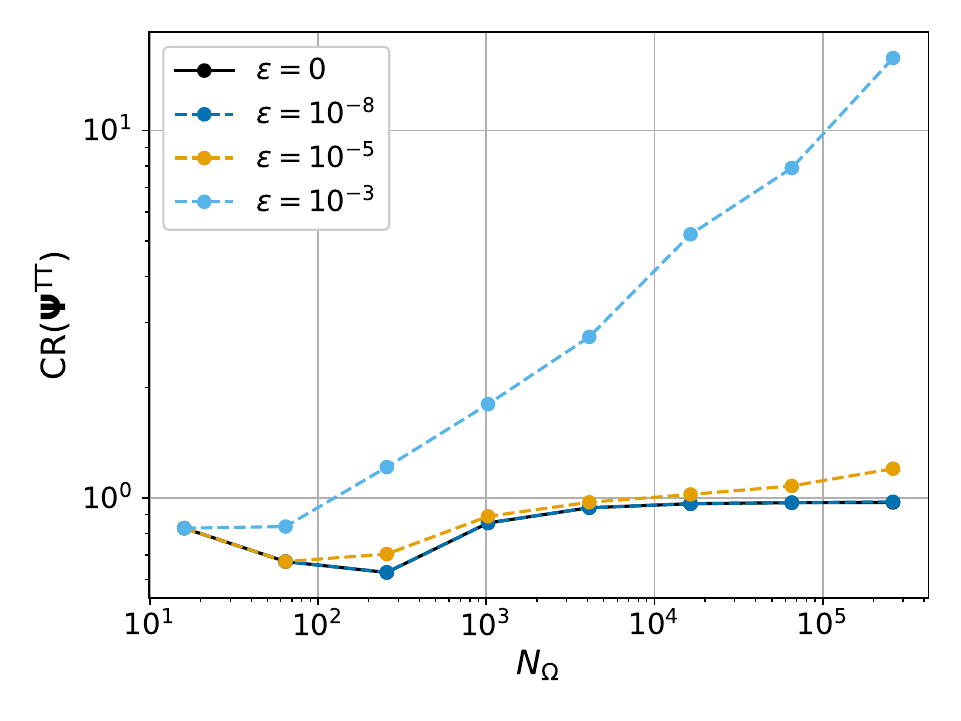}
    \caption{$p_{\hat x} = p_{\hat y} = 2$}
    \label{fig:fixed_circle_compression_psi}
\end{subfigure}
        
\caption{Compression ratio (CR) of the angular flux in tensor train (TT) format for increasing angular discretization. We present solutions to the fixed-source (a) square and (b) circular problems for $p_{\hat x} = p_{\hat y} = 2$ and varying truncation tolerance. These trends align with an increased degree of the basis functions.}
\label{fig:fixed_square_and_circle_compression_psi}
\end{figure}

\subparagraph{Angular Flux Compression in the TT Format}
We show the compression ratios for the angular flux decomposed into the TT format for the square and circle in \cref{fig:fixed_square_and_circle_compression_psi}. Across all truncation tolerances, we observe compression for $N_\Omega \ge 4096$ ordinates in the square, with $\epsilon = 10^{-3}$ always compressible. However, we note for $N_\Omega \ge 256$ with $p_{\hat x} = p_{\hat y} = 2$ we observe $r_{\text{max}}(\mathbf{\Psi}^{TT}) \ge 140$ for $\epsilon=10^{-8}$, $r_{\text{max}}(\mathbf{\Psi}^{TT}) \ge 110$ for $\epsilon=10^{-5}$, and $r_{\text{max}}(\mathbf{\Psi}^{TT}) \ge 40$ for $\epsilon=10^{-3}$. These ranks increase slightly with higher polynomial degrees. For the circular problem, as shown in \cref{fig:fixed_circle_compression_psi,fig:fixed_circle_compression_psi,fig:fixed_circle_compression_psi,fig:fixed_circle_compression_psi}, we see $\epsilon\in\{10^{-8}, 10^{-5}\}$ remain either completely incompressible in the TT format or slightly compressible at fine angular discretizations. This results in maximum ranks exceeding 1000, degrading the performance of TT adaptive rank solvers such as AMEn. 

\paragraph{Mesh Resolution Study}
Like the angular resolution study, below we present a discussion of operator compression in CSR and TT format; the effect of basis function polynomial degree and truncation tolerance on leakage fraction and scalar flux error; the computational scaling of CSR and TT products; and the compressability of the angular flux, all for varying mesh resolution. The main observations include:
\begin{itemize}
    \item Regular meshes again show no dependence of operator rank on mesh resolution, while meshes with curvilinear boundaries show variable operator rank with mesh resolution.
    \item Interior operators of meshes with curvilinear boundaries show a peak rank at coarser mesh discretizations, followed by a consistent decrease in rank as the mesh becomes increasingly regular locally. However, the boundary operator ranks increase with mesh resolution, but the rate of increase eventually plateaus, unlike the rapid growth observed with increasing angular resolution.
    \item TT again proves superior over CSR in interior operator compression, while boundary operators are best suited for CSR format. 
    \item Likely due to ray-effects for $N_\Omega = 256$, the square never sees a leakage fraction within $2\sigma$ of the MC reference solution while all basis function degrees get within $\sigma$ for the circle. 
    \item Coarser mesh discretizations see a larger benefit from increasing the basis function polynomial degree.
    \item The observations presented in the angular resolution study for the leakage fraction relative error of TT, TT (rounded), Mixed, and Mixed (rounded) to CSR apply for the mesh resolution study.
    \item For $N_\Omega = 256$ the scalar flux of the square never comes within $\epsilon_2\left(\mathbf{\Phi}^{\text{MC}} + 2\boldsymbol{\sigma}^{\text{MC}}, \mathbf{\Phi}^{\text{MC}}\right)$ with cell-by-cell scalar fluxes being potentially statistically significant. All polynomial degrees attain the same asymptotic error, indicating the angular discretization as the limiting factor for reducing the error to the reference MC solution. Again, coarser discretizations exhibit the largest improvement in error as the polynomial degree increases.
    \item The GPU again demonstrates superior performance over CPU in operator-vector products, with CSR consistently the fastest, followed by Mixed and Mixed (rounded). The increasing ranks of the outflow boundary operator make TT and TT (rounded) uncompetitive for patches with curvilinear boundaries. As in the angular resolution study, increasing the degree of the basis functions reduces the gap between CSR and Mixed/Mixed (rounded). For brevity, we do not explicitly show this result as it aligns closely with \cref{fig:fixed_square_and_circle_matvec}.
    \item The compressibility of the angular flux follows that observed in the angular resolution study, except the square with $\epsilon\in\{10^{-8}, 10^{-5}\}$ is far less or even incompressible for increasing mesh resolution.
    \item CR for the angular flux in TT format decreases with increasing NURBS basis function degree.
\end{itemize}
Consistent with the angular resolution study, TT offers better compression for interior operators as mesh resolution increases. For non-regular meshes, the mesh becomes increasingly regular with refinement, thereby decreasing the maximum rank of the interior operator. TT again falls short of CSR in time-to-solution, which is mitigated by leaving boundary operators in CSR format. 

\begin{figure}
\centering
\begin{subfigure}{0.47\textwidth}
    \includegraphics[width=\textwidth, trim=0.5cm 0.5cm 0.5cm 0.5cm, clip]{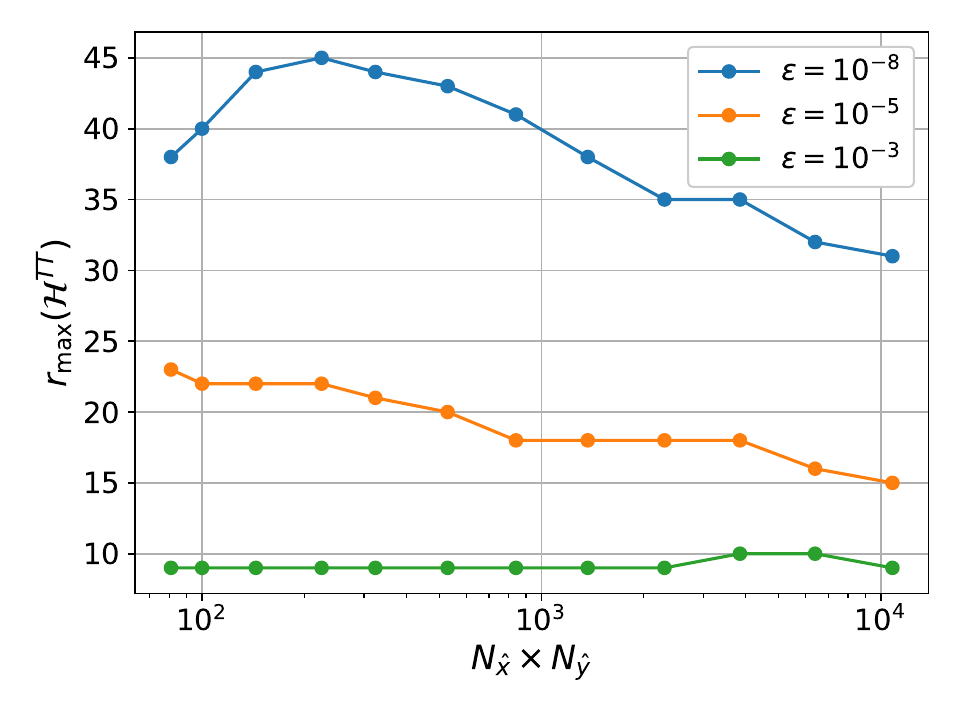}
    \caption{}
    \label{fig:fixed_circle_mesh_ranks_H}
\end{subfigure}
\begin{subfigure}{0.47\textwidth}
    \includegraphics[width=\textwidth, trim=0.5cm 0.5cm 0.5cm 0.5cm, clip]{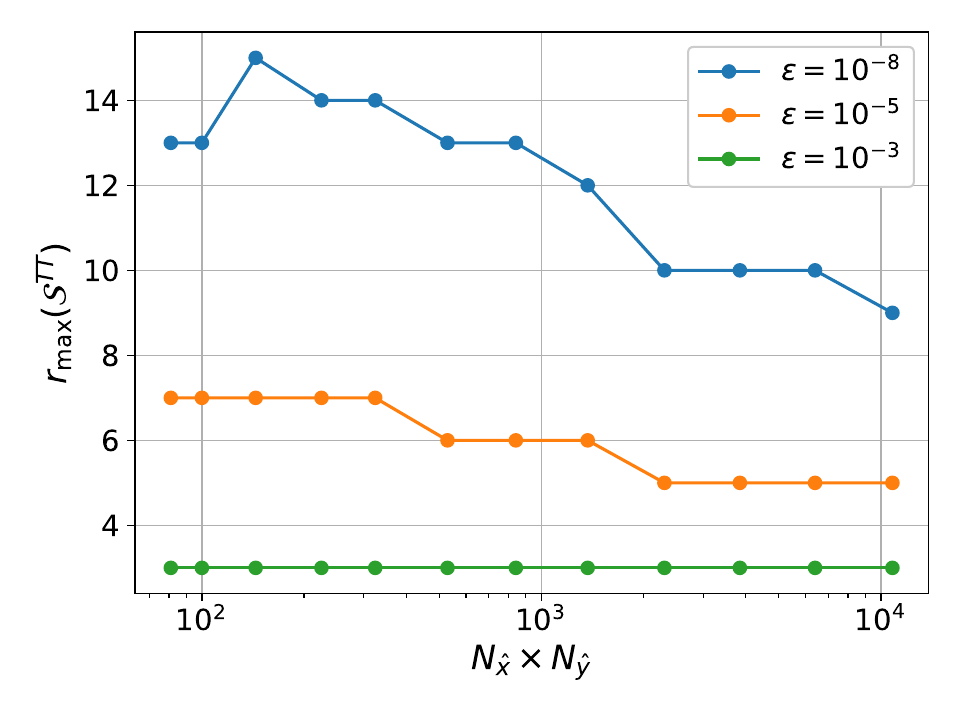}
    \caption{}
    \label{fig:fixed_circle_mesh_ranks_S}
\end{subfigure}
\begin{subfigure}{0.47\textwidth}
    \includegraphics[width=\textwidth, trim=0.5cm 0.5cm 0.5cm 0.5cm, clip]{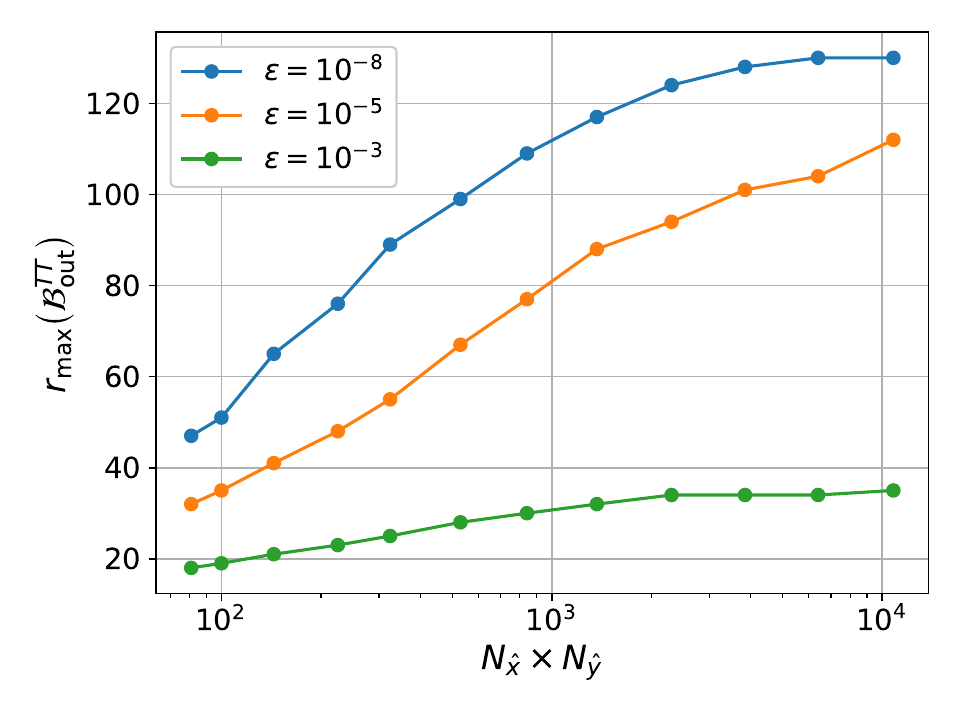}
    \caption{}
    \label{fig:fixed_circle_mesh_ranks_B_out}
\end{subfigure}
        
\caption{The maximum tensor train (TT) ranks versus the number of spatial degrees of freedom ($N_{\hat x}\times N_{\hat y}$) for the fixed source homogenized circle with $p_{\hat x} = p_{\hat y} = 4$ for the (a) streaming and collision, (b) scattering, and (c) outflow boundary operators.}
\label{fig:fixed_circle_mesh_ranks}
\end{figure}

\begin{figure}
\centering
\begin{subfigure}{0.47\textwidth}
    \includegraphics[width=\textwidth, trim=0.5cm 0.5cm 0.5cm 0.5cm, clip]{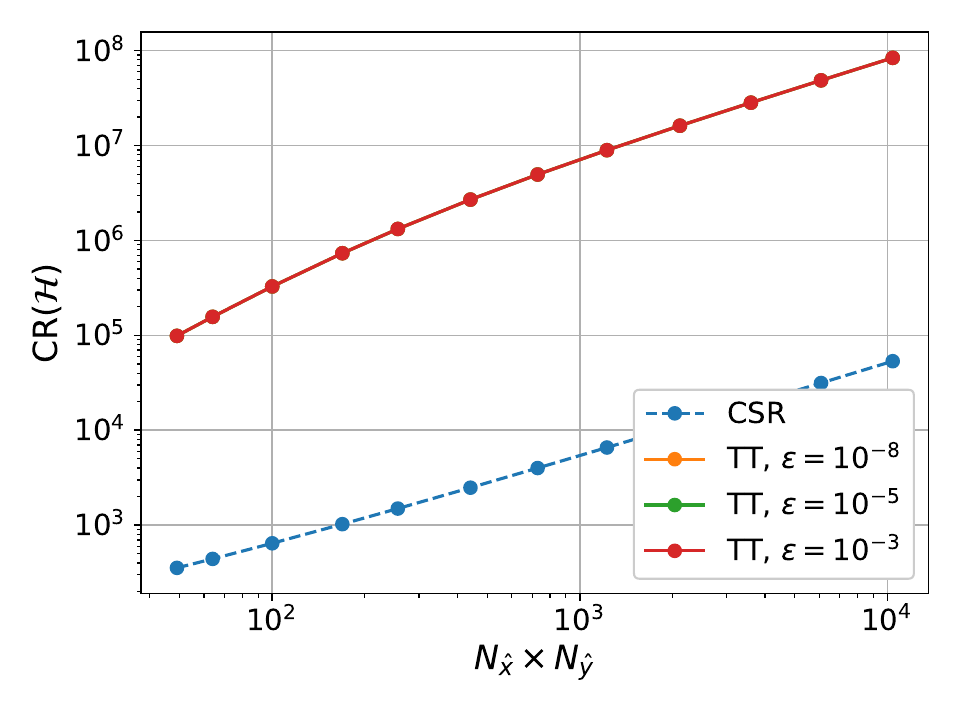}
    \caption{}
    \label{fig:fixed_square_mesh_comp_H}
\end{subfigure}
\begin{subfigure}{0.47\textwidth}
    \includegraphics[width=\textwidth, trim=0.5cm 0.5cm 0.5cm 0.5cm, clip]{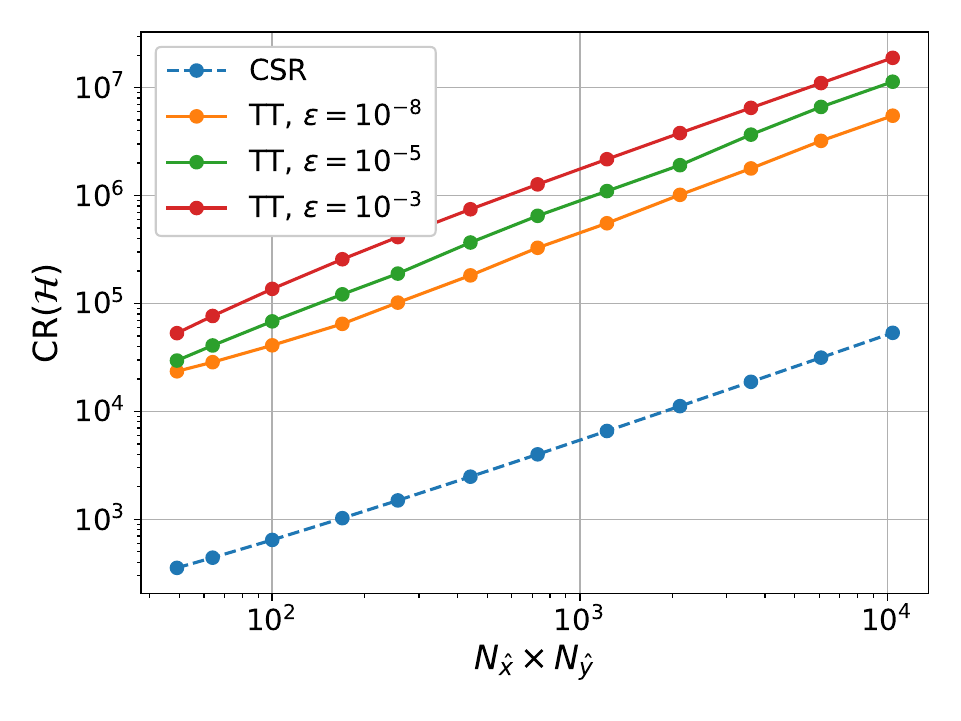}
    \caption{}
    \label{fig:fixed_circle_mesh_comp_H}
\end{subfigure}

\begin{subfigure}{0.47\textwidth}
    \includegraphics[width=\textwidth, trim=0.5cm 0.5cm 0.5cm 0.5cm, clip]{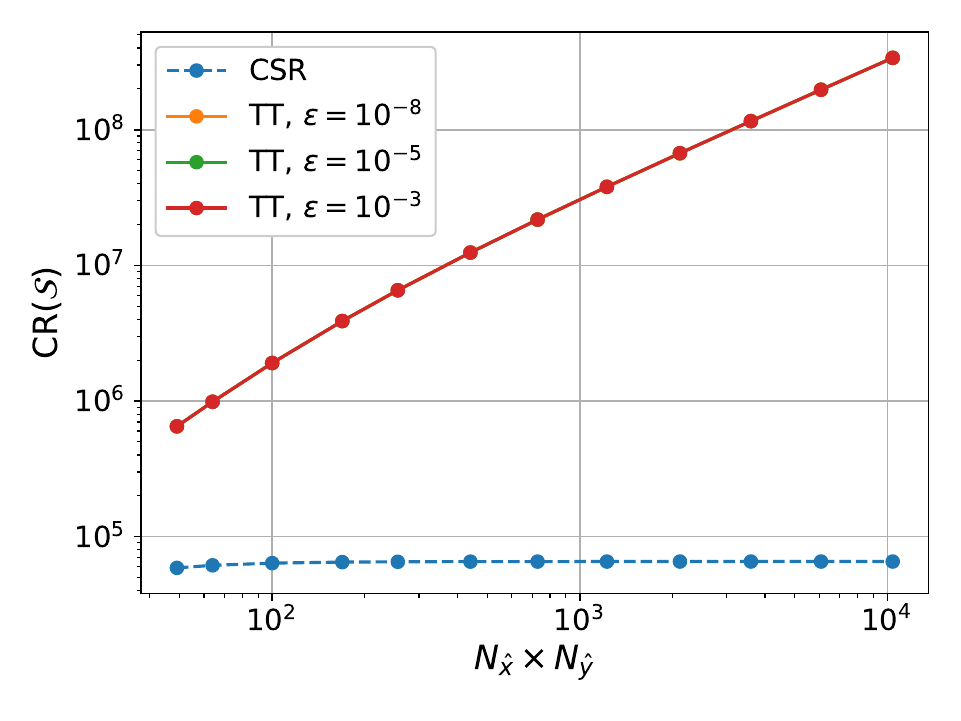}
    \caption{}
    \label{fig:fixed_square_mesh_comp_S}
\end{subfigure}
\begin{subfigure}{0.47\textwidth}
    \includegraphics[width=\textwidth, trim=0.5cm 0.5cm 0.5cm 0.5cm, clip]{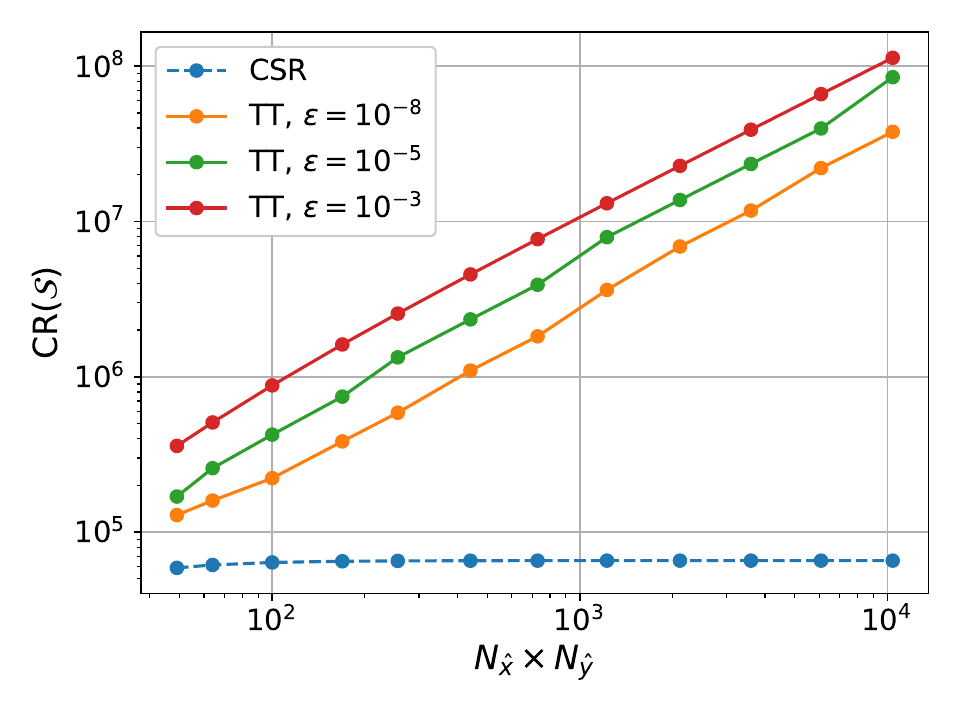}
    \caption{}
    \label{fig:fixed_circle_mesh_comp_S}
\end{subfigure}

\begin{subfigure}{0.47\textwidth}
    \includegraphics[width=\textwidth, trim=0.5cm 0.5cm 0.5cm 0.5cm, clip]{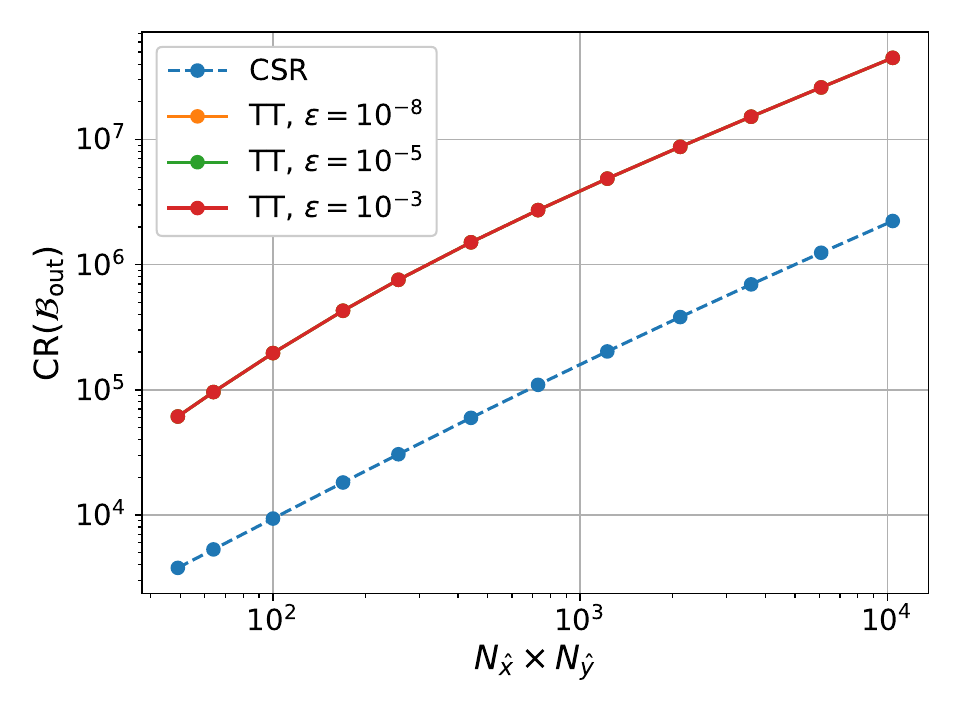}
    \caption{}
    \label{fig:fixed_square_mesh_comp_B_out}
\end{subfigure}
\begin{subfigure}{0.47\textwidth}
    \includegraphics[width=\textwidth, trim=0.5cm 0.5cm 0.5cm 0.5cm, clip]{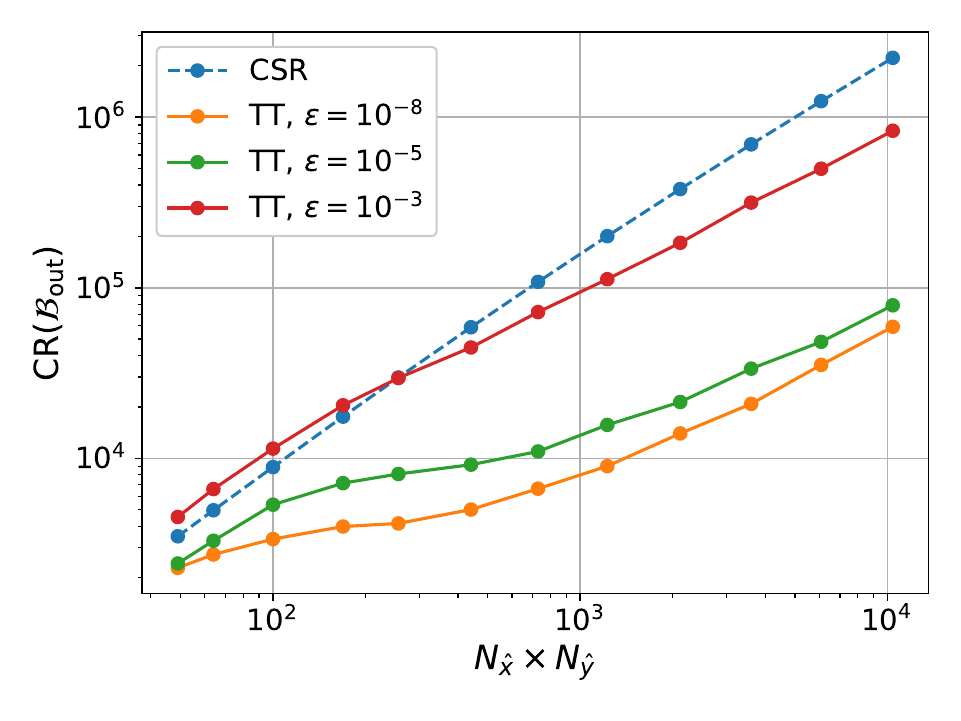}
    \caption{}
    \label{fig:fixed_circle_mesh_comp_B_out}
\end{subfigure}
        
\caption{Compression ratios (CRs) for the (a, b) streaming and collision, (c, d) scattering, and (e, f) outflow boundary operators for the fixed source homogenized square (left) and circle (right) problems. The above figures only show $p_{\hat x} = p_{\hat y} = 2$; however, the CR plots of $p_{\hat x} = p_{\hat y}\in\{3, 4, 6\}$ follow the same trends with the CR of the tensor train (TT) operators relative to Compressed Sparse Row (CSR) increasing for increasing basis function degree. We show TT truncation tolerances $\epsilon \in\{10^{-8}, 10^{-5}, 10^{-3}\}$ and CSR. Note the CR for the square shows no difference between truncation tolerances.}
\label{fig:fixed_square_and_circle_mesh_comp}
\end{figure}

\begin{table}
\centering
\caption{Exponent $\alpha$ for $\text{compression ratio (CR)}\propto \left(N_{\hat x} \times N_{\hat y}\right)^\alpha$ of each operator of the fixed source square and circle fit to the first and last three data points of \cref{fig:fixed_square_and_circle_mesh_comp} where $p_{\hat x} = p_{\hat y} = 2$. We show the exponent for the Compressed Sparse Row (CSR) and tensor train (TT) formats.}\label{tbl:fixed_square_and_circle_mesh_comp_p2_scale}
\begin{tabular}{ccccccccc} 
\toprule
\multirow{2}{*}{\textbf{Operator }} & \multicolumn{4}{c}{First Three Data Points}                                                                              & \multicolumn{4}{c}{Last Three Data Points}                                                                                \\ 
\cmidrule{2-9}
                                    & \textbf{CSR} & $\boldsymbol{\epsilon = 10^{-8}}$ & $\boldsymbol{\epsilon = 10^{-5}}$ & $\boldsymbol{\epsilon = 10^{-3}}$ & \textbf{CSR} & $\boldsymbol{\epsilon = 10^{-8}}$ & $\boldsymbol{\epsilon = 10^{-5}}$ & $\boldsymbol{\epsilon = 10^{-3}}$  \\ 
\hline\hline
\multicolumn{9}{c}{Homogenized Square}                                                                                                                                                                                                                                                     \\ 
\midrule
$\mathcal{H}$                       & 0.837        & 1.677                             & 1.677                             & 1.677                             & 0.985        & 1.025                             & 1.025                             & 1.025                              \\
$\mathcal{S}$                       & 0.110        & 1.504                             & 1.504                             & 1.504                             & 0.000        & 1.012                             & 1.012                             & 1.012                              \\
$\mathcal{B}_{\text{out}}$          & 1.272        & 1.632                             & 1.632                             & 1.632                             & 1.099        & 1.021                             & 1.021                             & 1.021                              \\ 
\midrule
\multicolumn{9}{c}{Homogenized Circle}                                                                                                                                                                                                                                                     \\ 
\midrule
$\mathcal{H}$                       & 0.837        & 0.780                             & 1.169                             & 1.321                             & 0.985        & 1.060                             & 1.064                             & 1.006                              \\
$\mathcal{S}$                       & 0.110        & 0.765                             & 1.269                             & 1.255                             & 0.000        & 1.100                             & 1.213                             & 1.004                              \\
$\mathcal{B}_{\text{out}}$          & 1.310        & 0.537                             & 1.110                             & 1.286                             & 1.101        & 0.980                             & 0.806                             & 0.914                              \\
\bottomrule
\end{tabular}
\end{table}

\subparagraph{Operator TT-Ranks and Compression}
For the homogenized square, the ranks of the operators match those of the angular-resolution operators and are constant. They are shown in \cref{tbl:fixed_source_constant_ranks}. We show the TT operator ranks for the circle in \cref{fig:fixed_circle_mesh_ranks} for $p_{\hat x} = p_{\hat y} = 4$. For the interior operators with $\epsilon \in\{10^{-8}, 10^{-5}\}$, there is an overall decreasing trend, with $\epsilon = 10^{-8}$ having a slight bump in rank for lower fidelity discretizations. This maximum rank and decreasing trend are observed in the rank coupling of the spatial parametric dimensions, indicating potential decoupling of the spatial dimensions. Locally, the mesh becomes more regular as fidelity increases. The interior operator ranks for $\epsilon = 10^{-3}$ remain essentially constant as mesh resolution increases. Like the angular resolution study, \cref{fig:fixed_circle_mesh_ranks_B_out} shows high coupling between angle and space as rank increases to $r_{\max}\left(\mathcal{B}_{\text{out}}^{\text{TT}}\right) > 100$ for $\epsilon\in\{10^{-8}, 10^{-5}\}$; however, the rank begins to plateau likely due to increased regularity in the high fidelity meshes. The $\epsilon = 10^{-3}$ reduces the ranks by more than $60$.

\par \Cref{fig:fixed_square_and_circle_mesh_comp} shows the CR of the operators for the square and circle for increasing spatial fidelity. The CR for $\mathcal{H}$ and $\mathcal{B}_{\text{out}}$ of the square follow that presented in the angular resolution study; however, as shown in \cref{tbl:fixed_square_and_circle_mesh_comp_p2_scale} the initial scaling of $\mathcal{H}^{\text{CSR}}$ is slightly sublinear while $\mathcal{B}^{\text{CSR}}_{\text{out}}$ is slightly superlinear but both approach linear with increasing mesh resolution. The CR of $\mathcal{S}^{\text{CSR}}$ is nearly constant for both the square and circle, as the CSR component only incorporates the spatial dimensions and not the angular. As in the angular resolution study, the difference in compression across varying truncation tolerances is observed only in the circle operators. Unlike the scaling of the circle interior operators in \cref{tbl:fixed_square_and_circle_comp_p2_scale}, the interior operators do not exhibit superlinear scaling for $\epsilon = 10^{-8}$ but do for increasing truncation tolerance, as shown in \cref{tbl:fixed_square_and_circle_mesh_comp_p2_scale}; however, they again approach linear scaling for higher-fidelity meshes. CSR exhibits superior compression for the outflow boundary operator in the circle problem as TT-rank increases on high-fidelity meshes. Again, TT compression relative to CSR increases with increasing basis-function degree.
\par Overall interior operators in TT format compress considerably more than CSR, with a decrease in ranks for increasing spatial resolution due to increasing regularity in the mesh. As in the angular resolution study, the coupling between angle and space in the outflow boundary operators yields high ranks in the TT format for curvilinear boundaries. 

\begin{figure}
\centering
\begin{subfigure}{0.47\textwidth}
    \includegraphics[width=\textwidth, trim=0.5cm 0.5cm 0.5cm 0.5cm, clip]{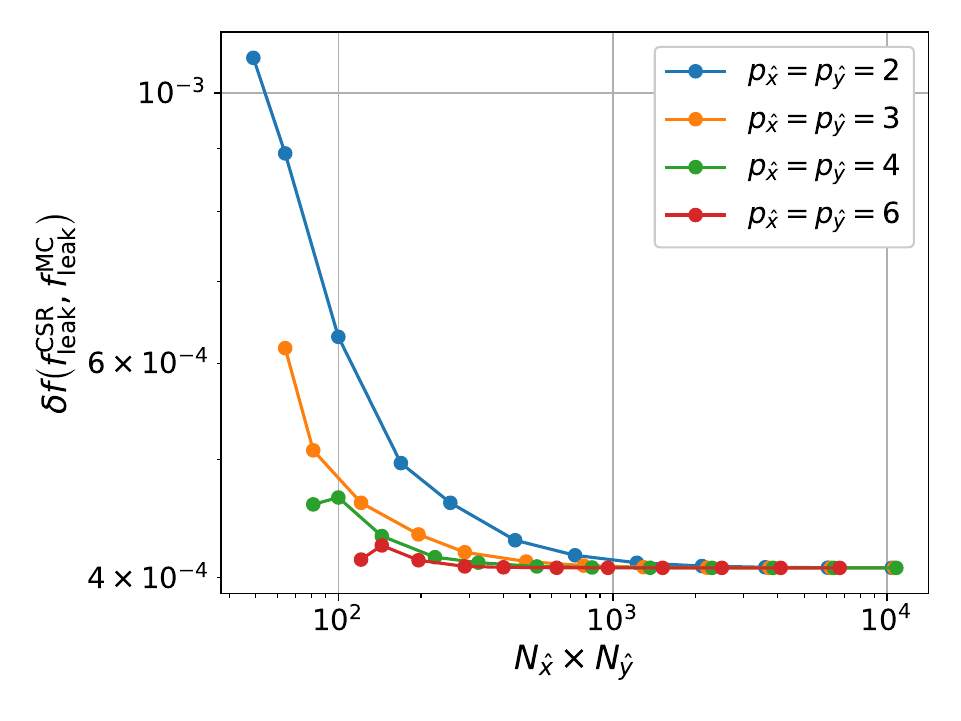}
    \caption{}
    \label{fig:fixed_square_mesh_leakage}
\end{subfigure}
\begin{subfigure}{0.47\textwidth}
    \includegraphics[width=\textwidth, trim=0.5cm 0.5cm 0.5cm 0.5cm, clip]{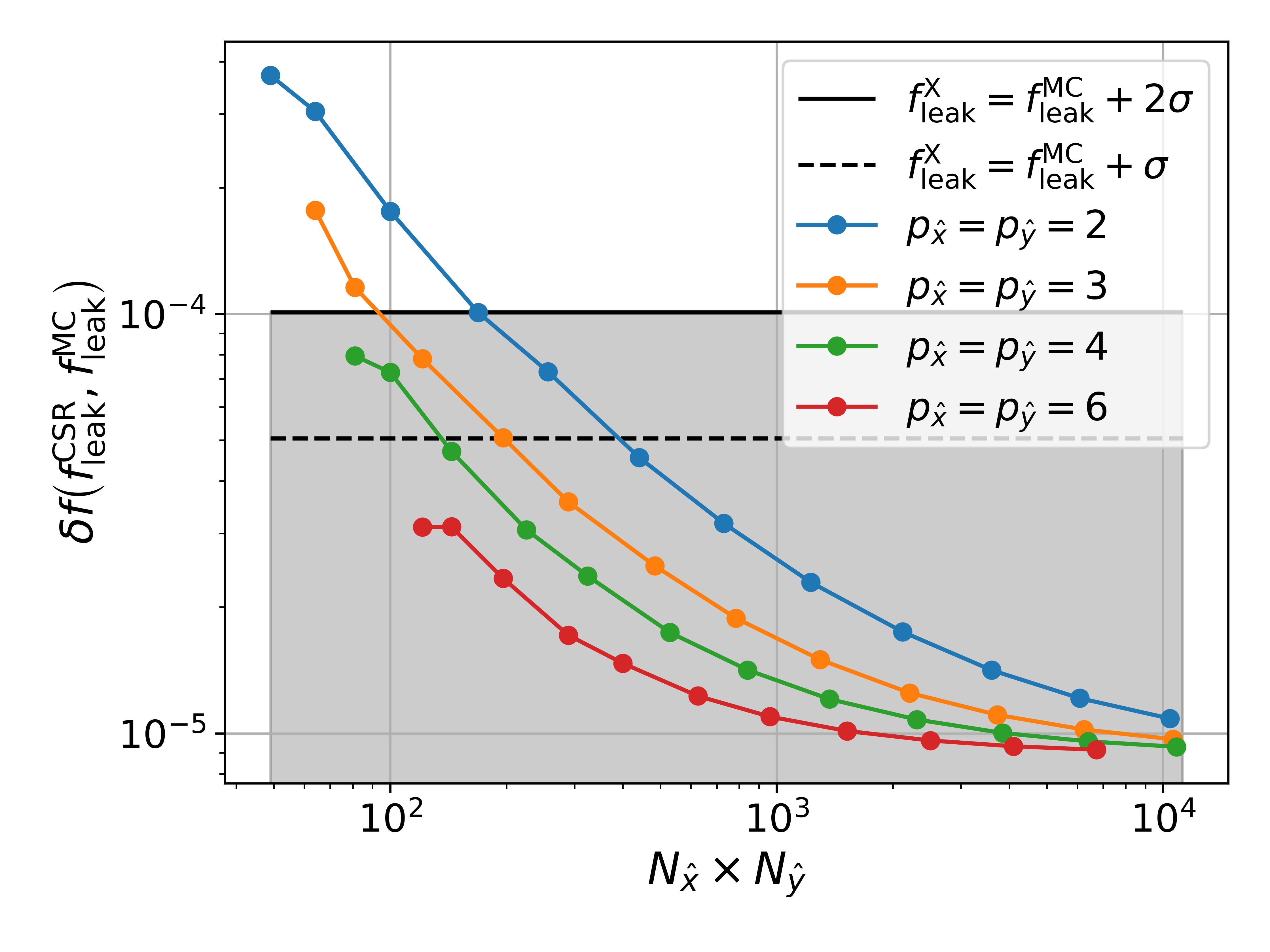}
    \caption{}
    \label{fig:fixed_circle_mesh_leakage}
\end{subfigure}
        
\caption{Leakage fraction relative error (\cref{eq:leakage_fraction_relative_error}) of CSR (\cref{eq:cases_csr}) to the Monte Carlo (MC) reference as a function of the number of spatial degrees of freedom ($N_{\hat x}\times N_{\hat y}$) for the fixed source homogenized (a) square and (b) circle. We show the basis function polynomial degrees $p_{\hat x} = p_{\hat y}\in\{2, 3, 4, 6\}$. For the circle we show $\sigma$ and $2\sigma$ from the reference MC solution; however, all polynomial degrees are $\delta f(f_{\text{leak}}^{\text{CSR}}, f_{\text{leak}}^{MC} + 2\sigma)\approx0.0003$ from $2\sigma$ for the square.}
\label{fig:fixed_square_and_circle_mesh_leakage}
\end{figure}

\subparagraph{Leakage Fraction Error}
In \cref{fig:fixed_square_and_circle_mesh_leakage}, we show the leakage fraction error of CSR to the MC reference versus spatial resolution for the square and circle. Likely due to ray-effects for $N_\Omega = 256$, the square solutions never come within $2\sigma$ of the MC solution, while all polynomial degrees come well within $\sigma$ for the circle. The circle shows much larger improvement in error for increasing polynomial degree for higher-fidelity mesh discretizations, while the square solutions plateau at $\delta f(f_{\text{leak}}^{\text{CSR}}, f_{\text{leak}}^{MC})\approx0.0004$ for $N_{\hat x}\times N_{\hat y} > 10^{3}$. We observe the most significant improvement in error with respect to the basis-function polynomial degree at the coarser discretizations for both geometries. The leakage fraction error of TT and Mixed relative to CSR follows that in \cref{fig:fixed_square_and_circle_leakage_to_csr} with greater variability. Again, the error increases as the degree of the basis functions increases.

\begin{figure}
    \centering
    \includegraphics[width=0.7\linewidth]{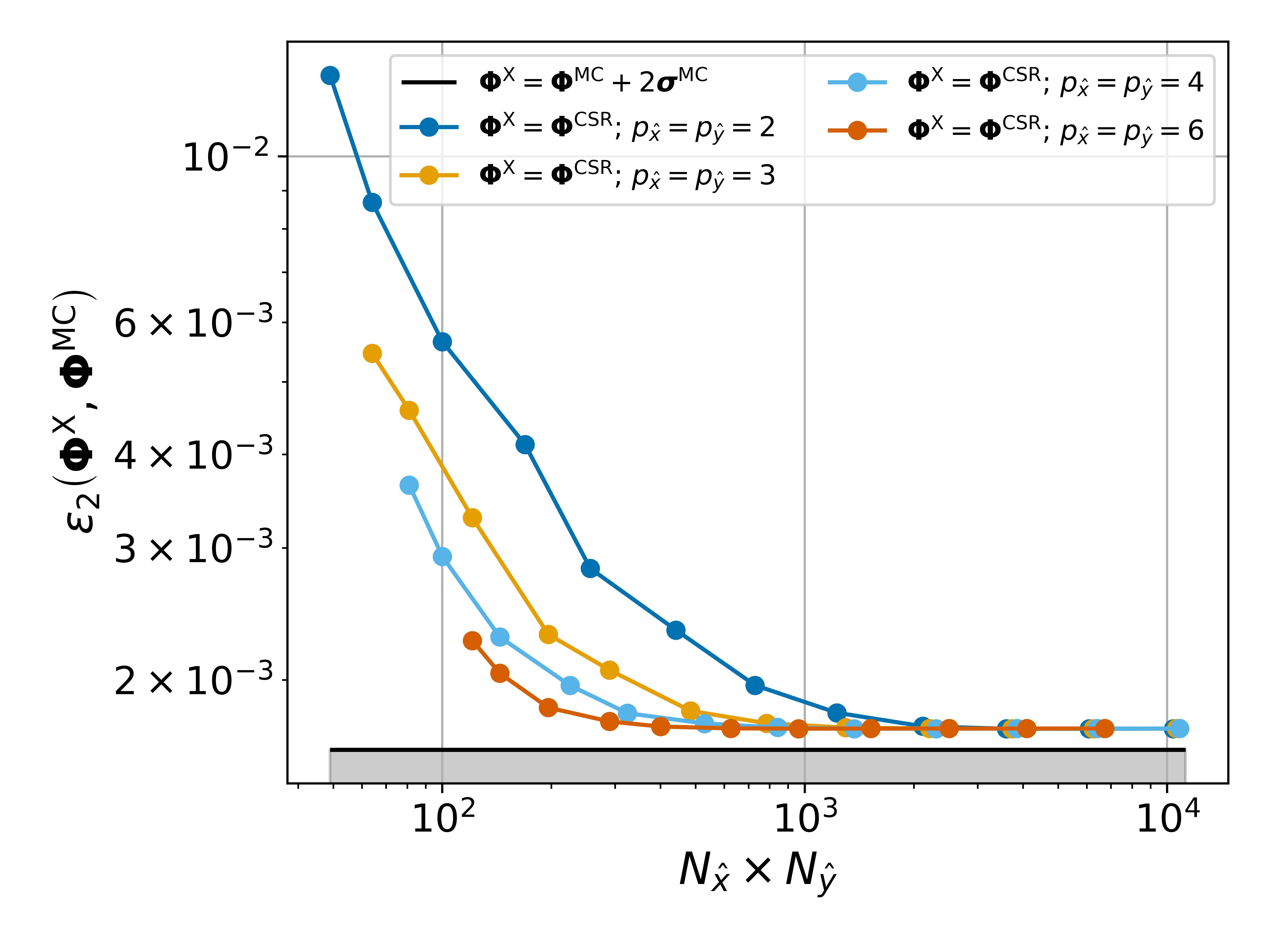}
    \caption{Scalar flux $L_2$ error (\cref{eq:eigenvector_error}) as a function of the number of spatial degrees of freedom ($N_{\hat x}\times N_{\hat y}$) for CSR (\cref{eq:cases_csr}) with $p_{\hat x} = p_{\hat y}\in\{2, 3, 4, 6\}$. We also show $\epsilon_2\left(\mathbf{\Phi}^{\text{MC}} + 2\boldsymbol{\sigma}^{\text{MC}}, \mathbf{\Phi}^{\text{MC}}\right)$ for reference.}
    \label{fig:fixed_square_mesh_l2}
\end{figure}

\subparagraph{Scalar Flux Error}
In \cref{fig:fixed_square_mesh_l2} we show the scalar flux $L_2$ error as a function of the number of spatial degrees of freedom for CSR with $p_{\hat x} = p_{\hat y}\in\{2, 3, 4, 6\}$. Again, we also provide $\epsilon_2\left(\mathbf{\Phi}^{\text{MC}} + 2\boldsymbol{\sigma}^{\text{MC}}, \mathbf{\Phi}^{\text{MC}}\right)$ for reference. All basis function degrees approach $\epsilon_2\approx 1.72\times 10^{-3}$, never coming within $\epsilon_2\left(\mathbf{\Phi}^{\text{MC}} + 2\boldsymbol{\sigma}^{\text{MC}}, \mathbf{\Phi}^{\text{MC}}\right)$. As the previous section suggests, the largest difference in basis function degrees with respect to the $L_2$ error occurs at the coarser discretizations. On a cell-by-cell basis, the $z$-score has a minimum of $3.974\times 10^{-4}~\sigma$, maximum of $19.362~\sigma$, mean of $1.816~\sigma$, $\text{Q}_1$ of $0.510~\sigma$, median of $1.127~\sigma$, and $\text{Q}_2$ of $2.203~\sigma$ for CSR with $N_{\hat x}\times N_{\hat y} = 4096$ and $p_{\hat x} = p_{\hat y} = 6$. Looking at the $L_2$ error between CSR and the other cases in \cref{eq:cases} follows the same trends observed in the angular resolution study; namely, all truncation tolerances closely follow each other, decreasing from $<10^{-7}$ in $L_2$ error for increasing mesh resolution, except for TT and TT (rounded) with $p_{\hat x} = p_{\hat y} = 6$ and $\epsilon = 10^{-3}$ which deviates from $\epsilon\in\{10^{-8}, 10^{-5}\}$ to $\epsilon_2\approx 10^{-5}$. 

\begin{figure}
\centering
\begin{subfigure}{0.47\textwidth}
    \includegraphics[width=\textwidth, trim=0.5cm 0.5cm 0.5cm 0.5cm, clip]{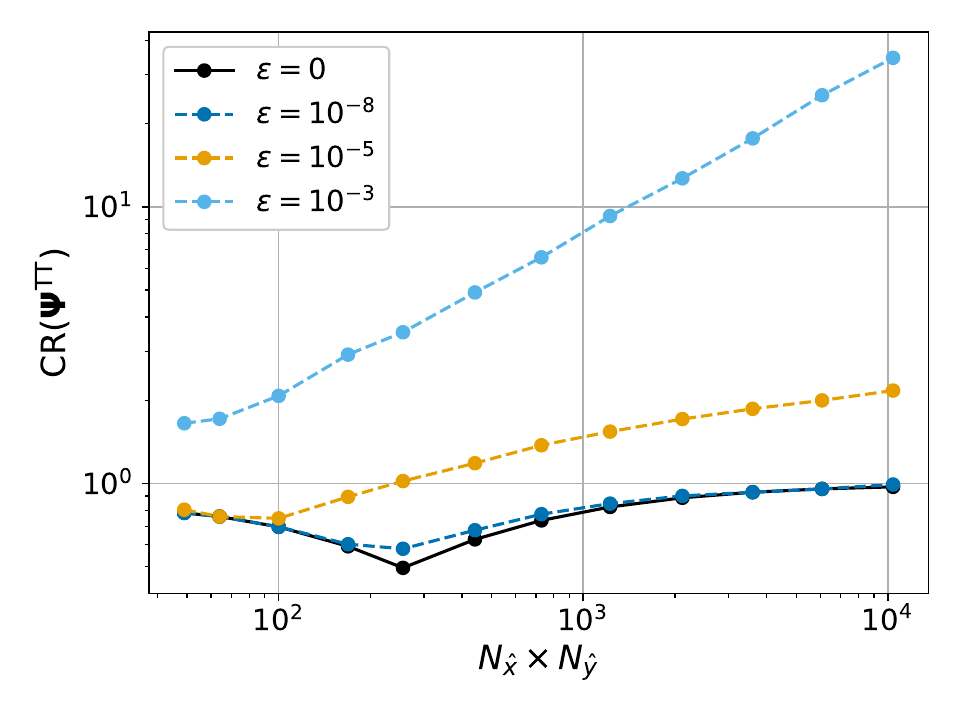}
    \caption{$p_{\hat x} = p_{\hat y} = 2$}
    \label{fig:fixed_square_mesh_compression_psi}
\end{subfigure}
\begin{subfigure}{0.47\textwidth}
    \includegraphics[width=\textwidth, trim=0.5cm 0.5cm 0.5cm 0.5cm, clip]{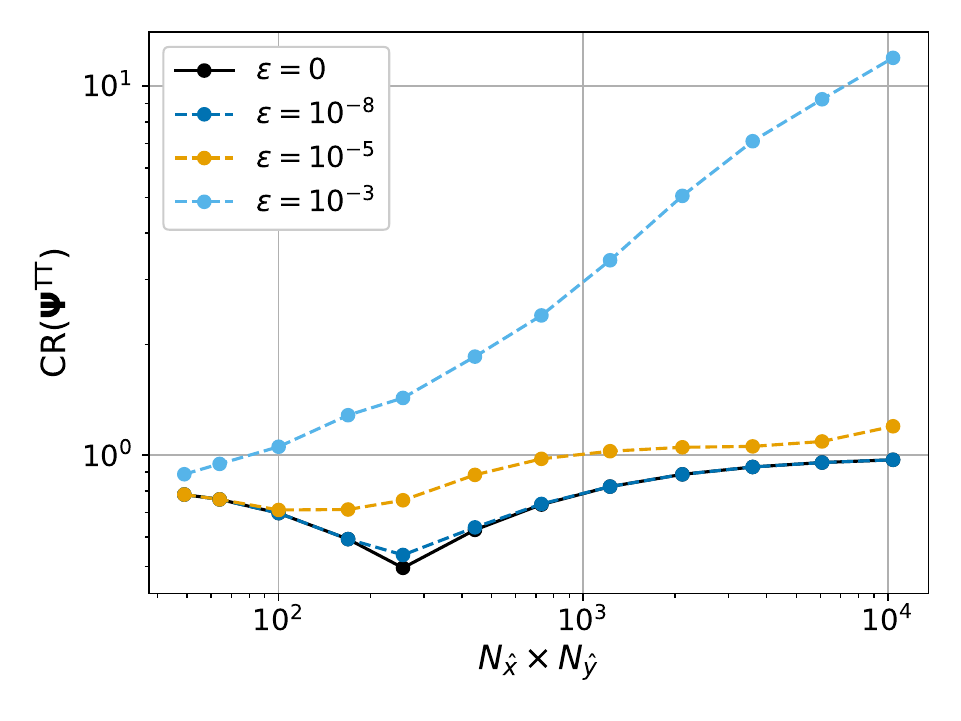}
    \caption{$p_{\hat x} = p_{\hat y} = 2$}
    \label{fig:fixed_circle_mesh_compression_psi}
\end{subfigure}
        
\caption{Compression ratio (CR) of the angular flux decomposed in the tensor train (TT) format for increasing number of spatial degrees of freedom ($N_{\hat x}\times N_{\hat y}$). We show $p_{\hat x} = p_{\hat y} = 2$ for the fixed source (a) square and (b) circle for full rank ($\epsilon = 0$) and truncation tolerances $\epsilon \in\{10^{-8}, 10^{-5}, 10^{-3}\}$.}
\label{fig:fixed_square_and_circle_mesh_compression_psi}
\end{figure}

\subparagraph{Angular Flux Compression in the TT Format}
\Cref{fig:fixed_square_and_circle_mesh_compression_psi} shows the compression ratios for the angular flux decomposed in the TT format for the square and circle for $p_{\hat x} = p_{\hat y} = 2$. The results for the circle, \cref{fig:fixed_circle_mesh_compression_psi}, closely follow those presented in the angular resolution study, as only $\epsilon = 10^{-3}$ is compressible in the TT format. However, the angular flux solution of the square appears only compressible for $\epsilon = 10^{-3}$, with $\epsilon = 10^{-5}$ marginally compressible, and $\epsilon = 10^{-8}$ not compressible. The angular flux becomes less compressible with increasing polynomial degree of the basis functions.

\subsubsection{Quarter circle source in void}\label{sec:fixed_quarter_circle}

\begin{figure}
\centering
\begin{subfigure}{0.45\textwidth}
    \includegraphics[width=\textwidth]{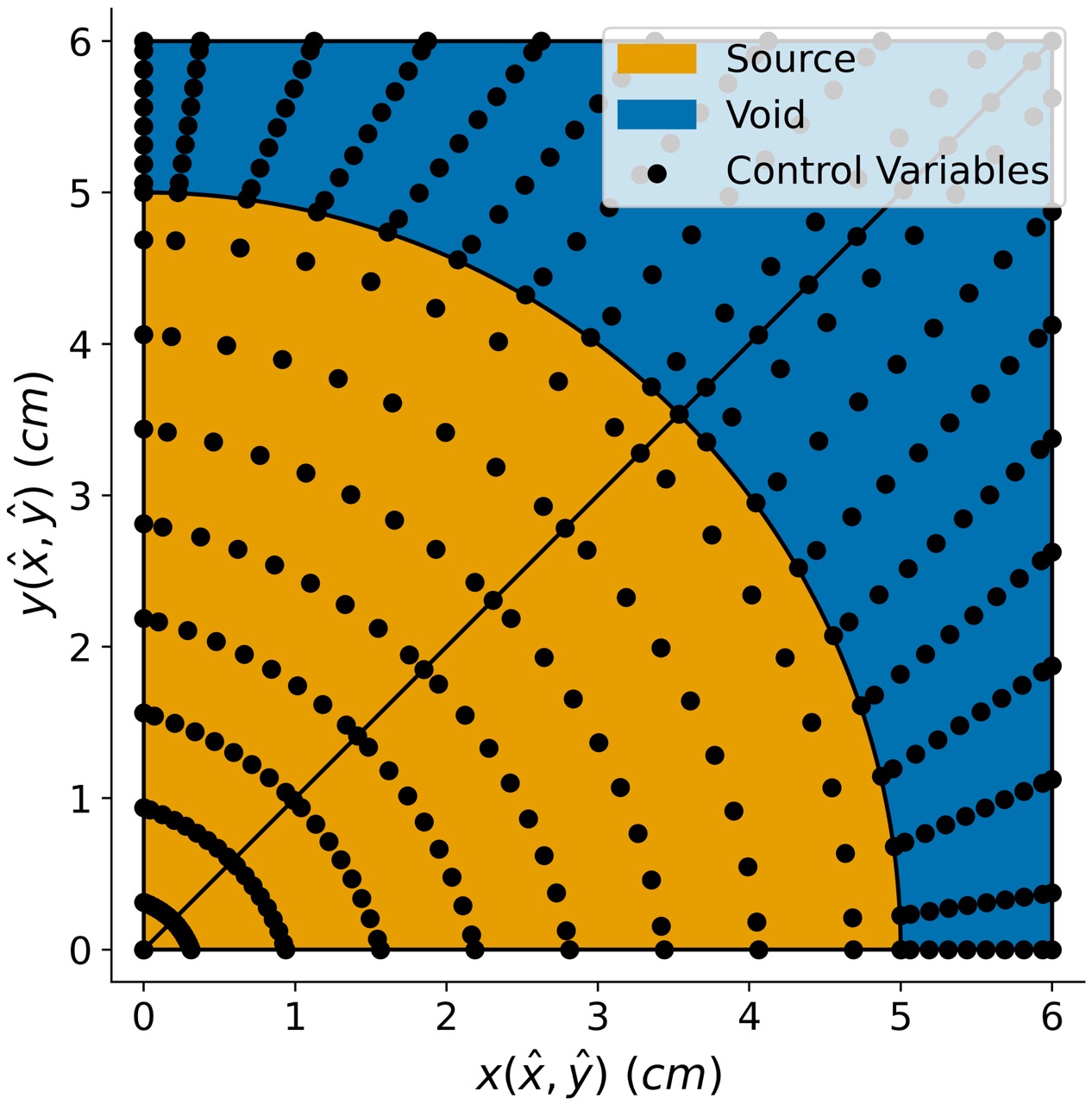}
    \caption{}
    \label{fig:fixed_quarter_circle}
\end{subfigure}
\begin{subfigure}{0.52\textwidth}
    \includegraphics[width=\textwidth]{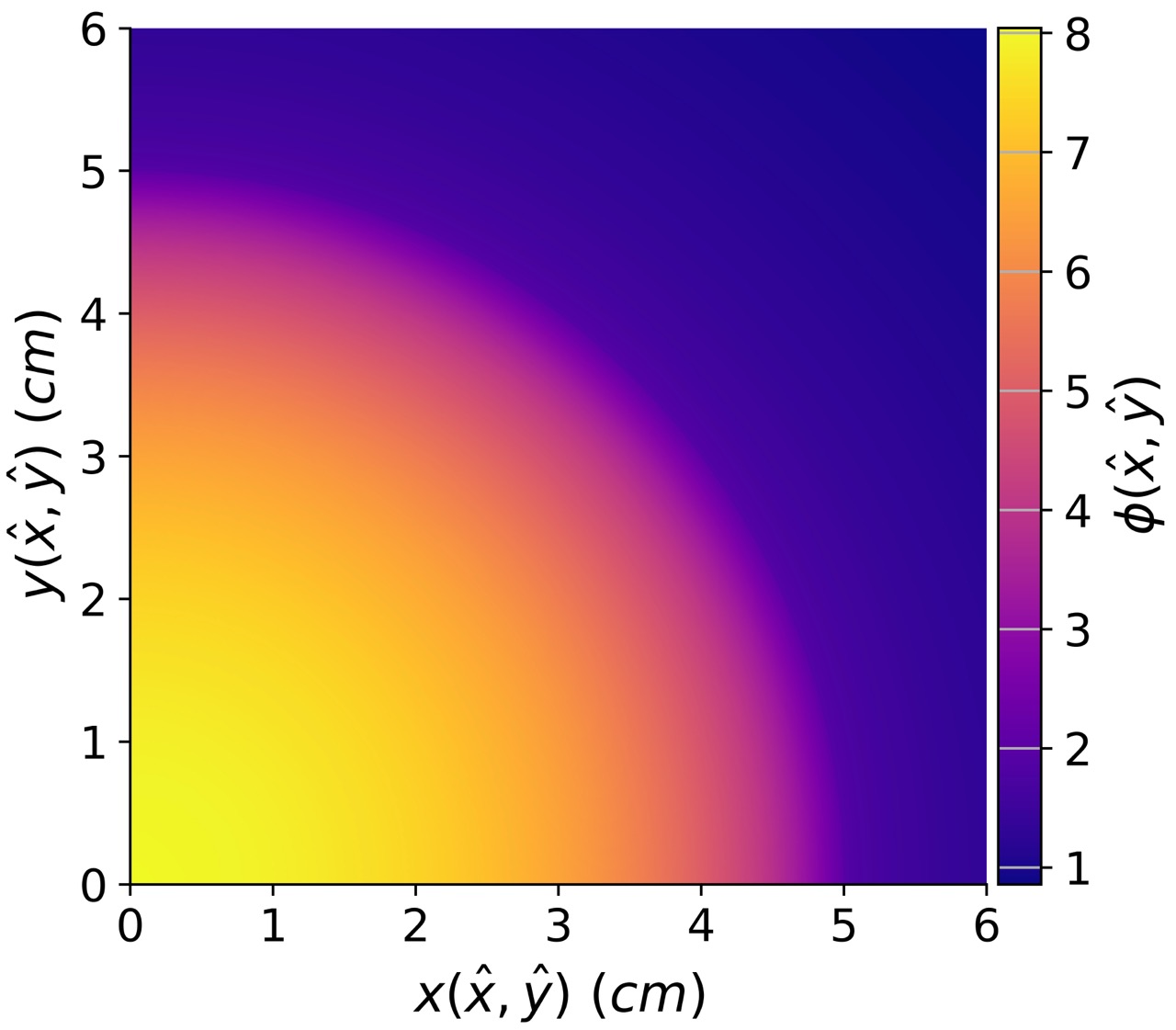}
    \caption{}
    \label{fig:fixed_quarter_circle_phi}
\end{subfigure}
        
\caption{Fixed quarter circular source surrounded by void with reflective boundary conditions along the left and bottom boundaries and vacuum on the rest. The (a) CAD geometry is represented by four patches (boundaries outlined in black) with $N_{\hat x}^u = N_{\hat y}^u = 8$ knot spans along each parametric direction and $p_{\hat x} = p_{\hat y} = 2$ NURBS basis functions for each patch. We show (b) an example scalar flux solution for this mesh solved with CSR (\cref{eq:cases_csr}) and $N_\Omega = 16384$ ordinates.}
\label{fig:fixed_quarter_circle_plots}
\end{figure}

\begin{table}
\centering
\caption{Cross sections for the fixed source quarter circle surrounded by void (\cref{sec:fixed_quarter_circle}) and cruciform source with annular shield (\cref{sec:fixed_cruciform}).}\label{tbl:fixed_quarter_cruciform_xs}
\begin{tabular}{cccc} 
\toprule
\textbf{Material} & $\boldsymbol{\Sigma_t~\left(cm^{-1}\right)}$ & $\boldsymbol{\Sigma_s~\left(cm^{-1}\right)}$ & $\boldsymbol{Q~\left(cm^{-3}s^{-1}\right)}$  \\ 
\hline\hline
\multicolumn{4}{c}{Quarter Circle Source in Void}                                                                                                              \\ 
\midrule
Source            & 1                                            & 0.9                                          & 1                                            \\
Void              & 0                                            & 0                                            & 0                                            \\ 
\midrule
\multicolumn{4}{c}{Cruciform Source with Shielding}                                                                                                            \\ 
\midrule
Source            & 0.01                                         & 0.008                                        & 1                                            \\
Void              & 0                                            & 0                                            & 0                                            \\
Shield            & 3                                            & 0.5                                          & 0                                            \\
\bottomrule
\end{tabular}
\end{table}

\par We next examine how multiple patches affect operator compression, error, operator-vector product scaling, and solution compression. The multi-patch test is the same as the homogeneous circular source presented in \cref{sec:fixed_homo}. Here, we represent the circle as a quarter-circle surrounded by a void, using $N_e = 4$ patches with reflective boundary conditions on the left and bottom faces, and vacuum on the remaining faces. \Cref{fig:fixed_quarter_circle} shows this CAD geometry with $N_{\hat x}^u = N_{\hat y}^u = 8$ knot spans along each parametric direction with $p_{\hat x} = p_{\hat y} = 2$ NURBS basis functions. The cross sections for this problem are shown in \cref{tbl:fixed_quarter_cruciform_xs}. We show a solution in \cref{fig:fixed_quarter_circle_phi} solved using CSR with $N_\Omega = 16384$ ordinates using the same spatial mesh in \Cref{fig:fixed_quarter_circle}. Our reference solution was generated using OpenMC with a leakage fraction of $f_{\text{leak}}^{\text{MC}} = 0.43995\pm 0.00002$ and a regular mesh solution $\mathbf{\Phi}^{\text{MC}}, \boldsymbol{\sigma}^{\text{MC}}\in\mathbb{R}^{1\times 128\times 128}$. 

\paragraph{Angular and Mesh Resolution Studies}

\begin{figure}
    \centering
    \includegraphics[width=0.7\linewidth]{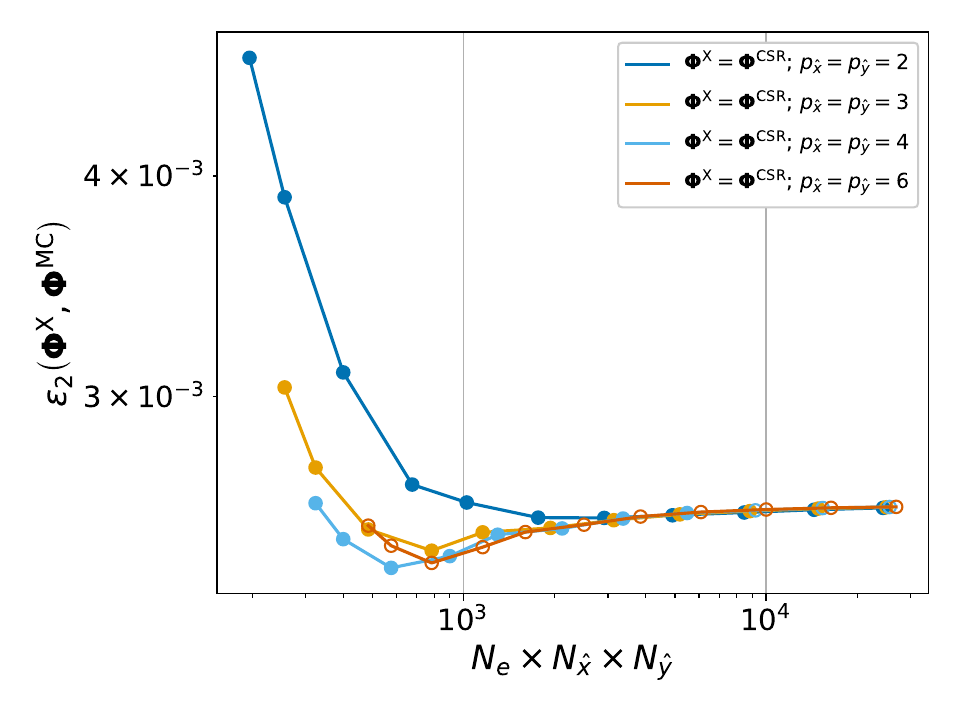}
    \caption{The scalar flux $L_2$ error (\cref{eq:eigenvector_error}) of CSR (\cref{eq:cases_csr}) to the Monte Carlo (MC) reference solution as a function of the number of spatial degrees of freedom ($N_e \times N_{\hat x}\times N_{\hat y}$) for NURBS basis function degrees $p_{\hat x} = p_{\hat y}\in\{2, 3, 4, 6\}$. The open circle indicates that the GMRES solver failed to converge within the allotted iterations and restarts.}
    \label{fig:fixed_quarter_circle_l2error}
\end{figure}

\par Many of the trends presented in \cref{sec:fixed_homo} are reproduced here, with the multi-patch quarter-circle representation closely following the homogenized circle results. For that reason, we only outline observations that deviate from \cref{sec:fixed_homo}:

\begin{itemize}
    \item For varying angular fidelity, the maximum rank of the interior operators remains consistent with the homogenized circle in \cref{tbl:fixed_source_constant_ranks} with the exception of $r_{\max}\left(\mathcal{H}^{TT}\right)\in [31, 33]$ for varying basis function degree. Additionally, at higher angular discretizations, the outflow and inflow boundary operator max rank for the quarter circle is lower than the single-patch circle in \cref{fig:fixed_square_ranks_B_out}.
    \item For increasing mesh resolution the streaming and collision operator features ranks $\sim 10$ lower than that shown in \cref{fig:fixed_circle_mesh_ranks_H} while the inflow and outflow boundary operators plateau at $r_{\max} = 143$ as opposed to $r_{\max} = 130$ for the outflow boundary operator of the homogenized circle show in \cref{fig:fixed_circle_mesh_ranks_B_out}.
    \item Compression for operators in TT format increased across angular and mesh resolutions while CSR maintained the same CR. The addition of the patch dimension provides further compression due to $\mathcal{O}(dN^2r^2)$ memory scaling for TT-operators. 
    \item Boundary operator compression in TT for $\epsilon\in\{10^{-8}, 10^{-5}\}$ was higher than CSR across all basis function degrees for all mesh discretizations, unlike that shown in \cref{fig:fixed_circle_mesh_comp_B_out}. This is likely due to the reduced number of curvilinear boundaries per patch.
    \item Like the relative error presented in \cref{fig:fixed_circle_leakage} all polynomial degrees got well within $2\sigma^{\text{MC}}$ and $p_{\hat x} = p_{\hat y}\in\{3, 4, 6\}$ got within $\sigma^{\text{MC}}$; however, for $p_{\hat x} = p_{\hat y} = 6$ for all angular and spatial discretizations GMRES failed to converge within $\epsilon_{\text{GMRES}} = 10^{-6}$ for the 1000 restarts and 100 iterations. Higher order basis functions increase the condition number, requiring more iterations or a preconditioner; this is especially apparent for multi-patch problems, as we discuss in \cref{sec:fixed_cruciform}. We also observe a failure for $p_{\hat x} = p_{\hat y} = 4$ to converge for $N_{\hat{x}}^{u} = N_{\hat y}^{u} \ge 44$ as well as $p_{\hat x} = p_{\hat y} = 3$ for $N_{\hat{x}}^{u} = N_{\hat y}^{u} \ge 58$.
    \item \Cref{fig:fixed_quarter_circle_l2error} shows the $L_2$ error of CSR to the MC reference solution for varying mesh fidelity. Like the homogenized square in \cref{fig:fixed_square_mesh_l2}, $\epsilon_2(\mathbf{\Phi}^{\text{CSR}}, \mathbf{\Phi}^{\text{MC}})$ converges to $0.0026$ for increasing mesh fidelity across all polynomial degrees. This is higher than $\epsilon_2(\mathbf{\Phi}^{\text{MC}} + 2\boldsymbol{\sigma}^{\text{MC}}, \mathbf{\Phi}^{\text{MC}}) = 0.0010$. Again, this indicates that the $N_\Omega = 256$ angular quadrature is the limiting factor, and that, in this case, higher-fidelity spatial meshes exacerbate ray effects at patch interfaces and boundaries, thereby increasing error.
    \item Compression for the angular flux in TT format followed the homogenized circle except $\text{CR}(\mathbf{\Psi}^{\text{TT}})$ approached $2$ for $\epsilon = 10^{-5}$ with increasing angular resolution. Additionally, for increasing mesh resolution, $\epsilon = 10^{-3}$ is only marginally compressible with $\text{CR}(\mathbf{\Psi}^{\text{TT}}) < 3$.
\end{itemize}

\subsubsection{Cruciform source with shielding}\label{sec:fixed_cruciform}

\begin{figure}
\centering
\begin{subfigure}{0.44\textwidth}
    \includegraphics[width=\textwidth]{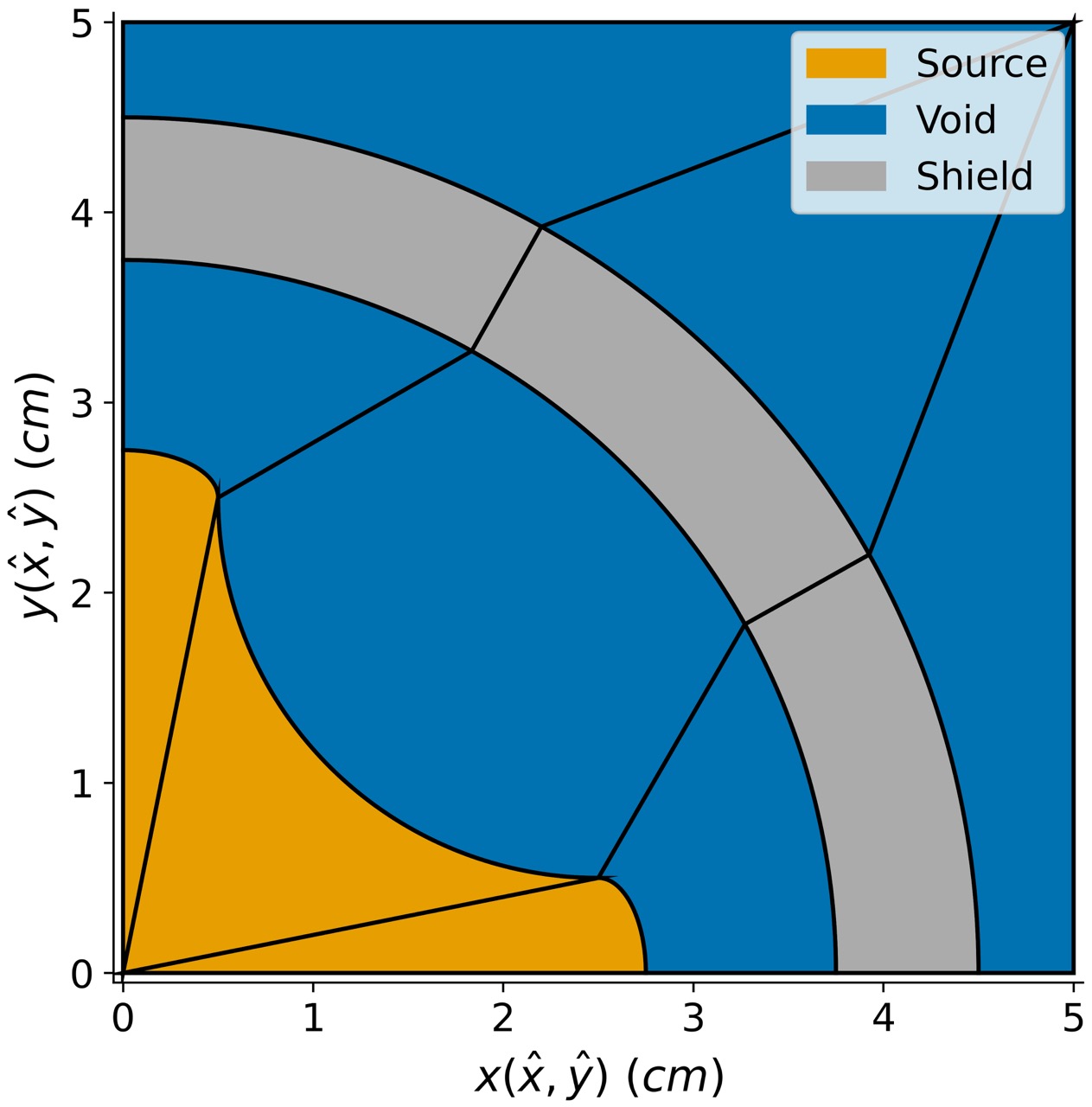}
    \caption{}
    \label{fig:fixed_cruciform}
\end{subfigure}
\begin{subfigure}{0.53\textwidth}
    \includegraphics[width=\textwidth]{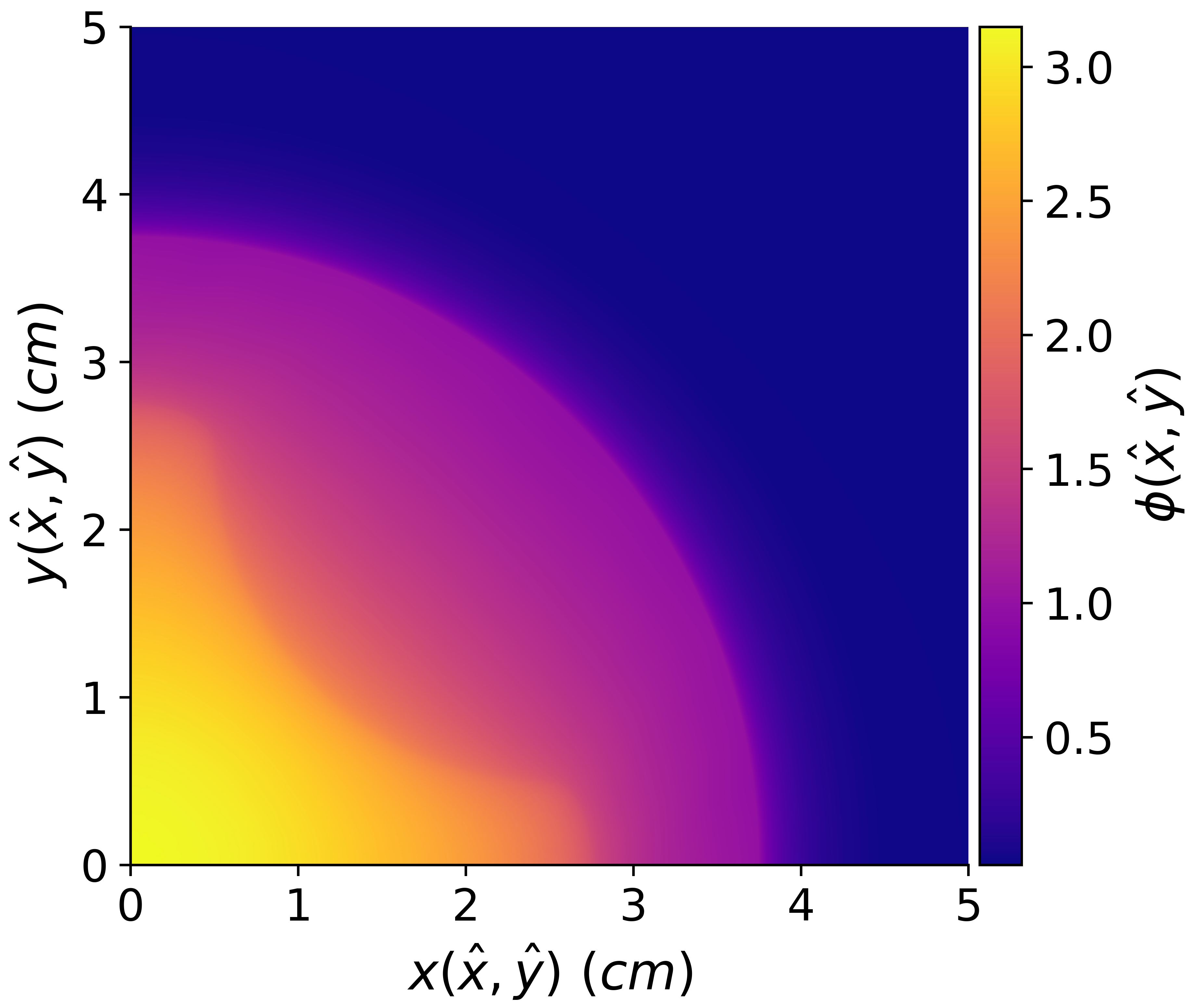}
    \caption{}
    \label{fig:fixed_cruciform_phi}
\end{subfigure}
        
\caption{Cruciform fixed source problem represented in (a) using quarter symmetry with $N_e = 12$ patches (boundaries outlined in black) each with $N_{\hat x}^u= N_{\hat y}^u = 13$ knot spans along each parametric direction with $p_{\hat x} = p_{\hat y} = 3$. We show the scalar flux solution in (b) computed with $N_{\Omega} = 4096$ ordinates using CSR (\cref{eq:cases_csr}).}
\label{fig:fixed_cruciform_plots}
\end{figure}

\par Our final and most challenging fixed source problem is a cruciform source surrounded by an annular shield in a vacuum. The single-group cross sections are given in \cref{tbl:fixed_quarter_cruciform_xs} with optically thin and thick material. The shield has a minimum optical depth of $2.25$ mean-free paths, with a scattering ratio $0.17$. The substantial variation in cross section produces a discontinuity in the transport solution at the void-shield interface. As shown in \cref{fig:fixed_cruciform}, we represent this geometry using quarter symmetry with $N_e = 12$ patches each with a $N_{\hat x}^u = N_{\hat y}^u = 13$ mesh with $p_{\hat x} = p_{\hat y} = 3$ NURBS basis functions. We use $N_\Omega = 4096$ ordinates and truncate all TTs to $\epsilon = 10^{-5}$. We solve the resulting linear system of $N_\Omega\times N_e\times \left(N_{\hat x}^u + p_{\hat x}\right)\times \left(N_{\hat y}^u + p_{\hat y}\right) = 12,582,912$ degrees of freedom with GMRES using $1000$ restarts and $100$ iterations. We show the scalar flux solution in \cref{fig:fixed_cruciform_phi} computed with CSR. 

\begin{table}
\centering
\caption{Compression ratios (CRs) and tensor train (TT) ranks for the TT and Compressed Sparse Row (CSR) formats for the cruciform fixed source problem. We also show the CRs for the left-hand side (LHS) for the CSR (\cref{eq:cases_csr}) and Mixed (rounded) (\cref{eq:cases_mixed_rounded}) cases.}\label{tbl:fixed_cruciform_crs}
\begin{tabular}{cccc} 
\toprule
\textbf{Operator}          & \textbf{Ranks}          & \textbf{TT CR}        & \textbf{CSR CR}       \\ 
\hline\hline
$\mathcal{H}$              & $\{3, 3, 3, 25, 17\}$   & $1.111\times 10^{9}$  & $1.590\times 10^{5}$  \\
$\mathcal{S}$              & $\{1, 1, 1, 6, 6\}$     & $1.157\times 10^{10}$ & $1.677\times 10^{7}$  \\
$\mathcal{B}_{\text{in}}$  & $\{6, 66, 110, 64, 6\}$ & $1.768\times 10^{7}$  & $5.262\times 10^{6}$  \\
$\mathcal{B}_{\text{out}}$ & $\{4, 64, 106, 61, 6\}$    & $1.923\times 10^{7}$  & $5.136\times 10^{6}$  \\
LHS of \cref{eq:cases_csr}                    & ---                     & ---                   & $1.548\times 10^{5}$  \\
LHS of \cref{eq:cases_mixed_rounded}~                    & ---                     & \multicolumn{2}{c}{$3.265\times 10^{6}$}      \\
\bottomrule
\end{tabular}
\end{table}

\par In \cref{tbl:fixed_cruciform_crs} we show the CRs and TT-ranks for the operators in TT and CSR formats. Notably, the largest ranks occur in the spatial cores of the interior operators, indicating the variety of mappings of the $12$ patches increases the ranks; however, $r_{\max}\left(\mathcal{H}^{\text{TT}}\right) = 25$ and $r_{\max}\left(\mathcal{S}^{\text{TT}}\right) = 6$ are considerably less than $r_{\max}\left(\mathcal{B}^{\text{TT}}_{\text{in}}\right) = 110$ and $r_{\max}\left(\mathcal{B}^{\text{TT}}_{\text{out}}\right) = 106$ and produce a significantly higher compression ratio $\text{CR}(\mathcal{H}^{\text{TT}}),\text{CR}(\mathcal{S}^{\text{TT}}) > 10^9$ three to four orders of magnitude higher than their CSR counterparts. The ranks of the boundary operators are distributed much more evenly than those of the interior operators, with the maximum ranks occurring at the rank coupling angle to space. We present the total operator compression for CSR and Mixed (rounded) cases, with Mixed (rounded) outperforming CSR; however, total CR for Mixed (rounded) is limited by the CSR boundary operators.

\begin{figure}
    \centering
    \includegraphics[width=0.7\linewidth]{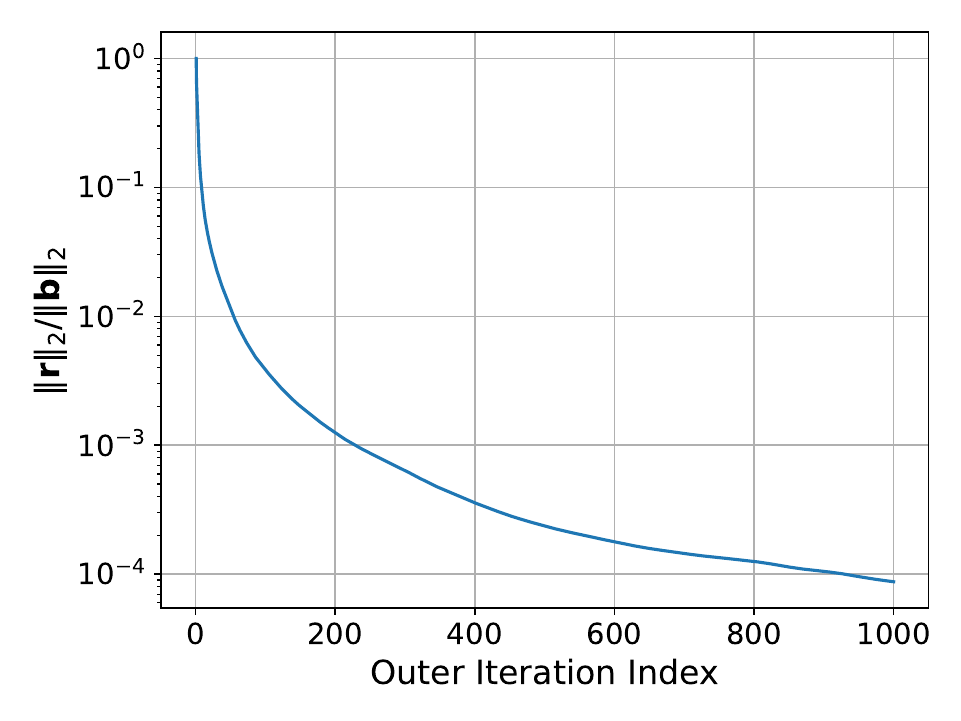}
    \caption{GMRES convergence of the fixed source cruciform problem with 1000 restarts (outer iteration index), each with 100 iterations.}
    \label{fig:fixed_cruciform_gmres}
\end{figure}

\par As shown in \cref{fig:fixed_cruciform_gmres}, restarted GMRES does not reach the desired relative residual norm of $\epsilon_{\text{GMRES}} = 10^{-6}$ within the prescribed 1000 restarts with 100 iterations. GMRES convergence begins with a steep descent in the early cycles, reducing the relative residual from $\sim 10^0$ to $\sim 10^{-2}$ in fewer than 100 restarts. We then transition to stagnation over the subsequent 900 restarts, reducing the relative residual to $\sim10^{-4}$. There was no difference in convergence behavior between CSR and Mixed (rounded), with a total solution time of $2.19~h$ and $6.22~h$, respectively. We clearly need a preconditioner to make GMRES a feasible solver for these complex multi-patch problems. Additionally, increasing the degree of the basis functions significantly degrades GMRES performance as the condition number increases.

\begin{figure}
\centering
\begin{subfigure}{0.47\textwidth}
    \includegraphics[width=\textwidth]{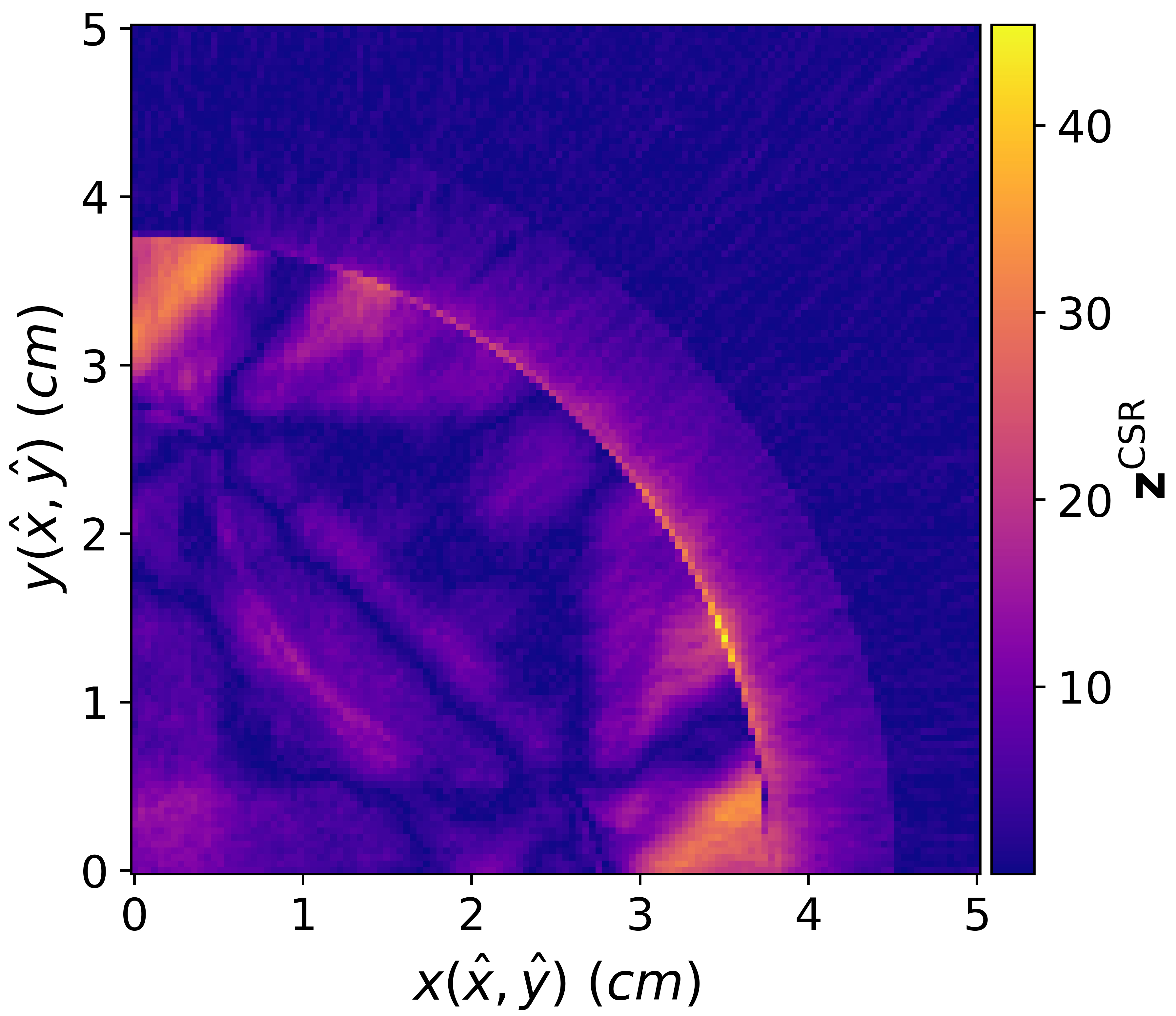}
    \caption{}
    \label{fig:fixed_cruciform_zscore_csr}
\end{subfigure}
\begin{subfigure}{0.47\textwidth}
    \includegraphics[width=\textwidth]{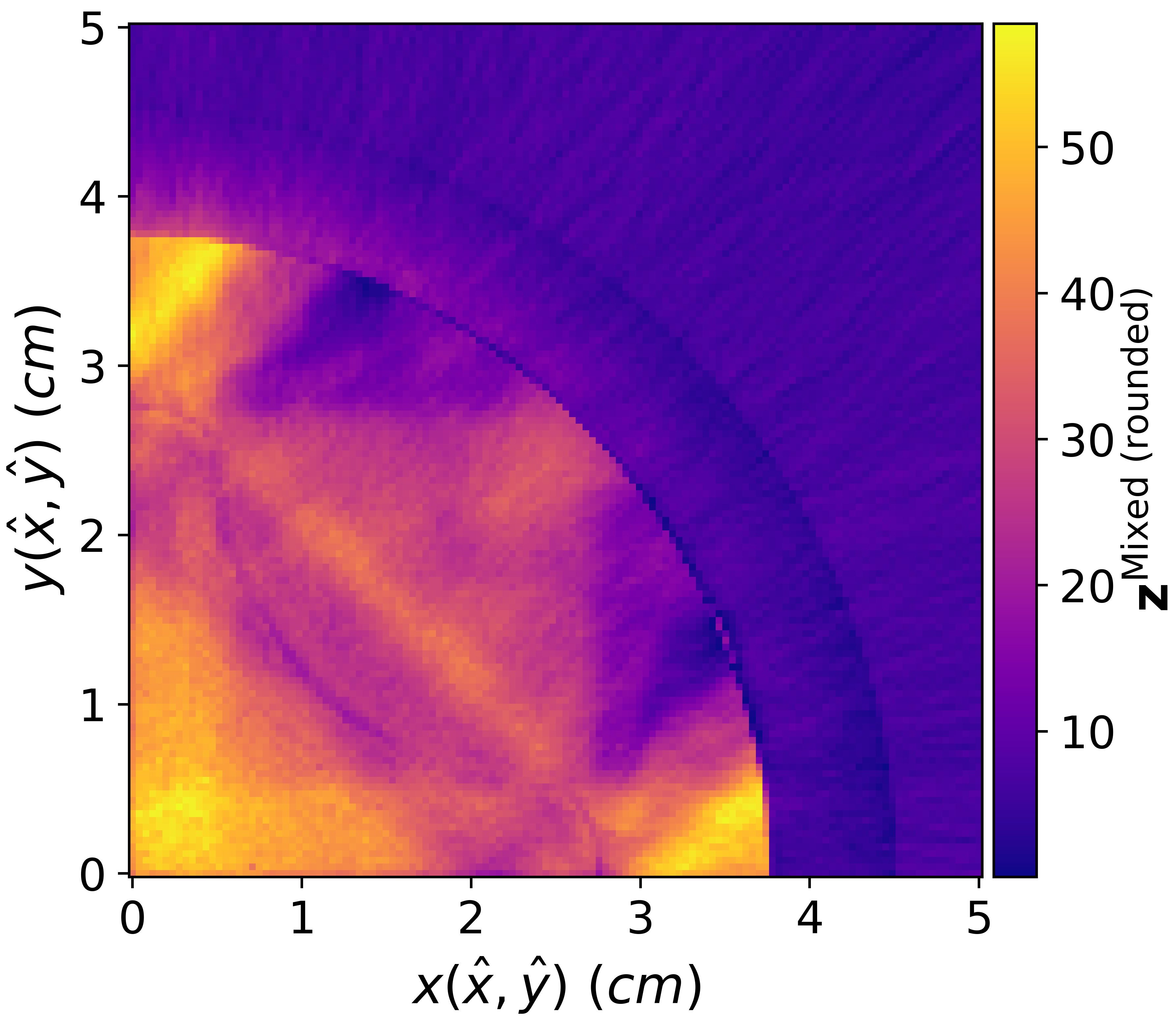}
    \caption{}
    \label{fig:fixed_cruciform_zscore_mixed_rounded}
\end{subfigure}
        
\caption{The $z$-score of (a) CSR (\cref{eq:cases_csr}) and (b) Mixed (rounded) (\cref{eq:cases_mixed_rounded}) for the cruciform fixed source problem.}
\label{fig:fixed_cruciform_zscore}
\end{figure}

\begin{table}
\centering
\caption{Statistics for the $z$-score and $L_2$ error for the fixed source cruciform source problem. The cases are given by \cref{eq:cases}.}\label{tbl:fixed_cruciform_stats}
\setlength{\tabcolsep}{3pt}
\resizebox{\columnwidth}{!}{
\begin{tabular}{cccccccc} 
\toprule
\textbf{Case}   & \textbf{Minimum~$\boldsymbol{(\sigma)}$} & \textbf{Q1~$\boldsymbol{(\sigma)}$} & \textbf{Median~$\boldsymbol{(\sigma)}$} & \textbf{Q2~$\boldsymbol{(\sigma)}$} & \textbf{Maximum~$\boldsymbol{(\sigma)}$} & \textbf{Mean~$\boldsymbol{(\sigma)}$} & $\boldsymbol{\epsilon_2(\mathbf{\Phi}^{\text{X}}, \mathbf{\Phi}^{\text{MC}})}$  \\ 
\hline\hline
CSR             & 0.000175                                 & 0.831                               & 2.46                                    & 6.85                                & 45.36                                    & 4.84                                  & 0.00310                                                                       \\
Mixed           & 0.000290                                 & 0.830                               & 2.46                                    & 6.85                                & 45.36                                    & 4.84                                  & 0.00310                                                                       \\
Mixed (rounded) & 0.036338                                 & 6.410                               & 9.27                                    & 28.03                               & 58.43                                    & 17.48                                 & 0.01723                                                                       \\
\bottomrule
\end{tabular}
}
\end{table}

\par The solutions of the unconverged solver found $f_{\text{leak}}^{\text{CSR}} = 6.914\times 10^{-2}$ and $f_{\text{leak}}^{\text{Mixed (rounded)}} = 6.800\times 10^{-2}$ which is $0.736\sigma$ and $-89.8\sigma$ off the MC reference. We note the leakage fraction and scalar flux solutions for Mixed recover to the CSR solution, indicating that the summation and rounding of $\mathcal{H}^{\text{TT}}$ and $\mathcal{S}^{\text{TT}}$ introduces a numerical diffusion in the patch boundaries and cross sections suppressing the flux especially within the source. This is apparent in \cref{fig:fixed_cruciform_zscore}, which shows colormaps of the $z$-score for CSR and Mixed (rounded) to the MC reference solution. Both plots show some slight oscillatory behavior with a lower error square artifact in $x, y\in[0, 3]$. We attribute the error oscillations to a combination of ray and geometric effects at patch boundaries, arising from higher continuity. In \cref{tbl:fixed_cruciform_stats} we show the statistics of the $z$-score as well as the $L_2$ error for each case. We again see significant differences between CSR and Mixed to Mixed (rounded), with all showing, on average, statistically significant deviations from the MC reference. 

\par A preconditioner is imperative to make solving multi-patch and high-fidelity problems with GMRES feasible; however, even for these multi-patch problems with large cross section discontinuities, TT provides significant compression in the interior operators. Further mesh refinement is needed to reduce scalar flux error to within statistical significance; however, the leakage fractions for CSR and Mixed were within $\sigma$ of OpenMC. 

\subsection{$k$-Eigenvalue}\label{sec:eig}

\par In the following we show three $k$-eigenvalue examples: a homogenized quarter circle in \cref{sec:eig_quarter_circle}, an infinite array of C5G7 fuel in \cref{sec:eig_c5g7}, and an infinite array of cruciform fuel in \cref{sec:eig_cruciform}. We solve all resulting linear systems with GMRES to $\epsilon_{\text{GMRES}} = 10^{-10}$ or a maximum number of restarts and iterations for each power iteration until $\epsilon_{\text{PI}} = 10^{-8}$ or we reach a maximum of $1000$ power iterations. All TTs are truncated to $\epsilon = 10^{-5}$. We compare the solution's eigenvalue and scalar flux with those of a reference analytical or OpenMC solution. To compare the eigenvalue, we use an error metric,
\begin{equation}\label{eq:eig_k_error}
    \delta k(k^{\text{X}}, k^{\text{ref}}) = k^{\text{X}} - k^{\text{ref}}.
\end{equation}
We present results only for the CSR and Mixed (rounded) cases on the GPU and normalize all scalar flux solutions to $\|\mathbf{\Phi}\|_2 = 1$ before computing errors.

\subsubsection{Quarter circle}\label{sec:eig_quarter_circle}

\begin{figure}
\centering
\begin{subfigure}{0.42\textwidth}
    \includegraphics[width=\textwidth]{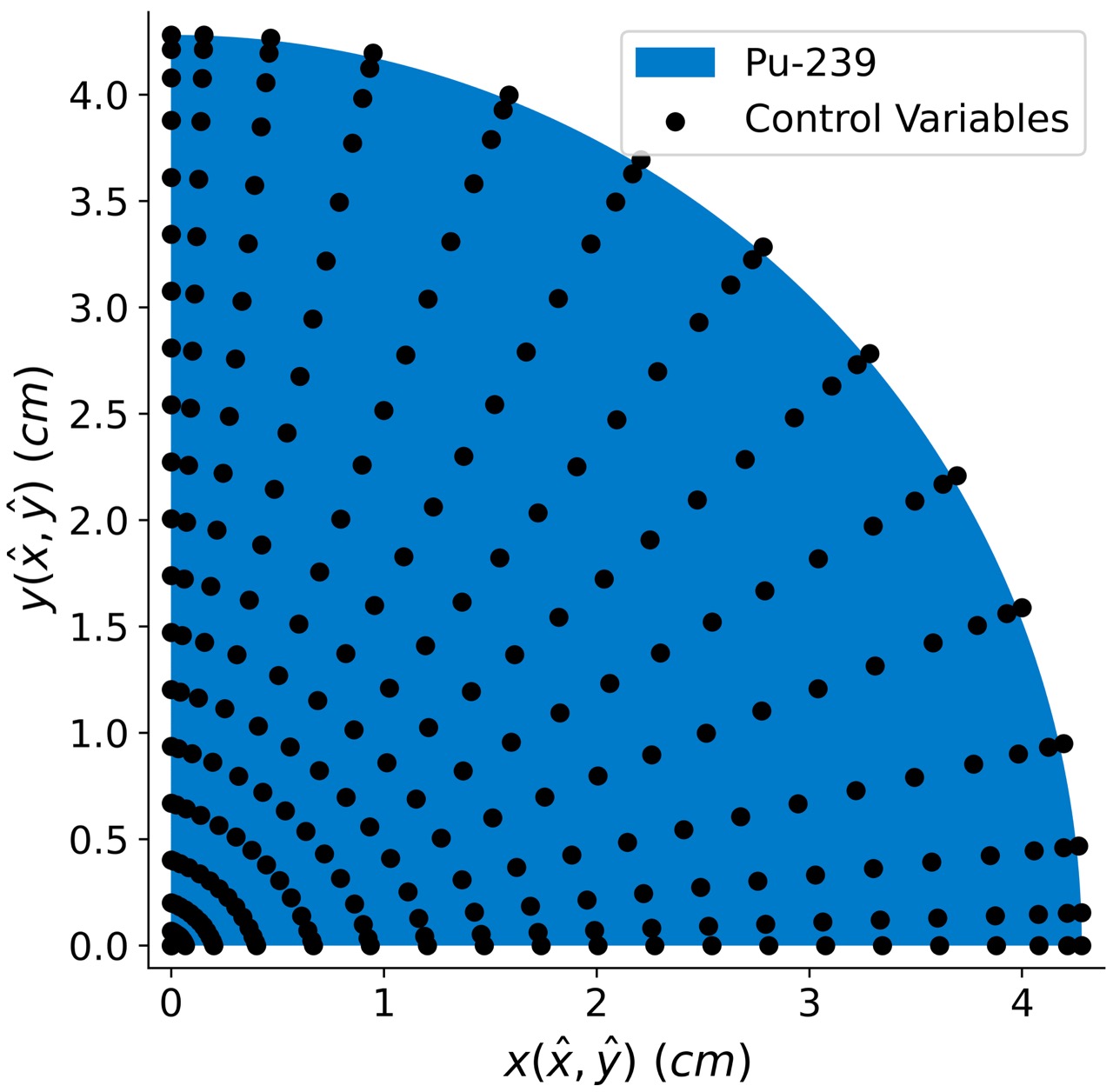}
    \caption{}
    \label{fig:eig_quarter_circle_geometry}
\end{subfigure}
\begin{subfigure}{0.53\textwidth}
    \includegraphics[width=\textwidth]{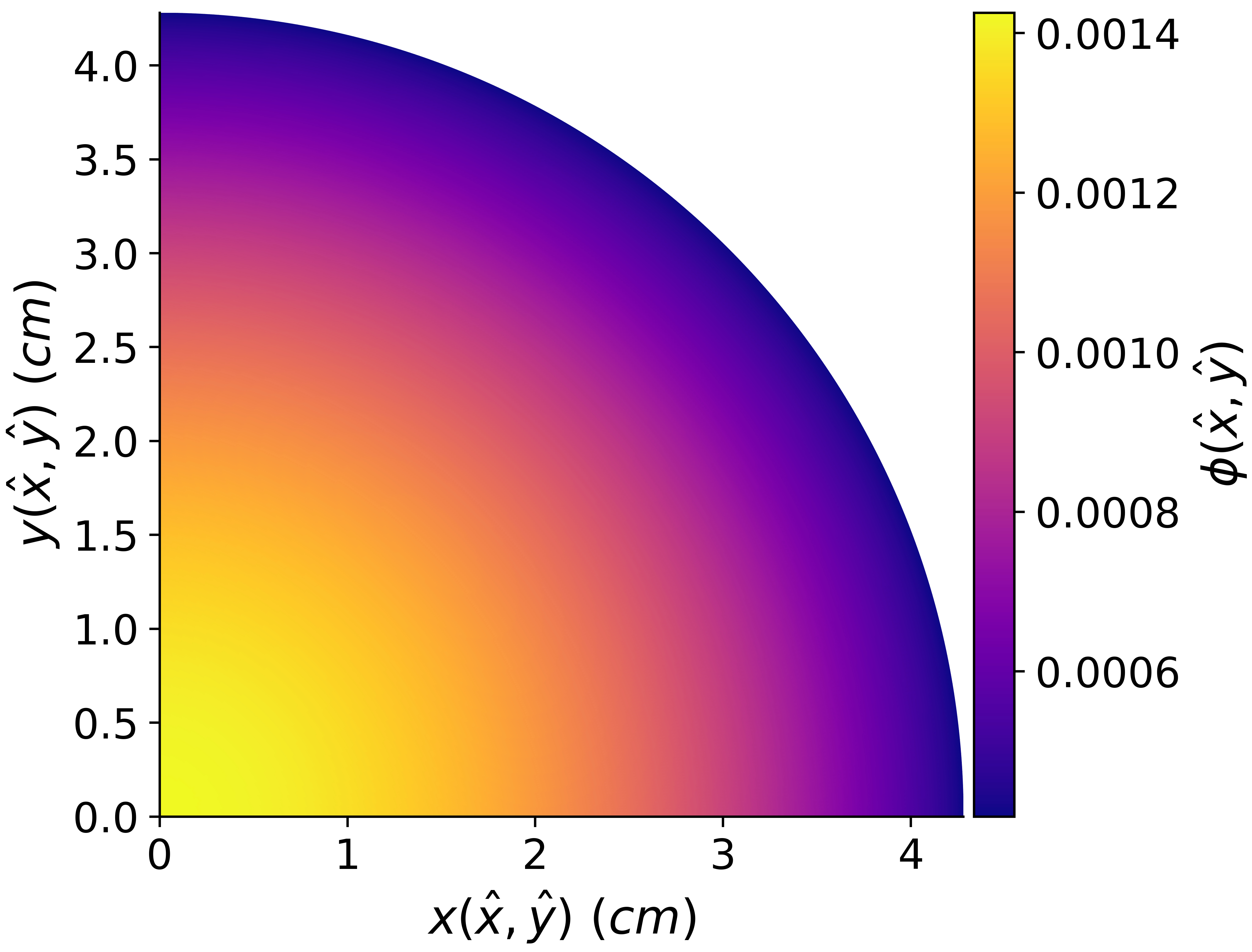}
    \caption{}
    \label{fig:eig_quarter_circle_phi}
\end{subfigure}
        
\caption{Eigenvalue critical circle problem from \cite{SOOD200355} with radius $r_c= 4.279960~cm$ represented in (a) as a quarter circle with reflecting boundaries on the left and bottom faces. We show the $N_{\hat x}^u = 10$ and $N_{\hat y}^u = 16$ mesh with $p_{\hat x} = p_{\hat y} = 4$ in (a) and the single-group scalar flux solution in (b).}
\label{fig:eig_quarter_circle}
\end{figure}

\par The first $k$-eigenvalue application is a homogenized critical circle from \cite{SOOD200355} with radius $r_c= 4.279960~cm$. This analytic benchmark has single-group ($N_E = 1$) cross sections $\s{t} = 0.32640~cm^{-1}$, $\nu\s{f} = 2.84 \cdot 0.081600~cm^{-1}$ and $\s{s} = 0.225216~cm^{-1}$. As shown in \cref{fig:eig_quarter_circle_geometry}, we model the geometry as a quarter-circle with reflective boundary conditions on the left and bottom faces and vacuum on the remaining faces. We discretize the LBTE with $N_\Omega = 4096$ ordinates and $N_e = 1$ patch with $N_{\hat x}^u =10$ and $N_{\hat y}^u = 16$ knot spans along each parametric dimension with $p_{\hat x} = p_{\hat y} = 4$ NURBS basis functions. We solve the resulting linear system of $N_\Omega \times N_E \times N_e\times (N_{\hat x}^u + p_{\hat x}) \times (N_{\hat y}^u + p_{\hat y}) = 1,146,880$ degrees of freedom using GMRES with $10$ restarts and $30$ iterations. The analytic benchmark provides reference solutions for the scalar flux in polar coordinates $\phi^{\text{ref}}(r)$: $\phi^{\text{ref}}(0.5r_c)/(\phi^{\text{ref}}(0))=0.8093$ and $\phi^{\text{ref}}(r_c)/(\phi^{\text{ref}}(0))=0.2926$. We define a relative error and an $L_2$ error,
\begin{subequations}\label{eq:eig_circle_error}
\begin{align}
    \delta\phi(r, \theta) &= \f{\phi^{\text{ref}}(0)}{\phi^{\text{ref}}(r)}\left(\f{\phi(r, \theta)}{\phi(0, \theta)} - \f{\phi^{\text{ref}}(r)}{\phi^{\text{ref}}(0)}\right),\label{eq:eig_circle_relerror}\\
    \epsilon_2(r) &= \f{\phi^{\text{ref}}(0)}{\phi^{\text{ref}}(r)}\sqrt{\frac{2}{\pi}\int_0^{\pi/2}\left(\f{\phi(r, \theta)}{\phi(0, \theta)} - \f{\phi^{\text{ref}}(r)}{\phi^{\text{ref}}(0)}\right)^2d\theta},\label{eq:eig_circle_l2error}
\end{align}
\end{subequations}
respectively.

\begin{table}
\centering
\caption{Compression ratios (CRs) and tensor train (TT) ranks for the TT and Compressed Sparse Row (CSR) formats for the $k$-eigenvalue problems. We show the CRs for the left-hand side (LHS) of the CSR (\cref{eq:cases_csr}) and Mixed (rounded) (\cref{eq:cases_mixed_rounded}) cases.}\label{tbl:eig_ranks}
\begin{tabular}{cccc} 
\toprule
\textbf{Operator}          & \textbf{TT-Ranks}             & \textbf{TT CR}        & \textbf{CSR CR}       \\ 
\hline\hline
\multicolumn{4}{c}{Homogeneous Quarter Circle}                                                          \\ 
\midrule
$\mathcal{H}$              & $\{3, 3, 3, 13\}$          & $4.199\times 10^{7}$  & $9.390\times 10^{3}$  \\
$\mathcal{S}$              & $\{1, 1, 1, 5\}$           & $2.608\times 10^{8}$  & $1.593\times 10^{7}$  \\
$\mathcal{B}_{\text{in}}$  & $\{2, 2, 2, 2\}$           & $1.341\times 10^{8}$  & $5.352\times 10^{5}$  \\
$\mathcal{B}_{\text{out}}$ & $\{4, 36, 27, 3\}$         & $1.133\times 10^{6}$  & $4.478\times 10^{5}$  \\
LHS of \cref{eq:cases_csr}                     & ---                        & ---                   & $9.297\times 10^{3}$  \\
LHS of \cref{eq:cases_mixed_rounded}                    & ---                        & \multicolumn{2}{c}{$3.067\times 10^{5}$}      \\ 
\midrule
\multicolumn{4}{c}{Infinite Array of C5G7 Fuel}                                                         \\ 
\midrule
$\mathcal{H}$              & $\{3, 3, 3, 4, 12, 17\}$   & $4.505\times 10^{8}$  & $9.949\times 10^{4}$  \\
$\mathcal{S}$              & $\{1, 1, 1, 2, 4, 9\}$     & $2.356\times 10^{9}$  & $1.049\times 10^{6}$  \\
$\mathcal{F}$              & $\{1, 1, 1, 1, 2, 7\}$     & $4.962\times 10^{9}$  & $1.049\times 10^{6}$  \\
$\mathcal{B}_{\text{in}}$  & $\{8, 32, 49, 49, 49, 8\}$ & $2.504\times 10^{7}$  & $1.775\times 10^{6}$  \\
$\mathcal{B}_{\text{out}}$ & $\{4, 28, 46, 46, 48, 7\}$ & $3.117\times 10^{7}$  & $1.783\times 10^{6}$  \\
LHS of \cref{eq:cases_csr}                    & ---                        & ---                   & $8.829\times 10^{4}$  \\
LHS of \cref{eq:cases_mixed_rounded}                    & ---                        & \multicolumn{2}{c}{$1.130\times 10^{6}$}      \\ 
\midrule
\multicolumn{4}{c}{Infinite Array of Cruciform Fuel (BA)}                                               \\ 
\midrule
$\mathcal{H}$              & $\{3, 3, 3, 6, 23, 9\}$    & $2.715\times 10^{9}$  & $2.985\times 10^{5}$  \\
$\mathcal{S}$              & $\{3, 3, 2, 4, 9, 5\}$     & $9.207\times 10^{9}$  & $5.243\times 10^{5}$  \\
$\mathcal{F}$              & $\{1, 1, 1, 2, 4, 4\}$     & $3.294\times 10^{10}$ & $1.049\times 10^{6}$  \\
$\mathcal{B}_{\text{in}}$  & $\{8, 38, 89, 93, 50, 4\}$ & $7.490\times 10^{7}$  & $5.156\times 10^{6}$  \\
$\mathcal{B}_{\text{out}}$ & $\{4, 34, 85, 89, 47, 3\}$ & $8.675\times 10^{7}$  & $5.181\times 10^{6}$  \\
LHS of \cref{eq:cases_csr}                    & ---                        & ---                   & $1.862\times 10^{5}$  \\
LHS of \cref{eq:cases_mixed_rounded}                    & ---                        & \multicolumn{2}{c}{$3.261\times 10^{6}$}      \\ 
\midrule
\multicolumn{4}{c}{Infinite Array of Cruciform Fuel (Gas)}                                              \\ 
\midrule
$\mathcal{H}$              & $\{3, 3, 3, 5, 22, 9\}$    & $3.006\times 10^9$    & $2.984\times 10^5$    \\
$\mathcal{S}$              & $\{3, 3, 2, 3, 9, 5\}$     & $1.005\times 10^{10}$ & $5.243\times 10^5$    \\
$\mathcal{F}$              & $\{1, 1, 1, 1, 3, 4\}$     & $4.631\times 10^{10}$ & $1.049\times 10^6$    \\
$\mathcal{B}_{\text{in}}$  & $\{8, 38, 89, 93, 50, 4\}$ & $7.490\times 10^7$    & $5.156\times 10^6$    \\
$\mathcal{B}_{\text{out}}$ & $\{4, 34, 85, 89, 47, 3\}$ & $8.675\times 10^7$    & $5.181\times 10^6$    \\
LHS of \cref{eq:cases_csr}                    & ---                        & ---                   & $1.862\times 10^{5}$  \\
LHS of \cref{eq:cases_mixed_rounded}                    & ---                        & \multicolumn{2}{c}{$3.261\times 10^6$}        \\
\bottomrule
\end{tabular}
\end{table}

\par \Cref{tbl:eig_ranks} shows the TT-ranks and compression ratios for the DG-LBTE operators in TT and CSR format. We also report the compression ratio for the combined operator for the CSR and Mixed (rounded) cases. The findings for the interior operators show four orders of magnitude more compression with TT than with CSR for $\mathcal{H}$. Notably, the reflective boundaries occur on axis-aligned boundaries. They are the only contribution to $\mathcal{B}_{\text{in}}$ which has $r_{\max}(\mathcal{B}_{\text{in}}) = 2$. In contrast, the outflow boundary, which incorporates the entire problem boundary including the curvilinear boundary, has $r_{\max}(\mathcal{B}_{\text{out}}) = 36$. Like the fixed source square discussed in \cref{sec:fixed_quarter_circle}, axis-aligned boundaries are low-rank due to the axis symmetry of the Chebyshev-Legendre angular quadrature and the normal direction falling along the axes. Additionally, the Mixed (rounded) left-hand side (LHS) operator is two orders of magnitude more compressed than CSR. 

\begin{table}
\centering
\caption{Metrics and solution time for the eigenvalue homogeneous circle where $\delta k^{\text{ref}} = 1$. We compute the eigenvalue error using \cref{eq:eig_k_error} and the scalar flux $L_2$ error with \cref{eq:eig_circle_l2error}. The cases presented are from \cref{eq:cases}.}\label{tbl:eig_circle_metrics}
\begin{tabular}{ccccc} 
\toprule
\textbf{Case}   & \textbf{Time $\boldsymbol{(min)}$} & $\boldsymbol{\delta k~(pcm)}$ & $\boldsymbol{\epsilon_2(0.5r_c)}$ & $\boldsymbol{\epsilon_2(r_c)}$  \\ 
\hline\hline
CSR             & 7.65                               & $-0.116$                        & $1.176\times 10^{-4}$             & $3.732\times 10^{-3}$           \\
Mixed (rounded) & 17.02                              & $-0.125$                        & $1.108\times 10^{-4}$             & $3.723\times 10^{-3}$           \\
\bottomrule
\end{tabular}
\end{table}

\begin{figure}
\centering
\begin{subfigure}{0.47\textwidth}
    \includegraphics[width=\textwidth, trim=0.5cm 0.5cm 0.5cm 0.5cm, clip]{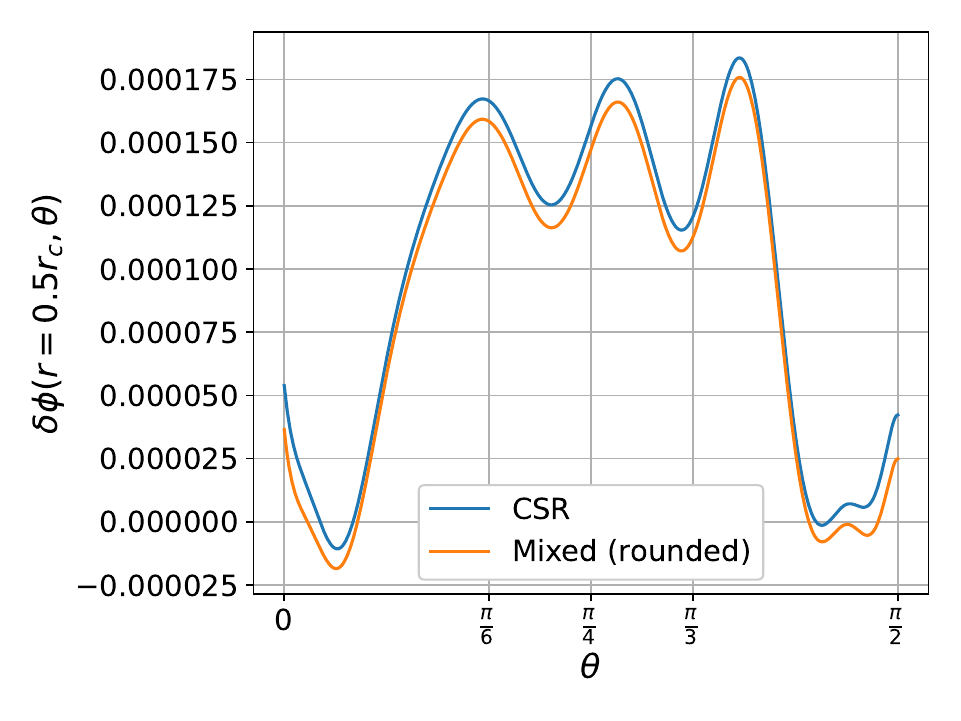}
    \caption{}
    \label{fig:eig_quarter_circle_error_0.5rc}
\end{subfigure}
\begin{subfigure}{0.47\textwidth}
    \includegraphics[width=\textwidth, trim=0.5cm 0.5cm 0.5cm 0.5cm, clip]{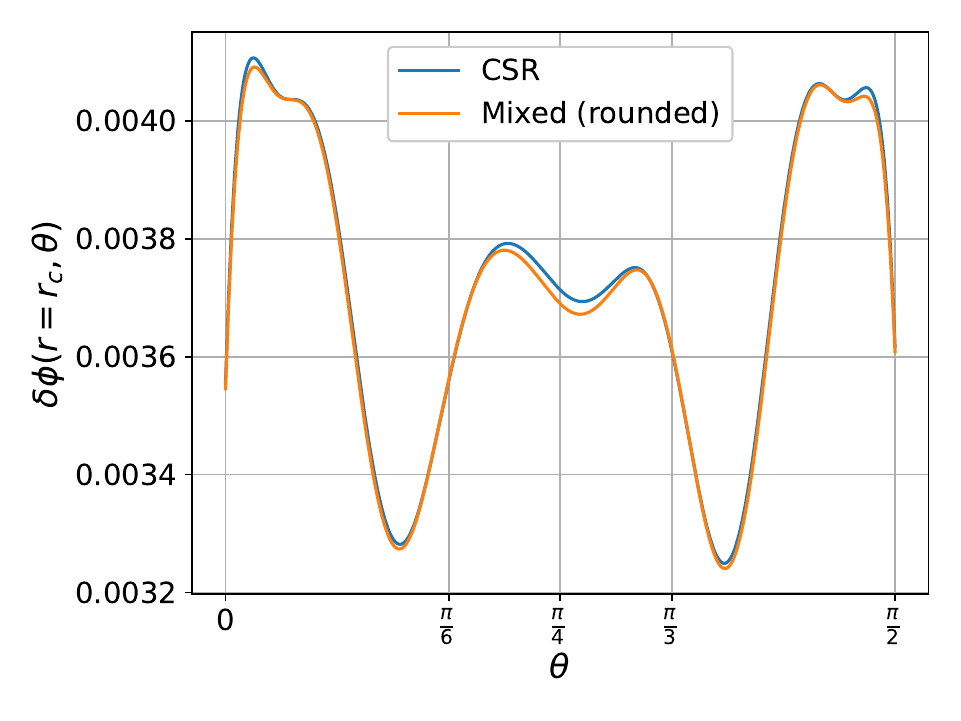}
    \caption{}
    \label{fig:eig_quarter_circle_error_rc}
\end{subfigure}
        
\caption{The relative error of the scalar flux computed with \cref{eq:eig_circle_relerror} as a function of polar angle for both CSR (\cref{eq:cases_csr}) and Mixed (rounded) (\cref{eq:cases_mixed_rounded}) cases for (a) $r = 0.5r_c$ and (b) $r = r_c$.}\label{fig:eig_circle_error}
\end{figure}

\par While the performance of TT compared to CSR in compression is significant, CSR is $2.22\times$ faster than Mixed (rounded) in solution time as shown in \cref{tbl:eig_circle_metrics}. Both yield an eigenvalue within a percent mille and minor $L_2$ errors, indicating good agreement with the analytic solution. The error increases radially with marginal differences between CSR and Mixed (rounded). We show the relative error of the scalar flux as a function of polar angle in \cref{fig:eig_circle_error}. The results align with \cref{tbl:eig_circle_metrics}, as Mixed (rounded) yields a slightly more accurate solution (from a scalar flux perspective). Although the expected solution should be independent of the polar angle, the NURBS mapping exhibits oscillations at a constant radius.

\par TDIGA yields a close approximation to the analytic problem in \cite{SOOD200355} with low $L_2$ errors and an eigenvalue within a percent mille, indicating that the exact geometric mapping of NURBS provides exceptional accuracy regardless of operator format. Again, the Mixed (rounded) case with interior operators in TT format achieves significant compression relative to CSR but incurs longer time-to-solution. 

\subsubsection{Infinite array of C5G7 fuel}\label{sec:eig_c5g7}

\begin{figure}
    \centering
    \includegraphics[width=0.5\linewidth]{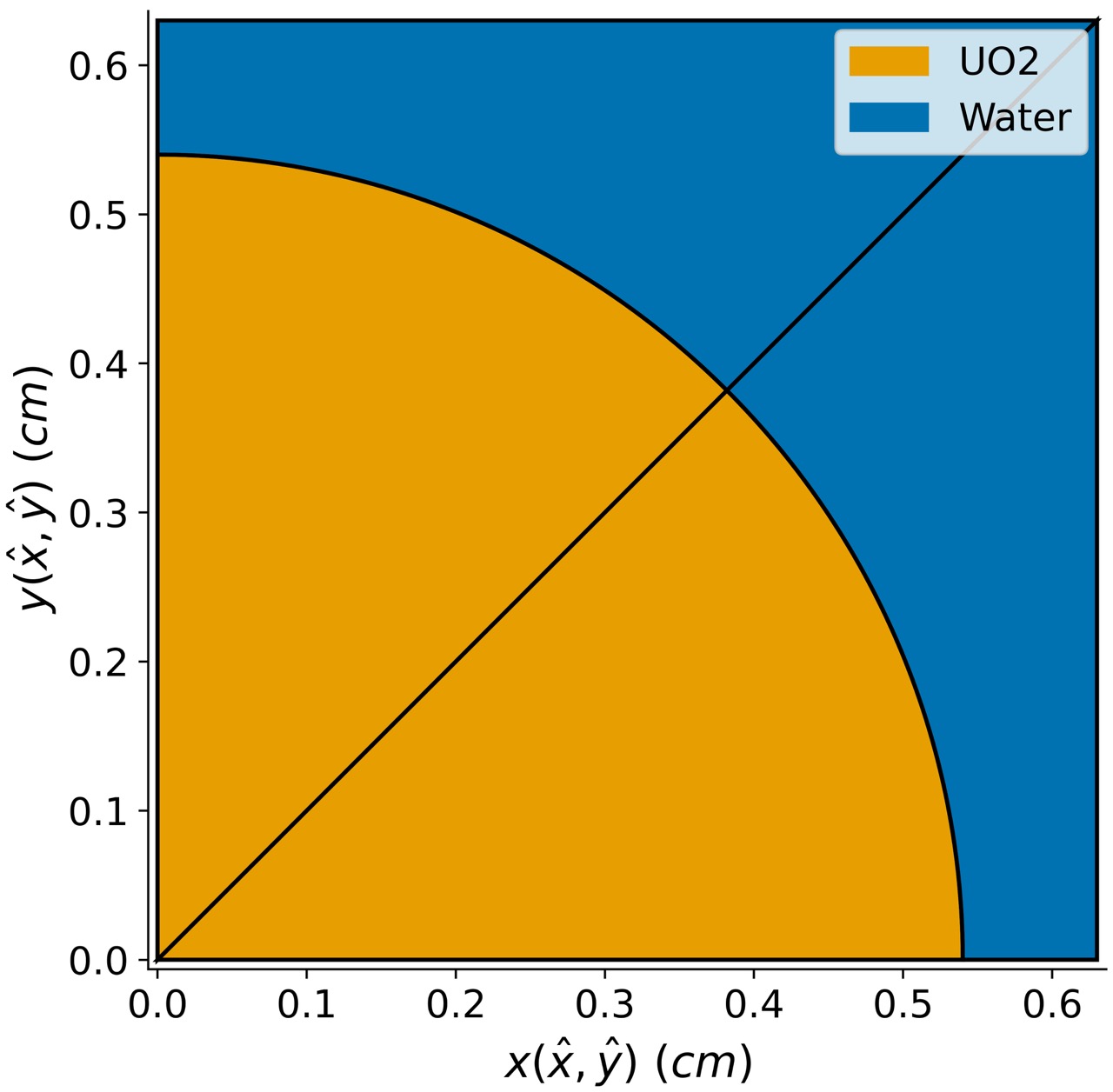}
    \caption{CAD geometry for the infinite array of C5G7 fuel. The problem is represented by a quarter-pin geometry with four patches (patch boundaries outlined in black) and reflective boundary conditions.}
    \label{fig:C5G7_pin}
\end{figure}

\begin{figure}
\centering
\begin{subfigure}{0.47\textwidth}
    \includegraphics[width=\textwidth, trim=0.5cm 0.5cm 0.5cm 0.5cm, clip]{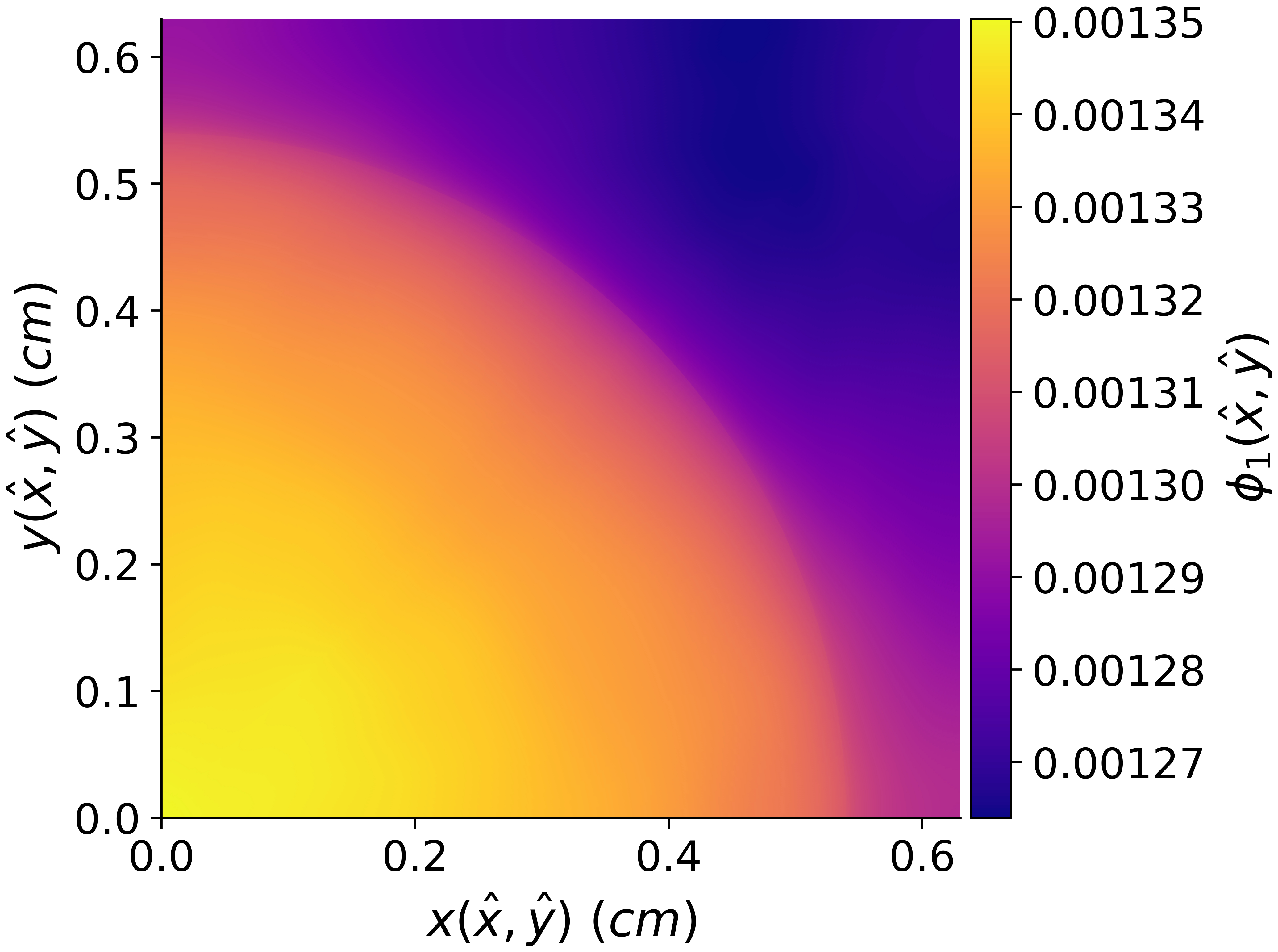}
    \caption{}
    \label{fig:eig_pincell_phi_1}
\end{subfigure}
\begin{subfigure}{0.47\textwidth}
    \includegraphics[width=\textwidth, trim=0.5cm 0.5cm 0.5cm 0.5cm, clip]{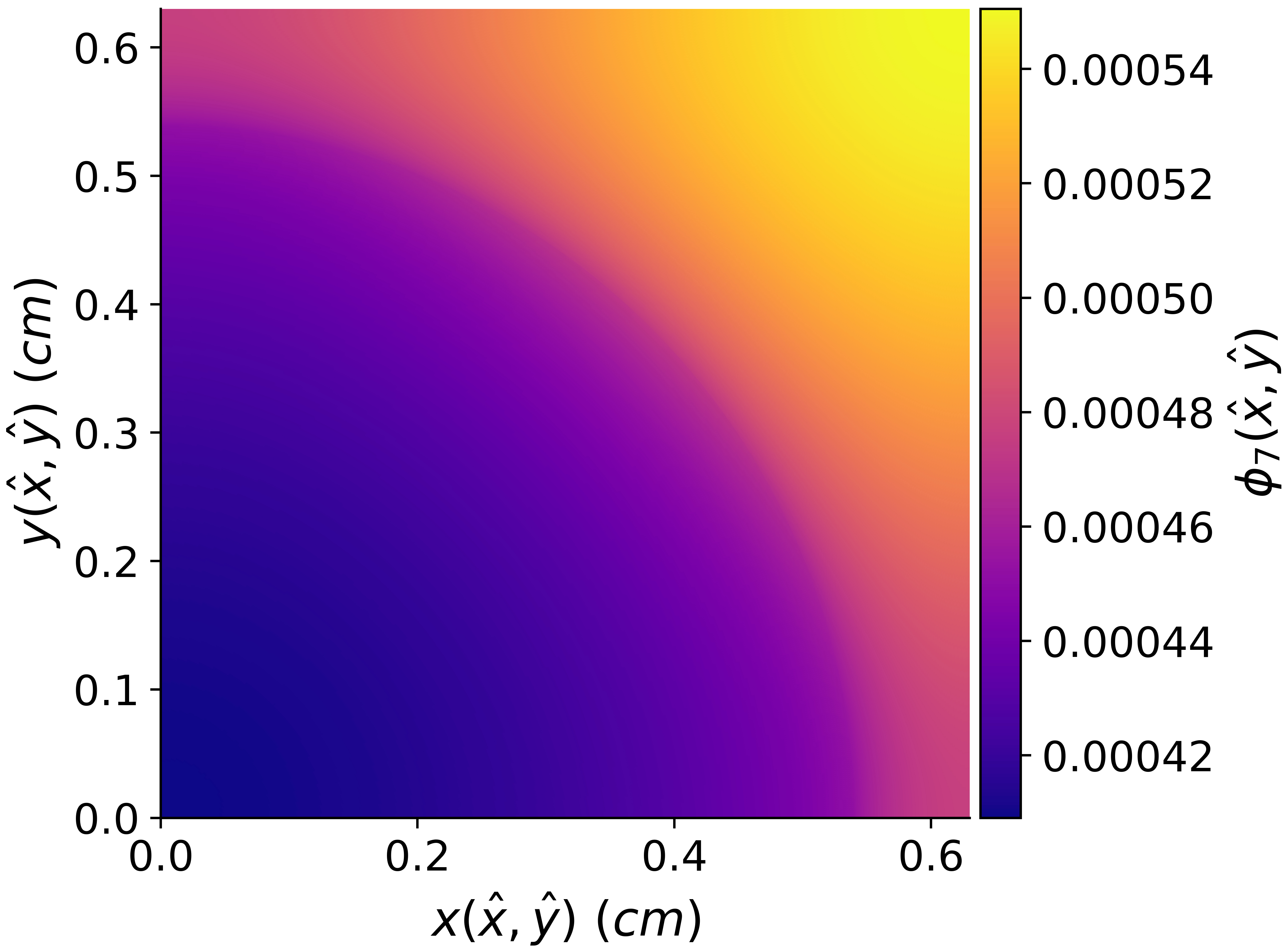}
    \caption{}
    \label{fig:eig_pincell_phi_7}
\end{subfigure}
        
\caption{The solutions of the (a) fastest ($i_E = 1$) and (b) slowest ($i_E = 7$) energy groups for the infinite array of C5G7 fuel solved with CSR (\cref{eq:cases_csr}).}\label{fig:eig_pincell_phi}
\end{figure}

\par The infinite array of C5G7 fuel is based on the C5G7 neutron transport benchmark \cite{C5G7} with seven-group cross sections and isotropic scattering. We represent an infinite array of $\text{UO}_2$ by the quarter circle shown in \cref{fig:C5G7_pin} with reflective boundary conditions on all sides. The reference OpenMC solution is $k^{\text{MC}} = 1.32559 \pm 0.00003$ and $\mathbf{\Phi}^{\text{MC}}, \boldsymbol{\sigma}^{\text{MC}}\in\mathbb{R}^{7\times 128\times 128}$. We used a discretization of $N_\Omega = 1024$, $N_E = 7$, $N_e = 4$, $N_{\hat x}^{u} = 10$, and $N_{\hat y}^u = 10$, with $p_{\hat x} = p_{\hat y} = 2$ resulting in $4,128,768$ degrees of freedom. Each power iteration used GMRES with a maximum of $10$ restarts and $75$ iterations. We show the solution of the fastest and slowest energy groups computed with CSR in \cref{fig:eig_pincell_phi}.

\par The TT-ranks and compression ratio of the operators are shown in \cref{tbl:eig_ranks} for CSR and TT formats. Again, interior operators see three to four orders of magnitude more compression in TT format than in the CSR format. Due to the separability in the interior operators and the isotropic scattering, we expect and observe rank-one coupling between angle, energy, and space. We observe significantly higher boundary operator ranks across dimensions, indicating a high degree of angular and spatial coupling; however, from a compression perspective, TT outperforms CSR. The LHS of the Mixed (rounded) case shows a similar compression boost to the homogeneous quarter-circle presented in the previous section.

\begin{table}
\centering
\caption{Solution times and error metrics for the infinite arrays of C5G7 fuel, cruciform fuel with a burnable absorber (BA) displacer, and cruciform fuel with a gas displacer. The eigenvalue error is computed with \cref{eq:eig_k_error} with $k^{\text{ref}} = k^{\text{MC}}$ and the $L_2$ error is computed with \cref{eq:eigenvector_error} with $\mathbf{\Phi}^{\text{ref}} = \mathbf{\Phi}^{\text{MC}}$. All cases refer to \cref{eq:cases}.}\label{tbl:eig_metrics}
\resizebox{\columnwidth}{!}{
\setlength{\tabcolsep}{3pt}
\begin{tabular}{ccccccccccc} 
\toprule
\textbf{Case}   & \textbf{Time $\boldsymbol{\left(min\right)}$} & $\boldsymbol{\delta k~\left(pcm\right)}$ & $\boldsymbol{\epsilon_2^1~\left(10^{-4}\right)}$ & $\boldsymbol{\epsilon_2^2~\left(10^{-4}\right)}$ & $\boldsymbol{\epsilon_2^3~\left(10^{-4}\right)}$ & $\boldsymbol{\epsilon_2^4~\left(10^{-4}\right)}$    & $\boldsymbol{\epsilon_2^5~\left(10^{-4}\right)}$    & $\boldsymbol{\epsilon_2^6~\left(10^{-4}\right)}$    & $\boldsymbol{\epsilon_2^7~\left(10^{-4}\right)}$    & $\boldsymbol{\epsilon_2~\left(10^{-4}\right)}$      \\ 
\hline\hline
\multicolumn{11}{c}{Infinite Array of C5G7 Fuel}                                                                                                                                                                                                                                                                                                       \\ 
\midrule
CSR                                                       & $10.89$                                         & $-21.34$                                 & $42.50$                                            & 4.853                                            & $5.157$                                            & $5.051$ & $5.628$ & $5.002$ & $4.275$ & $44.23$  \\
\begin{tabular}[c]{@{}c@{}}Mixed \\(rounded)\end{tabular} & $38.53$                                         & $-21.10$                                 & $42.49$                                            & 4.851                                            & 5.157                                            & 5.054 & 5.630 & 5.008 & 4.287 & 44.23  \\ 
\midrule
\multicolumn{11}{c}{Infinite Array of Cruciform Fuel (BA)}                                                                                                                                                                                                                                                                                             \\ 
\midrule
CSR                                                       & 250.48                                        & $-52.12$                                 & 31.99                                            & 5.261                                            & 5.683                                            & 5.985 & 9.240 & 12.96 & 12.71 & 39.17  \\
\begin{tabular}[c]{@{}c@{}}Mixed \\(rounded)\end{tabular} & 1584.29                                       & $-51.74$                                 & 31.99                                            & 5.261                                            & 5.692                                            & 5.997 & 9.223 & 12.93 & 12.68 & 39.15  \\ 
\midrule
\multicolumn{11}{c}{Infinite Array of Cruciform Fuel (Gas)}                                                                                                                                                                                                                                                                                            \\ 
\midrule
CSR                                                       & 291.74                                        & $-22.61$                                 & 39.98                                            & 7.166                                            & 6.782                                            & 8.779 & 8.682 & 7.611 & 10.67 & 44.95  \\
\begin{tabular}[c]{@{}c@{}}Mixed \\(rounded)\end{tabular} & 1352.34                                       & $-21.11$                                 & 39.98                                            & 7.164                                            & 6.786                                            & 8.787 & 8.687 & 7.563 & 10.60 & 44.92  \\
\bottomrule
\end{tabular}
}
\end{table}

\begin{figure}
\centering
\begin{subfigure}{0.47\textwidth}
    \includegraphics[width=\textwidth, trim=0.5cm 0.5cm 0.5cm 0.5cm, clip]{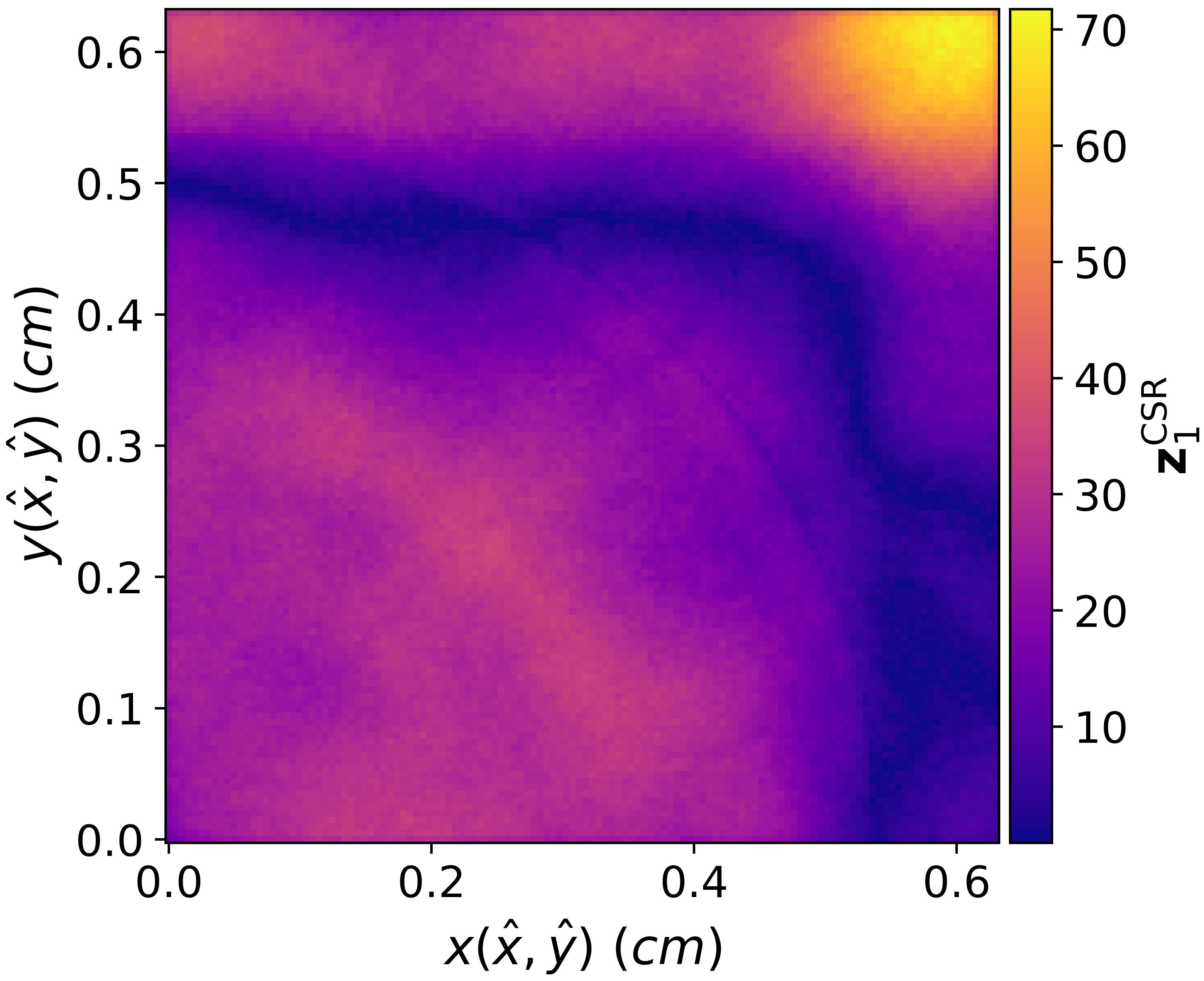}
    \caption{}
    \label{fig:eig_pincell_zscore_1}
\end{subfigure}
\begin{subfigure}{0.47\textwidth}
    \includegraphics[width=\textwidth, trim=0.5cm 0.5cm 0.5cm 0.5cm, clip]{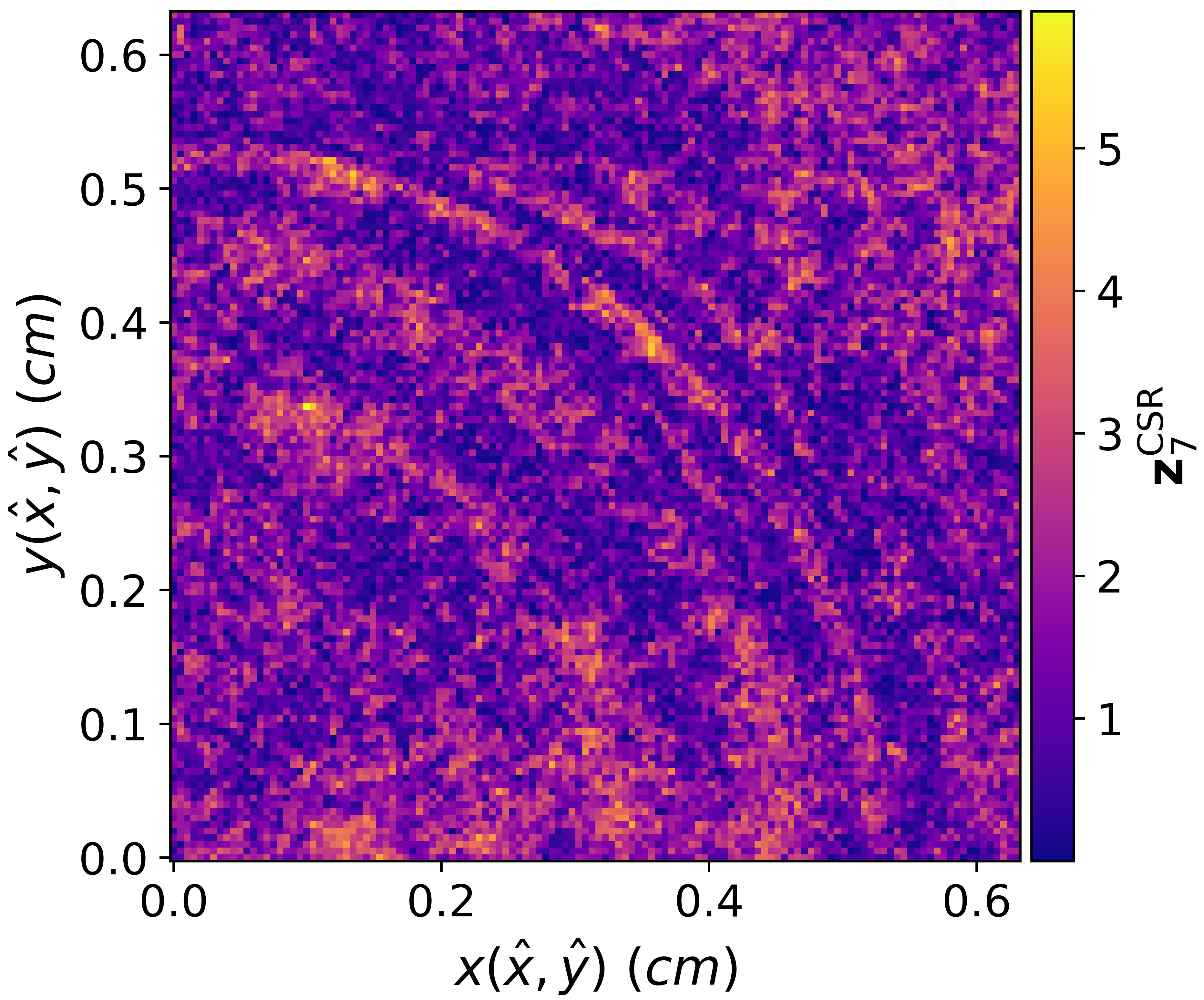}
    \caption{}
    \label{fig:eig_pincell_zscore_7}
\end{subfigure}
        
\caption{The $z$-score to the Monte Carlo (MC) reference solution for the (a) fastest ($i_E = 1$) and (b) slowest ($i_E = 7$) energy groups for the infinite array of C5G7 fuel solved with CSR (\cref{eq:cases_csr}).}\label{fig:eig_pincell_zscore}
\end{figure}

\par In \cref{tbl:eig_metrics} we show the solution time, eigenvalue error, and scalar flux $L_2$ errors for the CSR and Mixed (rounded) cases. The CSR again outperforms the Mixed (rounded) case by a factor of $3.53\times$ in time-to-solution. The errors are again negligible between CSR and Mixed (rounded), with the most significant error occurring in the fastest energy group. Again, a preconditioner may help the fastest-energy group converge more quickly; however, our approach of converging the global angular flux $L_2$ error neglects convergence behavior across energy groups. While the eigenvalue error is acceptable, it is statistically significant compared to the MC reference solution. We show the $z$-score of the fastest and slowest energy groups in \cref{fig:eig_pincell_zscore}. Note that these are from the CSR solution and are consistent with the Mixed (rounded) case. As expected, given the $L_2$ errors shown in \cref{tbl:eig_metrics}, the largest $z$-score occurs in the fastest energy group. We again observe slight oscillatory behavior in both groups and a low in the $z$-score following an axis-aligned square artifact, likely due to ray effects.

\subsubsection{Infinite array of cruciform fuel}\label{sec:eig_cruciform}

\begin{figure}
    \centering
    \includegraphics[width=0.5\linewidth]{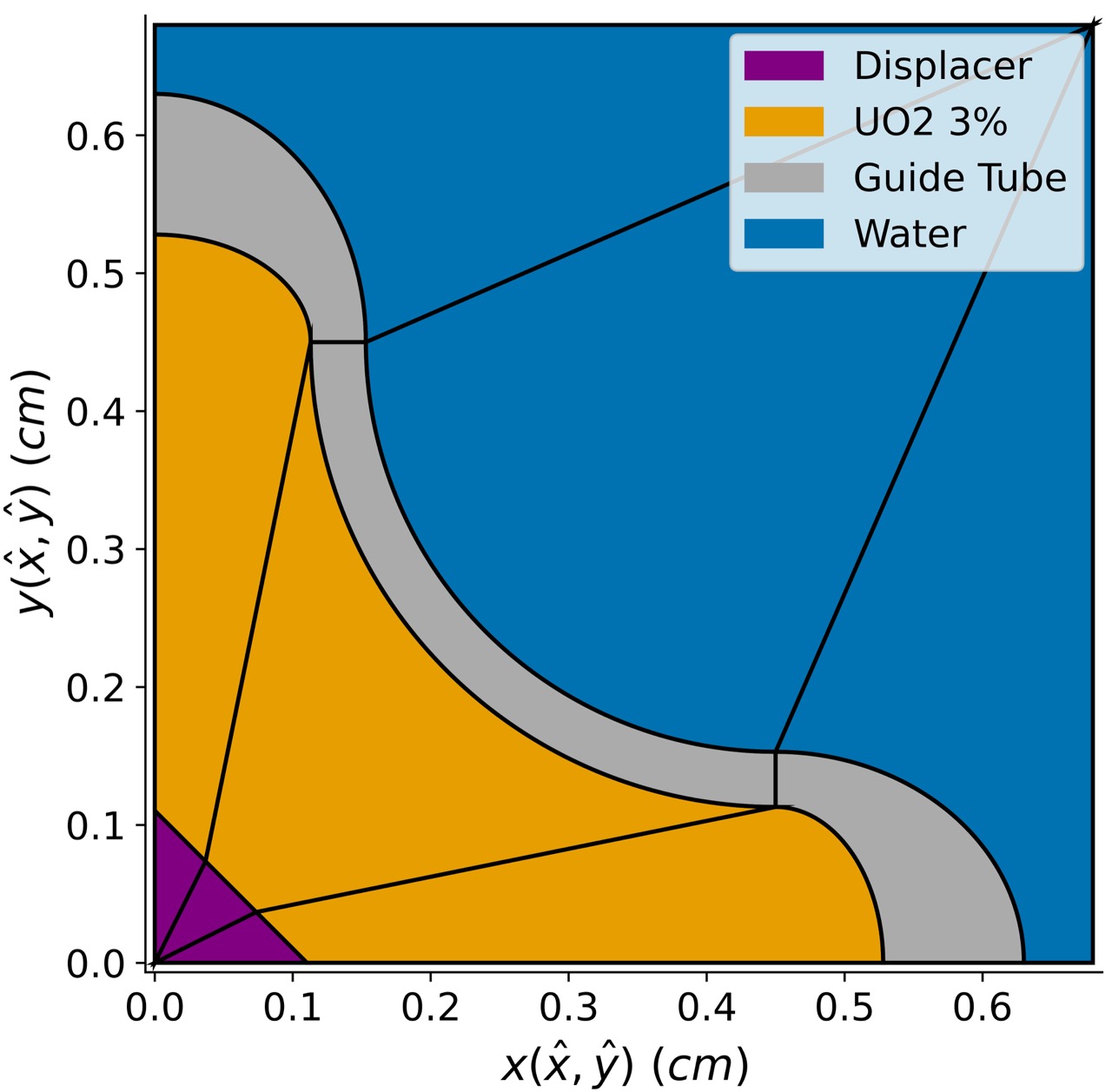}
    \caption{Geometry of the infinite array of cruciform fuel represented by $12$ patches (boundaries in black). The displacer is either a burnable absorber (BA) or a gas. All seven-group cross sections were taken from KAIST \cite{KAIST}.}
    \label{fig:eig_cruciform_geom}
\end{figure}

\begin{figure}
\centering
\begin{subfigure}{0.47\textwidth}
    \includegraphics[width=\textwidth, trim=0.5cm 0.5cm 0.5cm 0.5cm, clip]{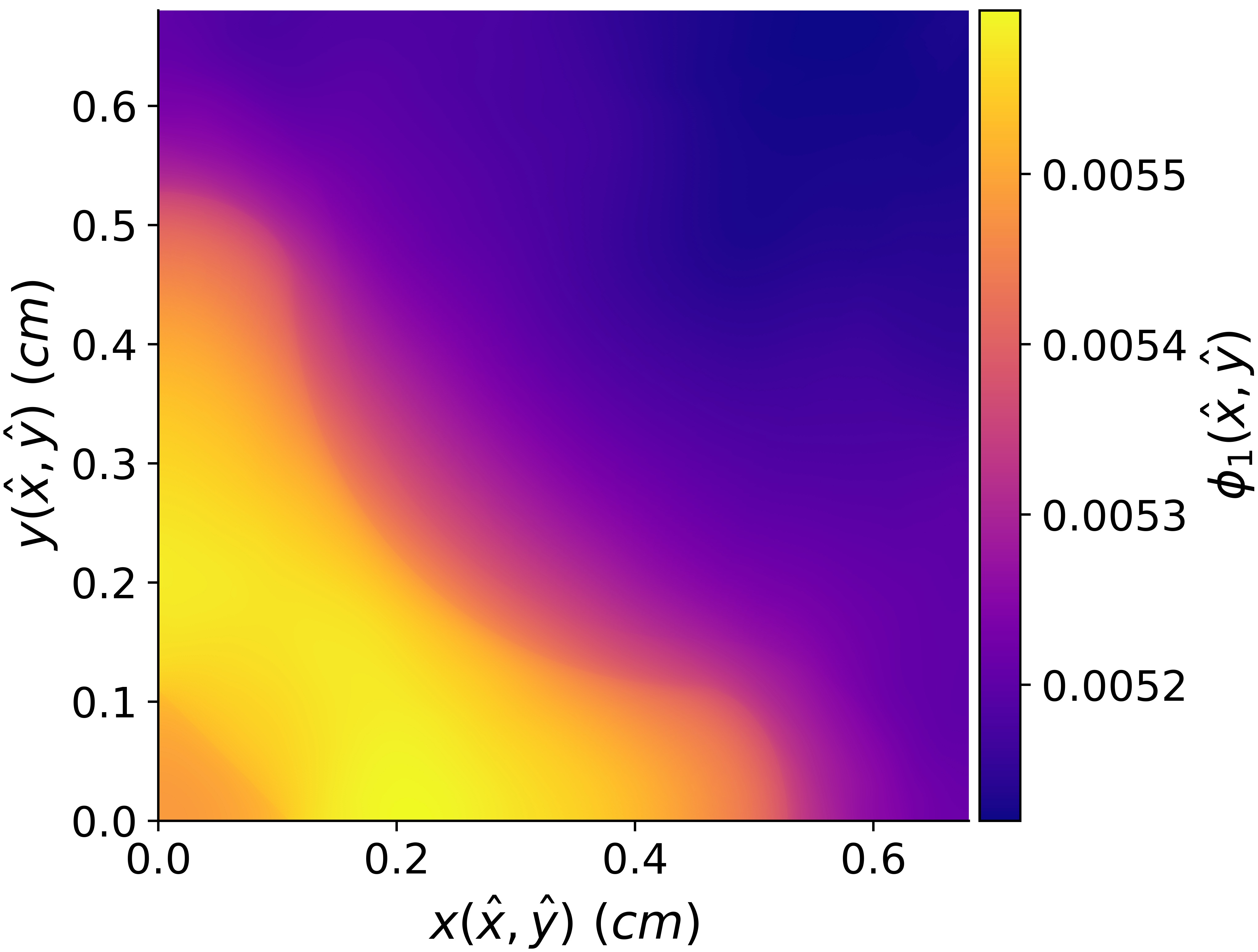}
    \caption{}
    \label{fig:eig_cruciform_phi_1_ba}
\end{subfigure}
\begin{subfigure}{0.47\textwidth}
    \includegraphics[width=\textwidth, trim=0.5cm 0.5cm 0.5cm 0.5cm, clip]{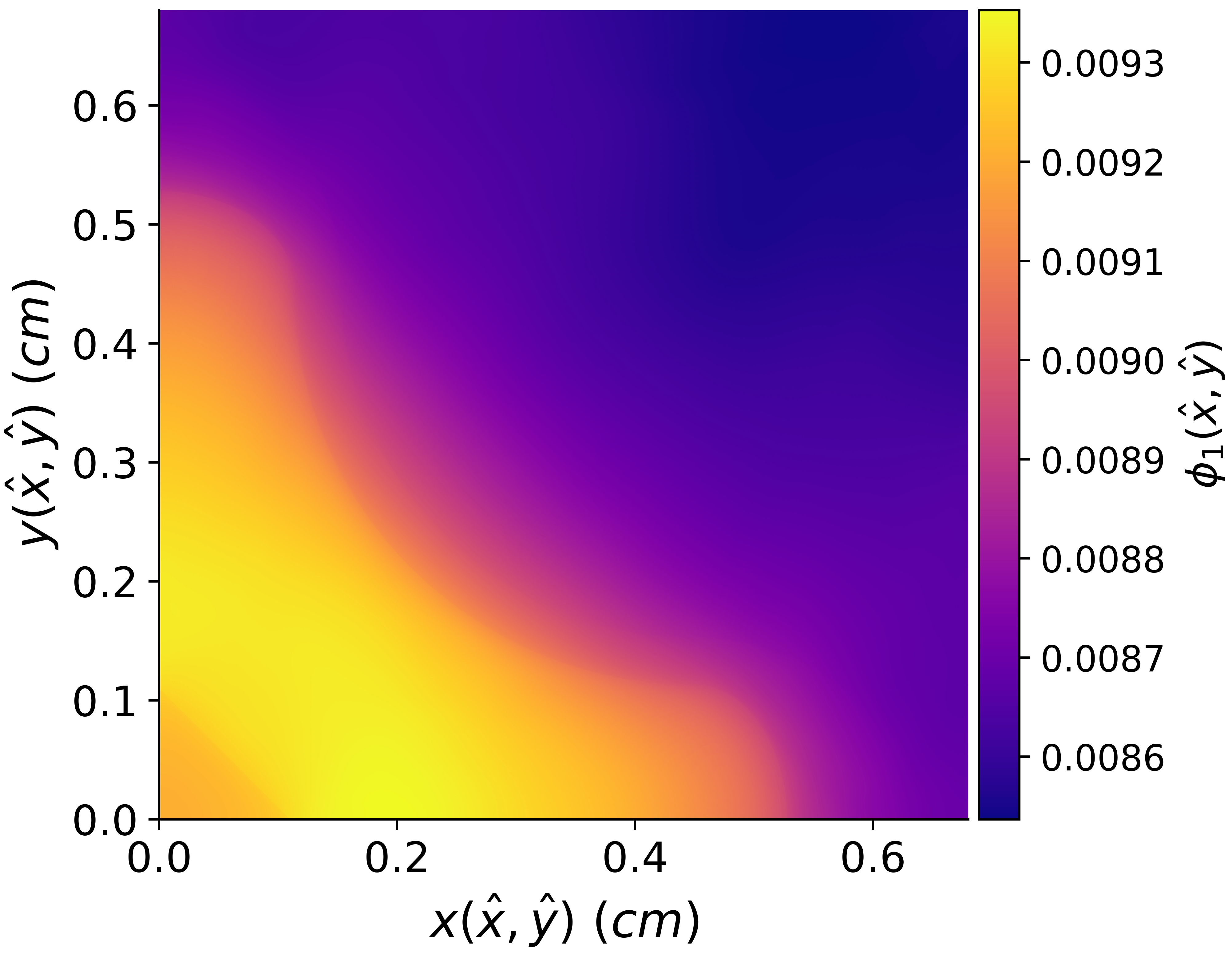}
    \caption{}
    \label{fig:eig_cruciform_phi_1_gas}
\end{subfigure}
\begin{subfigure}{0.47\textwidth}
    \includegraphics[width=\textwidth, trim=0.5cm 0.5cm 0.5cm 0.5cm, clip]{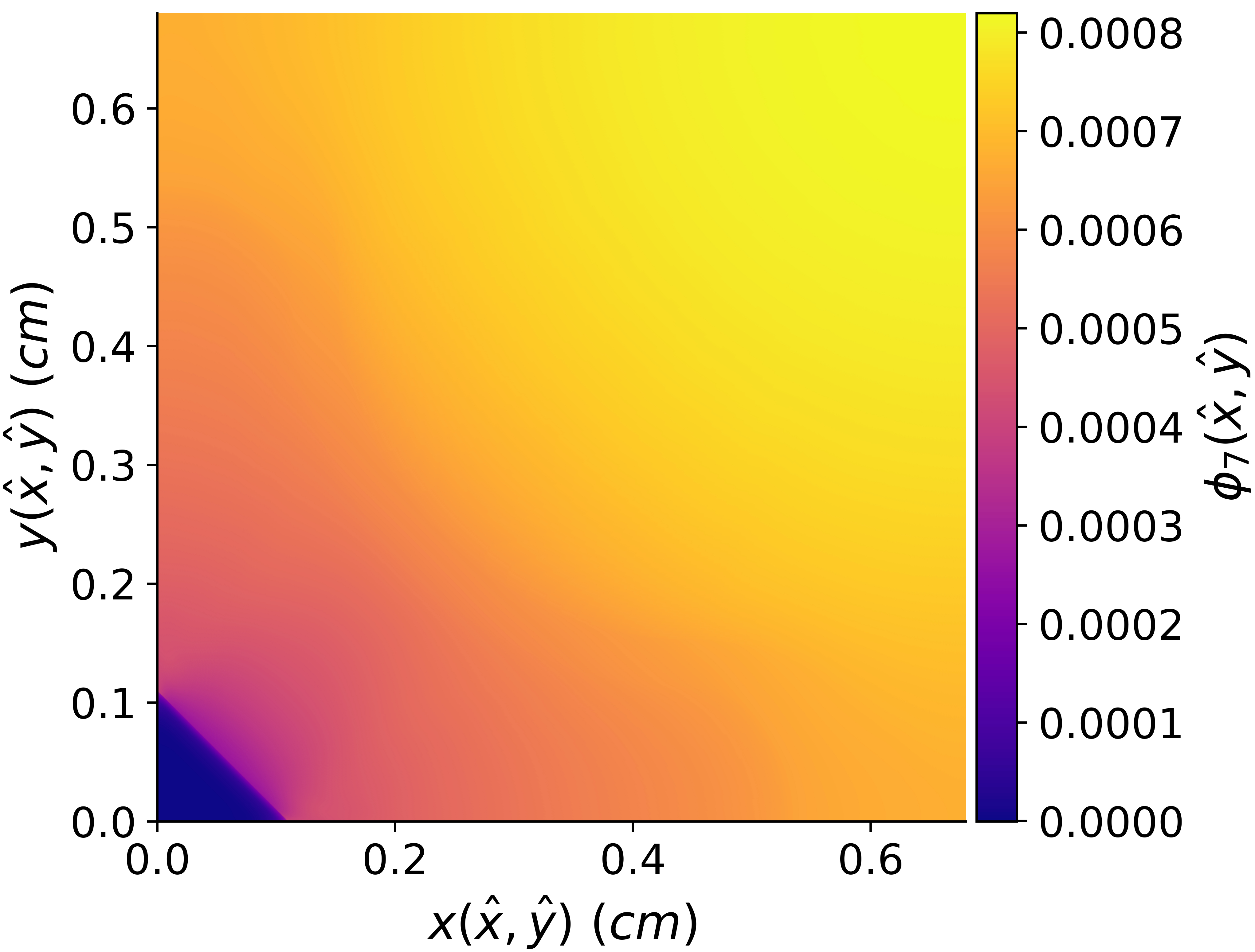}
    \caption{}
    \label{fig:eig_cruciform_phi_7_ba}
\end{subfigure}
\begin{subfigure}{0.47\textwidth}
    \includegraphics[width=\textwidth, trim=0.5cm 0.5cm 0.5cm 0.5cm, clip]{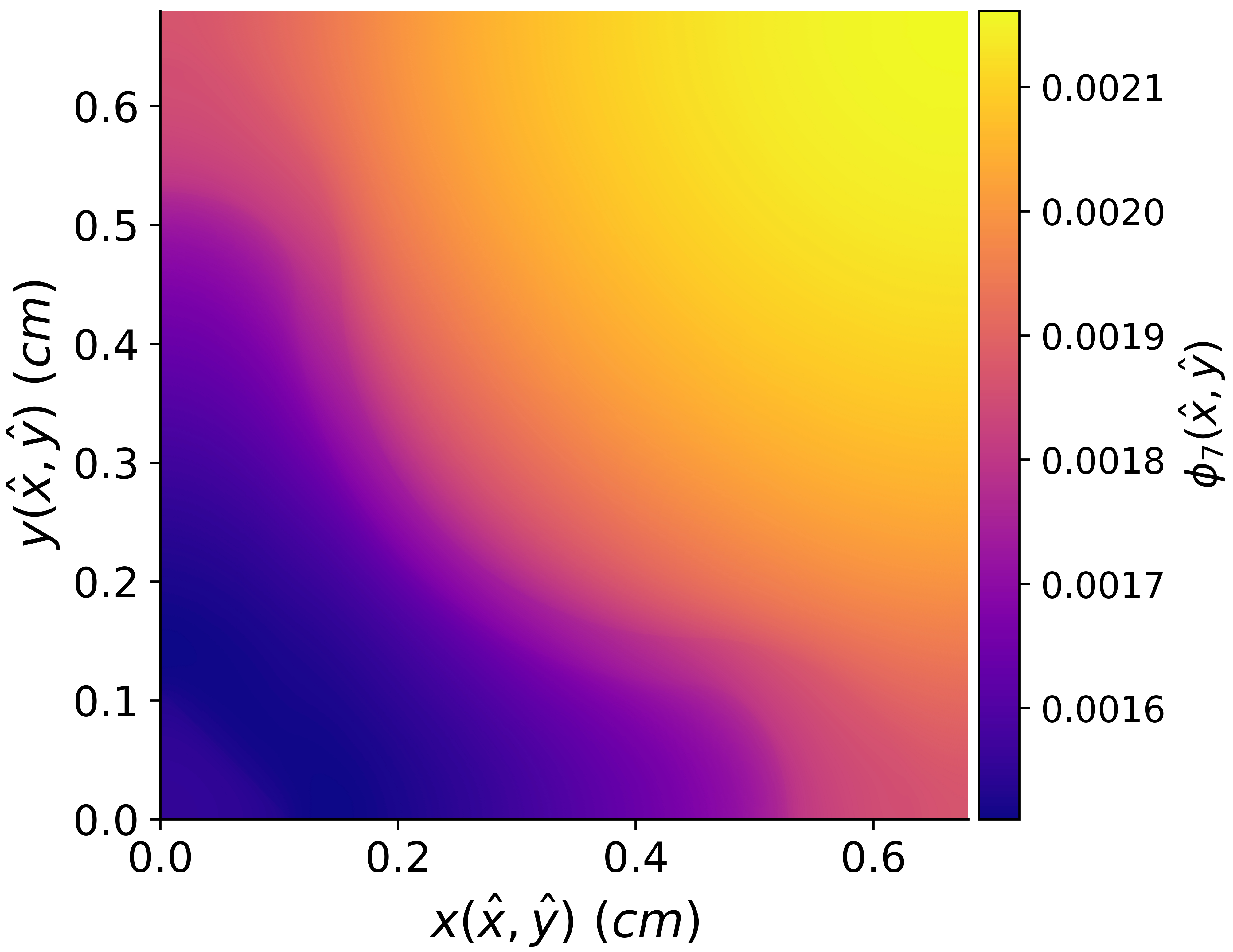}
    \caption{}
    \label{fig:eig_cruciform_phi_7_gas}
\end{subfigure}
        
\caption{The solutions for the (a, b) fastest ($i_E = 1$) and (c, d) slowest ($i_E = 7$) energy groups of the infinite array of cruciform fuel with a (left) burnable absorber (BA) or (right) gas displacer. These solutions were computed with CSR (\cref{eq:cases_csr}).}\label{fig:eig_cruciform_phi}
\end{figure}

\par \Cref{fig:eig_cruciform_geom} shows a quarter geometry of a four-lobe cruciform fuel pin modeled after similar designs published by Lightbridge \cite{Malone01122012}. We use all reflective boundary conditions and seven-group ($N_E = 7$) KAIST cross sections \cite{KAIST} with linearly anisotropic scattering. We solved the geometry in \cref{fig:eig_cruciform_geom} with burnable absorber (BA) and gas displacers, each having a reference OpenMC eigenvalue of $k^{\text{BA,ref}} = 0.83942\pm0.00006$ and $k^{\text{Gas,ref}} = 1.25669\pm0.00007$, respectively. We discretize the LBTE with $N_\Omega = 1025$ ordinates and $N_e = 12$ patches with polynomail degree $p_{\hat x} = p_{\hat y} = 2$ NURBS basis functions on $N_{\hat x}^u\times N_{\hat y}^u = 100$ knot spans where $N_{\hat x}^u= N_{\hat y}^u = 10$. The resulting linear system has $N_\Omega \times N_E\times N_e\times (N_{\hat x}^u + p_{\hat x}) \times (N_{\hat y}^u + p_{\hat y}) = 12,386,304$ degrees of freedom and is solved using GMRES with $10$ restarts and $75$ iterations. The solutions for the fastest ($i_E = 1$) and slowest ($i_E = 7$) energy groups are shown in \cref{fig:eig_cruciform_phi} for both the BA (left) and gas (right) displacers. All solutions were computed with CSR; however, as discussed later in this section, the Mixed (rounded) case closely matches the CSR solution.

\par We show the TT-ranks and CRs for the DG-LBTE operators in \cref{tbl:eig_ranks}, in both the TT and CSR formats, for both displacer types. The observations are consistent with prior findings, with TT demonstrating superior compression across all operators. Again, interior operators compress considerably more than boundary operators. Since this problem is linearly anisotropic, the scattering operator is no longer rank-one as the scattering cross section tensor and spherical harmonic operator are decomposed and rounded to the same $\epsilon = 10^{-5}$. We therefore observe an increase in the rank between angle and energy for $\mathcal{S}$ relative to the prior problems with isotropic scattering.

\begin{figure}
\centering
\begin{subfigure}{0.47\textwidth}
    \includegraphics[width=\textwidth, trim=0.5cm 0.5cm 0.5cm 0.5cm, clip]{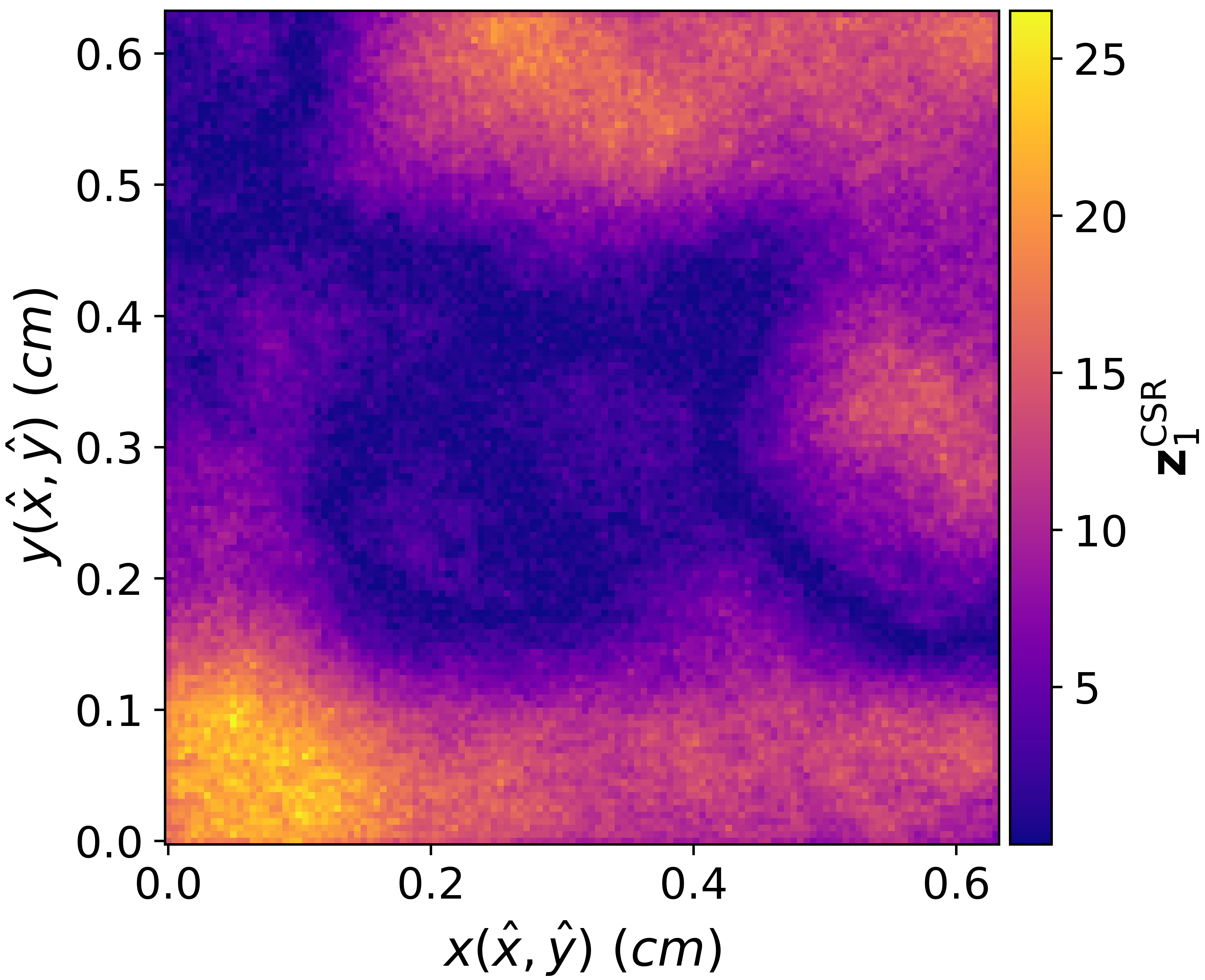}
    \caption{}
    \label{fig:eig_cruciform_zscore_1_ba}
\end{subfigure}
\begin{subfigure}{0.47\textwidth}
    \includegraphics[width=\textwidth, trim=0.5cm 0.5cm 0.5cm 0.5cm, clip]{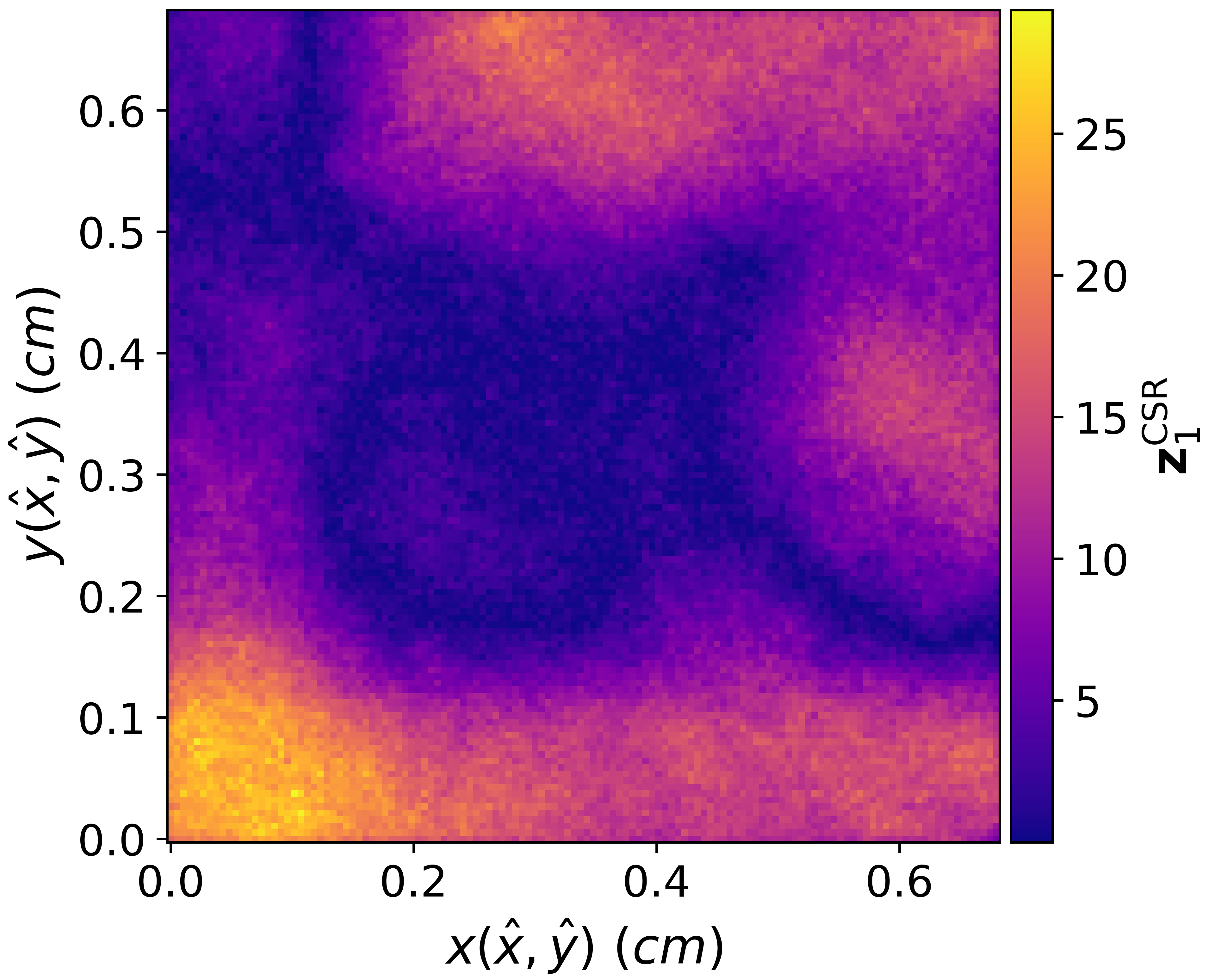}
    \caption{}
    \label{fig:eig_cruciform_zscore_1_gas}
\end{subfigure}
\begin{subfigure}{0.47\textwidth}
    \includegraphics[width=\textwidth, trim=0.5cm 0.5cm 0.5cm 0.5cm, clip]{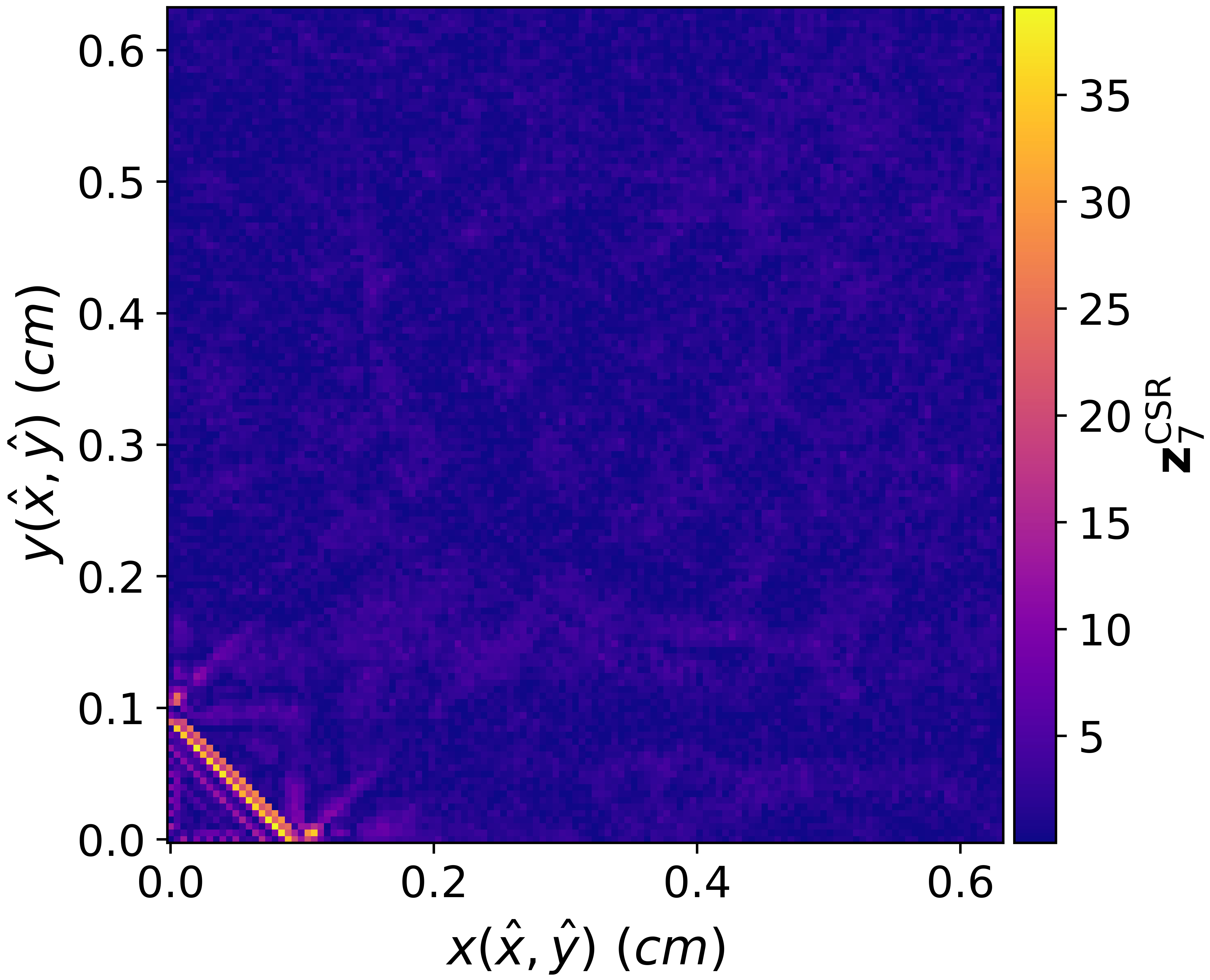}
    \caption{}
    \label{fig:eig_cruciform_zscore_7_ba}
\end{subfigure}
\begin{subfigure}{0.47\textwidth}
    \includegraphics[width=\textwidth, trim=0.5cm 0.5cm 0.5cm 0.5cm, clip]{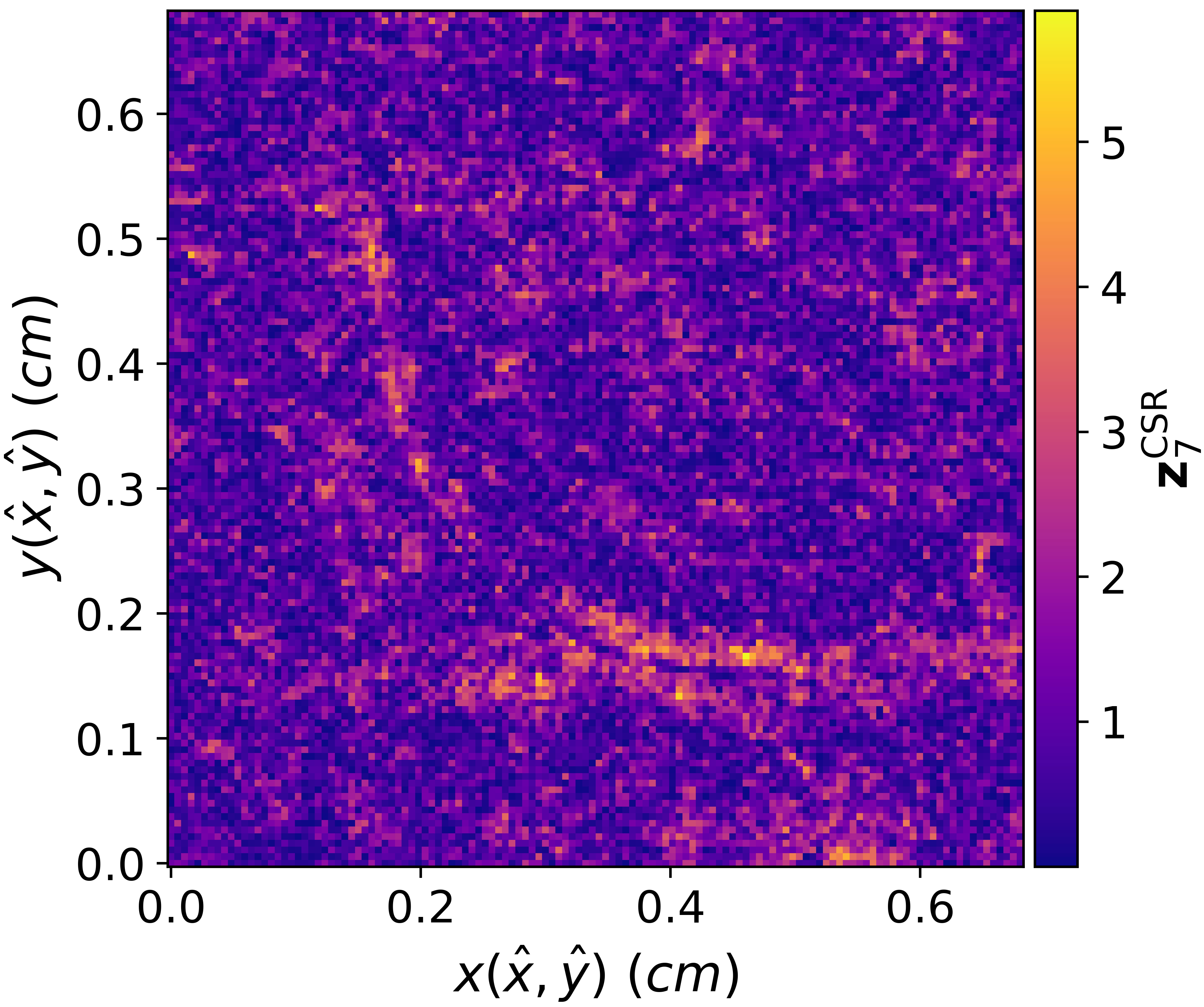}
    \caption{}
    \label{fig:eig_cruciform_zscore_7_gas}
\end{subfigure}
        
\caption{The $z$-score to the MC reference solution for the (a, b) fastest ($i_E = 1$) and (c, d) slowest ($i_E = 7$) energy groups for the infinite array of cruciform fuel with (left) burnable absorber (BA) or (right) gas displacers. All $z$-scores shown are those computed with CSR (\cref{eq:cases_csr}); however, the colormaps for Mixed (rounded) (\cref{eq:cases_mixed_rounded}) closely match CSR.}\label{fig:eig_cruciform_zscore}
\end{figure}

\par Again, in \cref{tbl:eig_metrics}, we present the solution time, eigenvalue error, and scalar flux $L_2$ error for the CSR and Mixed (rounded) cases for both the BA and gas displacers. CSR provides a $6.33\times$ and $4.64\times$ speedup over Mixed (rounded) for the BA and gas displacers, respectively. We attribute this difference to the slight increase in the rank of the BA interior operators relative to those of the gas. Additionally, Mixed (rounded) converged at 817 and 854 power iterations, compared with CSR's 825 and 963 power iterations for the BA and gas displacers, respectively. The errors between CSR and Mixed (rounded) are negligible; however, the eigenvalue and eigenvector errors of both to the MC reference are statistically significant. For the BA displacer, the eigenvalues are $-9~\sigma$ off the reference MC solution, while they are $-3~\sigma$ for the gas displacer problem. Both displacers exhibit a similar pattern: the majority of errors occur in the fastest energy group, followed by a decrease through groups five or six, after which the $L_2$ error increases slightly. We show the $z$-score for the fastest and slowest energy groups in \cref{fig:eig_cruciform_zscore}. Both the BA and gas displacers demonstrate matching oscillatory errors for the fastest energy group. However, for thermal energies, the considerable absorption cross section of the BA displacer produces a significant discontinuity in the scalar flux at the BA-$\text{UO}_2$ boundary, resulting in the largest $z$-score of all groups. The maximum $z$-score for the gas occurs in the fastest energy group, whereas those in the thermal range are significantly lower. The thermal range does show a collection of errors at the cladding-water interface. As shown in \cite{OWENS2017215}, further refinement near these material discontinuities will offer better agreement with the reference MC solution; however, the limitations of TT necessitate the same number of control points in each patch. 
\section{Conclusions and Future Work}\label{sec:conc}

\par We have presented the TDIGA method applied to the 2-D time-independent DG-LBTE discretized with discrete ordinates in angle, multigroup in energy, and IGA in space. Our approach builds on the PDG scheme developed in \cite{OWENS2016501} with a trivial generalization to their FDG scheme. We derived the 2-D time-independent DG-LBTE with IGA and higher-order scattering, and discussed an assembly procedure for the PDE operators in TT format. We solved the discretized equation using unpreconditioned restarted GMRES with operators in either TT or CSR format, and with an uncompressed solution vector. We applied the TDIGA method to fixed source and $k$-eigenvalue single-patch and multi-patch examples. We examined the behavior of operator compression, TT-ranks, operator application time, and solution error as angular and mesh resolutions increased and the NURBS basis function degree varied. We applied TDIGA to a shielded cruciform fixed-source problem involving optically thick and thin materials. The $k$-eigenvalue applications included a homogeneous circular source with an analytic reference solution, an infinite array of C5G7 fuel, and an infinite array of cruciform fuel with seven-group cross sections and linearly anisotropic scattering.

\par Across all problems presented, the TT representation of the DG-LBTE streaming and collision, scattering, and fission operators offers significant compression compared to CSR due to their explicit separability of space, angle, and energy. As angular resolution increases, the interior operator rank remains constant. Curvilinear boundaries yield a higher maximum rank for the interior operators due to the increased complexity of the Jacobian mapping. However, all interior operators had $r_{\max} < 50$ for $\epsilon = 10^{-8}$ and $r_{\max} < 25$ for $\epsilon=10^{-5}$ with little to no difference with increasing NURBS basis function degree or number of patches. However, the number of patches considered in this work is relatively low with $N_e \le 12$. Therefore, the effect of the number of patches with differing Jacobians on operator rank remains an open question. With increasing spatial mesh resolution, the maximum rank decreased as the Jacobian mapping became smoother and more regular, irrespective of the number of patches. In the homogeneous circle presented in \cref{sec:fixed_homo}, the worst-case scenario for TT compression of interior operators, TT compressed the scattering and collision operator from 11.4 PB to 9.2 MB for $N_\Omega = 262144$, $N_{\hat x}^u = N_{\hat y}^u = 10$, and $p_{\hat x} = p_{\hat y} = 2$ while CSR managed 12.5 GB. However, due to high angular and spatial coupling, the boundary operator ranks explode to $r_{\max} > 500$ for non-axis-aligned meshes for $\epsilon\in\{10^{-8}, 10^{-5}\}$ with increasing angular resolution but plateau to $100 <r_{\max} < 150$ at the finest spatial discretization we considered.

\par We note all but the infinite array of cruciform source (\cref{sec:eig_cruciform}) demonstrate problems with isotropic scattering for which the scattering operator is rank-one between angle, energy, and space. However, for higher-order scattering, the scattering operator loses its explicit separability between angle and energy because the spherical-harmonic operator and the group-to-group scattering cross section tensor must be decomposed in an approximate TT format. The dependence of the rank of the scattering operator on scattering order remains an open research question and is likely problem-dependent. 

\par As stated in \cite{gorodetsky2025thermalradiationtransporttensor}, the truncation tolerance $\epsilon$ introduces a new axis for error convergence; however, given that the solution vector is uncompressed, its effects may be less than those discussed \cite{gorodetsky2025thermalradiationtransporttensor}. DG transport locally conserves particle balance; however, TT-rounding introduces an approximation that may potentially break particle balance. The GMRES convergence discussed in the shielded cruciform fixed-source problem (\cref{sec:fixed_cruciform}) demonstrated no differing behavior between the three cases presented, while the infinite array of cruciform fuel with gas required 109 fewer power iterations for summed and rounded interior operators in TT format compared to CSR. The ideal $\epsilon$ is problem dependent as for example in the shielded cruciform fixed-source problem the summation and rounding of $\mathcal{H}^{\text{TT}}$ and $\mathcal{S}^{\text{TT}}$ with significant variations in material properties produces a diffusion of the shield cross section outside of the shield, suppressing the flux while the infinite array of C5G7 fuel and cruciform fuel problems (\cref{sec:eig_c5g7,sec:eig_cruciform}) showed little to no difference using the same approach compared to CSR. We note this discrepancy may be recovered by applying $\mathcal{H}^{\text{TT}}$ and $\mathcal{S}^{\text{TT}}$ separately. The angular and mesh resolution scaling studies of the fixed source problems (\Cref{sec:fixed_homo,sec:fixed_quarter_circle}) showed either decreasing or constant error between $\epsilon\in\{10^{-8}, 10^{-5}, 10^{-3}\}$ and CSR for increasing angular or mesh resolution. This discrepancy proved to be insignificant in the eigenvalue problems (\cref{sec:eig}) compared to the error of the MC or analytic reference solutions.

\par Our decision to solve for the uncompressed solution vector proved correct as the angular flux was marginally compressible in the TT format for $\epsilon \in\{10^{-5}, 10^{-3}\}$ while $\epsilon = 10^{-8}$ was close to full rank for problems with curvalinear boundaries. However, this uncompressed representation resulted in operator-vector scaling at $\mathcal{O}(dN^dr^2\log(N))$ as opposed to $\mathcal{O}(dNr^3)$ if the angular flux was in TT format. This is significantly less favorable than $\mathcal{O}(\text{nnz})$ for CSR, resulting in longer time-to-solution for TT operators, especially those with high ranks from curvalinear boundaries. This is mitigated for mixed representations with low-rank interior operators in TT format, while the high-rank boundary operators are in CSR; however, this representation never fully recovers the CSR time-to-solution. Increasing the degree of the NURBS basis functions also reduces the gap between TT and CSR as the number of nonzeros per degree of freedom increases. Future work may identify more fruitful alternative TN formats beyond TT that compress both the PDE operators and the solution vector. Solver development for TN topologies may also yield greater efficiency.

\par Additionally, this multi-patch TT representation necessitates that there be at least the same number of control points in each patch, which significantly hampers the efficiency of potential AMR strategies. In future work, treating each patch independently with its own TT representation can restore local AMR and exploit the sparse structure within unrounded TT-cores, potentially reducing the $\mathcal{O}(dN^dr^2\log(N))$ scaling. However, the number of TTs required then scales as with the number of patches. Additionally, this eliminates the potential for compression in lattice structures such as fuel assemblies, where we can treat assembly and pin locations as additional dimensions in the TT operators.

\par As demonstrated by the multi-patch numerical examples discussed in \cref{sec:fixed_cruciform,sec:eig_c5g7,sec:eig_cruciform}, the number of GMRES and power iterations to converge is intractable for real-world applications without a preconditioner. Therefore, extending this work to more complex nuclear systems requires the development of efficient tensorized preconditioners. Given the exceptional compression of the interior operators in the TT format and the refinement-invariant, exact geometric representation offered by the NURBS basis, geometric multigrid (GMG) with TT operators may be a promising approach. An example of GMG with IGA was applied to the neutron diffusion equation in \cite{10.1115/ICONE26-81316}. However, traditional FEA spatial preconditioners are less effective for high-order meshes. Additionally, DG methods are commonly used in domain-decomposition frameworks, which enable scalability across multiple nodes or GPUs. Preconditioners in angle and energy will also be necessary.

\par Future work would also extend this to 3-D space; however, this presents some difficulties, particularly in the angular cores of the TT operators. Moving to 3-D adds another TT in the streaming operator. Along with expanding the quadrant cores to octant cores, to preserve rank one within the angular terms of the streaming operator, we require octant symmetry, which is not supported in the spherical-coordinate product quadrature of Chebyshev-Legendre. Alternatively, octant-symmetric quadrature sets, or absorbing the octant core into the polar and azimuthal cores, are options, the latter at the cost of one dimension. This also affects the boundary operators and complicates the spherical harmonic operator,
\begin{subequations}
    \begin{align}
        Y_l(\bo, \bo') &= Y_{e, l}^0(\bo)Y_{e, l}^0(\bo') + 2\sum_{m = 1}^lY_{e, l}^m(\bo)Y_{e, l}^m(\bo') + Y_{o, l}^m(\bo)Y_{o, l}^m(\bo'),\\
        Y_{o, l}^m(\bo) &= Y_{o, l}^m(\mu, \gamma) = \left[(2l + 1)\frac{(l - |m|)}{(l + |m|)}\right]^{1/2}\sin(m\gamma)(1 - \mu^2)^{|m|/2}\frac{d^{|m|}}{d\mu^{|m|}}P_l(\mu),
    \end{align}
\end{subequations}
with the presence of another term for the odd spherical harmonic function $Y_{o, l}^m(\bo)$. Finally, all TTs would gain a dimension for the control points along the $\hat z$-axis.

\section*{Acknowledgments}
\par This work was funded by the University Nuclear Leadership Program (UNLP) fellowship awarded to the first author by the U.S. Department of Energy Office of Nuclear Energy.

\section*{Data Availability}

\par Scripts and figures for the results presented in this paper can be found on our repository \texttt{ttnte\_tdiga\_jcp2025} (\href{https://github.com/myerspat/ttnte\_tdiga\_jcp2025}{https://github.com/myerspat/ttnte\_tdiga\_jcp2025}). There, we provide documentation on installing and setting up our transport code \texttt{ttnte} (\href{https://github.com/myerspat/ttnte}{https://github.com/myerspat/ttnte}) to reproduce the results. The raw data is hosted on Zenodo \cite{myers_2026_18294213}, and we provide documentation on installing it in \texttt{ttnte\_tdiga\_jcp2025}.

\section*{CRediT Author Statement}

\begin{itemize}
    \item \textbf{Patrick A. Myers}: Conceptualization, Data Curation, Formal Analysis, Funding Acquisition, Investigation, Methodology, Software, Validation, Visualization, Writing --- Original Draft
    \item \textbf{Joseph A. Bogdan}: Software, Validation, Writing --- Review and Edit
    \item \textbf{Majdi I. Radaideh}: Conceptualization, Project Administration, Resources, Supervision, Writing --- Review and Edit 
    \item \textbf{Brian C. Kiedrowski}: Conceptualization, Formal Analysis, Funding Acquisition, Project Administration, Resources, Supervision, Writing --- Review and Edit
\end{itemize}

\appendix
\setcounter{figure}{0}
\setcounter{table}{0}
\section{How to Read Tensor Network Diagrams}\label{app:ball_and_stick}

\begin{figure}
\centering
\begin{subfigure}[b]{0.3\textwidth}
\centering
\begin{tikzpicture}[
    thick,
    every node/.style={font=\small},
    tensor/.style={
        circle,
        draw=black,
        fill=blue!20,
        minimum size=7mm,
        inner sep=0pt
    },
    bigtensor/.style={
        circle,
        draw=black,
        fill=red!20,
        minimum size=10mm,
        inner sep=1pt
    },
    leg/.style={line width=1.2pt},
    freeindex/.style={},
    arrowstyle/.style={->, thick, line width=1pt},
    dotstyle/.style={font=\Large}
]

\node[tensor] (G1) at (0,0) {$\mathcal{X}$};

\node[freeindex] (N1) at (-1.2, 0) {$i_1$};

\draw[leg] (G1) -- (N1);

\end{tikzpicture}
\caption{$\mathcal{X}_i$}
\label{fig:app_vector}
\end{subfigure}
\begin{subfigure}[b]{0.3\textwidth}
\centering
\begin{tikzpicture}[
    thick,
    every node/.style={font=\small},
    tensor/.style={
        circle,
        draw=black,
        fill=blue!20,
        minimum size=7mm,
        inner sep=0pt
    },
    bigtensor/.style={
        circle,
        draw=black,
        fill=red!20,
        minimum size=10mm,
        inner sep=1pt
    },
    leg/.style={line width=1.2pt},
    freeindex/.style={},
    arrowstyle/.style={->, thick, line width=1pt},
    dotstyle/.style={font=\Large}
]

\node[tensor] (G1) at (0,0) {$\mathcal{A}$};

\node[freeindex] (N1) at (-1.2, 0) {$i_1$};
\node[freeindex] (N2) at (1.2, 0) {$i_2$};

\draw[leg] (G1) -- (N1);
\draw[leg] (G1) -- (N2);

\end{tikzpicture}
\caption{$\mathcal{A}_{i_1, i_2}$}
\label{fig:app_matrix}
\end{subfigure}
\begin{subfigure}[b]{0.3\textwidth}
\centering
\begin{tikzpicture}[
    thick,
    every node/.style={font=\small},
    tensor/.style={
        circle,
        draw=black,
        fill=blue!20,
        minimum size=7mm,
        inner sep=0pt
    },
    bigtensor/.style={
        circle,
        draw=black,
        fill=red!20,
        minimum size=10mm,
        inner sep=1pt
    },
    leg/.style={line width=1.2pt},
    freeindex/.style={},
    arrowstyle/.style={->, thick, line width=1pt},
    dotstyle/.style={font=\Large}
]

\node[tensor] (G1) at (0,0) {$\mathcal{T}$};

\node[freeindex] (N1) at (1.2, 0) {$i_2$};
\begin{scope}[rotate around={120:(G1)}]
    \node[freeindex] (N2) at (1.2,0) {$i_3$};
\end{scope}
\begin{scope}[rotate around={240:(G1)}]
    \node[freeindex] (N3) at (1.2,0) {$i_4$};
\end{scope}

\draw[leg] (G1) -- (N1);
\draw[leg] (G1) -- (N2);
\draw[leg] (G1) -- (N3);

\end{tikzpicture}
\caption{$\mathcal{T}_{i_2,i_3,i_4}$}
\label{fig:app_3d_tensor}
\end{subfigure}
\begin{subfigure}[b]{0.45\textwidth}
\centering
\resizebox{\columnwidth}{!}{
\begin{tikzpicture}[
    thick,
    every node/.style={font=\small},
    tensor/.style={
        circle,
        draw=black,
        fill=blue!20,
        minimum size=7mm,
        inner sep=0pt
    },
    bigtensor/.style={
        circle,
        draw=black,
        fill=red!20,
        minimum size=10mm,
        inner sep=1pt
    },
    leg/.style={line width=1.2pt},
    freeindex/.style={},
    arrowstyle/.style={->, thick, line width=1pt},
    dotstyle/.style={font=\Large}
]

\node[tensor] (G1) at (0,0) {$\mathcal{X}$};
\node[tensor] (G2) at (1.5,0) {$\mathcal{A}$};

\node[freeindex] (N1) at (2.7, 0) {$i_2$};


\draw[leg] (G2) -- (N1);
\draw[leg] (G1) -- (G2) node[midway, above] {$N_1$};

\draw[arrowstyle] (3,0) -- (3.5,0);

\node[tensor] (G3) at (4, 0) {$\mathcal{Y}$};

\node[freeindex] (N2) at (5.2, 0) {$i_2$};

\draw[leg] (G3) -- (N2);

\end{tikzpicture}
}
\caption{Tensor network diagram for \cref{eq:app_matvec}.}
\label{fig:contraction_2d}
\end{subfigure}
\begin{subfigure}[b]{0.45\textwidth}
\centering
\resizebox{\columnwidth}{!}{
\begin{tikzpicture}[
    thick,
    every node/.style={font=\small},
    tensor/.style={
        circle,
        draw=black,
        fill=blue!20,
        minimum size=7mm,
        inner sep=0pt
    },
    bigtensor/.style={
        circle,
        draw=black,
        fill=red!20,
        minimum size=10mm,
        inner sep=1pt
    },
    leg/.style={line width=1.2pt},
    freeindex/.style={},
    arrowstyle/.style={->, thick, line width=1pt},
    dotstyle/.style={font=\Large}
]

\node[tensor] (G1) at (0,0) {$\mathcal{T}$};
\node[tensor] (G2) at (1.5,0) {$\mathcal{A}$};

\node[freeindex] (N1) at (2.7, 0) {$i_1$};
\begin{scope}[rotate around={120:(G1)}]
    \node[freeindex] (N2) at (1.2,0) {$i_3$};
\end{scope}
\begin{scope}[rotate around={240:(G1)}]
    \node[freeindex] (N3) at (1.2,0) {$i_4$};
\end{scope}

\draw[leg] (G2) -- (N1);
\draw[leg] (G1) -- (N2);
\draw[leg] (G1) -- (N3);
\draw[leg] (G1) -- (G2) node[midway, above] {$N_2$};

\draw[arrowstyle] (3,0) -- (3.5,0);

\node[tensor] (G3) at (4, 0) {$\mathcal{G}$};

\node[freeindex] (N4) at (5.2, 0) {$i_1$};
\begin{scope}[rotate around={120:(G3)}]
    \node[freeindex] (N5) at (5.2,0) {$i_3$};
\end{scope}
\begin{scope}[rotate around={240:(G3)}]
    \node[freeindex] (N6) at (5.2,0) {$i_4$};
\end{scope}

\draw[leg] (G3) -- (N4);
\draw[leg] (G3) -- (N5);
\draw[leg] (G3) -- (N6);

\end{tikzpicture}
}
\caption{Tensor network diagram for \cref{eq:app_tensor_contract}.}
\label{fig:contraction_3d}
\end{subfigure}

\caption{Example tensor network diagrams for a vector $\mathcal{X}\in\mathbb{R}^{N_1}$, matrix $\mathcal{A}\in\mathbb{R}^{N_1\times N_2}$, and a three-dimensional tensor $\mathcal{T}\in\mathbb{R}^{N_2\times N_3\times N_4}$. We also show the diagram for a matrix vector product of $\mathcal{A}$ and $\mathcal{X}$ over $i_1$ and for a contraction of $\mathcal{A}$ and $\mathcal{T}$ over $i_2$.}
\label{fig:app_tn_diagrams}
\end{figure}
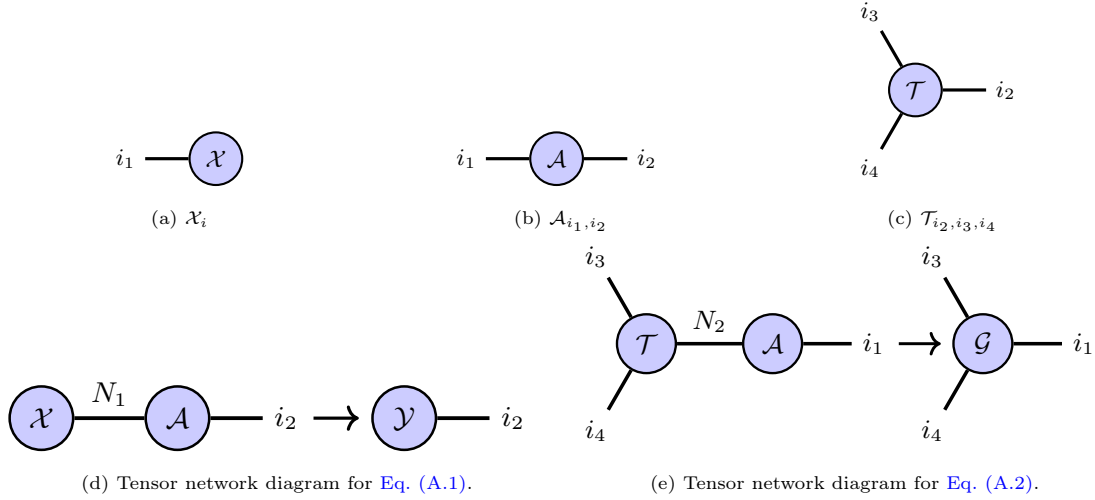

\par Tensor network diagrams are a useful tool for understanding multidimensional tensors and their operations. In tensor network diagrams, we represent multidimensional tensors as solid circles, and an annotated line emanating from each circle corresponds to a dimension. As examples we show a vector $\mathcal{X}\in\mathbb{R}^{N_1}$ with index $i_1\in\{1, 2, \dots,N_1\}$ in \cref{fig:app_vector}, a matrix $\mathcal{A}\in\mathbb{R}^{N_1\times N_2}$ with indices $(i_1, i_2)\in\{1, 2,\dots,N_1\}\times \{1, 2, \dots, N_2\}$ in \cref{fig:app_matrix}, and a three-dimensional tensor $\mathcal{T}\in\mathbb{R}^{N_2\times N_3\times N_4}$ with indices $(i_2, i_3, i_4)\in\{1, 2, \dots, N_2\}\times \{1, 2, \dots, N_3\}\times \{1, 2, \dots, N_4\}$ in \cref{fig:app_3d_tensor}. The dangling lines are \textit{free indices}, which we can connect to other tensors to represent a tensor contraction over those indices. As an example we show the matrix-vector product of $\mathcal{A}$ applied to $\mathcal{X}$ over $i_1$ in \cref{fig:contraction_2d} or mathematically,
\begin{equation}\label{eq:app_matvec}
    \mathcal{Y}_{i_2} = \sum_{i_1 = 1}^{N_1}\mathcal{A}_{i_1, i_2}\mathcal{X}_{i_1}.
\end{equation}
In \cref{fig:contraction_3d} we show the contraction of $\mathcal{T}$ and $\mathcal{A}$ over $i_2$ which represents the tensor $\mathcal{G}\in\mathbb{R}^{N_1\times N_3\times N_4}$,
\begin{equation}\label{eq:app_tensor_contract}
    \mathcal{G}_{i_1, i_3, i_4} = \sum_{i_2 = 1}^{N_2}\mathcal{T}_{i_2, i_3, i_4}\mathcal{A}_{i_1, i_2},
\end{equation}
with indices $(i_1, i_3, i_4)\in\{1, 2, \dots, N_1\}\times \{1, 2, \dots, N_3\}\times \{1, 2, \dots,N_4\}$. When the contraction happens, only the non-contracted indices remain in the new tensor.

\par As stated in \cref{sec:tt}, the purpose of tensor networks is to represent a full tensor in a compressed contraction network of smaller tensors. In the tensor train (TT) format, this means we do not contract the rank dimensions. For products such as those in \cref{eq:intg_span,eq:b_spans} with tensor network diagrams in \cref{fig:VR,fig:b_spans}, respectively, we contract across the bonded free indices but not the ranks, except for tensors without free indices, such as the inner products over the spatial quadrature (cores $4$ and $5$ in \cref{fig:VR} for example). This factorization implies some degree of separability, with the degree of coupling determined by the ranks of the factorized tensors. High rank dimensions imply a high degree of coupling between tensors and an inseparability between free indices and the physical dimensions they represent. 

\bibliographystyle{elsarticle-num}
\bibliography{references}

\end{document}